\documentclass[twoside,11pt]{Classes/CUEDthesisPSnPDF}

\usepackage{lscape}
\usepackage{etaremune}
\usepackage{longtable}

\def\ltr{$L_{\rm X}-T$ }

\title{The XMM-Newton/SDSS Galaxy Cluster Survey}

\author{Ali Said Ahmed Takey}

\collegeordept{Leibniz-Institute for Astrophysics Potsdam (AIP)}

\university{}

\crest{\includegraphics[bb = 0 0 255 280, width=30mm]{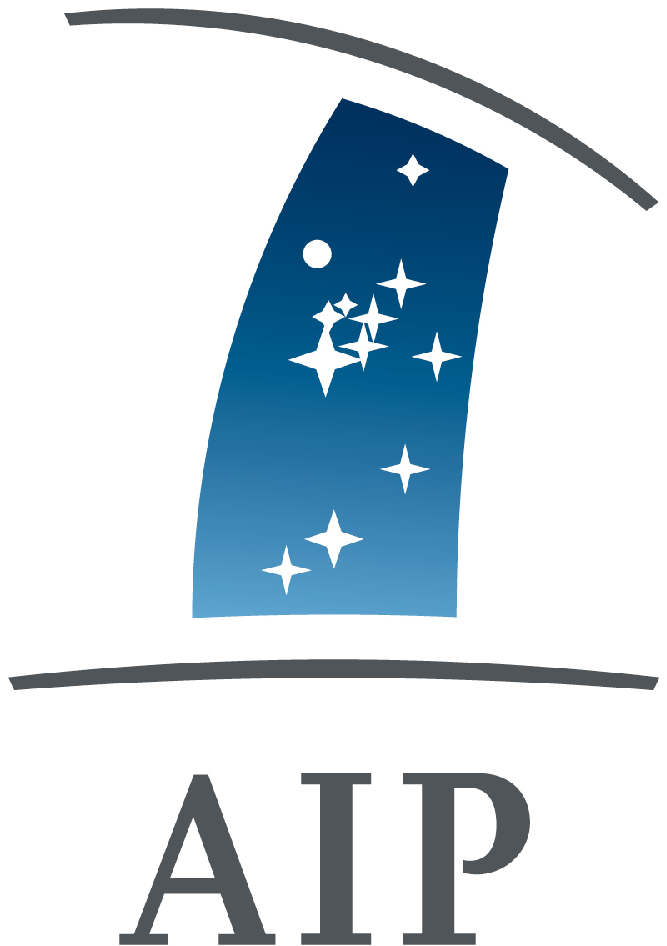}
\hspace{2cm}   
\includegraphics[bb = 0 0 91 96, width=30mm]{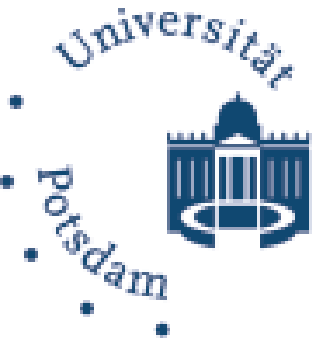} }

\renewcommand{\submittedtext}{A thesis submitted to Postdam University \\
for the degree of \\}
\degree{Doctor of Natural Science}
\degreedate{2013}

\usepackage{StyleFiles/watermark}

\begin{document}

\maketitle

\setcounter{secnumdepth}{3}
\setcounter{tocdepth}{3}

\frontmatter 
\pagenumbering{roman}


\begin{dedication} 


\vspace{7cm}

{\bf Supervisors}
\begin{enumerate}
 \item Prof.~Dr.~Matthias Steinmetz (Leibniz-Institut f{\"u}r Astrophysik Potsdam (AIP), Potsdam, Germany)
 \item PD~Dr.~Axel Schwope (AIP, Potsdam, Germany)
 \item Dr.~Georg Lamer (AIP, Potsdam, Germany) 
\end{enumerate}

\vspace{1cm}

{\bf Examiners}

\begin{enumerate}
 \item Prof.~Dr.~Matthias Steinmetz (AIP, Potsdam, Germany)
 \item Prof.~Dr.~Thomas Reiprich (Argelander-Institut f{\"u}r Astronomie, Bonn, Germany) 
 \item Prof.~Dr.~Joseph Mohr (Ludwig-Maximilians-Universit{\"a}t, M{\"u}nchen, Germany)

\end{enumerate}

\end{dedication}




\begin{abstracts}        

Galaxy clusters are the largest known gravitationally bound objects, their 
study is important for both an intrinsic understanding of their systems and  
an investigation of the large scale structure of the universe. The 
multi-component nature of galaxy clusters offers multiple observable 
signals across the electromagnetic spectrum. At X-ray wavelengths, galaxy 
clusters are simply identified as X-ray luminous, spatially extended, and 
extragalactic sources.
X-ray observations offer the most powerful technique for constructing cluster
catalogues. The main advantages of the X-ray cluster surveys are their excellent 
purity and completeness and the X-ray observables are tightly correlated with 
mass, which is indeed the  most fundamental parameter of clusters. 
In my thesis I have conducted the 2XMMi/SDSS galaxy cluster survey, which  
is a serendipitous search for galaxy clusters based on the X-ray extended 
sources in the XMM-Newton Serendipitous Source Catalogue (2XMMi-DR3). 
The main aims of the survey are to identify new X-ray galaxy clusters,
investigate their X-ray scaling relations, identify distant cluster candidates, 
and  study the correlation of the X-ray and optical properties. The 
survey is constrained to those extended sources that are in the footprint 
of the Sloan Digital Sky Survey (SDSS) in order to be able to identify the 
optical counterparts as well as to measure their redshifts that are mandatory 
to measure their physical properties. The overlap area between the XMM-Newton 
fields and the SDSS-DR7 imaging, the latest SDSS data release at the 
starting of the survey, is 210 deg$^2$. The survey comprises 1180 X-ray 
cluster candidates with at least 80 background-subtracted photon counts, 
which passed the quality control process.            

To measure the optical redshifts of the X-ray cluster candidates, I used three 
procedures; (i) cross-matching these candidates with the recent and largest 
optically selected cluster catalogues in the literature, which yielded the 
photometric redshifts of about a quarter of the X-ray cluster candidates. 
(ii) I developed a finding algorithm to search for overdensities of galaxies 
at the positions of the X-ray cluster candidates in the photometric redshift 
space and to measure their redshifts from the SDSS-DR8 data, which provided 
the photometric redshifts of 530 groups/clusters.
(iii) I developed an algorithm to identify the cluster candidates associated with 
spectroscopically targeted Luminous Red Galaxies (LRGs) in the SDSS-DR9 and to 
measure the cluster spectroscopic redshift, which provided 324 groups and clusters 
with spectroscopic confirmation based on spectroscopic redshift of at least 
one LRG. In total, 
the optically confirmed cluster sample comprises 574 groups and clusters with 
redshifts ($0.03 \leq z \leq 0.77$), which is the largest X-ray selected cluster 
catalogue to date based on observations from the current X-ray observatories 
(XMM-Newton, Chandra, Suzaku, and Swift/XRT). Among the cluster sample, about 
75 percent are newly X-ray discovered groups/clusters and 40 percent are new 
systems to the literature.              

To determine the X-ray properties of the optically confirmed cluster sample,  
I reduced and analysed their X-ray data in an automated way following the 
standard pipelines of processing the XMM-Newton data. In this analysis, 
I extracted the cluster spectra from EPIC(PN, MOS1, MOS2) images within an 
optimal aperture chosen to maximise the signal-to-noise ratio. The spectral 
fitting procedure provided the X-ray temperatures $kT$ (0.5 - 7.5 keV) for 345 
systems that have good quality X-ray data. For all the optically confirmed 
cluster sample,  I measured the physical properties $L_{500}$  
($0.5 \times 10^{42} - 1.2 \times 10^{45}$\ erg\ s$^{-1}$) and $M_{500}$ 
($1.1 \times 10^{13} - 4.9 \times 10^{14}$\ M$_{\odot}$) from an iterative 
procedure using published scaling relations. The present X-ray detected 
groups and clusters are in the low and intermediate luminosity regimes apart 
from few luminous systems, thanks to the XMM-Newton sensitivity and the 
available XMM-Newton deep fields

The optically confirmed cluster sample with measurements of redshift and X-ray 
properties can be used for various astrophysical applications. 
As a first application, I investigated the \ltr relation for the first 
time based on a large cluster sample of 345 systems with X-ray spectroscopic 
parameters drawn from a single survey. The current sample includes groups 
and clusters with wide ranges of redshifts, temperatures, and luminosities. 
The slope of the relation is consistent with the published ones of nearby 
clusters with higher temperatures and luminosities. The derived relation is 
still much steeper than that predicted by self-similar evolution. 
I also investigated the evolution of the slope and the scatter of the \ltr 
relation with the cluster redshift. After excluding the low luminosity groups, 
I found no significant changes of the slope and the intrinsic scatter of the 
relation with redshift when dividing the sample into three redshift bins. When 
including the low luminosity groups in the low redshift subsample, I found its 
\ltr relation becomes flatter than  the relation of the intermediate and high 
redshift subsamples.

As a second application of the optically confirmed cluster sample from our 
ongoing survey, I investigated the correlation between the cluster X-ray 
and the optical parameters that have been determined in a homogenous way. 
Firstly, I investigated the correlations between the BCG properties (absolute 
magnitude and optical luminosity) and the cluster global properties (redshift 
and mass). Secondly, I computed the richness and the optical luminosity within 
$R_{500}$ of a nearby subsample ($z \le 0.42$, with a complete membership 
detection from the SDSS data) with measured X-ray temperatures from 
our survey. The relation between the estimated optical luminosity and 
richness is also presented. Finally, the correlation between the cluster 
optical properties (richness and luminosity) and the cluster global 
properties (X-ray luminosity, temperature, mass) are investigated. 

\end{abstracts}


\begin{Zusammenfassung}        

Im Rahmen dieser Arbeit habe ich die 2XMMi/SDSS Galaxienhaufendurchmusterung erstellt (2XMMi/SDSS galaxy cluster survey), eine Suche nach
Galaxienhaufen welche auf der Detektion ausgedehnter R\"ontgenquellen im \emph{XMM-Newton} Quellenkatalog (2XMMi-DR3) basiert. Die Hauptziele dieser Suche
sind die Identifizierung bisher unbekannter r\"ontgenheller Galaxienhaufen, die Erforschung ihrer Beziehungen zwischen R\"ontgenleuchtkraft und
Temperatur (X-ray scaling relation), eine Entdeckung von m\"oglichen weit entfernten Galaxienhaufen und die Beziehung zwischen Eigenschaften im
Optischen und R\"ontgenbereich. Die Durchmusterung ist f\"ur alle Quellen der Himmelsregionen ausgelegt, die vom \emph{Sloan Digital Sky Survey} (SDSS) erfasst werden. Das Ziel besteht darin, ihre optischen Gegenst\"ucke zu finden und deren Rotverschiebungen zu bestimmen. Die gemeinsamen Himmelsareale
zwischen \emph{XMM-Newton} und dem Bildmaterial vom \emph{SDSS-DR7} umfassen 210~deg$^2$. Meine Durchmusterung enth\"alt 1180 m\"ogliche
Galaxienhaufen mit wenigstens 80 vom Hintergrund bereinigten Photonen im R\"ontgenbereich, die einer Qualit\"atskontrolle erfolgreich standgehalten
haben.

Um die Rotverschiebungen der m\"oglichen Galaxienhaufen im optischen Bereich zu bestimmen nutzte ich drei Vorgehensweisen: (i) Ein Abgleich jener
Kandidaten mit den neuesten und umfangreichsten Katalogen optisch ausgew\"ahlter Galaxienhaufen, die in der Literatur verf\"ugbar sind. (ii) Ich
entwickelte einen Algorithmus, um Rotverschiebungen der optischen Gegenst\"ucke aus Daten vom \emph{SDSS-DR8} zu ermitteln, welches zu photometrischen
Rotverschiebungen von 530 Galaxiengruppen-/haufen f\"uhrte. (iii) Ein weiterer von mir entwickelter Algorithmus nutzte die spektroskopischen Rotverschiebung von roten leuchtkr\"aftigen Galaxien (LRGs) in den Daten des \emph{SDSS-DR9} und ergab 324 Gruppen und Haufen.
Zusammengefasst enth\"alt diese Probe 574 auch im optischen nachgewiesener Galaxiengruppen und -haufen mit bekannten Rotverschiebungen
($0.03 \leq z \leq 0.77$) - der zur Zeit umfangreichste Katalog von im R\"ontgenbereich ausgew\"ahlten Galaxienhaufen basierend auf aktuellen
R\"ontgenbeobachtungen. Unter jenen Haufen waren ca. 75\% im R\"ontgenbereich nicht bekannt und 40\% fanden in der bisherigen Literatur noch keine
Erw\"ahnung.

Um die R\"ontgeneigenschaften der im Optischen best\"atigten Haufen zu bestimmen, war eine automatische Reduktion und Analyse der R\"ontgendaten
unverzichtbar. Die Prozedur, welche Modelle an die R\"ontgenspektren anpasste, ergab Temperaturen $kT$ von $0.5 - 7.5$ keV f\"ur 345 Kandidaten. F\"ur alle Haufen, die auch im optischen auffindbar waren, bestimmte ich die physikalischen Eigenschaften $L_{500}~(0.5 \times 10^{42} - 1.2 \times 10^{45}$ erg~s$^{-1}$) und $M_{500}~(1.1 \times 10^{13} - 4.9 \times 10^{14}$ M$_{\odot}$).

Die Probe optisch best\"atigter Galaxienhaufen mit gemessenen Rotverschiebungen und R\"ontgeneigenschaften kann f\"ur viele astrophysikalische Anwendungen
genutzt werden. Als eine der ersten Anwendungen betrachtete ich die Beziehung zwischen \(L_{X} - T \); das erste Mal f\"ur eine so grosse Anzahl von
345 Objekten. Der aktuelle Katalog enth\"alt Gruppen und Haufen, die einen grossen Bereich in Rotverschiebung, Temperatur und Helligkeit abdecken.
Der Anstieg jener Beziehung ist im Einklang mit bereits publizierten Werten f\"ur nahegelegene Galaxienhaufen von hoher Temperatur und Helligkeit.
Nach dem Ausschluss leuchtschwacher Gruppen und der Einteilung der Daten in drei nach Rotverschiebung geordneter Gruppen, waren keine signifikanten
\"Anderungen von Anstieg und intrinsischer Streuung zu beobachten.

Als zweite Anwendung unserer Durchmusterung, untersuchte ich die Haufen bez\"uglich deren Eigenschaften im Optischen und im R\"ontgenbereich. Zuerst
betrachtete ich den Zusammenhang zwischen den Eigenschaften (absolute Helligkeit und optische Leuchkraft) der hellsten Haufengalaxie (BCG) mit denen
des Haufens als Ganzem (Rotverschiebung und Masse). Danach berechnete ich die Reichhaltigkeit der Galaxienhaufen und deren optische Leuchtkraft
innerhalb von \(R_{500}\) f\"ur eine Stichprobe nahegelegener Haufen (\(z \leq 0.42\), hier sind SDSS Daten noch empfindlich genug um den Grossteil
der Haufengalaxien abzubilden) mit gemessenen R\"ontgentemperaturen. \\ Schlussendlich konnten die Wechselwirkungen zwischen den optischen
Eigenschaften (Reichhaltigkeit und Leuchtkraft) und den globalen Eigenschaften (R\"ontgenleuchtkraft, Temperatur und Masse) n\"aher untersucht werden.

\end{Zusammenfassung}


\begin{acknowledgements}      

\begin{onehalfspace}
  
Foremost, I would like to express my thanks to my advisor PD~Dr.~Axel Schwope 
for the continuous support of my Ph.D study and research, for his patience, 
motivation, enthusiasm, and immense knowledge. His guidance helped me in all 
the time of research and writing of this thesis. 

Besides my advisor, I am also thankful to Dr.~Georg Lamer for his scientific 
advice and knowledge and many insightful discussions and suggestions. I am 
very grateful to Prof.~Dr.~Matthias Steinmetz, scientific chairman of the 
Leibniz Institute for Astrophysics Potsdam (AIP), for being my   
supervisor at the university and for reviewing my thesis.

I gratefully acknowledge the funding sources that made my Ph.D work possible. 
I was funded by the the Egyptian Ministry of Higher Education and Scientific 
Research (MHESR). The partial financial support of the AIP is gratefully 
acknowledged. 

My sincere thanks are due to the anonymous referees of the published papers 
from my Ph.D thesis for their detailed review, constructive criticism and 
excellent advice that improved the discussion of the results. 

I warmly thank the remaining members of the X-ray Astronomy group (Iris 
Traulsen, Gabriele Sch{\"o}nherr, Valentina Scipione, Adriana Pires, Sabine 
Thater, Andreas Rabitz, Arjen de Hoon, Alexey Mints, Robert Schwarz, Rene 
Heller, Ada Nebot Gomez-Moran, Jose Ramirez, Justus Vogel, Daniele Facchino) 
for the kind hospitality during my stay at the AIP. 

My warm thanks are due to my professors and colleagues at the National 
Research Institute of Astronomy and Geophysics (NRIAG), Egypt especially 
Prof. Issa Issa, Hamed Ismail, and Gamal Ali for their encouragement.    

Lastly, I would like to thank my family for all their love and encouragement. 
For my parents who raised me with a love of science and supported me in all 
my pursuits. My special gratitude is due to my wife, children, sisters, 
brothers, and friends for their loving support.

\end{onehalfspace}

\end{acknowledgements}

\begin{publications}        


\section*{\normalsize Refereed Publications}

\begin{etaremune}

\item {\bf Takey, A}., Schwope, A., Lamer, G. 2013, The 2XMMi/SDSS Galaxy 
Cluster Survey. III. Clusters associated with spectroscopically targeted 
LRGs in SDSS-DR10, accepted for publication in A\&A.

\item {\bf Takey, A}., Schwope, A., Lamer, G. 2013, The 2XMMi/SDSS Galaxy 
Cluster Survey. II. The optically confirmed cluster sample and the \ltr 
relation, A\&A, 558, A75.
 
\item de Hoon, A., Lamer, G., Schwope, A., ......., {\bf Takey, A}. 2013, Distant 
galaxy clusters in a deep XMM-Newton field within the CFTHLS D4, A\&A, 551, A8.
 
\item {\bf Takey, A}., Schwope, A., Lamer, G. 2011, The 2XMMi/SDSS Galaxy 
Cluster Survey. I. The first cluster sample and X-ray luminosity-temperature 
relation, A\&A, 534, A120.
   
\item {\bf Takey, A. S}., Ismail, H. A., Isaa, I. A., Bakry, A. A., and  
Ismail, M. N.  2006, Frequency Distribution of HII Regions Radii as a Distance 
Indicator in Seyfert galaxies, Al-Azhar Bulletin of Science, vol. 17, 
No. 1(June): pp. 121-127, in Egypt.

\item Ismail, H. A., Alawy, A. E., {\bf Takey, A. S}., Isaa, I. A., 
Selim, H. H. 2005, Frequency Distribution of HII Regions Radii as a Distance 
Indicator, JKAS, 38, 71.

\end{etaremune}


\section*{\normalsize Non-refereed Publications}

\begin{etaremune}

\item {\bf Takey, A}., Schwope, A., Lamer, G. 2013, Clusters associated 
with spectroscopically targeted LRGs in SDSS-DR9, ``THE MASS PROFILE OF GALAXY 
CLUSTERS from the core to the outskirts'', 8-22 March 2013, Madonna di Campiglio,
Italy (Poster)

\item {\bf Takey, A}., Schwope, A., Lamer, G. 2012, The 2XMMi/SDSS Galaxy Cluster Survey,
``IAU XVIII General Assembly``, 20-31 August 2012, Beijing, China (Poster)

\item {\bf Takey, A}., Schwope, A., Lamer, G. 2012, The 2XMMi/SDSS Galaxy Cluster Survey, 
Proceedings of a workshop ``Galaxy Clusters as Giant Cosmic Laboratories``,  
21-23 May 2012, Madrid, Spain., p.57 (Oral talk) 

\item {\bf Takey, A}., Schwope, A., Lamer, G. 2011, XMM-Newton/SDSS Galaxy Cluster Survey,
''The X-ray Universe 2011``, 27-30 June 2011, Berlin, Germany (Poster)

\item {\bf Takey, A}., Schwope, A., Lamer, G. 2010, XMM-SDSS Galaxy Cluster Survey,
''First Azarquiel School of Astronomy ``, 4-11 July 2010, Granada, Spain (Oral talk)

\item {\bf Takey, A}., Schwope, A., Lamer, G., de Hoon, A, 2010, XMM-SDSS Cluster Survey,
''Evolution of galaxies, their central black holes and their large-scale environment``, 
20-24 September 2010, Potsdam, Germany  (Poster)

\end{etaremune}


\end{publications}

\tableofcontents
\listoffigures
\listoftables

\mainmatter 


\chapter{Introduction}



\section{Clusters of Galaxies}

Charles Messier and William Herschel constructed the first systematic 
catalogues of nebulae, which led to the discovery of the tendency of 
galaxies to clusters. In the nineteenth and early twentieth century, larger 
samples of nebulae were compiled that made the tendency of galaxy to cluster 
becomes more apparent. These spiral and elliptical nebulae were proved by 
\citet{Hubble26} to be true galaxies like the Milky Way located at 
large distances. This implied that clusters of galaxies are systems with 
enormous size and mass. Hubble's proof revolutionized the study of the 
most prominent clusters of galaxies. Assuming a virial equilibrium of 
galaxy motions and using the velocity measurements of cluster galaxies, 
\citet[][]{Zwicky33, Zwicky37} and \citet{Smith36} found the total gravitating cluster 
masses for the Coma and Virgo clusters enormously large. The masses measured in that 
way were much larger than the stellar masses of the cluster galaxies by a factor 
of $\sim 200-400$, which led to the postulation of the existence of large amounts 
of dark matter. Clusters of galaxies were first identified as large 
concentrations in the projected galaxy distribution containing hundreds to 
thousands of galaxies, clustering over a small region on the sky 
\citep[][]{Abell58, Zwicky61, Abell89}.

The extended X-ray emission of the hot intracluster medium (ICM) was 
discovered in the Coma cluster by \citep[e.g,][]{Meekins71, Gursky71, 
Cavaliere71, Forman72, Kellogg72}. The ICM was found to be 
smoothly filling the intergalactic space and emitting at X-rays via  
thermal bremsstrahlung. The discovery of the ICM provided a part of the missing 
mass in the Coma cluster and led to the detection of clusters up to high 
redshifts later.  In addition, the temperature measurements of the ICM 
provided another confirmation that the gravitational potential of clusters 
requires a dark component of matter. The hot plasma in ICM distorts the 
cosmic microwave background (CMB) photons through inverse Compton 
scattering \citep[][]{Sunyaev70, Sunyaev72, Sunyaev80}, which was called the 
Sunyaev-Zel'dovich (SZ) effect. Recently, this technique led to the discovery 
of hundreds of galaxy clusters \citep[see the review article by][]{Kravtsov12}.

\begin{figure}
\centering{
  \resizebox{150mm}{!}{\includegraphics[viewport=0  40  500 228, clip]{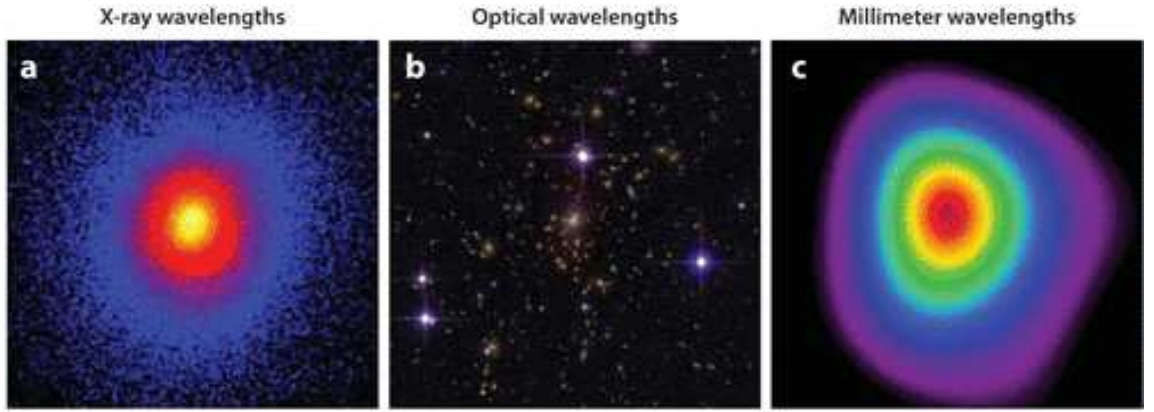}}}
  \caption[Observations of Abell 1835 at X-rays (a), optical (b), and millimeter 
(c) wavelengths]{Observations of Abell 1835 at X-rays (a), optical (b), and millimeter 
(c) wavelengths. The images have a size of $5.2'\times 5.2'$ 
(1.2 Mpc $\times$ 1.2 Mpc at the cluster redshift, $z = 0.25$) centered on the 
X-ray emission peak position. Image from \cite{Allen11}. } 
  \label{f:Abell1835}
\end{figure}

Figure~\ref{f:Abell1835} shows the observations of a dynamically relaxed 
cluster, Abell 1835 ($z = 0.25$), at various wavelength bands. The intracluster 
gas is observed at X-ray and millimeter wavelengths as shown in images (a) and 
(c), respectively. The different colours  indicate the different temperatures 
and densities of the intracluster gas. Cluster galaxies are observed 
in the optical band as shown in image (b). The figure is taken from 
\cite{Allen11}.

According to the hierarchical structure formation scenario of our universe, clusters of 
galaxies are the largest building blocks after stars and galaxies. X-ray 
observations reveal that clusters are well defined and connected structural 
entities since the cluster X-ray emission traces its whole structure in 
a contiguous way as shown in Figure~\ref{f:Abell1835}(a). Galaxy 
clusters are formed essentially by gravitational collapse and have had time 
to reach their dynamical equilibrium. Thus, clusters appear only in a relative 
late epoch in the cosmic history, i.e.~very massive clusters exist only up to 
a certain yet to be determined redshift.  The characteristic form of galaxy 
clusters can be well assessed by observations as well as described by 
theoretical models \citep{Boehringer06}.       
%

Clusters of galaxies are the largest gravitationally bound objects consisting   
mostly of dark matter ($\sim 78 \% $), hot thermal plasma ($\sim 11 \% $), and 
galaxies ($\sim 2 \% $). The total mass of galaxy clusters lies in the range of 
$10^{14} - 10^{15}$\ M$_\odot$. The depth of the cluster potential can be 
measured using the velocity dispersion of cluster galaxies or the plasma 
temperature. The 
velocity dispersions of galaxies is in the range from 300 km s$^{-1}$ in galaxy 
groups to 1500 km s$^{-1}$ in the most massive galaxy clusters. The plasma 
temperature $kT$ of clusters ranging from 2 to 15 keV gives information about the 
cluster potential since the hot ICM is in the form of an approximately 
hydrostatic equilibrium \citep{Boehringer06}.

{\bf Importance of galaxy clusters:}
They form the largest astrophysical laboratories that are suitable for 
various studies of astrophysical processes. Among the astrophysical studies 
of galaxy clusters are the following:        

\begin{enumerate}

\item Investigation of galaxy evolution within a dense and well defined 
environment.  

\item Study the evolution of the dynamical and thermal structure of galaxy 
clusters.

\item Investigation of the chemical enrichment of ICM and the interaction 
between the intracluster gas and cluster galaxies.

\item Study faint and high redshift galaxies beyond the cluster redshift since 
massive clusters act as natural telescopes, extremely powerful gravitational 
lenses. 

\item They probe the high density tail of the cosmic density field since they 
arise from the gravitational collapse of rare high peaks of primordial density 
perturbations.

\item The evolution of the cluster population is tightly connected to the 
evolution of the large-scale structure and the universe as a whole. Therefore, 
the number density of galaxy clusters is highly sensitive to specific 
cosmological scenarios.

\item The space density of clusters has been used to measure the amplitude of 
density perturbations. In addition its evolution depends on the value of the 
matter density parameter.

\item  The space density of distant galaxy clusters can be used as a powerful 
cosmological diagnostics.

\item Probe the nature and amount of dark matter.

\item Constraining the Dark Energy equation of state. 

\end{enumerate}

{\bf  Structure of the introduction:} 
as mentioned above the astrophysical studies of clusters are too broad, 
therefore I will briefly summarize information and cluster studies that are 
relevant to the thesis work. In Section 1.2 the observational techniques used 
to detect galaxy clusters in multi-wavelengths approach are presented. I also 
provide more information about X-ray observations of galaxy clusters as well 
as the current X-ray telescopes. In Section 1.3 I present the constructed 
cluster catalogues based on multi-wavelengths data with more details about 
X-ray selected galaxy clusters. 
In Section 1.4 I present the main observable parameters obtained from cluster 
signals across the electromagnetic spectrum. The correlations between the 
cluster properties in same and different electromagnetic bands are presented 
in Section 1.5. Finally the aims and the outlines of the thesis are presented 
in Section 1.6.         



\section{Observational Techniques}

The multi-component nature of galaxy clusters offers various observable 
signals across the electromagnetic spectrum \citep[e.g,][]{Sarazin88, Allen11}. 
Clusters have been discovered firstly at optical and NIR wavelengths due to 
the stellar emission from their galaxies and intracluster light. The total 
matter distribution in clusters can be measured through studying the 
gravitational lensing.
At radio wavelengths, Synchrotron emission from relativistic electrons 
is visible. The hot ionized ICM is observed at X-ray and mm wavelengths.
In the following, I will briefly present the 
observational techniques of galaxy clusters in optical/NIR, mm, and X-ray 
wavelengths. I will also present the current X-ray observatories especially 
XMM-Newton since the thesis work is based mainly on archival observations 
of this satellite observatory. 


\subsection{Optical and NIR observations}
The optical and NIR emission of galaxy clusters is radiated mainly as starlight. 
Cluster galaxies are clustered in three spatial dimensions, R.A, Dec, and their 
position along the line of sight. The uncertainty of their distances leads to 
projection effects which contaminate the cluster richness and confuse cluster 
detection. Therefore any cluster finding algorithm faces the challenge of 
discriminating cluster galaxies from the field galaxies \citep{Hao10}.
The ability of determining the position of galaxies along the line of sight 
is limited by available data. When spectroscopic redshifts of galaxies are 
available, cluster finding techniques give a secure cluster detection and 
an accurate richness measurement \citep[e.g,][]{Miller05, Berlind06, Yang07}.

Over the past six decades, the photometric data of galaxies were the main  
database to optically detect clusters of galaxies. The finding algorithms 
depend on the type of the available photometric data. When only single band 
data were available, the cluster detecting algorithms were based on magnitudes 
\citep[][]{Abell58, Shectman85, Abell89, Postman96, Gal00, Kim02, Goto02, Gal03, 
Lopes04}. 
This method was successful in detecting massive clusters but can not provide good 
purity and completeness for less massive clusters. In addition, the estimated
richness causes a large scatter in the richness-mass relation that was derived 
from the cluster sample constructed from this procedure \citep{Hao10}.

The availability of multi-band imaging greatly reduces the projection effects 
that plagued the optical cluster detections. There are two procedures to 
identify galaxy clusters based on multi-colour data. First; use the colours of 
galaxies to obtain their photometric redshifts that were used to discriminate 
the cluster galaxies. There are many algorithms to assign the photometric 
redshifts for galaxies based on their magnitudes and colours 
\citep[e.g,][]{Bolzonella00, Oyaizu08, Gerdes10}. These methods are limited 
by the available template sets of spectroscopic redshifts. Many cluster finding 
algorithms were based on the photometric redshifts of galaxies \citep[][]{Li08, 
Wen09, Wen12}. The estimated cluster redshifts based on these algorithms have 
an uncertainty of about 0.02 \citep{Hao10}. 

The second method is searching for galaxy clusters based on clustering of their 
galaxies in the colour space. 
The galactic content of the regular clusters is entirely dominated 
by early type galaxies (E, S0) while the irregular clusters usually contains 
early and late type galaxies \citep{Bahcall77, Allen11}. In addition the early 
type galaxies are dominant in the central regions of clusters. The existence of 
early type galaxies (with a homogeneous nature of stellar population) in many 
clusters gives them similar spectral energy distributions (SEDs) that include 
the strong feature of $4000\, \AA$ break. Therefore, cluster galaxies are tightly 
clustered in colour as well as space and exhibit a narrow colour scatter in 
colour-magnitude diagram. Since the $4000\, \AA$ break shifts across the optical 
filters with increasing cluster redshift, there is a strong correlation 
between cluster galaxy colour and cluster redshift. 

The red-sequence of cluster 
galaxies is a very prominent feature of clusters, therefore it provides a 
powerful technique to discriminate between the cluster galaxies and the field 
galaxies. For more information about red-sequence galaxies in clusters and 
detected clusters using this technique, we refer to \citep[e.g,][]{Bower92, 
Smail98, Gladders98, Lopez-Cruz04, Mullis05, Gladders05, Koester07, Stott09, 
Hao09, Mei09, Hao10}.


\subsection{Gravitational lensing}

The gravity associated with a mass concentration will bend light rays passing 
near to it, according to general relativity. This phenomenon is known as 
gravitational lensing, which can magnify and distort the images of background 
galaxies. Galaxy clusters produce the largest observed angular deflections of 
light rays in the universe since they have the deepest gravitational potentials 
on large scale.  Observations reveal strong distortions and multiple images of 
individual background galaxies caused by the more massive and compact galaxy 
clusters, that is called strong lensing. Gravitational lensing can be detected 
clearly in the statistical appearance of background galaxies observed through 
clusters (weak lensing) and in the filed.  
\citet{Zwicky37} proposed that cluster masses could be measured through 
the gravitational lensing of background galaxies. Recently, the techniques 
of weak and strong lensing were applicable and provided unbiased mass 
measurements.  
The mass 
measured by this technique is an independent measurement that is used for 
the comparison with the mass measurements obtained by other methods based on 
X-ray, SZ, and optical observables 
\citep[e.g.][]{Boehringer06, Bartelmann10, Allen11}.            

   
\subsection{Sunyaev-Zel'dovich (SZ) effect}

The hot intracluster gas can be observed at millimeter wavelengths through its 
inverse Compton scatter of the cosmic microwave background (CMB), which has 
a perfect blackbody spectrum. This scattering boosts the photon energy and gives 
rise to a small shift in the CMB spectrum as the CMB photons pass through the hot 
gas in the ICM. This effect was predicted by \cite{Sunyaev72} and is called the 
thermal Sunyaev-Zel'dovich (SZ) effect. The magnitude of the thermal SZ effect 
is proportional to the line of sight integral of the product of the gas density 
$n_{\rm e}$ and temperature $T$ as $Y\,\propto\,\int\,n_{\rm e}\,T\,dV$. 
The thermal SZ effect is independent of the cluster distance,  
which benefits the cosmological applications based on SZ selected clusters and 
leads to discover clusters up to high redshifts.
Precise observations of the CMB distortion provides measurements of the cluster 
temperature in addition they provide a mass proxy of galaxy clusters.
The motion of clusters with respect to the CMB produces an additional smaller 
distortion that is know as kinetic SZ effect. Its magnitude is proportional 
to the peculiar velocity of clusters \citep[e.g.][]{Carlstrom02, Voit05, 
Allen11}.




\subsection{X-ray observations}

\cite{Limber59} suggested that diffuse gas must be present among galaxies, 
and galaxy clusters are filled with a hot intracluster diffuse gas because 
galaxy formation can not be 100 percent efficient. \citet{Felten66} was 
inspired by a spurious X-ray detection of the Coma cluster and attributed 
that X-ray emission to thermal bremsstrahlung. The first detections of the 
extended X-ray emission from the ICM in the Coma cluster was done by 
\citet{Meekins71} and \citet{Gursky71}, who suggested that most of rich 
clusters include an X-ray emission with luminosity in the range of 
$10^{43} - 10^{44}$ erg s$^{-1}$.   \cite{Cavaliere71} suggested that many 
extragalactic X-ray sources are probably associated with galaxy clusters. 
The nature of the diffuse cluster X-ray emission was first established by 
\citet{Solinger72}. The diffuse X-ray emission of clusters originates in a hot 
ICM plasma with temperatures in the range $10^7 - 10^8$ K, which radiates most 
of its thermal emission in the soft X-ray regime. For more information about 
the history of the investigation of galaxy clusters, see the review paper by 
\citet{Biviano00}.  

The ICM behaves as a fully ionized plasma with its emissivity dominated by 
thermal bremsstrahlung, i.e. radiation from free-free transitions of electrons
being accelerated in the coulomb potential of the nuclei (mainly hydrogen). 
In addition to the bremsstrahlung emission, there are free-bound emission 
(recombination) and bound-bound emission (line radiation). The emissivities 
of these emission processes in ICM are proportional to the square of the 
gas density that ranges from $\sim 10^{-1}$ cm$^{-3}$ in the cluster center to    
$\sim 10^{-5}$ cm$^{-3}$ in the cluster outskirt \citep{Allen11}. 

The emissivity of the thermal bremsstrahlung process at the frequency $\nu$ 
scales as 
\begin{equation}
\epsilon_\nu \propto\,n_en_i g(\nu,T)\,T^{-1/2}\exp{\left(-h\nu/ k_BT\right)}
\label{eq:emissivity}
\end{equation}
where $n_e$ and $n_i$ are the number density of electrons and ions, respectively.
$g(\nu,T)$ is the Gaunt factor and it scales with $\ln(k_BT/h\nu)$. The X-ray 
luminosity of the cluster is obtained by integrating Equation~\ref{eq:emissivity} 
over the energy range of the X-ray emission and over the gas distribution. 
The powerful X-ray luminosity of clusters ($L_{\rm X} \sim 10^{43} - 
10^{45}$ erg s$^{-1}$) places them among the most luminous 
extragalactic X-ray sources in the universe beside Active Galactic Nuclei 
(AGNs) and allow for cluster identification up to high redshifts. The main 
difference between the appearance of X-ray emission from clusters and AGNs 
is the extent of the emission since clusters appear as spatially extended 
sources while AGNs appear as point-like sources, which makes ease of cluster 
identification. 
For clusters with X-ray temperatures of $T_{\rm X} \ge 3$ keV, the pure 
bremsstrahlung emissivity gives a good approximation for their X-ray emission 
while for cooler systems the contribution from metal emission lines should be 
taken into account \citep{Rosati02}.


\subsubsection{Advantages of X-ray surveys}

The hot ICM plasma is tracing the shape of the cluster, therefore the X-ray 
appearance provides information about the cluster structure. Using the recent 
advanced X-ray observatories (e.g. Chandra and XMM-Newton) as well as previous 
missions (e.g. ROSAT), X-ray observations (images and spectra) of galaxy 
clusters provided a wealth of detailed knowledge on their structure, 
composition, formation history, and their population in the sky. In addition 
to understanding the intrinsic properties of clusters, X-ray observations 
have been used to investigate the link between the formation of the large 
scale structure and the underlying cosmological models \citep{Boehringer06}. 

The X-ray selection of clusters has several advantages for cosmological 
surveys: the observable X-ray luminosity and temperature  of a cluster is 
tightly correlated with its total mass, which is indeed its most fundamental 
parameter \citep{Reiprich02}. These relations provide the ability to measure 
both the mass function \citep{Boehringer02} and power spectrum 
\citep{Schuecker03}, which directly probe the cosmological models.  Since the 
cluster X-ray emission is strongly peaked on the dense cluster core, X-ray 
selection is less affected by projection effects than optical surveys and 
clusters can be identified efficiently over a wide redshift range.  
 
To determine the evolution of the space density of galaxy clusters, it is 
required to count the number of clusters of a given mass per unit volume at
different redshifts. There are three essential tools to achieve that, (i) an 
efficient method to identify galaxy clusters over a wide redshift range, (ii) 
an observable estimator of the cluster mass, and (iii) a method to 
compute the cluster survey volume or the selection function. X-ray 
observations of clusters provide an efficient method of identification, 
which fulfills the requirements to be used as a cosmological test. Therefore, 
X-ray studies of clusters provide; (i) an efficient way of mapping the large 
scale structure and the evolution of the universe and (ii) valuable means of 
understanding their internal structure as well as the history of cosmic baryons 
\citep{Rosati02}.


\subsubsection{X-ray telescopes}

At the beginning of the 1990s, studies of X-ray selected clusters made 
substantial progress by the advent of new X-ray missions. The main mission 
was the ROSAT satellite, which provided all-sky survey and deep pointed 
observations in the energy band 0.1-2 keV. Its angular resolution was less 
than 5 arcsec at half energy width. ROSAT had a field-of-view with 
a diameter of $\sim$ 2 degree. ASCA and Beppo-SAX satellites provided follow-up 
observations for some of the newly discovered ROSAT clusters that revealed 
hints of the complex physics of the ICM. The new generation of X-ray satellites, 
Chandra and XMM-Newton, have provided significantly deep surveys for relatively 
small sky areas as well as follow-up observations that were used to understand 
the properties of the ICM in addition to unveiling the interplay between ICM and 
baryon physics \citep{Rosati02}. The main characteristics of the current X-ray 
telescope, Chandra and XMM-Newton, are summarised below in a nutshell.

{\bf The Chandra X-ray Observatory (CXO)} was launched on July 23, 1999 by 
the Space Shuttle Columbia. Chandra is the X-ray component of the NASA's Great 
Observatories program, which provides unprecedented capabilities for sub-arcsecond 
imaging, spectrometric imaging, and for high resolution spectroscopy over 
the energy band 0.08 - 10 keV. The superior resolution of Chandra shows many 
important details about the nature of the ICM distribution like cold fronts 
and shock waves as well as X-ray cavities produced by AGN radio lobes 
\citep{Boehringer06}.

The science instruments comprise two imaging/readout devices (HRC and ACIS) 
and two gratings (LETG and HETG), which are described in the following; (1) High 
Resolution Camera (HRC) that comprises two detectors, one for imaging with 
field-of-view (FOV) of $30' \times 30'$ and pixel size of $0.13''$, the other 
detector is used with a grating for spectroscopy, (2) Advanced CCD Imaging Spectrometer 
(ACIS) that is used for imaging with FOV of $16.9' \times 16.9'$ with pixel size 
of $0.492''$ as well as for spectroscopy when used as a readout instrument, 
(3) Low-Energy Transmission Grating (LETG) that provides high-resolution 
spectroscopy in the energy range 0.08 - 2 keV with a resolving power of 
$E/\bigtriangleup E > 1000$, (4) High-Energy Transmission grating (HETG), 
which provides high-resolution spectroscopy from 0.4 to 4 keV 
(using the Medium-Energy Grating, MEG) and from 0.8 to 8 keV (using the 
High-Energy Grating, HEG) \citep{Weisskopf00}.

\begin{figure}
\centering{
  \resizebox{100mm}{!}{\includegraphics{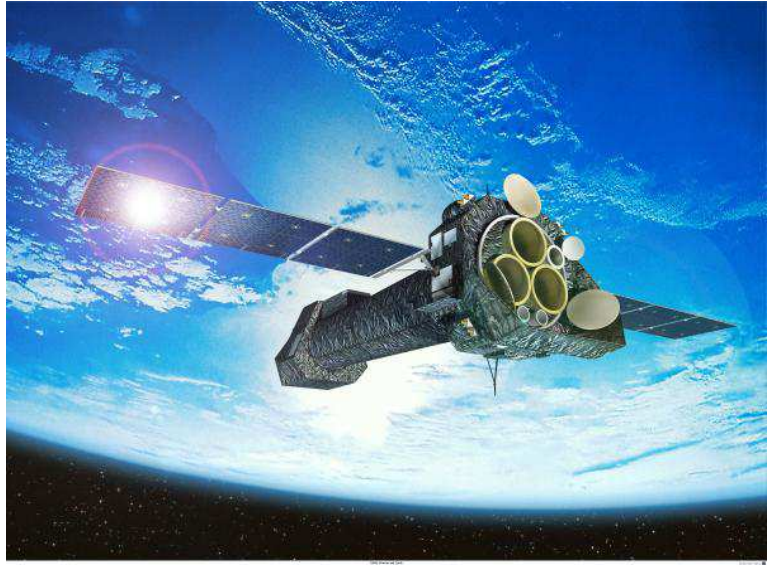}}}
  \caption[Artist view of the XMM-Newton observatory]{Artist view of the XMM-Newton observatory, the ESA's second 
cornerstone of the Horizon 2000 Science Programme. Image courtesy of ESA. } 
  \label{f:XMM_med}
\end{figure}

\begin{figure}
\centering{
  \resizebox{100mm}{!}{\includegraphics[viewport=0  100  1280 1000, clip]{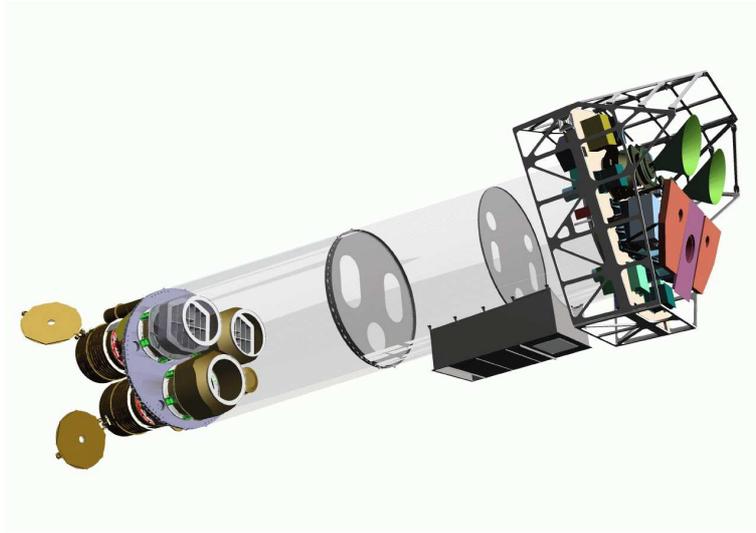}}}
  \caption[Sketch of the XMM-Newton payload]{Sketch of the XMM-Newton payload. 
The three mirror modules are visible at the lower part. The two mirrors at the 
lower left are equipped with Reflection Grating Arrays. The focal X-ray 
instruments are shown at the right end, which are the MOS cameras with their 
radiators (black/green horns), the radiator of the PN camera (violet) and 
the RGS detectors (in pink). The OM telescope is not seen in the payload 
since it is obscured by the lower mirror module. Image courtesy of ESA. } 
  \label{f:XMM_payl}
\end{figure}

{\bf XMM-Newton}, X-ray Multi-Mirror Mission, is the second of the European Space 
Agency's (ESA) four cornerstone missions defined in the Horizon 2000 program. 
It was launched by Ariane 504 on December 10th 1999 into a highly elliptical 
orbit (period $\sim$ 48 hour), with an apogee of about 115,000 km and a perigee 
of ca.~6000 km. The high elliptical orbit offers continuous target visibility 
of up to about 40 hours, with a minimum height for science observations of 
46,000 km. The XMM-Newton spacecraft is the largest scientific satellite ever 
launched by the ESA. Its length and weight are a bout 10 m and 4 tons, 
respectively. An artist image of the XMM-Newton observatory and its payload are 
displayed in Figure~\ref{f:XMM_med} and Figure~\ref{f:XMM_payl}, respectively.

XMM-Newton carries two types of telescopes, first; three Wolter type X-ray 
telescopes, second; a 30-cm optical/UV telescope.  The science instruments 
on board XMM-Newton are (1) European Photon Imaging Camera (EPIC), (2) 
Reflection Grating Spectrometer (RGS), and (3) Optical Monitor (OM). EPIC 
has three CCD cameras (MOS1, MOS2, PN) for X-ray imaging, moderate resolution 
spectroscopy, and X-ray photometry. RGS has two identical spectrometers 
(RGS1 and RGS2) for high-resolution X-ray spectroscopy and spectro-photometry. 
Both EPIC and RGS reside in the focal planes of the X-ray telescopes. OM is 
a co-aligned telescope for optical/UV imaging and spectroscopy. The six 
instruments can be operated simultaneously or independently with the 
possibility to work in different modes of data acquisition. Thus, XMM-Newton 
provides simultaneous observations at X-ray and optical/UV wavelengths.

EPIC has two types cameras (MOS: Metal Oxide Semiconductor and PN) that are 
fundamentally different in their geometry as well as other properties like the 
readout time and the quantum efficiency. The MOS chip arrays consists of 7 
individual identical front-illuminated chips that are not co-planar but 
offset with respect to each other. The MOS cameras receive 44 $\%$ of the 
reflected light by two X-ray telescopes while the rest of the reflected light
is received by the two RGS. The EPIC PN camera is a single silicon wafer 
consisting of 12 CCDs that are back-illuminated. It resides at the focal 
plane of one of the X-ray telescopes and receives the total reflected beam. 
Both MOS and PN cameras allow several modes of data acquisition.

The main characteristics of XMM-Newton are listed in 
Table~\ref{tbl:XMM_Char} that is obtained from \citet{Piconcelli12}. 
The performance of X-ray telescopes is characterised by the image quality, 
effective area, and straylight rejection efficiency. The quality of an X-ray 
mirror module is determined by its ability to focus photons. One of the 
XMM-Newton’s major strong points is that the core of its on-axis point-spread 
function (PSF) is narrow and varies little over a wide energy range (0.1-6 keV) 
but the PSF becomes slightly more energy dependent above 6 keV. The mirror 
performance is also characterised by the effective area, which reflects 
the ability of the mirrors to collect radiation at different photon energies.    
XMM-Newton carries X-ray telescopes with the largest effective area of 
a focusing telescope ever since the total geometric effective area of 
their mirrors at 1.5 keV energy is 4650 cm$^{2}$.

Both the shape of the X-ray PSF and the effective area is a function of 
the off-axis angle within the mirrors field-of-view. The PSF at large 
off-axis angles is elongated due to off-axis aberration (astigmatism). The 
effective area declines as a function of off-axis angle, which is called 
vignetting.  
The other important characteristic of X-ray telescope performance is the 
efficiency to reject the straylight that produces a contaminated image by 
a diffuse background light emitted by X-ray sources located outside the 
field of view. XMM-Newton's X-ray telescopes have X-ray 
baffles to reduce such effects. For more information about XMM-Newton, 
instruments, and it characteristics, we refer to the XMM-Newton Users 
Handbook by \citet{Piconcelli12}.

\begin{table}
\caption{The main characteristics of the XMM-Newton. The table is from \citet{Piconcelli12}. }
\label{tbl:XMM_Char}     
\centering 
 {\footnotesize     
\begin{tabular}{c c c c c c  }         
\hline\hline                        
Instrument           & EPIC MOS        & EPIC PN         &  RGS                    &  OM \\
\hline 
Bandpass             & 0.15-12 keV     & 0.15-12 keV     & 0.35-2.5 keV            & 180-600 nm   \\
Orbital target visibility & 5-135 ks        & 5-135 ks        & 5-135 ks                & 5-145 ks      \\
Sensitivity (erg s$^{-1}$ cm$^{-2}$)   & $\sim 10^{-14}$ & $\sim 10^{-14}$ & $\sim 8 \times 10^{-5}$ & 20.7 mag       \\
Field of view (FOV)  & 30$'$           & 30$'$           & $\sim$ 5$'$             & 17$'$          \\
PSF (FWHM/HEW)       & 5$''$/14$''$    & 6$''$/15$''$    & N/A                     & 1.4$''$-2.0$''$ \\
Pixel size           & 40 $\mu$m (1.1$''$) & 150 $\mu$m (4.1$''$)  & 81 $\mu$m ($9 \times 10^{-3} \AA $) & 0.476513$''$ \\
Timing resolution    & 1.75 ms             & 0.03 ms               & 0.6 s                             & 0.5 s   \\
Spectral resolution  & $\sim$ 70 eV        & $\sim$ 80 eV          & 0.04/0.025 $\AA$                   & 350 \\
\hline
\end{tabular}
}
\end{table}



\section{Constructing Cluster Catalogues}

\subsection{Optically selected clusters}

The first cluster catalogue based on optical data was compiled by 
\citet{Abell58}. He visually inspected the photographic plates from the Palomar
Observatory Sky Survey. To include a cluster in his catalogue he required
a concentration of 50 galaxies or more in a magnitude range $m_3 + 2$, 
where $m_3$ is the third brightest galaxy, within a circle with a radius 
of $\sim$ 2 Mpc in the redshift range $0.02 \le z \le 0.20$. The distances 
were estimated based on the magnitude of the tenth brightest galaxy. The 
cluster catalogue was updated and enlarged to comprise about 4000 clusters 
by \citep{Abell89}. Based on the same optical data with less strict criteria, 
\citet{Zwicky61} constructed a large catalogue of 9700 galaxy clusters extending 
to poor clusters and high redshift ones. 
In addition to these main catalogues many group and cluster samples were 
detected over smaller areas on the sky and up to higher redshifts 
\citep[e.g,][]{Huchra82, Gunn86, Postman96, Gladders05}. The SDSS 
photometric and spectroscopic data allowed to compile large samples 
of optically selected galaxy groups and clusters.

The SDSS saw the first light in 1998 and has started the routine survey in 2000 
\citep{York00}. It uses a dedicated 2.5 m telescope at Apache Point Observatory
(APO) in Southern new Mexico. The telescope is equipped with a large-format 
mosaic CCD  camera to image the sky in five optical bands ({\tt u, g, r, i, z})
and two digital spectrographs to obtain spectra of galaxies, quasars, and stars.  
The first, SDSS-I (2000-2005), and second, SDSS-II (2005-2008), phases of the 
survey provided imaging covering more than a quarter of the sky, photometric  
catalogues, spectra, and redshifts. These data have been made public in  
yearly releases, the final one of those projects was DR7 in 2009. 
The SDSS has extended the previous phases with a new project called SDSS-III, 
which started in 2008 and will carry out observations for six years. 
The present data release of the SDSS is DR9, which includes imaging data of 
14555 deg$^2$ as well as the first spectroscopic data from the SDSS-III  
Baryon Oscillation Spectroscopic Survey (BOSS) that covers a spectroscopic 
footprint area on the sky of 3275 deg$^2$.
Among the scientific goals of SDSS are to create a well-calibrated and 
contiguous imaging and spectroscopic survey of northern Galactic cap at 
high latitudes, in addition to imaging of a series of stripes in Southern
Galactic cap to understand the Galaxy and to explore Type Ia 
supernovae \citep{Ahn12}.
 
In the last 13 years, SDSS data was extensively used to detect galaxy clusters 
through applying various detection techniques that are described in Section 
1.2. \citet{Hao10} listed in Table 1 most of the optical cluster surveys and 
the detection methods using different optical surveys. Here I only summarise 
the optical cluster surveys, which were based on the SDSS data. 
Table~\ref{tbl:SDSS_surveys} lists the main properties of the largest cluster 
catalogues extracted from various releases of SDSS. It lists the catalogue 
name, finding algorithm, used data, number of detected clusters (Nr.CLGs), 
redshift range, and references in columns 1, 2, 3, 4, 5, and 6, respectively.

\begin{table}
\caption{A list of the largest cluster catalogues selected optically based on 
SDSS data using different finding algorithms.}
\label{tbl:SDSS_surveys}     
\centering 
 {\scriptsize                                  
\begin{tabular}{c c c c c c c c}         
\hline\hline                        

Catalogue & Algorithm &  Data &  Nr.CLGs  & Redshift  & References \\  
\hline                                 
SDSS CE      & Cut and enhance (CE) & Magnitudes& 4,638   & 0.01-0.44 & \cite{Goto02}\\
 C4          & C4                   & Colours   &  748    & 0.02-0.17 & \cite{Miller05}\\
 Berlind2006 & Friends-of-friends   & Spec-z    & 8,148   & 0.05-0.1  & \cite{Berlind06}\\
 MaxBCG      & Red-sequence         & Colours   & 13,823  & 0.1-0.3   & \cite{Koester07}\\
 WHL09       & Friends-of-friends   & Photo-z   & 39,688  & 0.05-0.6  & \cite{Wen09}\\
 GMBCG       & Red-sequence         & Colours   & 55,000  & 0.1-0.55  & \cite{Hao10}\\
 AMF         & Adaptive-Matched-    & Magnitudes& 69,173  & 0.05-0.78 & \cite{Szabo11}\\
             & Filter               & + Photo-z &         &           &   \\
 WHL12       & Friends-of-friends   & Photo-z   & 132,684 & 0.05-0.8  & \cite{Wen12}\\
\hline
\end{tabular}
 }
\end{table}


\subsection{X-ray selected clusters}

The statistical studies of clusters of galaxies provide complementary 
and powerful constraints on the cosmological parameters \citep[e.g.][]{Voit05, 
Allen11}. Therefore the X-ray surveys of galaxy clusters, which provide pure 
and clean cluster samples, are an important tool for cosmology and large 
scale structure. Many clusters have been detected in X-ray observations 
taken by the previous X-ray missions e.g.  
Uhuru, HEAO-1, Ariel-V, Einstein, and EXOSAT, which have allowed a more 
accurate characterization of  their physical properties (for a review, 
see \cite{Rosati02}). The ROSAT All Sky Survey \citep[RASS,][]{Voges99} 
and its deep pointed observations  have led to the discovery of hundreds 
of clusters. Based on ROSAT observations, 1743 clusters have been 
identified, which are compiled in a meta-catalogue called MCXC by
\cite{Piffaretti11}. The MCXC catalogue is based on published  RASS-based  
(NORAS, REFLEEX, BCS, SGP, NEP, MACS, and CIZA) and serendipitous 
(160D, 400D, SHARC, WARPS, and EMSS) cluster catalogues.

The current generation of X-ray satellites XMM-Newton, Chandra, and Suzaku  
provided follow-up observations of individual clusters allowing a precise 
determination of their spatially resolved spectra \citep[e.g.][]{Vikhlinin09b, 
Pratt10, Arnaud10}. Several other projects are being conducted to detect 
galaxy clusters from the observations of the XMM-Newton, Chandra, and the 
Swift/X-ray telescopes that provided contiguous surveys for small areas 
\citep[e.g.][]{Finoguenov07, Finoguenov10, Adami11, Suhada12} and pointed 
observations that cover slightly large areas on the sky \citep[e.g.][]
{Boschin02, Barkhouse06, Kolokotronis06, Peterson09, Fassbender11, Takey11, 
Mehrtens12, Clerc12, Tundo12, de-Hoon13, Takey13a, Takey13b}.
So far these surveys provided a substantial cluster sample of few hundreds 
up to redshift of 1.57. \citet{Tundo12} summarised in Table 1 the ongoing 
surveys of galaxy clusters, here I regenerated the table and added the completed
ROSAT surveys as well as our survey as listed in Table~\ref{tbl:X-ray_surveys}. 
It lists the satellite, survey name, type (contiguous or serendipitous), 
survey area, flux limit in 0.5-2 keV, number of detected clusters (Nr. CLGs), 
and its reference in columns 1, 2, 3, 4, 5, 6, and 7, respectively.  

Although, the current X-ray telescopes provide contiguous surveys covering small 
solid angles, they provided thousands of pointed observations. The summed 
field-of-views of these individual observations is covering slightly large area 
on the sky. The pointed observations were the main data base for serendipitous 
cluster surveys that led to the identification of hundreds of galaxy clusters 
as shown in the table above. 
The main X-ray serendipitous sources catalogues to date are the XMM-Newton 
Serendipitous Source Catalogue, 2XMMi-DR3 \citep{Watson09}, and the Chandra 
Source Catalogue \citep{Evans10}. Both catalogues provided thousands of 
extended sources, which are the main resources to discover new clusters of 
galaxies. Unfortunately, these catalogues have a large fraction of 
likely spurious detections that need to be identified and excluded through 
visual inspection of their X-ray images.

As listed in Table~\ref{tbl:X-ray_surveys}, there are only a few thousands of 
clusters that have been identified at X-ray wavelengths. Since most of the 
current X-ray archives are already explored, there is no expected significant 
increase of X-ray detected galaxy clusters from Chandra or XMM-Newton. The 
current status will be 
changed dramatically by the launch of eROSITA, which is scheduled in 2014. It will 
perform the first imaging all-sky survey in the energy band [0.5-10] keV with 
an unprecedented spectral and angular resolution. This will led to detect about 
100,000 clusters of galaxies up to redshift $z \sim 1.3$ in order to study 
the large scale structure and test cosmological models \citep{Predehl10}.

\begin{table}
\caption{Main properties of the completed  and ongoing X-ray cluster surveys.}
\label{tbl:X-ray_surveys}     
\centering 
 {\tiny                                  
\begin{tabular}{c c c c c c c c}         
\hline\hline                        

Satellite &  Name  & Type &  Area          & Flux limit    & Nr.      & References \\  
          &        &      &  deg$^2$       & (0.5-2.0 keV) & CLGs     &            \\
\hline                                 
ROSAT     & MCXC & All Sky Survey  &  &   &  1743 & \cite{Piffaretti11} \\ 
          &      & + Serendipitous &  &   &       & \\ 

\hline
Chandra   & DCS   & Serendipitous  & 5.55 & $0.6 \times 10^{-14}$ & 36 & \cite{Boschin02} \\
          & ChaMP & Serendipitous  & 13.0 & $1.0 \times 10^{-14}$ & 49 & \cite{Barkhouse06}\\    
\hline
XMM-Newton& SEXCLAS  & Serendipitous & 2.1   & $0.6 \times 10^{-14}$ & 19  & \cite{Kolokotronis06}\\ 
          & COSMOS   & Contiguous    & 2.1   & $0.2 \times 10^{-14}$ & 72  & \cite{Finoguenov07}\\
          & SXDF     & Contiguous    & 1.3   & $0.2 \times 10^{-14}$ & 57  & \cite{Finoguenov10}\\
          & XMM-LSS  & Contiguous    & 11.0  & $\sim  10^{-14}$      & 66  & \cite{Adami11}\\
          & XDCP     & Serendipitous & 76.0  & $\sim 10^{-14}$  & 22  & \cite{Fassbender11}\\
          & XCS      & Serendipitous & 410.0 & $>$ 300 net cts       & 504 & \cite{Mehrtens12}\\
          & XCLASS   & Serendipitous & 90.0  & $2 \times 10^{-14}$   & 347 & \cite{Clerc12}\\
          & XMM-BCS  & Contiguous    & 6.0   & $0.6 \times 10^{-14}$ & 46  & \cite{Suhada12}\\
          &{\bf 2XMMi/SDSS} & Serendipitous & 210   & $>$ 80 net cts  & 574 & {\bf This thesis}\\
          &                 &               &       &                 &     & \cite{Takey11, Takey13a, Takey13b} \\   
\hline

XMM+Chandra & Peterson09 & Serendipitous & 163.4 & $0.3 \times 10^{-14}$ & 462 & \cite{Peterson09}\\
\hline
Swift/XRT   & SXCS       & Serendipitous & 40.0  & $1.0 \times 10^{-14}$ & 72  & \cite{Tundo12}\\
\hline
\end{tabular}
}
\end{table}


\subsection{SZ selected clusters}

Recently, several galaxy cluster surveys have been conducted at mm wavelengths 
through the Sunyaev-Zel'dovich (SZ) effect. Although, mm survey covering large 
area was proposed in the nineties by \cite{Barbosa96} to constrain the 
cosmological parameters, it is only recently achieved to have such large 
surveys that provided several hundreds of galaxy clusters. These surveys 
were done by either ground-based telescopes e.g the South Pole Telescope 
\citep[SPT,][]{Reichardt13} and the Atacama Cosmology Telescope \citep[ACT,]
[]{Hasselfield13} or the Planck Satellite \citep[e.g.][]{Planck11, Planck13}. 
The number of confirmed clusters detected by the Plank, SPT, and ACT are 861, 
158, and 68, respectively. The SZ surveys provide a complementary 
tool to determine the cluster mass, the most fundamental parameter of clusters, 
through the tight correlations between the SZ signal with the mass. The main 
advantage of this technique is a redshift-independent signal that led to discover 
clusters up to high redshifts.             



\section{Observable parameters}

In addition to identification of clusters through the optical surveys they also 
provide several observables. The main observables are the richness, net cluster 
galaxies after subtracting the expected background galaxies, and the optical 
luminosity of the clusters. Both observables can be used to estimate the cluster 
mass but with a large uncertainty. Follow-up 
of individual clusters provided data that can be used to measure their masses 
with much better accuracy through the galaxy number density, luminosity, and 
velocity dispersion profiles. Another possibility to infer the cluster mass 
with high precision is through gravitational lensing of background galaxies. 
       
Assuming the ICM is an isothermal sphere, the gas density profile of the 
intracluster gas can be derived by the King model that was originally applied 
for globular star clusters by \citet{King62} as
\begin{equation}
\rho_{\rm g}(r)\, \propto \,\Bigg[1 + \Big(\frac{r}{r_{\rm c}} \Big)^2 \Bigg]^{-3\beta/2},
\end{equation}  
and the X-ray surface brightness profile of clusters can be described by the 
$\beta$ model \citep{Cavaliere76} : 
\begin{equation}
 S_{\rm X}(r)\, \propto \,\Bigg[1 + \Big(\frac{r}{r_{\rm c}} \Big)^2 \Bigg]^{-3\beta + 1/2},
\end{equation}
where r$_{\rm c}$ is the core radius and $\beta$ value equals approximately 2/3.
The $\beta$ model provides a good approximation to the observed surface 
profiles except for the cool core clusters that have an extra central brightness 
enhancement and it also shows a deviation in the far cluster outskirts 
\citep[e.g.,][]{Boehringer06}. The parameters of the $\beta$ model are the 
$r_{\rm c}$, $\beta$, and the normalization. 

Measuring the plasma temperature and abundance requires high quality X-ray 
data because photons must be distributed among multiple energy bins. The global 
ICM temperature is obtained by fitting a single-temperature emission model to 
a cluster spectrum that contains multiple temperature. Having enough data, 
the temperature and density profiles can be measured. Assuming the ICM is 
a spherically symmetric system in hydrostatic equilibrium, the measured gas 
density, $\rho$(r), and temperature, $T(r)$, profiles can be related to the 
overall mass of the clusters, 
\begin{equation}
M(r) = - \frac{rkT(r)}{G \mu m_{\rm p}} \Bigg[\frac{d\, \ln\, \rho(r)}{d\, \ln\, r} + \frac{d\, \ln\, T(r)}{d\, \ln\, r} \Bigg].
\end{equation}
where M(r) is the mass within radius $r$, G is Newton's constant, $k$ is the 
Boltzmann constant, and $\mu m_{\rm p}$ is the mean molecular weight 
\citep[e.g.,][]{Sarazin88, Voit05, Allen11}. For those clusters with 
low-quality X-ray data, the X-ray observables provide good mass proxies as 
described in the next section.



\section{Scaling relations}

Various correlations are found among the observable parameters as well as 
relations between the observables with the cluster mass. In the following 
subsections, I will briefly show the relation for X-ray and optical 
observables and their use to infer the cluster mass.


\subsection{X-ray scaling relations}

If clusters are formed through the gravitational collapse of a homogeneous 
spherical over-density of non-interacting dark matter, it is expected that 
the collapse process and the produced dark matter halos are self-similar. 
This means that clusters with lower mass are scaled down versions of clusters 
with higher masses. 
N-body simulations suggested that the dark matter halo has a density 
distribution, known as NFW profile, with the form
\begin{equation}
 \rho_{\rm g}(r)\ \propto\ r^{-p}\ (r + r_{\rm s})^{p - q} 
\end{equation}
where $r_{\rm s}$ is the scale radius and the most common representations are 
$p = 1$ and $q = 3$ \citep{Navarro97}. The characteristic radius depends on the
total mass and the formation time. The NFW profile fits the mass profile 
of the most well-relaxed clusters but it shows some deviations of clusters 
with ongoing mergers \citep{Boehringer06}.

If clusters are approximately self-similar systems, their global parameters are 
expected to follow tight relations and to scale with mass, which are called 
scaling relations. Assuming clusters are spherically symmetric systems 
in a hydrostatic equilibrium and are formed at the same epoch, the theoretical 
scaling relations for the cluster parameters (temperature, luminosity, and 
mass) can be derived. The expected relations for self-similar systems M-T, 
L-T, and L-M at redshift = 0 and the evolved ones are given in Col. 2 and 3 
of Table~\ref{tbl:X-ray_relations}. Since the scaling relations are 
theoretically expected for clusters at redshift = 0, a correction factor, 
$E(z)$, should be applied to the derived parameters for higher redshift 
clusters \citep[e.g.~][]{Kotov05}. The evolution factor, $E(z)$, is dimensionless 
and defined as $\bigl[\Omega_{\rm M} (1+z)^{3} + \Omega_{\Lambda}\Bigr]^{1/2}$.                 

These scaling relations have been calibrated observationally by several groups
\citep[e.g.,][]{Reiprich02, Pratt09}. These studies showed that the observed 
relations have a partly differing power law exponent from the expected one from 
the scaling relations according to the self-similar evolution. This break from 
the self-similarity is clearly shown in the L-T relation \citep[e.g.][]
{Markevitch98, Pratt09, Mittal11, Eckmiller11,  Reichert11, Takey11, 
Maughan12, Hilton12}. These investigations showed that the observed L-T 
relation is much steeper than the slope predicted by self-similar evolution. 
This indicates  that the ICM is heated by an additional source of energy, 
which comes mainly from AGNs \citep{Blanton11}. The 
inclusion of AGN-feedback at high redshifts in cosmological evolution models 
indeed gives better agreement between simulated and observed L-T relation 
\citep{Hilton12}.

\begin{table}
\caption{X-ray scaling relations that are theoretically expected from the 
self-similar model.}
\label{tbl:X-ray_relations}     
\centering 
 \begin{tabular}{c c c c c c c c}         
\hline\hline                        

relations & self-similar relations &  evolved self-similar relations \\
          & redshift = 0           &    redshift = z                  \\
\hline \\                                
M-T       & $M\ \propto\ T_{\rm X}^{3/2}$   &  $M\ \propto\ E^{-1}(z)\ .\ T_{\rm X}^{3/2}$ \\\\
L-T       & $L_{\rm Xbol}\ \propto\ T_{\rm X}^{2}$     &  $L_{\rm Xbol}\ \propto\ E(z)\ .\ T_{\rm X}^{2}$        \\\\
L-M       & $L_{\rm Xbol}\ \propto\ M^{4/3}$   &  $L_{\rm Xbol}\ \propto\ E^{7/3}(z)\ .\ M^{4/3}$ \\\\
\hline
\end{tabular}
\end{table}


\subsection{X-ray-optical relations}

As discussed above, the most important parameter of a galaxy cluster is the mass, 
which can determined through the X-ray observables, SZ signals, and gravitational 
lensing. Although, these techniques provide accurate mass measurements, they 
are too expensive in terms of observations and they also provided small cluster 
samples. The alternative and the simple way to obtain mass measurements for 
a large cluster sample is to find an unbiased photometric proxy such as optical 
richness or optical luminosity. Establishing and calibrating such 
mass-observable relations are essential for determining the cluster mass 
function \citep{Lopes09}.       

Solid observational evidence indicates a strong interaction between the two 
baryonic components of galaxy clusters. The evolution of galaxies in clusters 
is influenced by the hot diffuse gas in the ICM. The observed metal abundance in 
the ICM is produced by the pollution metals expelled from galaxies via galactic 
winds \citep[][]{Finoguenov01, DeGrandi02}. To understand the complex physics 
of galaxy cluster baryon components, it is required to combine X-ray and optical 
observations of a large sample of these systems and to compare their optical 
and X-ray appearance \citep{Popesso04}.        

Several studies have presented the correlations between the X-ray 
observables such as luminosity and temperature as well as the cluster mass  
with both optical richness and luminosity \citep[e.g.][]{Popesso04, Popesso05, 
Lopes06, Rykoff08, Lopes09, Wen09,Szabo11} based on known massive X-ray galaxy 
clusters in the literature apart from RASS-SDSS sample that comprised galaxy 
groups and clusters by \cite{Popesso04}. These correlations show the ability 
to predict the X-ray properties of galaxy clusters, the most expensive to 
observe, as well as the cluster masses from the optical properties, and vice 
versa within a certain accuracy.   
In addition several groups investigated the relation between the optical 
luminosity of the BCGs and the cluster masses \citep[e.g.][]{Popesso07, 
Mittal09}. They found a relation with inconsistent slopes as $L_{\rm BCG}\, 
\propto \, M^{0.1...0.6 }$. The absolute magnitude of the BCGs is also found 
to be dependent on the cluster redshift \citep{Wen12}.





\section{Aims and outline of this work}

In this thesis I am going to explore the XMM-Newton archival data in order to 
identify a large sample of new X-ray detected galaxy groups and clusters. 
The constructed catalogue will allow us to investigate the properties of these 
new systems, to trace the evolution of the X-ray scaling relations and the 
correlations between the X-ray and optical properties. To measure the X-ray 
luminosities and temperatures of the new constructed sample, I need to 
measure their redshifts. Therefore I constrained the survey on those XMM-Newton 
fields that are in the foot print of the SDSS in order to measure their optical 
redshifts. The thesis is divided into six chapters. Each chapter is almost 
self-contained, therefore I will present each chapter as a paper format. The 
second chapter is already published in the {\it Astronomy \& Astrophysics 
Journal}. The work of the third chapter is accepted for publication 
in the same journal while the fourth and fifth chapters will be submitted in 
the near future. 

The thesis is organised as follows: in Chapter 2 I will describe the 
cluster survey, the X-ray cluster candidates list, the X-ray data reduction 
and analysis, the cluster sample with available redshifts in the literature, 
the first cluster sample with X-ray luminosities and temperatures, and the 
X-ray luminosity-temperature relations. In Chapter 3 I will present our 
strategy to measure the photometric redshifts of the clusters, the optically 
confirmed cluster sample and their X-ray properties, and the X-ray 
luminosity-temperature relation that is derived from a large sample 
including groups and clusters over a wide redshift range. 

In Chapter 4 I will enlarge the optically confirmed cluster sample using the 
recent data of the SDSS that includes spectra of the luminous red galaxies 
(LRGs) in BOSS. In Chapter 5 I will investigate the relations between 
the BCGs properties (absolute magnitude, optical luminosity) and the cluster 
properties (redshift, mass), in addition to the correlation between 
the X-ray properties (temperature, luminosity, mass) and the optical 
properties (richness and optical luminosity) of a subsample with X-ray 
temperature measurements. I finally summarise the the work of my thesis 
as well as the main contributions from this study in Chapter 6.



%



\chapter[The 2XMMi/SDSS Galaxy Cluster Survey: I. The first cluster sample]{The 2XMMi/SDSS Galaxy Cluster Survey: I. The first cluster sample \footnote{This chapter is published in the {\it Astronomy \& Astrophysics Journal}, 2011A\&A...534A..120T} }

%
%



\section*{Abstract}
We present a catalogue of X-ray selected galaxy clusters and groups as 
a first release of the 2XMMi/SDSS Galaxy Cluster Survey. The survey is a search
for galaxy clusters detected serendipitously in observations with XMM-Newton
in the footprint of the Sloan Digital Sky Survey (SDSS). 
The main aims of the survey are to identify new X-ray galaxy clusters,
investigate their X-ray scaling relations, identify distant cluster candidates, 
and  study the correlation of the X-ray and optical properties. In this paper, 
we describe the basic strategy to identify and characterize the X-ray cluster
candidates that currently comprise 1180 objects selected from the second
XMM-Newton serendipitous source catalogue (2XMMi-DR3). 
Cross-correlation of the initial catalogue with recently published optically
selected SDSS galaxy cluster catalogues yields photometric redshifts for 275
objects. Of these, 182 clusters have at least one member with a spectroscopic
redshift from existing public data (SDSS-DR8). We developed an automated
method to reprocess the XMM-Newton X-ray observations, determine the optimum
source extraction radius, generate source and background spectra, and derive
the temperatures and luminosities of the optically confirmed clusters.
Here we present the X-ray properties of the first cluster sample, which
comprises 175 clusters, among which 139 objects are new X-ray discoveries while
the others were previously known as X-ray sources. 
For each cluster, the catalogue provides: two identifiers, coordinates,
temperature,  flux [0.5-2] keV, luminosity [0.5-2] keV extracted from an
optimum aperture, bolometric luminosity $L_{500}$, total mass $M_{500}$,
radius $R_{500}$, and the optical properties of the counterpart.  The 
first cluster sample from the survey covers a wide range of redshifts from 0.09 to 0.61,
bolometric luminosities $L_{500} = 1.9 \times 10^{42} - 1.2 \times
10^{45}$\,erg s$^{-1}$, and masses $M_{500} = 2.3 \times 10^{13} - 4.9 \times
10^{14}$\,M$_\odot$.  We extend the relation between the X-ray bolometric
luminosity $L_{500}$ and the X-ray temperature towards significantly lower 
$T$ and $L$ and still find that the slope of the linear $L-T$ relation is  
consistent with values published for high luminosities.


\section{Introduction}
Galaxy clusters are the most visible tracers of large-scale structure. They
occupy very massive dark matter halos and are observationally  accessible by
a wide range of means. Their locations are found to corresponding to large numbers of
tightly clustered galaxies, pools of hot X-ray  emitting gas, and relatively
strong features in the gravitational lensing shear field. Precise observations
of large numbers of clusters  provide an important tool for testing our
understanding of cosmology and structure formation. Clusters are also
interesting laboratories  for the study of galaxy evolution under the
influence of extreme environments \citep{Koester07}. 

The baryonic matter of the clusters is found in two forms: first, individual
galaxies within the cluster, which are most effectively studied through  optical and NIR
photometric and spectroscopic surveys; and second, a hot, ionized intra-cluster
medium (ICM), which can be studied by means of its X-ray emission and the Sunyaev-Zeldovich
(SZ) effect \citep[][]{Sunyaev72,Sunyaev80}. The detection of clusters using SZ effect is 
a fairly new and highly promising  technique for which tremendous progress has been made 
in finding high redshift clusters and measuring the total cluster mass  
\citep[e.g.][]{Planck11, Vanderlinde10, Marriage11}.   

The X-ray selection of clusters has several advantages for cosmological surveys: the
observable X-ray luminosity and temperature  of a cluster is tightly
correlated with its total mass, which is indeed its most fundamental parameter
\citep{Reiprich02}. These relations provide the ability to measure both the mass
function \citep{Boehringer02} and power spectrum \citep{Schuecker03}, which directly 
probe the cosmological models.  Since the cluster X-ray emission is
strongly peaked on the dense cluster core, X-ray selection is less affected by projection 
effects than optical surveys and clusters can be identified efficiently over
a wide redshift range.  

Many clusters have been found in X-ray observations with Uhuru, HEAO-1, Ariel-V,
Einstein, and EXOSAT, which have allowed a more accurate characterization of  their
physical proprieties (for a review, see \cite{Rosati02}). The ROSAT All Sky
Survey \citep[RASS,][]{Voges99} and the deep pointed observations  have led to
the discovery of hundreds of clusters. In ROSAT observations, 1743 clusters have been 
identified, which are compiled in a meta-catalogue called MCXC by
\cite{Piffaretti11}. The MCXC catalogue is based on published  RASS-based  
(NORAS, REFLEEX, BCS, SGP, NEP, MACS, and CIZA) and
serendipitous (160D, 400D, SHARC, WARPS, and EMSS) cluster catalogues.  

The current generation of X-ray satellites XMM-Newton, Chandra, and Suzaku have
provided follow-up observations of statistical samples of ROSAT  clusters for
cosmological studies \citep{Vikhlinin09b} and detailed information on the
structural proprieties of the cluster population
\citep[e.g.][]{Vikhlinin06,Pratt10,Arnaud10}. Several projects are ongoing to
detect new clusters of galaxies from XMM-Newton and Chandra observations
(e.g. the XSC \citep{Romer01}, XDCP \citep{Fassbender07}, XMM-LSS
\citep{Pierre06}, COSMOS \citep{Finoguenov07}, SXDS \citep{Finoguenov10}, and
ChaMP \citep{Barkhouse06}). 

In this paper, we present the 2XMMi/SDSS galaxy cluster survey, a search
for galaxy clusters based on extended sources in the 2XMMi catalogue
\citep{Watson09}  in the field of view the Sloan Digital Sky Survey (SDSS). The
main aim of the survey is to build a large catalogue of new X-ray clusters  
in the sky coverage of SDSS. The catalogue will allow us to investigate the
correlation between the X-ray and optical properties of the clusters. 
One of the long term goals of the project is to improve the X-ray scaling
relations, and to prepare for the eROSITA cluster surveys,  a mid term goal is
the selection of the cluster candidates beyond the SDSS-limit for studies of
the distant universe. Here we present a first cluster sample of the survey which
comprises 175 clusters found by cross-matching the 2XMMi sample with published
SDSS based optical cluster catalogues. 

The paper is organized as follows: in Sect.~2 we describe the procedure of
selecting the X-ray cluster candidates as well as  their possible
counterparts in SDSS data. In Sect.~3 we describe the X-ray data the reduction
and analysis of the optically confirmed clusters. The discussion of the
results is described in Sect.~4. Section 5 concludes the paper. The cosmological
parameters  $\Omega_{\rm M}=0.3$, $\Omega_{\Lambda}=0.7$ and
$H_0=70$\ km\ s$^{-1}$\ Mpc$^{-1}$ were used throughout this paper.


\section{Sample construction}
We describe our basic strategy for identifying clusters among the 
extended X-ray sources in the 2XMMi catalogue. We then proceed
by cross-matching the initial catalogue with those of optically
selected galaxy clusters from the SDSS thus deriving a catalogue of X-ray
selected and optically confirmed clusters with measured redshifts, whose X-ray
properties are analysed in Sect.~3.

\subsection{X-ray cluster candidate list}
X-ray observations provide a robust method for the initial identification of
galaxy clusters as extended X-ray sources. A strategy to create a clean galaxy
cluster sample is to construct a catalogue of  X-ray cluster candidates
followed by optical observations. XMM-Newton archival observations provide the
basis for creating catalogues of serendipitously identified point-like and extended
X-ray sources. The largest X-ray source  catalogue ever produced is the 
second XMM-Newton source catalogue \citep{Watson09}.  The latest edition of
this catalogue is 2XMMi-DR3, which was released on 2010 April 28.  
The 2XMMi-DR3 covers 504 deg$^{2}$ and contains $\sim 3$ times as many discrete
sources as either the ROSAT survey or pointed catalogues. The catalogue
contains 353191 X-ray source detections corresponding to 262902 unique X-ray
sources detected in 4953 XMM-Newton EPIC (European Photon Imaging Camera)
observations made between 2000 February 3 and 2009 October 8.  

The 2XMMi-DR3 contains 30470 extended source detections, which form the primary
database for our study. This initial sample contains a very significant number
 of spurious detections caused by the clustering of unresolved point sources, edge
effects, the shape of the PSF (point spread function) of the X-ray mirrors,
large extended sources consisting of several minor sources, and other
effects such as X-ray ghosts and similar. 

We applied several selection steps to obtain a number of X-ray extended
sources that were then visually inspected individually. In our study, we considered only
sources at high Galactic latitudes, $|b| > 20^{\circ}$, and discarded those 
that were flagged as spurious in the 2XMMi-DR3 catalogue by the screeners of
the XMM-Newton SSC (Survey Science Centre). The source detection pipeline
used for the creation of the 2XMMi catalogues allows for a maximum core radius
of extended sources of 80 arcsec. Sources with extent parameters equalling
that boundary were discarded, screening of a few examples shows that those
sources are spurious or large extended sources (the targets of the
observations) that were discarded anyhow. This initial selection reduced the
number to 4027 detections. 

Since our main aim is the generation of a serendipitous cluster sample, we
removed sources that were the targets of the XMM-Newton observation. We also
discarded fields containing large extended sources and selected only those
fields within the footprint of SDSS, which left 1818 detections.  After
removing multiple detections of the same extended sources (phenomena caused by either
problematic source geometries in a single XMM-Newton observation or 
duplicate detections in a re-observed field), the catalogue was reduced to
1520 extended sources that were regarded as unique.

This list still contains spurious detections for a number of reasons:
(a) point-source confusion, (b) resolution of one asymmetric extended source
into several symmetric extended sources, (c) the ill-known shape of the PSF
leads to an excess of sources near bright point-sources, for both 
point-like and extended sources, and (d) edge effects/low exposure times.
To remove the obvious spurious cases, we visually inspected the X-ray images 
of the initial 1520 detections using the FLIX upper limit 
server\footnote{http://www.ledas.ac.uk/flix/flix.html}. As a result, we were 
left with 1240 confirmed extended X-ray sources. 

We then made use of the SDSS to remove additional non-cluster sources. 
We downloaded the XMM-Newton EPIC X-ray images from the XMM-Newton Science
Archive \citep[XSA:][]{Arviset02} and created summed EPIC (PN+MOS1+MOS2)
images in the energy band $0.2 - 4.5$\,keV. Using these, we created smoothed
X-ray contours, which were overlaid onto co-added $r$, $i$, and $z-$band SDSS
images. Visual inspection of those optical multi-colour images with X-ray
contours overlaid, allowed us to remove extended sources corresponding to
nearby field galaxies, as well as those objects that are likely spurious 
detections. The resulting list which passes these selection criteria 
contains 1180 cluster candidates, about 75 percent of which  
are newly discovered.

\begin{figure}[t]
 \centering{ 
   \resizebox{100mm}{!}{\includegraphics{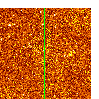}}}
  \caption[The X-ray-optical overlay of the representative cluster, 2XMM
    J104421.8+213029 at photometric redshift = 0.4975]{The X-ray-optical overlay of the representative cluster 2XMM
    J104421.8+213029 at photometric redshift = 0.4975. The X-ray contours are
    overlaid on the SDSS co-added image obtained in  $r$, $i$, and
    $z$-bands. The field of view is $4'\times 4'$ centred on the X-ray
    cluster position. The cross-hair indicates the cluster stellar mass centre 
    as given by \cite{Szabo11}.}
    \label{f:overlay}
\end{figure}

Figure~\ref{f:overlay} shows the X-ray-optical overlay of a new X-ray 
cluster, which has a counterpart in SDSS at photometric redshift = 0.4975 and
a stellar mass centre indicated by the cross-hair as given by \cite{Szabo11}
(see Section 2.2). We use this cluster to illustrate the main steps of our
analysis in the following sections. In Appendix A, Figures A.1 to A.4 show the 
X-ray-optical overlays and the extracted  X-ray spectra for four clusters 
illustrating results for various redshifts covered by our sample and different 
X-ray fluxes.

About one quarter of the X-ray selected cluster candidates have no plausible
optical counterpart. These are regarded as high-redshift candidates 
beyond the SDSS limit at $z \geq 0.6$, and suitable targets for dedicated
optical/near-infrared follow-up observations \citep[see e.g.~][]{Lamer08}.


\subsection{The cross-matching with optical cluster catalogues}
The SDSS offers the opportunity to produce large galaxy-cluster
catalogues. Several techniques were applied to identify likely clusters from
multiband imaging and SDSS spectroscopy. We use those published catalogues to
cross-identify common sources in our X-ray selected and those optical
samples. All these optical catalogues give redshift information per cluster,
which we use in the following to study the X-ray properties of our sources
($z_{p}$ indicates a photometric, $z_{s}$ a spectroscopic redshift). 

Table~\ref{t:sdss} lists the main properties of the optical cluster catalogues 
that we used to confirm our X-ray selection. Below we provide a very brief 
description of each of these together with the acronym used by us:

\begin{itemize}

\item{\bf GMBCG} The {\it Gaussian Mixture Brightest Cluster
  Galaxy} catalogue  \citep{Hao10}  
  consists of more than 55,000 rich clusters across the redshift
  range $0.1 < z_{p} < 0.55$ identified in SDSS-DR7.  The galaxy clusters were 
  detected by identifying the cluster red-sequence plus a brightest cluster
  galaxy (BCG). The cross-identification of X-ray cluster candidates with the 
  GMBCG within a radius of 1 arcmin yields 136 confirmed clusters.  

\item{\bf WHL} The catalogue of Wen, Han \& Liu \citep{Wen09} consists of
  39,668 clusters of galaxies drawn from SDSS-DR6 and covers the redshift
  range $0.05 < z_{p} < 0.6$. A cluster was identified if more than eight
  member galaxies of $ M_{r} \le -21$ were found within a radius of 0.5\,Mpc and
  within a photometric redshift interval $z_{p} \pm 0.04(1 + z_{p})$. We
  confirm 150 X-ray clusters by cross-matching within 1 arcmin. 

\item{\bf MaxBCG} The {\it max Brightest Cluster
  Galaxy} catalogue \citep[][]{Koester07} lists 13,823 clusters in
  the redshift range $0.1 < z_{p} < 0.3$ from SDSS-DR5. The clusters were
  identified using maxBCG red-sequence technique, which uses the clustering 
  of galaxies on the sky, in both magnitude and colour, to identify groups and 
  clusters of bright E/S0 red-sequence galaxies. The cross-match with our X-ray 
  cluster candidate list reveals 54 clusters in common within a radius of 
  one arcmin. 

\item{\bf AMF} The {\it Adaptive Matched Filter} catalogue of \cite{Szabo11}
  lists 69,173 likely galaxy clusters in the redshift range $0.045 < z_{p} <
  0.78$ extracted from SDSS-DR6 using an adaptive matched filter (AMF) cluster
  finder. The cross-match yields 127 confirmed X-ray galaxy clusters. 
\end{itemize}

In the AMF-catalogue, the cluster centre is given as the anticipated centre of
the stellar mass of the cluster, while in the other three catalogues the  cluster 
centre is the position of the brightest galaxy cluster (BCG).

Many of our X-ray selected clusters have counterparts in several optical
cluster catalogues within our chosen search radius of one arcmin. In these
cases, we use the redshift of the optical counterpart, which has minimum spatial
offset from the X-ray position. Table~\ref{t:sdss} 
lists in the second to last column the number of matching X-ray sources per 
optical catalog individually and in the last column the final number after 
removal of duplicate identifications. 
%

\begin{table}
\caption[Main properties of the cluster catalogues with optically (SDSS-based)
  selected entries]{Main properties of the cluster catalogues with optically (SDSS-based)
  selected entries. The last two columns give the number of matching X-ray
  selected clusters individually and cumulatively.}
\label{t:sdss}     
\centering                                    
\begin{tabular}{c c c c c c}         
\hline                       
CLG         & Nr.       & Redshift   & SDSS      & X-ray      & Nr.CLG  \\  
catalogue   & CLG       & range      &           & CLG ($1'$) & sample \\
\hline                                  
    GMBCG   & 55,000  & 0.1 - 0.55 &  DR7      &  136  & 123 \\     
    WHL     & 39,688  & 0.05 - 0.6 &  DR6      &  150  & 72 \\
    MaxBCG  & 13,823  & 0.1 - 0.3  &  DR5      &  54   & 20 \\
    AMF     & 69,173  &0.045 - 0.78&  DR6      &  127  & 60 \\
\hline
   Total    &         &            &           &       & 275 \\     
\hline 
\end{tabular}
\end{table}

The unique optically confirmed X-ray cluster sample obtained by cross-matching
with the four catalogues consists of 275 objects having at least photometric 
redshifts. 
After cross-identification, we found 120 clusters of the optically confirmed cluster 
sample with a spectroscopic redshift for the brightest galaxy cluster (BCG) from the 
published optical catalogues. Since the latest data release, SDSS DR8, provides more 
spectroscopic redshifts, we searched for additional spectra of BCGs 
and other member galaxies. We ran SDSS queries searching for galaxies 
with spectroscopic redshifts $z_{s(g)}$ within 1 Mpc from the X-ray 
centre. We considered a galaxy as a member of a cluster if
$|z_{p} - z_{s(g)}| < 0.05$.  

\begin{figure}
\centering{
  \resizebox{100mm}{!}{\includegraphics[viewport=15  5  525 400, clip]{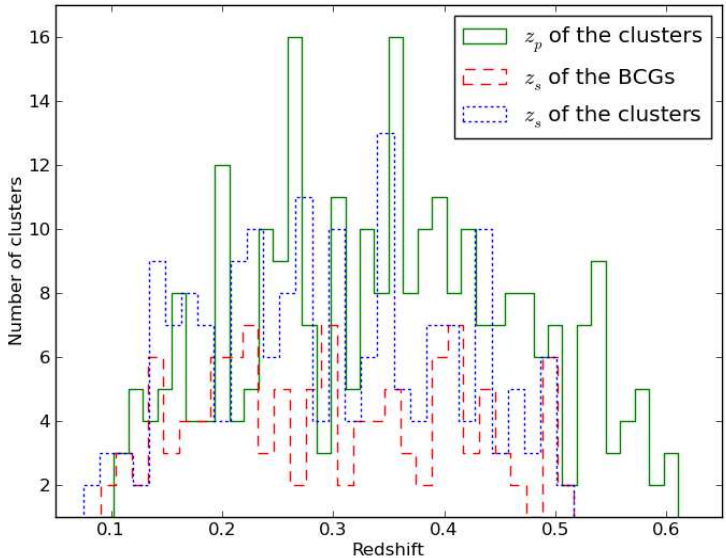}}}
  \caption[The distribution of the optical redshifts for the confirmed
    clusters sample]{The distribution of the optical redshifts for the confirmed
    clusters sample. The distribution includes the cluster photometric
    redshifts $z_{p}$ (solid line) with a median 0.36, spectroscopic redshifts
    of the BCGs $z_{s}$ (dashed line) with a median 0.3 from the optical
    cluster catalogues and the cluster  spectroscopic redshifts $z_{s}$
    (dotted line) with a median 0.3 from the SDSS data.} 
  \label{f:zdist}
\end{figure}

\begin{figure}
\centering{
  \resizebox{100mm}{!}{\includegraphics[viewport=15  5  525 400, clip]{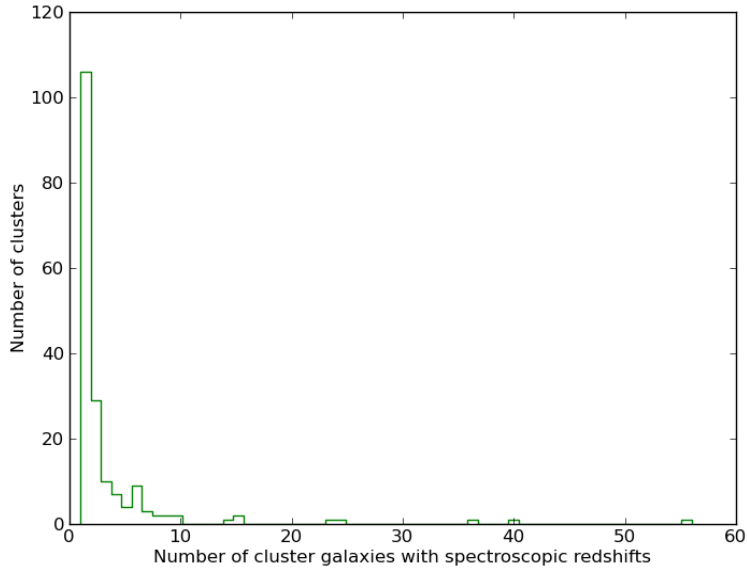}}}
  \caption{The distribution of the spectroscopically confirmed cluster members
    per cluster.}
  \label{f:numspec}
\end{figure}

\begin{figure}
\centering{
  \resizebox{100mm}{!}{\includegraphics[viewport=15  10  525 400, clip]{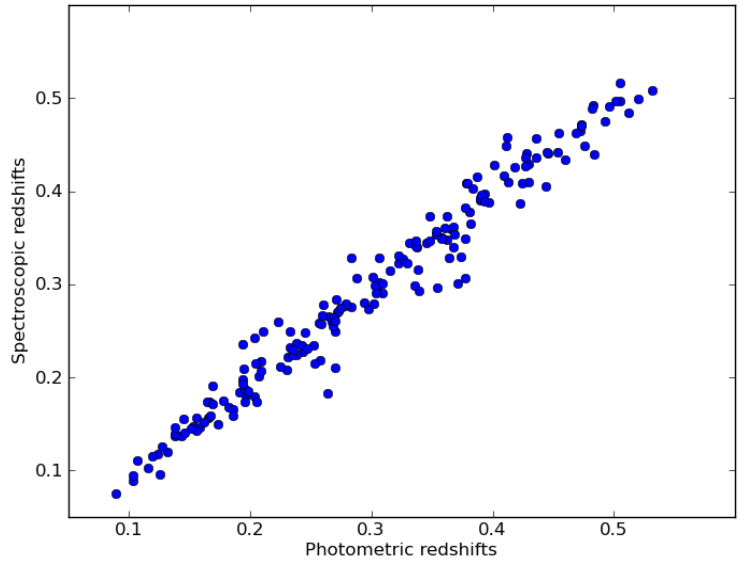}}}
  \caption{The relation between the photometric and spectroscopic redshifts of
    the confirmed cluster sample.} 
  \label{f:zpzs}
\end{figure}

The spectroscopic redshift of the cluster was calculated as the average
redshift for the cluster galaxies with spectroscopic redshifts.  The confirmed
cluster sample with spectroscopic redshifts for at least one galaxy 
includes 182 objects. Therefore, the unique optically confirmed X-ray cluster
 sample has the photometric redshifts for all of them, 120 spectroscopic 
 redshifts for 120 BCGs from the optical cluster catalogues, and 182 clusters 
with one or more members with spectroscopic redshifts from the SDSS database. 
Figure~\ref{f:zdist} shows the distribution of the cluster photometric 
redshift $z_{p}$, the distribution of spectroscopic redshifts $z_{s}$ 
of the BCGs as given in the various optical cluster catalogues, and the average
spectroscopic redshift of the cluster members (which we refer to as the cluster
spectroscopic redshift) for the confirmed cluster sample. Figure~\ref{f:numspec} 
shows the distribution of the number of cluster galaxies that have a spectroscopic 
redshift in the SDSS database. The relation between the photometric and spectroscopic 
redshifts of the cluster sample is shown in Figure~\ref{f:zpzs}. Since this relation 
was found to be tight (where the Gaussian distribution of ($z_{s}$ - $z_{p}$) has 
$\sigma$ = 0.02), we were able to rely on the photometric redshifts for the cluster 
with no spectroscopic information.

\begin{figure}
\centering{
  \resizebox{100mm}{!}{\includegraphics[viewport=15  10  535 400, clip]{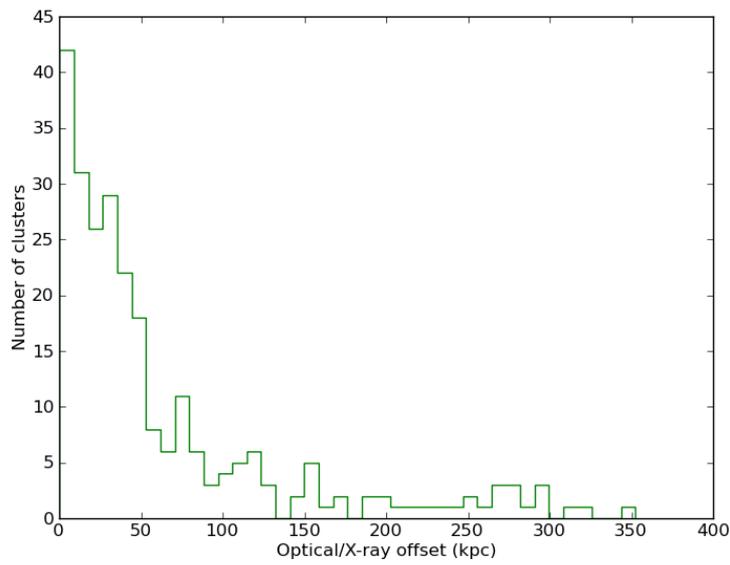}}} 
  \caption{ The distribution of the linear separation between the position of
    the BCG and the X-ray cluster position.} 
  \label{f:xoff}
\end{figure}

We used an angular separation of one arcmin to cross-match the X-ray
cluster candidates with the optical cluster catalogues. The corresponding
linear separation was calculated using the spectroscopic redshift, if
available, or the photometric redshift. Figure~\ref{f:xoff} shows the
distribution of the linear separation between the X-ray centre and the BCG
position. For AMF clusters (60 objects), we identified the BCGs of 40 
systems within one arcmin and computed their offsets, which are included in this 
aforementioned distribution. The BCGs were selected as the brightest galaxies with 
$|z_{p} - z_{p(BCG\ cand.)}| < 0.05$ among the three 
BCG candidates given for each AMF cluster published by 
\cite{Szabo11}. The other 20 AMF clusters are not included in 
Figure~\ref{f:xoff}, because their BCG is outside one arcmin. It is not always the
case that the BCG lies exactly on the X-ray peak. \citet{Rykoff08} model the
optical/X-ray offset distribution by matching a sample of maxBCG clusters to
known X-ray sources from the ROSAT survey. They found a large excess of X-ray
clusters associated with the optical cluster centre. There is a tight core in
which the BCG is within $\sim$ 150\ h$^{-1}$\ kpc  of any X-ray source, as
well as a long tail extending to  $\sim$ 1500\ h$^{-1}$\ kpc. It is shown in 
Figure~\ref{f:xoff} that the majority of the confirmed sample have the BCG 
within a radius ($\sim$ 150 kpc), as well as a tail extending to 352 kpc 
that is consistent with the optical/X-ray offset distribution of   
\cite{Rykoff08}.  

We searched the Astronomical Database SIMBAD and the NASA/IPAC
Extragalactic Database (NED) to check whether they had been identified and catalogued
previously. We used a search radius of one arcmin. About 85 percent of the
confirmed sample are new X-ray clusters, while the remainder had been
previously studied using ROSAT, Chandra, or XMM-Newton data.


\section{X-ray data analysis}
\begin{figure}
\centering{
  \resizebox{100mm}{!}{\includegraphics[viewport=15  0  530 400, clip]{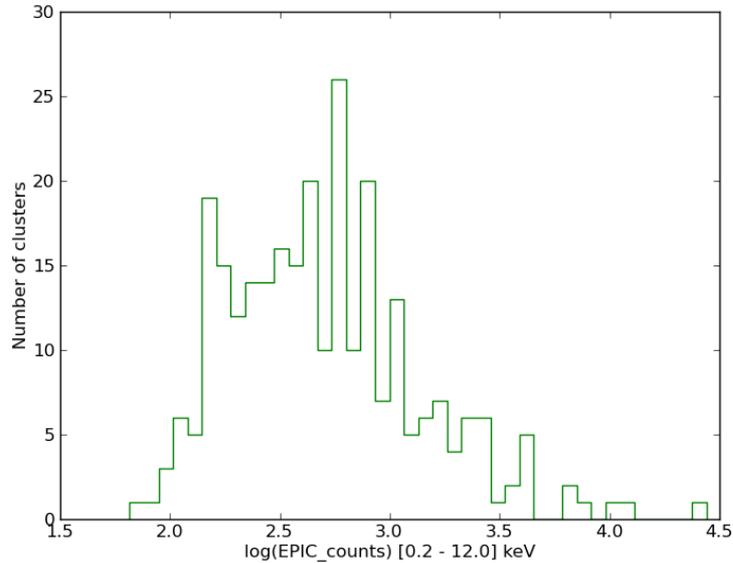}}}
  \caption{The distribution of EPIC counts in [0.2-12] keV as given in the 2XMMi-DR3 
catalogue for the confirmed cluster sample.} 
  \label{f:ctsdist}
\end{figure}

The optically confirmed clusters have a wide range of source counts (EPIC counts in
the broad band energy 0.2-12 keV from the 2XMMi-DR3 catalogue) from 80 to 28000 counts 
as shown in Figure~\ref{f:ctsdist}. To analyse
the X-ray data, we have to determine the optical redshifts, except for some   
candidates with more than 1000 net photons for which it is possible to estimate the 
X-ray redshift \citep[e.g.][]{Lamer08,Yu11}. In this paper, we use the cluster 
spectroscopic redshifts where available or the photometric redshifts 
that we obtained from the cross-matching as described in the previous section.

The data reduction and analysis of the optically confirmed sample was carried
out using the XMM-Newton Science Analysis Software (SAS) version 10.0.0.

\subsection{Standard pipelines}
The raw XMM-Newton data were downloaded using the Archive InterOperability
System (AIO), which provides access to the XMM-Newton Science Archive (XSA). The raw
data were provided in the form of a bundle of files known as observation
data files (ODF), which contain uncalibrated event files, satellite attitude
files, and calibration information. The main steps in the data reduction were:
(i) the generation of calibrated event lists for the EPIC (MOS1, MOS2, and PN)
cameras using the latest calibration data. This was done using the SAS packages
{\tt cifbuild},  {\tt odfingest}, {\tt epchain}, and {\tt emchain}. (ii) The
creation of background light curves to identify time intervals with poor 
quality data. (iii) The filtering of the EPIC event lists to exclude periods of
high background flaring and bad events. (iv) To create a sky image of the
filtered data set. The last three steps were performed using SAS packages {\tt evselect},
{\tt tabgtigen}, and {\tt xmmselect}. 


\subsection{Analysis of the sample}
We now describe the procedure to determine the source and
background regions for each cluster, extract the  source and background
spectra, fit the X-ray spectra, and finally measure the X-ray parameters
(e.g.~temperature, flux, and luminosity). As input to the task generating the X-ray
spectra, we used the filtered event lists as described in the previous
section. 
 

\subsubsection{Optimum source extraction radius}
The most critical step in generating the cluster X-ray spectra is to determine
the source extraction radius. We developed a method to  optimize the
signal-to-noise ratio (SNR) of the spectrum for each cluster.  To
calculate the extraction radius with the highest integrated SNR we created
radial profiles of each cluster in the energy band $0.5 -
2.0$\,keV. Background sources, taken from the EPIC PPS source lists, were
excluded and the profiles were exposure corrected using the EPIC exposure
maps. Since we did not perform a new source detection run, the SNR was 
calculated as a function of radius taking into account the background 
levels as given in the 2XMMi catalogue.

\begin{figure}
\centering{
  \resizebox{100mm}{!}{\includegraphics[viewport=15  10  530 393,clip]{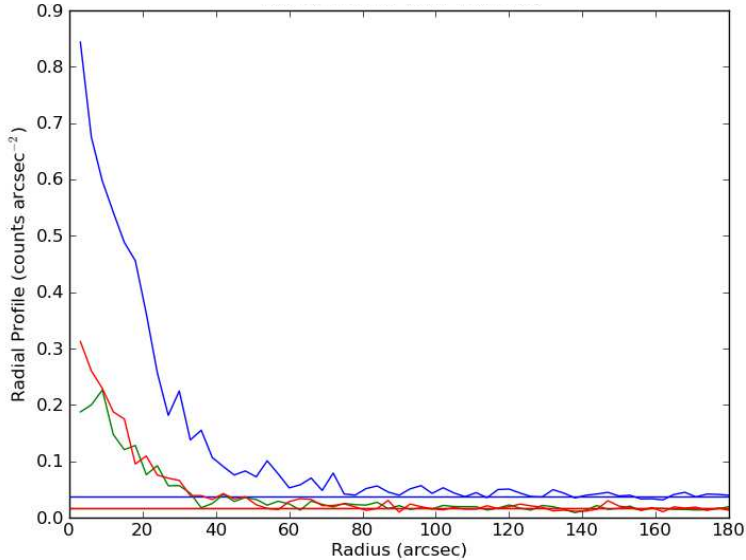}}}
  \caption[The radial profile of 2XMM J104421.8+213029 MOS1(green), MOS2(red),
    and PN (blue) images]{The radial profile of 2XMM J104421.8+213029 MOS1(green), MOS2(red),
    and PN (blue) images. The horizontal lines indicate the background values
    for MOS1, MOS2 and PN with the same colour as the profile. } 
  \label{f:rprof}
\end{figure}

\begin{figure}
\centering{
  \resizebox{100mm}{!}{\includegraphics[viewport=10  10 365 270,clip]{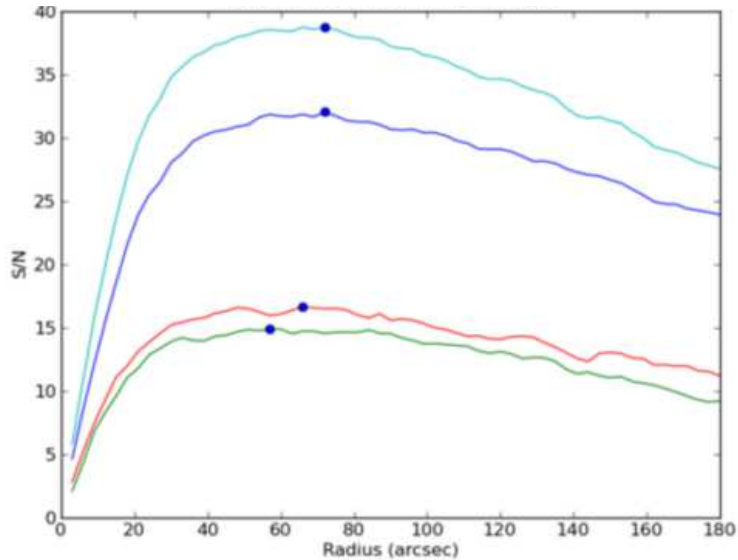}}}
  \caption[The signal-to-noise ratio (SNR) profiles of 2XMM J104421.8+213029]{The 
    signal-to-noise ratio (SNR) profiles of 2XMM J104421.8+213029 in
    MOS1 (green), MOS2 (red), PN (blue) and EPIC (MOS1+MOS2+PN) (cyan) data.
    The cluster optimum extraction radius ($72''$) is corresponding to the
    highest SNR as indicated by a point in the EPIC SNR profile.} 
\label{f:snr}
\end{figure}

The radial profiles of the X-ray surface brightness of the representative
cluster in MOS1, MOS2 and PN data are shown in Figure~\ref{f:rprof}. The
background values of the cluster in the EPIC images are indicated by the
horizontal line with the same colours as the profiles. Figure~\ref{f:snr} shows
the SNR profiles of the representative cluster in MOS1, MOS2, PN and EPIC
(MOS1+MOS2+PN) data as a function of the radius from the cluster centre. The
optimum extraction radius ($72''$) is determined from the maximum value in the
EPIC SNR plot, which is indicated by a point in Figure~\ref{f:snr}. 


\subsubsection{Spectral extraction} 
\begin{figure}[h]
\centering{
  \resizebox{100mm}{!}{\includegraphics{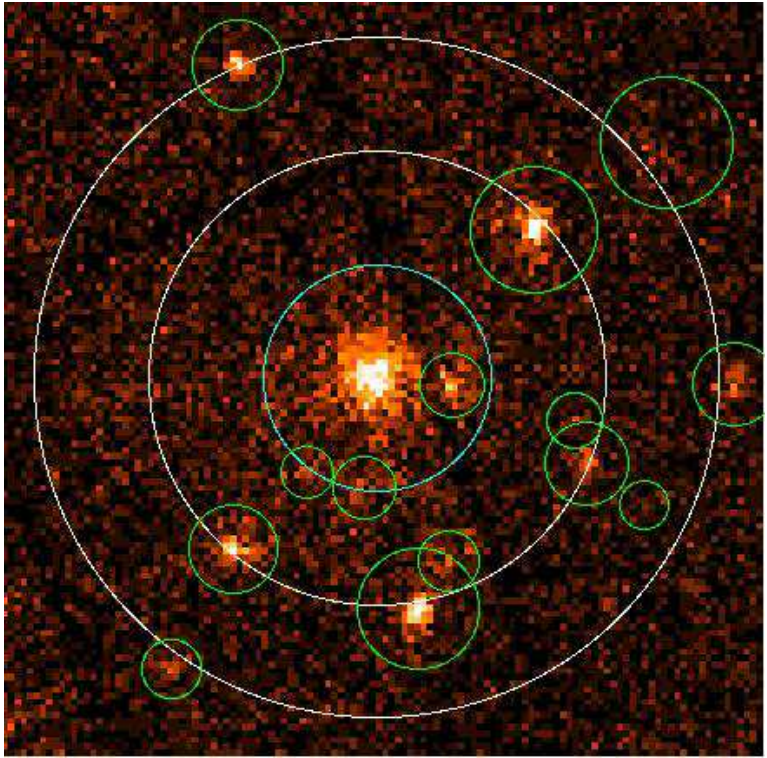}}}
  \caption[The representative cluster extraction and background region]{The 
    representative cluster extraction region is the inner circle
    with colour cyan. The background region is the annulus with white colour.
    The excluding field sources are indicated by green circles. The field of
    view is $8'\times 8'$ centred at the cluster position. } 
  \label{f:mask}
\end{figure}

The EPIC filtered event lists were used to extract the X-ray spectra of the
cross-correlated X-ray and optical cluster sample. The spectra of each cluster candidate 
were extracted from a region with an optimum extraction radius as described
in the previous section. The background spectra were extracted from a
circular annulus around the cluster with inner and outer radii equalling two
and three times the optimum radius, respectively. Other unrelated nearby
sources were masked and excluded from the source and background regions that
were finally used to extract the X-ray spectra. 
Figure~\ref{f:mask} shows the cluster and background regions, as well as
the excluded regions of field sources for the representative cluster. The SAS
task {\tt especget} was used to generate the cluster and background spectra and
to create the response matrix files (redistribution  matrix file (RMF) and
ancillary response file (ARF)) required to perform the X-ray spectral
fitting with XSPEC.

\subsubsection{Spectral fitting}
The photon counts of each cluster spectrum were grouped into bins with at least
one count per bin before a fit of a spectral model was applied to the data
using the Ftools task {\tt grppha}. The spectral fitting was carried out using
XSPEC software version 12.5.1 \citep{Arnaud96}. Before executing the algorithm to fit
the spectra, the Galactic HI column (nH) was derived from the HI map from the
Leiden/Argentine/Bonn (LAB) survey \citep{Kalberla05}. This parameter  was fixed
while fitting the X-ray spectrum. The redshift of the spectral
model was fixed to the optical cluster redshift either the spectroscopic redshift
for 182 clusters or the photometric redshift for the remainder cluster sample.  

For each cluster, the available EPIC spectra were fitted simultaneously. The
employed fitting model was a multiplication of a $\tt TBABS$ absorption model 
\citep{Wilms00} and a single-temperature optically thin thermal plasma
component \citep[the $\tt MEKAL$ code   in XSPEC terminology,][]{Mewe86} 
to model the X-ray plasma emission from the ICM. 
The metallicity was fixed at 0.4 $Z_\odot$.
 This value is the mean of the metallicities of  95 galaxy clusters
in the redshift range from 0.1 to 0.6 (the same redshift range of the confirmed
sample) observed by Chandra \citep{Maughan08}. The free parameters are the
X-ray temperature and the spectral normalization. The fitting was done using
the Cash statistic with one count per bin  following the recommendation of
\cite{Krumpe08} for small count statistics.  

\begin{figure}
\centering{
  \resizebox{100mm}{!}{\includegraphics[angle=-90, viewport=30  0  520 660 ,clip]{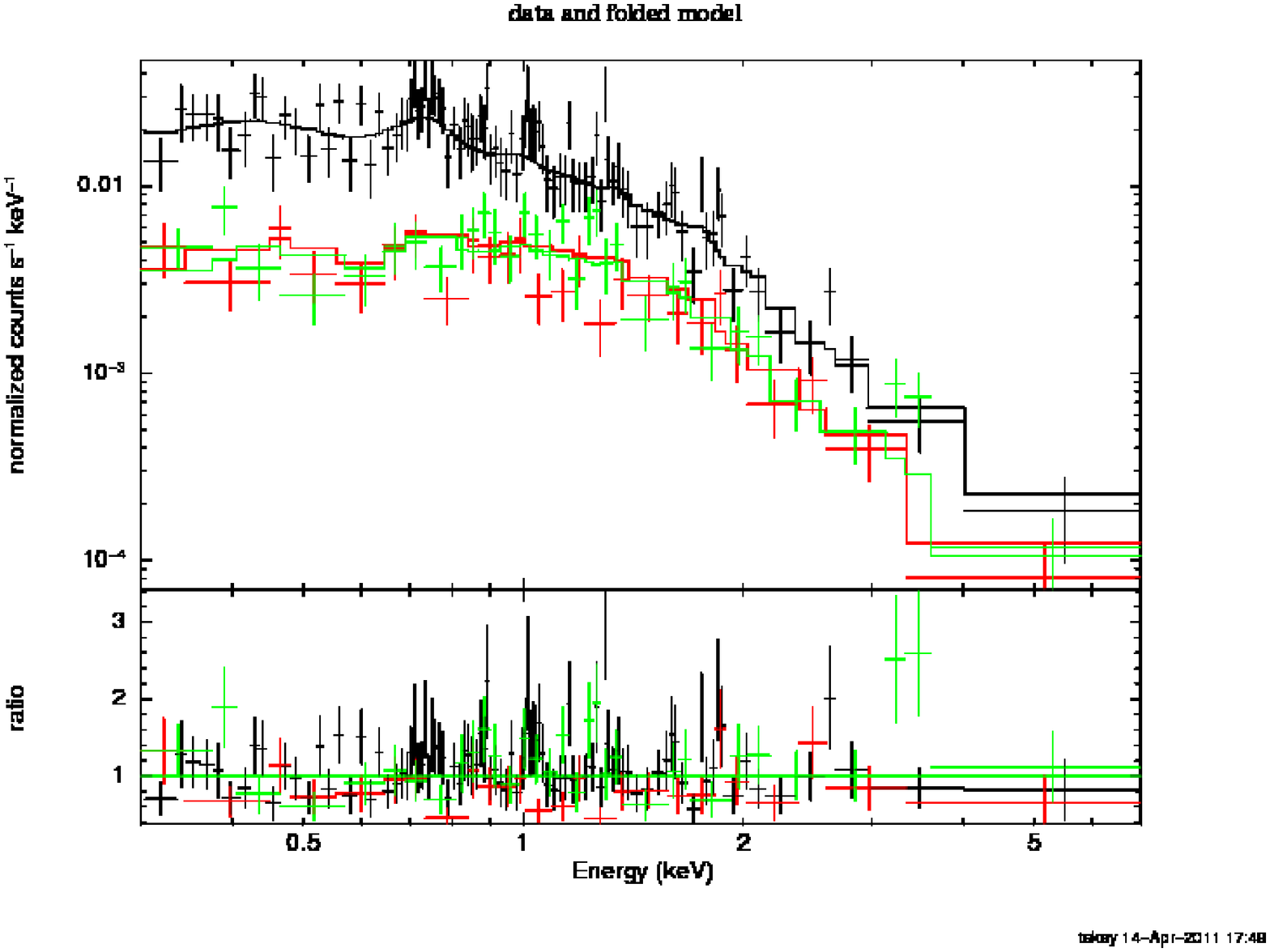}}} 
  \caption{The EPIC PN (black), MOS1 (green) and MOS2 (red) data with the best-fit
 MEKAL model for the representative cluster. }  
  \label{f:xfit}
\end{figure}

To avoid the fitting algorithm converging to a local minimum
of the fitting statistics, we ran series of fits stepping from 0.1
to 15 keV  with a step size = 0.05 using the $\tt{steppar}$ command within
XSPEC. The cluster temperature, its flux in the [0.5-2]\,keV band, its X-ray 
luminosity in the [0.5-2]\,keV band, the bolometric luminosity, and the corresponding 
errors were derived from the best-fitting model. We assumed that the fractional error
in the bolometric luminosity was the same as the fractional error in the aperture
luminosity [0.5-2] keV (within an aperture defined by the optimum extraction radius). 
Figure~\ref{f:xfit} shows the fits to the EPIC (MOS1, MOS2, and PN)
spectra and the models for the representative cluster. Figures A.1 to A.4
 in the Appendix A show the fitted spectra of four clusters with different X-ray 
surface brightnesses  and data qualities at different redshifts covering the whole 
redshift range of the confirmed clusters.


\section{Analysis of a cluster sample with reliable X-ray parameters}

We analysed the X-ray data of the optically and X-ray 
confirmed clusters to measure the global temperature of the hot ICM. 
We developed an optimal extraction method for the X-ray spectra
maximising the SNR. The cluster spectra 
were fitted with absorbed thin thermal plasma emission models with
pre-determined redshift and interstellar column density to determine 
the aperture X-ray temperature ($T_{ap}$), flux ($F_{ap}$) [0.5-2] keV, 
luminosity ($L_{ap}$) [0.5-2] keV, and their errors.  
We accepted the measurements of $T_{ap}$ and  $L_{ap}$ if the fractional
errors were smaller than 0.5. About 80 percent of the confirmed clusters passed
this fractional error filter. For these clusters, another visual screening of the
spectral fits (Figure~\ref{f:xfit}) and the X-ray images (Figure~\ref{f:mask}) 
was done.  When the spectral extraction of a given cluster was strongly affected
by the exclusion of field sources within the  extraction radius or a poor 
determination of the background spectrum, it was also excluded from the final 
sample, which comprises 175 clusters. For a fraction of 80 percent, this is 
the first X-ray detection and the first temperature measurement. 

\begin{figure}
\centering{
  \resizebox{100mm}{!}{\includegraphics[viewport=15  5  535 400, clip]{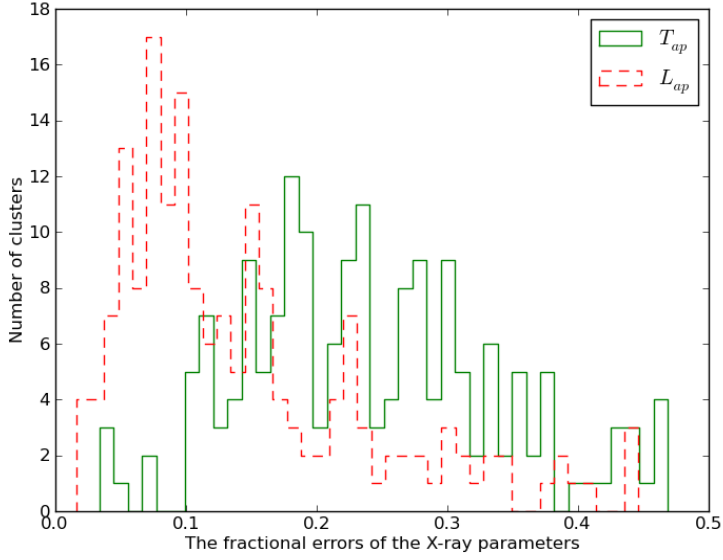}}}
  \caption[The distribution of the fractional errors in the X-ray temperatures
    and luminosities]{The distribution of the fractional errors in the X-ray temperatures
    (solid) and luminosities (dashed) derived from spectra extracted within the  
optimum aperture for the energy range 0.5-2 keV of the first cluster sample.}
  \label{f:reler}
\end{figure}

Our subsequent presentation of our analysis and discussion refers to those 175 
objects with reliable X-ray
parameters. The distribution of the $T_{ap}$ and $L_{ap}$ [0.5-2] keV fractional
errors for the first cluster sample is shown in Figure~\ref{f:reler}. It is
clearly evident that the  cluster luminosity is more tightly constrained than the temperature. 
For about 86 percent of the sample,  the fractional errors 
are smaller than 0.25. Therefore, we estimated  several physical parameters
for each cluster based on the bolometric luminosity $L_{bol}$ within the
optimal aperture. The median correction factor between aperture bolometric  luminosities 
and aperture luminosities in the energy band [0.5-2] keV ( $L_{bol}$ / $L_{ap}$ ) 
was found to be 1.7. We assumed that the fractional error in $L_{bol}$ was identical 
to that of $L_{ap}$ [0.5-2] keV. The estimated parameters are $R_{500}$
the radius at which the mean mass density is 500 times the critical density of
the Universe (see Eq. 2) at the cluster redshift, $L_{500}$ the bolometric 
luminosity within $R_{500}$, and $M_{500}$ the cluster mass within
$R_{500}$. We used an iterative procedure to estimate the physical parameters
using published $L_{500} - T_{500}$ and $L_{500} - M_{500}$ relations  
\citep[][their orthogonal fit for $M_{500}$ with Malmquist bias correction]{Pratt09}.
Our procedure is similar to that used by \cite{Piffaretti11} and \cite{Suhada10},  
which consists of the following steps:

\begin{itemize}
\item[(i)] We estimate $M_{500}$ using the $L - M$ relation
\begin{equation}
 M_{500} = 2 \times 10^{14} M_{\odot}\ \bigl(
 \frac{h(z)^{-7/3}\ L_{bol}}{1.38\times10^{44}\ erg\ s^{-1}} \Bigr)^{1/2.08},  
\end{equation}
where $h(z)$  is the Hubble constant normalised to its present-day value, $
h(z) = \bigl[\Omega_{\rm M} (1+z)^{3} + \Omega_{\Lambda}\Bigr]^{1/2} $. We 
approximate $L_{500}$ as the aperture bolometric luminosity 
$L_{bol}$,  which we correct in an iterative way. 

\item[(ii)] We compute $R_{500}$ 
\begin{equation}
 R_{500} = \sqrt[3]{3 M_{500} / 4\pi 500 \rho_{c}(z)},
\end{equation} 
where the critical density is  $\rho_{c}(z) = h(z)^{2} 3 H^{2} / 8 \pi G$ .

\item[(iii)] 
We compute the cluster temperature within $R_{500}$ using the $L - T$ relation 
\begin{equation}
 T = 5 \mbox{keV} \ \bigl(\frac{h(z)^{-1}\ L_{bol}}{7.13 \times 10^{44}\ erg\ s^{-1}}\Bigr)^{1/3.35}.
\end{equation} 

\item[(iv)] We calculate the core radius $r_{core}$ and $\beta$  using scaling
  relations from \cite{Finoguenov07} 
\begin{equation}
 r_{core}  = 0.07 \times R_{500} \times \bigl(\frac{T}{1\ keV}\bigr)^{0.63},
\end{equation} 
\begin{equation}
 \beta  = 0.4\ \bigl(\frac{T}{1\ keV}\bigr)^{1/3}.
\end{equation} 

\item[(v)] We calculate the enclosed flux within $R_{500}$ and the optimum
  aperture by extrapolating the $\beta$-model. The ratio of the two
  fluxes is calculated, i.e.  $\gamma = F_{500} / F_{\rm bol} $. 

\item[(vi)]
We finally compute a corrected value of $L_{500} = \gamma \times L_{\rm bol}$.
\end{itemize}

We then considered $L_{500}$ as input for another iteration and all
computed parameters were updated. We repeated this iterative procedure until
 converging to a final solution.  At this stage, the $L_{500}$, $M_{500}$, 
and $R_{500}$ were determined. The median correction factor between extrapolated 
luminosities and  aperture bolometric luminosities ($L_{500}$/$L_{\rm bol}$) 
was 1.5. To calculate the errors in Eqs. 1 and 3,  we 
included the measurement errors in the aperture bolometric luminosity $L_{bol}$   
, the intrinsic scatter in  the $L - T$ and $L - M$ relations, and the 
propagated errors caused by the uncertainty in their slopes and intercepts. 
For Eq. 4 and Eq. 5, we included only the propagated errors of their 
independent parameters since their intrinsic scatter had not been published. 
Finally, all the measured errors were taken into account when computing the errors 
in $L_{500}$ and $M_{500}$ in the last iteration. The errors 
in $L_{500}$ and $M_{500}$ were still underestimated because of the possible scatter 
in the relations in Eq. 4 and Eq. 5.       

We investigated the $L - T$ relation in the first cluster sample using $T_{ap}$ and
$L_{500}$.  Figure~\ref{f:L-T} shows the relation between $L_{500}$ (corrected for the
redshift evolution) and $T_{ap}$ (uncorrected for cooling flows).  Here we
assumed that $T_{ap}$ did not differ significantly from $T_{500}$ and its
error was derived from the spectral fits. The best-fit linear relation (solid line) 
derived from an orthogonal distance  regression fit (ODR) \citep[][which takes 
into account measurement errors in both variables]{Boggs90} between their logarithm, is 

\begin{equation}
\log\ (h(z)^{-1}\ L_{500}) = (0.57 \pm 0.05)  + (3.41 \pm 0.15)\ \log\ (T_{ap}).
\end{equation} 

The best-fit power law relation derived from a BCES orthogonal fit to the
$L_{500}$ - $T_{500}$ relation published by \cite{Pratt09}  for the REXCESS
sample is plotted as the dashed line in Figure~\ref{f:L-T} . 
The ODR slope (present work), $3.41 \pm 0.15$, is consistent with the BCES 
orthogonal slope \citep{Pratt09} of the REXCESS
sample, $3.35 \pm 0.32$. In addition, the present slope is consistent with the BCES 
orthogonal slope (3.63 $\pm$ 0.27) of the $L - T$ relation derived from a sample of 
114 clusters (without excluding the core regions) observed 
with Chandra across a wide range of temperature (2 $<$ kT $<$ 16 keV) 
and redshift (0.1 $<$ z $<$ 1.3) by \cite{Maughan11}. 

We tested the corresponding uncertainty in the error budget of $L_{500}$ caused 
by the above-mentioned unknown scatter in Eq. 4 and 5: for example, a $\sigma = \pm 0.1$ 
scatter in the $\beta$ value results in a fractional error in   $L_{500}$ of 17$\%$. 
If we take into account the newly estimated errors in $L_{500}$ when fitting the $L - T$ relation, 
the revised slope of 3.32 is within the error in the original slope as in Eq. 6 and still 
consistent with the published ones.
   
\begin{figure}
\centering{
  \resizebox{100mm}{!}{\includegraphics[viewport=20  10  535 400, clip]{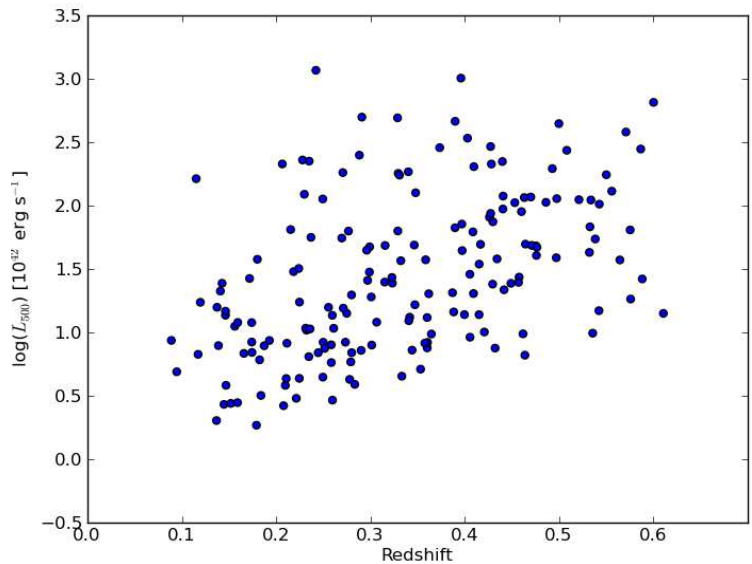}}}
  \caption{The distribution of the X-ray bolometric luminosities $L_{500}$ as a 
function of optical redshifts of the first cluster sample.}
  \label{f:L-z}
\end{figure}

Our sample represents cluster temperatures ranging from 0.45 to 5.92 keV and values of 
bolometric luminosity in the $L_{500}$ range $1.9 \times 10^{42} - 1.2 \times
10^{45}$\ erg\ s$^{-1}$  in a wide redshift range  0.1 - 0.6.  Most of the
published $L - T$ relations were derived from   local cluster samples with
temperatures higher than 2 keV. The current relation is derived for our
sample, which includes clusters and  groups with low temperatures 
and luminosities in a wide redshift range up to $z = 0.6$. The distribution of
luminosity as a function of redshift is shown in Figure~\ref{f:L-z}.  



\begin{figure}
\centering{
  \resizebox{100mm}{!}{\includegraphics[viewport=20  10  535 400, clip]{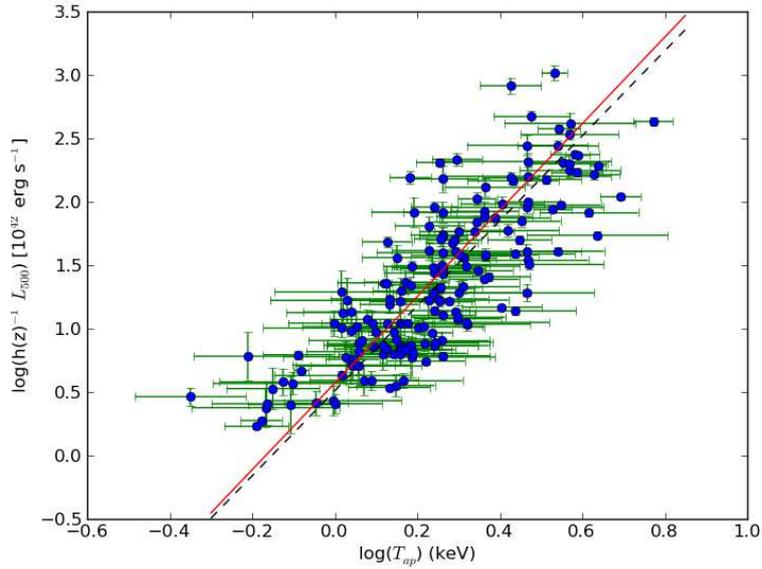}}}
  \caption[The relation between the X-ray bolometric luminosities $L_{500}$ and 
aperture temperatures $T_{ap}$ of the first cluster sample]{The relation 
between the X-ray bolometric luminosities $L_{500}$ and 
aperture temperatures $T_{ap}$ of the first cluster sample.  The solid line indicates
 the best fit of the sample using orthogonal distance regression (ODR). The dashed 
line is the extrapolated relation for REXCESS sample \citep{Pratt09} using 
a BCES orthogonal fit.}
  \label{f:L-T}
\end{figure}

Table~\ref{t:Cat}, available at the CDS,  represents the first cluster sample containing as many as  
175 X-ray clusters. In addition, the first cluster sample with the X-ray-optical overlay and fitted spectra 
for each cluster is publicly available from\\ \url{http://www.aip.de/groups/xray/XMM_SDSS_CLUSTERS/17498.html}.
 In the catalogue, we provide the cluster identification number 
(detection Id, detid) and its name (IAUNAME) in (cols. [1] and [2]), the right 
ascension and declination of X-ray emission in equinox J2000.0 (cols. [3] and [4]), 
the XMM-Newton observation Id (obsid) (col. [5]), the optical redshift (col. [6]), 
the scale at the cluster redshift in kpc/$''$ (col. [7]), 
the aperture and $R_{500}$ radii in kpc (col.[8] and [9]), the cluster aperture  
X-ray temperature $T_{ap}$ and its positive and negative errors in keV (cols. [10], 
[11] and [12], respectively), the aperture X-ray flux $F_{ap}$ [0.5-2] keV 
and its positive and negative errors in units of $10^{-14}$\ erg\ cm$^{-2}$\ s$^{-1}$ 
(cols. [13], [14] and [15], respectively), the aperture  
X-ray luminosity $L_{ap}$ [0.5-2] keV and its positive and negative errors in units 
of $10^{42}$\ erg\ s$^{-1}$ (cols. [16], [17] and [18], respectively),
the cluster bolometric luminosity $L_{500}$ and its error in units of 
$10^{42}$\ erg\ s$^{-1}$ (cols. [19] and [20]), the cluster mass $M_{500}$ and its error
in units of $10^{13}$\ M$_\odot$ (cols. [21] and [22]), the Galactic HI column in 
units $10^{22}$\ cm$^{-2}$ (col.[23]), the identification number of the cluster
in optical catalogue (col.[24]), the BCG right ascension and declination in equinox 
J2000.0 (cols. [25] and [26]) although for AMF catalogue they represent the 
cluster stellar mass centre, the cluster photometric redshift (col.[27]), the average 
spectroscopic redshift of the cluster galaxies with available spectroscopic
redshifts and their number (cols.[28] and [29]), the linear offset between the 
cluster X-ray position and the cluster optical position (col.[30]), 
the optical cluster catalogue names that identify the cluster (col.[31]) 
(Note: the optical parameters are extracted from the first one), 
and the alternative name of the  X-ray clusters previously identified using ROSAT, Chandra, or 
XMM-Newton data and its reference in NED and SIMBAD databases (col.[32] and [33]).


\section{Summary and outlook}
We have presented the first sample of X-ray galaxy clusters from the 2XMMi-Newton/SDSS 
Galaxy Cluster Survey. The survey comprises 1180 cluster candidates selected as X-ray 
serendipitous sources from the second XMM-Newton serendipitous source catalogue
(2XMMi-DR3) that had been observed by the SDSS. A quarter of the candidates 
are identified as distant cluster candidates beyond z = 0.6, because there is no apparent 
overdensity of galaxies in the corresponding 
SDSS images. Another quarter of the candidates had been previously identified
in optical cluster catalogues extracted from SDSS data.    
Our cross-correlation of the X-ray cluster candidates with four optical cluster catalogues within 
a matching radius of one arcmin confirmed 275 clusters and provided us with 
the photometric redshifts for all of them and the spectroscopic redshifts for 120 BCGs. 
We extracted all available spectroscopic redshifts 
for the cluster members from recent SDSS data. Among the confirmed cluster sample, 182 clusters 
have spectroscopic redshifts for at least one galaxy member.   
More than 80 percent of the confirmed sample are newly identified X-ray clusters and the 
others had been previously identified using ROSAT, Chandra, or XMM-Newton data.
 We reduced and analysed the X-ray data of the confirmed sample in an automated way. 
The X-ray temperature, flux and luminosity of
the confirmed sample and their errors were derived from  spectral fitting.
 The analysed sample in the present work contains 175 X-ray galaxy clusters with acceptable
 measurements of X-ray parameters (ie. with fractional errors smaller than 0.5) 
from reasonable quality fitting (139 objects being newly discovered in X-rays).
 In addition, we derived the physical properties ($R_{500}$, $L_{500}$ and $M_{500}$) 
of the study sample from an iterative procedure using the published scaling relations. 
The relation between the X-ray bolometric luminosity $L_{500}$ and aperture temperature of 
the sample is investigated. The slope of the relation agrees with the slope of the 
same relation in the  REXCESS sample \citep{Pratt09}. The present relation is derived from
 a large sample with low luminosities and temperatures  across a wide redshift range 0.09 - 0.61.   
 
As one extension to this project, we intend to obtain SDSS photometric redshifts
 of all 2XMMi-DR3 X-ray cluster candidates that have been detected in  SDSS imaging. This 
will significantly increase the sample size and the identified fraction of the 2XMMi cluster sample.
Further improvements in the accuracy of the X-ray parameters for about  10 percent of the confirmed 
sample will be made by analysing repeated observations of those clusters.

\clearpage
\begin{landscape}
\begin{table}
\caption{The first 25 entries of the first cluster sample}
\label{t:Cat}
{\tiny
\begin{tabular}{|c|l|r|r|c|r|c|r|r|r|r|r|r|r|r|r|r|r|}
\hline
  \multicolumn{1}{|c|}{detid\tablefootmark{a}} &
  \multicolumn{1}{c|}{Name\tablefootmark{a}} &
  \multicolumn{1}{c|}{ra\tablefootmark{a}} &
  \multicolumn{1}{c|}{dec\tablefootmark{a}} &
  \multicolumn{1}{c|}{obsid\tablefootmark{a}} &
  \multicolumn{1}{c|}{z\tablefootmark{b}} &
  \multicolumn{1}{c|}{scale} &
  \multicolumn{1}{c|}{$R_{ap}$} &
  \multicolumn{1}{c|}{$R_{500}$} &
  \multicolumn{1}{c|}{$T_{ap}$} &
  \multicolumn{1}{c|}{$+ eT_{ap}$} &
  \multicolumn{1}{c|}{$- eT_{ap}$} &
  \multicolumn{1}{c|}{$F_{ap}$\tablefootmark{c}} &
  \multicolumn{1}{c|}{$+ eF_{ap}$} &
  \multicolumn{1}{c|}{$- eF_{ap}$} &
  \multicolumn{1}{c|}{$L_{ap}$\tablefootmark{d}} &
  \multicolumn{1}{c|}{$+ eL_{ap}$} &
  \multicolumn{1}{c|}{$- L_{ap}$} \\

  \multicolumn{1}{|c|}{} &
  \multicolumn{1}{c|}{IAUNAME} &
  \multicolumn{1}{c|}{(deg)} &
  \multicolumn{1}{c|}{(deg)} &
  \multicolumn{1}{c|}{} &
  \multicolumn{1}{c|}{} &
  \multicolumn{1}{c|}{kpc/$''$} &
  \multicolumn{1}{c|}{(kpc)} &
  \multicolumn{1}{c|}{(kpc)} &
  \multicolumn{1}{c|}{(keV)} &
  \multicolumn{1}{c|}{(keV)} &
  \multicolumn{1}{c|}{(keV)} &
  \multicolumn{1}{c|}{} &
  \multicolumn{1}{c|}{} &
  \multicolumn{1}{c|}{} &
  \multicolumn{1}{c|}{} &
  \multicolumn{1}{c|}{} &
  \multicolumn{1}{c|}{} \\

(1)  &  (2)  &  (3)  & (4)  &   (5)   &  (6)  &   (7)   &   (8)   &  (9)   &  (10)    &  (11) &  (12) & (13) & (14) & (15) &  (16)  & (17)  &  (18)   \\
\hline 

005825 &     2XMM J003917.9+004200 &   9.82489 &   0.70013 & 0203690101 & 0.2801 & 4.24 &  89.14 &  474.51 & 1.07 & 0.33 & 0.42 &   0.51 & 0.17 & 0.22 &   1.27 &  0.46 &  0.40 \\
005901 &     2XMM J003942.2+004533 &   9.92584 &   0.75919 & 0203690101 & 0.4152 & 5.50 & 247.28 &  566.70 & 1.83 & 0.42 & 0.18 &   2.19 & 0.23 & 0.19 &  12.70 &  1.32 &  1.07 \\
006920 &     2XMM J004231.2+005114 &  10.63008 &   0.85401 & 0090070201 & 0.1468 & 2.57 & 107.87 &  464.68 & 1.41 & 0.28 & 0.33 &   1.09 & 0.41 & 0.30 &   0.62 &  0.19 &  0.17 \\
007340 &     2XMM J004252.6+004259 &  10.71952 &   0.71650 & 0090070201 & 0.2595 & 4.02 & 385.77 &  535.23 & 1.98 & 0.64 & 0.32 &   3.29 & 0.40 & 0.27 &   6.39 &  0.71 &  0.55 \\
007362 &     2XMM J004253.7-093423 &  10.72397 &  -9.57311 & 0065140201 & 0.4054 & 5.42 & 259.99 &  553.50 & 1.49 & 0.52 & 0.18 &   2.25 & 0.58 & 0.24 &  12.64 &  2.86 &  2.43 \\
007881 &     2XMM J004334.0+010106 &  10.89197 &   1.01844 & 0090070201 & 0.1741 & 2.95 & 203.84 &  549.63 & 1.34 & 0.15 & 0.13 &   4.97 & 0.27 & 0.46 &   4.15 &  0.44 &  0.34 \\
008026 &     2XMM J004350.7+004733 &  10.96148 &   0.79263 & 0090070201 & 0.4754 & 5.94 & 445.43 &  576.17 & 2.32 & 0.64 & 0.47 &   2.93 & 0.23 & 0.24 &  22.38 &  1.37 &  1.94 \\
008084 &     2XMM J004401.3+000647 &  11.00567 &   0.11323 & 0303562201 & 0.2185 & 3.53 & 307.39 &  622.06 & 1.83 & 0.40 & 0.20 &   8.71 & 1.28 & 0.83 &  11.73 &  1.69 &  1.58 \\
021508 &     2XMM J015917.2+003011 &  29.82169 &   0.50328 & 0101640201 & 0.2882 & 4.33 & 259.91 &  838.95 & 1.98 & 0.30 & 0.24 &  32.62 & 2.76 & 2.07 &  79.81 &  6.20 &  6.81 \\
021850 &     2XMM J020342.0-074652 &  30.92533 &  -7.78128 & 0411980201 & 0.4398 & 5.68 & 341.05 &  752.50 & 3.71 & 1.24 & 0.85 &  10.50 & 0.75 & 0.82 &  62.67 &  5.59 &  6.38 \\
030611 &     2XMM J023150.5-072836 &  37.96059 &  -7.47683 & 0200730401 & 0.1791 & 3.02 & 208.60 &  406.59 & 0.65 & 0.16 & 0.08 &   0.95 & 0.06 & 0.09 &   0.88 &  0.06 &  0.07 \\
030746 &     2XMM J023346.9-085054 &  38.44543 &  -8.84844 & 0150470601 & 0.2799 & 4.24 & 254.58 &  561.31 & 1.78 & 0.55 & 0.45 &   3.26 & 0.74 & 0.55 &   7.49 &  1.57 &  1.13 \\
034903 &     2XMM J030637.1-001803 &  46.65469 &  -0.30096 & 0201120101 & 0.4576 & 5.81 & 331.40 &  531.81 & 2.05 & 0.83 & 0.45 &   1.73 & 0.24 & 0.18 &  11.24 &  1.49 &  1.25 \\
042730 &     2XMM J033757.5+002900 &  54.48959 &   0.48351 & 0036540101 & 0.3232 & 4.69 & 253.05 &  566.44 & 1.80 & 0.34 & 0.37 &   3.00 & 0.40 & 0.33 &   9.22 &  0.93 &  1.00 \\
080229 &     2XMM J073605.9+433906 & 114.02470 &  43.65179 & 0083000101 & 0.4282 & 5.60 & 386.12 &  752.33 & 3.87 & 0.65 & 0.46 &  11.22 & 0.65 & 0.47 &  58.15 &  3.49 &  2.67 \\
083366 &     2XMM J075121.7+181600 & 117.84073 &  18.26679 & 0111100301 & 0.3882 & 5.28 & 316.50 &  501.14 & 1.20 & 0.30 & 0.20 &   1.65 & 0.28 & 0.25 &   8.44 &  1.27 &  1.55 \\
083482 &     2XMM J075427.4+220949 & 118.61452 &  22.16371 & 0110070401 & 0.3969 & 5.35 & 304.78 &  643.87 & 2.18 & 0.55 & 0.43 &   5.70 & 0.70 & 0.58 &  26.81 &  4.35 &  3.06 \\
089735 &     2XMM J082746.9+263508 & 126.94574 &  26.58556 & 0103260601 & 0.3869 & 5.26 & 236.90 &  530.40 & 1.69 & 1.05 & 0.46 &   1.63 & 0.55 & 0.33 &   7.93 &  2.19 &  2.51 \\
089885 &     2XMM J083146.1+525056 & 127.94516 &  52.84719 & 0092800201 & 0.5383 & 6.34 & 513.76 &  565.42 & 3.47 & 1.08 & 0.72 &   2.26 & 0.09 & 0.09 &  20.16 &  1.23 &  0.56 \\
090256 &     2XMM J083454.8+553422 & 128.72859 &  55.57287 & 0143653901 & 0.2421 & 3.82 & 286.38 & 1102.80 & 3.40 & 0.25 & 0.24 & 165.21 & 3.59 & 4.71 & 258.02 &  5.58 &  6.62 \\
090966 &     2XMM J083724.7+553249 & 129.35324 &  55.54712 & 0143653901 & 0.2767 & 4.21 & 315.60 &  677.20 & 1.83 & 0.43 & 0.24 &  11.90 & 1.67 & 1.61 &  26.22 &  4.04 &  2.91 \\
092117 &     2XMM J084701.9+345114 & 131.75794 &  34.85384 & 0107860501 & 0.4643 & 5.86 & 369.31 &  582.95 & 2.73 & 1.20 & 0.63 &   2.72 & 0.16 & 0.17 &  18.71 &  1.01 &  1.04 \\
092718 &     2XMM J084847.8+445611 & 132.19968 &  44.93637 & 0085150101 & 0.5753 & 6.55 & 353.92 &  567.36 & 1.92 & 0.38 & 0.17 &   2.53 & 0.20 & 0.14 &  30.39 &  3.25 &  1.00 \\
097911 &     2XMM J092545.5+305858 & 141.43996 &  30.98303 & 0200730101 & 0.5865 & 6.61 & 357.18 &  713.02 & 3.57 & 0.75 & 0.70 &   7.55 & 0.51 & 0.48 &  86.75 &  6.62 &  7.56 \\
098728 &     2XMM J093205.0+473320 & 143.02097 &  47.55565 & 0203050701 & 0.2248 & 3.61 & 259.96 &  567.30 & 1.36 & 0.22 & 0.23 &   5.05 & 0.77 & 0.80 &   7.46 &  1.32 &  1.05 \\

\hline
\end{tabular}
}
\tablefoot{ Full Table~\ref{t:Cat} is available at the CDS via anonymous ftp to cdsarc.u-strasbg.fr (130.79.128.5) or via 
\url{http://cdsarc.u-strasbg.fr/viz-bin/qcat?J/A+A/534/A120}. In addition, the cluster catalogue with the X-ray-optical overlay and fitted spectra 
for each cluster is publicly available from \url{http://www.aip.de/groups/xray/XMM_SDSS_CLUSTERS/17498.html}.\\ 
\tablefoottext{a}{All these parameters are extracted from the 2XMMi-DR3 catalogue.} \\
\tablefoottext{b}{The cluster redshift from col. (27) or col. (28).}\\     
\tablefoottext{c}{Aperture X-ray flux $F_{ap}$ [0.5-2] keV and its positive and negative errors in units of $10^{-14}$\ erg\ cm$^{-2}$\ s$^{-1}$.}\\
\tablefoottext{d}{Aperture X-ray luminosity $L_{ap}$ [0.5-2] keV and its positive and negative errors in units of $10^{42}$\ erg\ s$^{-1}$.} \\
\tablefoottext{e}{X-ray bolometric luminosity $L_{500}$ and its error in units of $10^{42}$\ erg\ s$^{-1}$.} \\
\tablefoottext{f}{The cluster mass $M_{500}$ and its error  in units of $10^{13}$\ M$_\odot$.}\\
\tablefoottext{g}{The Galactic HI column in units $10^{22}$\ cm$^{-2}$.}\\
\tablefoottext{h}{These parameters are obtained from first catalogue in col. (31).} \\
\tablefoottext{i}{These parameters are extracted from SDSS-DR8 data.}  \\
\tablefoottext{j}{The names of the optical catalogues which detected the cluster.} \\
\tablefoottext{k}{The known X-ray cluster names from NED or SIMBAD.} \\  
   }
\end{table}
\end{landscape}

\begin{landscape}
\addtocounter{table}{-1}
\begin{table}
\caption{\label{} The first 25 entries of the first cluster sample, continued.}
{\tiny
\begin{tabular}{|c|r|r|r|r|r|l|r|r|r|r|r|r|l|l|r|}
\hline
  \multicolumn{1}{|c|}{detid\tablefootmark{a}} &
  \multicolumn{1}{c|}{$L_{500}$\tablefootmark{e}} &
  \multicolumn{1}{c|}{$\pm eL_{500}$} &
  \multicolumn{1}{c|}{$M_{500}$\tablefootmark{f}} &
  \multicolumn{1}{c|}{$\pm eM_{500}$} &
  \multicolumn{1}{c|}{nH\tablefootmark{g}} &
  \multicolumn{1}{c|}{objid\tablefootmark{h}} &
  \multicolumn{1}{c|}{RA\tablefootmark{h}} &
  \multicolumn{1}{c|}{DEC\tablefootmark{h}} &
  \multicolumn{1}{c|}{$z_{p}$\tablefootmark{h}} &
  \multicolumn{1}{c|}{$z_{s}$\tablefootmark{i}} &
  \multicolumn{1}{c|}{$N_{z_{s}}$\tablefootmark{i}} &
  \multicolumn{1}{c|}{offset} &
  \multicolumn{1}{c|}{opt-cats\tablefootmark{j}} &
  \multicolumn{1}{c|}{known X-ray\tablefootmark{k}} &
  \multicolumn{1}{c|}{ref.} \\

  \multicolumn{1}{|c|}{} &
  \multicolumn{1}{c|}{} &
  \multicolumn{1}{c|}{} &
  \multicolumn{1}{c|}{} &
  \multicolumn{1}{c|}{} &
  \multicolumn{1}{c|}{} &
  \multicolumn{1}{c|}{} &
  \multicolumn{1}{c|}{} &
  \multicolumn{1}{c|}{} &
  \multicolumn{1}{c|}{} &
  \multicolumn{1}{c|}{} &
  \multicolumn{1}{c|}{} &
  \multicolumn{1}{c|}{(kpc)} &
  \multicolumn{1}{c|}{} &
  \multicolumn{1}{c|}{CLG} &
  \multicolumn{1}{c|}{} \\

  (1)  &  (19)  & (20)  &  (21)  &  (22)   &  (23)    &  (24)  &   (25)   &  (26) & (27) & (28)  & (29)& (30)& (31)& (32)& (33)  \\
\hline

005825 &    6.93 &   1.96 &  4.05 &  1.03 & 0.0198 & J003916.6+004215     &   9.82500 &   0.69981 & 0.2942 & 0.2801 &  1 &   5.10 & WHL,maxBCG               & -                              &  - \\
005901 &   34.56 &   3.06 &  8.04 &  1.62 & 0.0195 & 587731187282018514   &   9.92728 &   0.76163 & 0.3873 & 0.4152 &  4 &  56.11 & GMBCG,WHL,AMF            & -                              &  - \\
006920 &    3.82 &   0.76 &  3.29 &  0.79 & 0.0179 & 588015510347776148   &  10.63096 &   0.85021 & 0.1532 & 0.1468 &  6 &  36.04 & GMBCG,maxBCG             & -                              &  - \\
007340 &   13.64 &   1.53 &  5.68 &  1.20 & 0.0178 & J004252.7+004306     &  10.71964 &   0.71845 & 0.2676 & 0.2595 &  5 &  28.19 & WHL                      & [PBG2005] 03                   &  1 \\
007362 &   28.75 &   6.27 &  7.40 &  1.66 & 0.0270 & 587727226768523625   &  10.72131 &  -9.57365 & 0.4054 & 0.0000 &  0 &  52.21 & GMBCG,WHL                & -                              &  - \\
007881 &   11.95 &   0.89 &  5.61 &  1.17 & 0.0182 & 588015510347907142   &  10.89606 &   1.01972 & 0.1661 & 0.1741 &  7 &  45.59 & GMBCG,WHL,AMF            & [PBG2005] 1                    &  1 \\
008026 &   48.17 &   4.51 &  9.06 &  1.82 & 0.0179 & 587731187282477303   &  10.95832 &   0.78838 & 0.4929 & 0.4754 &  1 & 113.06 & GMBCG,WHL                & -                              &  - \\
008084 &   30.15 &   4.29 &  8.52 &  1.77 & 0.0168 & 588015509274165366   &  11.00568 &   0.11512 & 0.2580 & 0.2185 &  3 &  24.13 & GMBCG,WHL                & -                              &  - \\
021508 &  249.85 &  29.20 & 22.56 &  4.45 & 0.0234 & 588015509819293880   &  29.82170 &   0.50343 & 0.2882 & 0.0000 &  0 &   2.25 & GMBCG                    & [VMF98] 021                    &  2 \\
021850 &  222.60 &  29.56 & 19.37 &  3.86 & 0.0187 & 587727884967346553   &  30.92570 &  -7.78196 & 0.4834 & 0.4398 &  1 &  16.02 & GMBCG,AMF                & -                              &  - \\
030611 &    1.85 &   0.11 &  2.28 &  0.53 & 0.0318 & J023149.7-072834     &  37.95694 &  -7.47613 & 0.2039 & 0.1791 &  2 &  40.11 & WHL                      & -                              &  - \\
030746 &   19.75 &   3.68 &  6.70 &  1.47 & 0.0300 & 587724240688316594   &  38.44672 &  -8.84924 & 0.2799 & 0.0000 &  0 &  23.01 & GMBCG,WHL,maxBCG,AMF     & -                              &  - \\
034903 &   27.30 &   3.53 &  6.98 &  1.46 & 0.0627 & 588015508752892483   &  46.65299 &  -0.30216 & 0.4120 & 0.4576 &  1 &  43.63 & GMBCG                    & -                              &  - \\
042730 &   24.46 &   2.52 &  7.22 &  1.48 & 0.0614 & 588015509830107502   &  54.48798 &   0.48583 & 0.3220 & 0.3232 &  5 &  47.71 & GMBCG,WHL                & [PBG2005] 13                   &  1 \\
080229 &  212.68 &  16.29 & 19.09 &  3.67 & 0.0549 & J073605.8+433908     & 114.02408 &  43.65224 & 0.4011 & 0.4282 &  2 &  12.85 & WHL                      & -                              &  - \\
083366 &   14.50 &   2.56 &  5.39 &  1.20 & 0.0520 & 8718.0.0.8718        & 117.83460 &  18.26690 & 0.3969 & 0.3882 &  1 & 110.47 & AMF                      & -                              &  - \\
083482 &   71.54 &  10.89 & 11.54 &  2.37 & 0.0617 & J075427.2+220941     & 118.61320 &  22.16150 & 0.3937 & 0.3969 &  1 &  48.61 & WHL                      & -                              &  - \\
089735 &   20.56 &   6.14 &  6.38 &  1.58 & 0.0356 & 588016841246441787   & 126.94684 &  26.58608 & 0.4225 & 0.3869 &  2 &  21.17 & GMBCG,WHL,AMF            & -                              &  - \\
089885 &   54.55 &   3.55 &  9.23 &  1.82 & 0.0411 & J083146.4+525057     & 127.94344 &  52.84933 & 0.5383 & 0.0000 &  0 &  54.34 & WHL                      & -                              &  - \\
090256 & 1167.06 & 156.57 & 48.71 & 10.09 & 0.0429 & 587737808499048612   & 128.72875 &  55.57253 & 0.2033 & 0.2421 &  2 &   4.84 & GMBCG                    & 4C +55.16                      &  3 \\
090966 &   62.97 &   8.97 & 11.72 &  2.39 & 0.0405 & 8222                 & 129.35330 &  55.54786 & 0.2780 & 0.2767 &  2 &  11.21 & maxBCG                   & -                              &  - \\
092117 &   49.68 &   3.40 &  9.27 &  1.83 & 0.0292 & 587732482731737444   & 131.75750 &  34.85367 & 0.4725 & 0.4643 &  1 &   8.47 & GMBCG,WHL                & -                              &  - \\
092718 &   64.29 &   5.19 &  9.74 &  1.93 & 0.0279 & 23315.0.0.23315      & 132.20100 &  44.93770 & 0.5753 & 0.0000 &  0 &  38.44 & AMF                      & [VMF98] 060                    &  2 \\
097911 &  279.44 &  31.21 & 19.60 &  3.85 & 0.0178 & 18874.0.0.18874      & 141.43440 &  30.98410 & 0.5865 & 0.0000 &  0 & 116.35 & AMF                      & -                              &  - \\
098728 &   17.34 &   2.73 &  6.51 &  1.40 & 0.0129 & 1896                 & 143.01990 &  47.55527 & 0.2350 & 0.2248 &  1 &  10.65 & maxBCG,AMF               & -                              &  - \\

\hline
\end{tabular}
}
\tablebib{1- \cite{Plionis05}; 2- \cite{Vikhlinin98}; 3- \cite{Cavagnolo08}; 4- \cite{Finoguenov07}; 5-\cite{Horner08}; 
 6-\cite{Dietrich07}; 7- \cite{Burenin07}; 8- \cite{Romer00}; 9- \cite{Basilakos04}; 10- \cite{Schuecker04}; 11- \cite{Sehgal08}
}
\end{table}
\end{landscape}

\chapter[II. The optically confirmed cluster sample and the \ltr relation]{II. The optically confirmed cluster sample and the \ltr relation\footnote{This chapter is published in the {\it Astronomy \& Astrophysics Journal}, 2013A\&A...558A..75T}}


\section*{Abstract}
 
   {\it Aims.} We compile a sample of X-ray selected galaxy groups and clusters 
 from the XMM-Newton serendipitous source catalogue (2XMMi-DR3) with optical 
confirmation and redshift measurement from the Sloan Digital Sky Survey (SDSS).
We present an analysis of the X-ray properties of this new sample with
particular emphasis on the X-ray luminosity-temperature (\ltr) relation.

   {\it Methods.} The X-ray cluster candidates were selected from the 2XMMi-DR3 catalogue in 
the footprint of the SDSS-DR7. We developed a finding algorithm to search for 
overdensities of galaxies at the positions of the X-ray cluster candidates in 
the photometric redshift space and to measure the redshifts of the clusters 
from the SDSS data. 
For optically confirmed clusters with good quality X-ray data we derive
the X-ray flux, luminosity  and temperature from proper spectral fits, while
the X-ray flux for clusters with low quality X-ray data is obtained from the
2XMMi-DR3 catalogue.

   {\it Results.} The developed detection algorithm provides the photometric redshift of 530 
galaxy clusters. Among them, 310 clusters have a spectroscopic redshift for at 
least one member galaxy. About 75 percent of the optically confirmed cluster 
sample are newly discovered X-ray clusters. Also, 301 systems are known as optically 
selected clusters in the literature while the remainder are new discoveries in 
X-ray and optical bands. The optically confirmed cluster sample spans a wide 
redshift range 0.03-0.70 (median $z$=0.32). 
In this paper, we present the catalogue of X-ray selected galaxy groups and 
clusters from the 2XMMi/SDSS galaxy cluster survey. The catalogue has two 
subsamples: (i) a cluster sample comprising 345 objects with their X-ray 
spectroscopic temperature and flux from the spectral fitting; 
(ii) a cluster sample consisting 185 systems with their X-ray flux from the 
2XMMi-DR3 catalogue since their X-ray data is not sufficient to do spectral 
fitting. For each cluster, the catalogue also provides the X-ray bolometric 
luminosity and the cluster mass at $R_{500}$ based on scaling relations and 
the position of the likely brightest cluster galaxy (BCG). 
The updated \ltr relation of the current sample with X-ray spectroscopic 
parameters is presented. We find the slope of the \ltr relation is consistent
with published ones. We see no evidence for evolution in the slope and 
intrinsic scatter of the \ltr relation with redshift when excluding the 
low luminous groups.



\section{Introduction}

   Galaxy clusters are the largest known gravitationally bound objects, their 
study is important for both an intrinsic understanding of their systems and  
an investigation of the large scale structure of the universe. The multi-component 
nature of galaxy clusters offers multiple observable signals across the 
electromagnetic spectrum \citep[e.g.][]{Sarazin88, Rosati02}. 
The hot, ionized intra-cluster 
medium (ICM) is investigated  at X-ray wavelengths and the Sunyaev-Zeldovich 
(SZ) effect \citep[][]{Sunyaev72,Sunyaev80}. The cluster galaxies are most 
effectively studied through  optical and NIR photometric and spectroscopic 
surveys. The statistical studies of clusters of galaxies provide complementary 
and powerful constraints on the cosmological parameters \citep[e.g.][]{Voit05, 
Allen11}.
 
X-ray observations offer the most powerful technique for constructing cluster
catalogues. The main advantages of the X-ray cluster surveys are their excellent 
purity and completeness and the X-ray observables are tightly correlated with 
mass \citep[e.g.][]{Reiprich02, Allen11}. 
Reliable measurements of cluster masses allow us to measure 
both the mass function \citep{Boehringer02} and power spectrum 
\citep{Schuecker03}, which directly probe the cosmological models. 

At X-ray wavelengths, galaxy clusters are simply identified as X-ray luminous, 
continuous, spatially extended, extragalactic sources \citep{Allen11}. 
Several X-ray cluster samples have been constructed from previous X-ray 
missions and used for a variety of 
astrophysical studies \citep[e.g.][]{Romer94, Forman78, Scharf97, Vikhlinin98, 
Boehringer00, Borgani01, Boehringer04, Burenin07}. The current generation of 
X-ray satellites XMM-Newton, Chandra, and Suzaku  provided follow-up observations 
of individual clusters allowing a precise determination of their spatially 
resolved spectra \citep[e.g.][]{Vikhlinin09b,Pratt10,Arnaud10}. Several other 
projects are being conducted to detect galaxy clusters from the observations of 
the XMM-Newton, Chandra, and the X-ray Telescope on board of the Swift 
satellite \citep[e.g.][]{Barkhouse06, Kolokotronis06, Finoguenov07, 
Finoguenov10, Adami11, Fassbender11, Takey11, Mehrtens12, Clerc12, Tundo12}.

We have started a serendipitous search for galaxy clusters based on extended 
sources in the 2XMMi-DR3 catalogue, the second XMM-Newton source 
catalogue \citep{Watson09}, in the footprint of the SDSS-DR7. The main aim of 
the survey is to construct a large catalogue of newly discovered X-ray selected 
groups and clusters from XMM-Newton archival observations. The catalogue will 
allow us to investigate the evolution of X-ray scaling relations as well as 
the correlation between the X-ray and optical properties of the clusters. 

The survey comprises 1180 X-ray selected cluster candidates. 
A cross-correlation of these 
with recently published optically selected SDSS galaxy cluster catalogues 
yielded photometric redshifts for 275 objects. 
Of these, 175 clusters were published by \citep[][Paper I hereafter]{Takey11} 
together with their X-ray luminosity, temperature and mass.
The first cluster sample from the survey covers a wide range of redshifts from 
0.09 to 0.61.
We extended the relation between the X-ray bolometric 
luminosity at $R_{500}$  (the radius at which the cluster mean density 
is 500 times the critical density of the Universe at the cluster redshift)
and the X-ray temperature towards significantly lower 
luminosities than reported in the literature and found that the slope of the
linear \ltr relation was consistent with that for more luminous clusters.


In the present paper, we expand the optically confirmed sample from the survey 
by searching for the optical counterparts of cluster candidates that had been 
missed by previous cluster finding algorithms and their members detected in the 
SDSS imaging (see Paper I for a sample of X-ray and optically selected groups 
and clusters).
We present the algorithm used to identify the optical counterparts of the 
X-ray cluster candidates and to estimate the cluster redshifts using SDSS data. 
As a result, we present a catalogue of X-ray selected galaxy groups and clusters
(including the objects in Paper I) from the ongoing 
2XMMi/SDSS galaxy cluster survey.
The catalogue provides the X-ray properties (e.g. temperature, flux, luminosity, 
and mass) and the cluster photometric redshift and, if available, the 
cluster spectroscopic redshift and the position of the likely brightest cluster 
galaxy (BCG) of the optically confirmed cluster sample. 

The X-ray luminosity-temperature (\ltr) relation was investigated by several 
authors \citep[e.g.][]{Markevitch98, Pratt09, Mittal11, Eckmiller11,  Reichert11, 
Takey11, Maughan12, Hilton12}. 
These studies showed that the observed \ltr relation is much steeper than that  
predicted by self-similar evolution. This indicates that the ICM 
is heated by an additional source of energy, which comes mainly from Active 
Galactic Nuclei (AGNs) \citep{Blanton11}. 
The inclusion of AGN-feedback in cosmological evolution models indeed gives
better agreement between simulated and observed \ltr under certain
circumstances \citep{Hilton12}.
Here, we present an updated \ltr relation based on the largest sample of 
X-ray selected groups and clusters to date drawn from a single survey based on 
XMM-Newton observations. The sample spans a wide redshift range from 
0.03 to 0.67. 

The format of this paper is as the follows. In Section 2, we describe  
the construction of the X-ray cluster candidates list and the optically 
confirmed cluster sample with their redshift estimations. In Section 3, 
we present the X-ray data reduction and analysis of the constructed sample. In 
Section 4, the results and discussion of the cluster sample is 
presented. We summarise our results in Section 5. The cosmological 
parameters  $\Omega_{\rm M}=0.3$, $\Omega_{\Lambda}=0.7$ and 
$H_0=70$\ km\ s$^{-1}$\ Mpc$^{-1}$ were used throughout this paper.



\section{Sample construction}
We started our search based on the XMM-Newton 
serendipitous sources followed by searching of overdensities of galaxies in 
3D space. In the following subsections, we present the strategy to create 
the X-ray cluster candidates list. To derive the X-ray properties of these 
candidates, we need to determine their redshift either from the X-ray data, 
which is only possible for the X-ray brightest clusters, or from the optical 
data, which is used in the current work. 
We also present the algorithm which is used to detect the clusters in the optical 
band and to estimate their redshifts from the SDSS data. The comparison of the 
measured redshifts with the published ones is presented.


\subsection{X-ray cluster candidates}

The survey comprises X-ray cluster candidates selected as serendipitous 
sources from the 2XMMi-DR3 catalogue in the footprint of SDSS-DR7.
The number of XMM-Newton fields that were used in constructing the 
2XMMi-DR3 catalogue in the footprint of SDSS-DR7 at high galactic latitude 
$|b| > 20^{\circ}$ is 1200 fields after excluding the multiple observations 
of the same field. We also excluded those fields that were flagged as bad 
(the whole field) and not suitable for source detection according to the  
manual flag given in the 2XMMi-DR3 catalogue. The total area of those fields that 
were included in our survey is 210 deg$^2$ taking into account the overlap areas 
among the fields.  

The cluster candidate selection was based on X-ray extended sources that 
passed the quality assessment during the construction of the
catalogue  by the XMM-Newton Survey Science Center (SSC).
The extent parameter of each extended source in the 2XMMi-DR3 catalogue is 
determined by the SAS task {\tt emldetect} by fitting a convolution of 
a $\beta$ model ($\beta$=2/3) and the instrument PSF (point spread function) 
to each input source. The source is classified as extended if the extent 
parameter varies between 6 to 80 arcsec and if the extent likelihood is 
larger than 4 \citep{Watson09}.

The completeness of the 2XMMi-DR3 extended source catalogue is not easy to 
assess since the 2XMMi-DR3 catalogue was constructed from 4953 
observations that have different exposure times. The wide range of exposure 
times yields various flux limits.  
\cite{Muehlegger10} simulated two fields (LBQS and SCSA with exposure time 
52 ks and 8.8 ks, respectively) in the XMM-Newton Distant Cluster project 
(XDCP) in order to test the detection probability. They used a source 
detection technique that is similar to the one used in detecting the 2XMMi-DR3
sources. 
According to their simulations, the higher detection probability was 
for clusters with intermediate core radii in the range of 15 to 25 arcsec. The
probability goes down with the decrease of the photon counts as well as the 
core radius ($< 7$ arcsec) due to the difficulty to discriminate extended sources 
from point sources. The detection probability of sources with large core radii 
($ > 75 $ arcsec) and low number of photon was low because these systems 
disappear in the background due to their low surface brightness. 
The detection probability decreases beyond the off-axis angle of 12 arcmin, 
caused by vignetting.
Based on these results clusters with low photon counts or large core radii 
might be missed in the 2XMMi-catalogue or might be listed with incorrect 
source parameters.

The selected extended sources were visually inspected by us
in two steps to  
exclude likely spurious detections. The first visual inspection was done 
using the X-ray images through the FLIX upper limit 
server\footnote{http://www.ledas.ac.uk/flix/flix.html}. The second one was 
done using the X-ray-optical overlays, where the X-ray flux contours were overlaid 
onto the co-added SDSS images in $r$, $i$, and $z-$bands. 
The former inspection allowed us the remove the obvious spurious cases 
due to point source confusion, X-ray artifacts, and near very bright
sources. Extended sources were also rejected if they were found within another
extended source or at the very edge of the CCDs. The latter inspection 
enabled us also to remove the X-ray extended sources corresponding to low redshift  
galaxies. The resulting list which passed these selection criteria includes 
1180 X-ray cluster candidates with at least 80 net photon counts. More than 
75 percent are new X-ray detections of galaxy groups and clusters.  

\begin{figure}[t]
\centering{
  \resizebox{100mm}{!}{\includegraphics{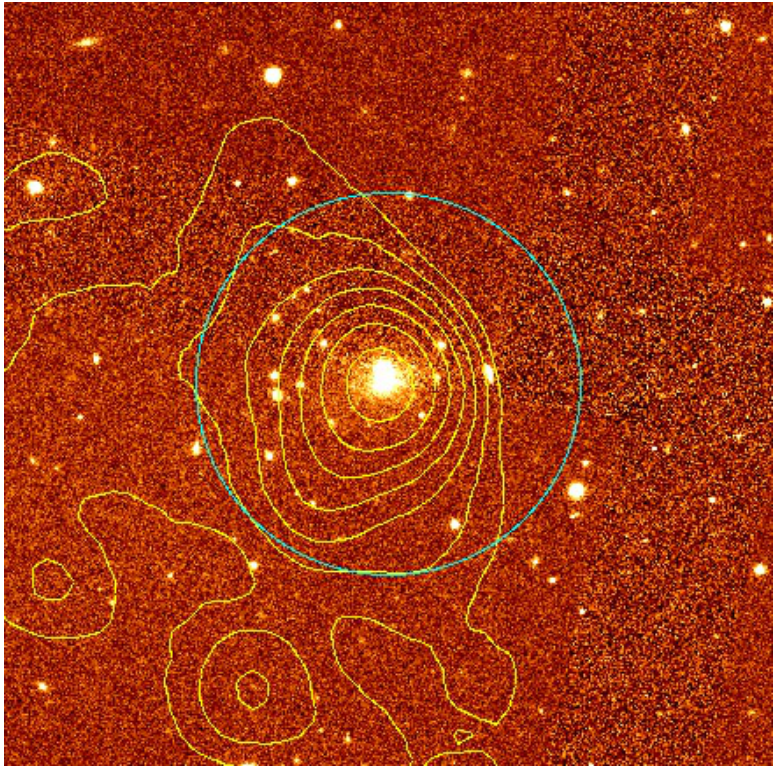}}}
  \caption[The X-ray-optical overlay of the example cluster 2XMM J102133.2+213752 
at spectroscopic redshift = 0.1873]{The X-ray-optical overlay of the example cluster
2XMM J102133.2+213752 at spectroscopic redshift = 0.1873. The X-ray flux 
contours (0.2 - 4.5 keV) are overlaid on combined image from $r$, $i$, and 
$z-$bands SDSS images. The plotted cyan circle is a circle with one arcmin radius 
around the X-ray emission peak. The field of view is $4'\times 4'$ centred on 
the X-ray cluster position.} 
  \label{f:108143_overlay}
\end{figure}

Figure~\ref{f:108143_overlay} shows the X-ray-optical overlay of a newly 
discovered galaxy cluster in X-ray and optical observations at 
redshift = 0.1873. This cluster has been serendipitously detected (at off-axis 
angle of about 11 arcmins) in XMM-Newton EPIC observations of the galaxy NGC 3221.
We use this cluster as an example for illustrating the main steps of 
estimating the cluster redshift and the X-ray analyses in the following sections.

\subsection{Construction of the optically confirmed cluster sample}

Various methods have been developed to define the cluster membership of 
galaxies from the data provided by the SDSS. They are based on different 
properties of the clusters and their galaxy members, e.g.~using the cluster
red-sequence, or the E/S0 ridge-line 
\citep[e.g.][]{Koester07, Hao10}, or an overdensity of galaxies in the
photometric redshift space \citep{Wen09}. Also, clusters of  
galaxies are identified by convolving the optical galaxy survey with a set of 
filters in position, magnitude, and redshift space based on a modeling of the 
cluster and field galaxy distributions \citep{Szabo11}. 

In Paper I, we have optically confirmed about a quarter of the X-ray cluster 
candidates through cross-correlation with previously identified clusters in
four optical cluster catalogues \citep[][]{Hao10, Wen09, Koester07, Szabo11}. 
The remainder of the X-ray cluster candidates are either distant cluster 
candidates beyond the SDSS detection limits i.e. z $\ge$ 0.6, which need 
follow-up imaging and spectroscopic confirmation or there are overdensities 
of galaxies at the X-ray cluster positions that were not recognized by any
previous optical cluster finders (see e.g.~Figure~\ref{f:108143_overlay}).
Therefore, for those clusters with members detected in the SDSS imaging, 
we developed our own algorithm to search for optical counterparts
and determine their redshifts from photometric redshifts in the SDSS database.


\subsubsection{Estimation of the cluster redshifts}

Since we have prior information about the cluster position, the position of 
the X-ray emission peak, we can use this information to simplify the cluster
finding procedure. We searched for an overdensity of galaxies
around the X-ray position of the cluster candidates within a certain
redshift interval. 
We created a galaxy sample for each X-ray cluster candidate by selecting all
galaxies from the SDSS-DR8 in an area with 10 arcmins radius centred on the 
X-ray source position. This radius corresponds to a physical radius of 500\,kpc
at redshift redshift 0.04, which is about our low redshift limit.

The galaxies were selected from the {\tt galaxy} view table in the SDSS-DR8, which 
contains the photometric parameters measured for resolved primary objects, 
classified as galaxies.  Also, the photometric redshifts and, if available, the 
spectroscopic redshifts of the galaxy sample were selected from the 
{\tt Photoz} and {\tt Specz} tables, respectively, in the SDSS-DR8. 
The extracted parameters of the galaxy 
sample include the coordinates, the model magnitudes in $r-$band, 
the photometric redshifts, and, if available, the spectroscopic redshifts. 
When spectroscopic redshifts of galaxies are available, we use those 
instead of the photometric redshifts.

\begin{figure}
\centering{
  \resizebox{100mm}{!}{\includegraphics{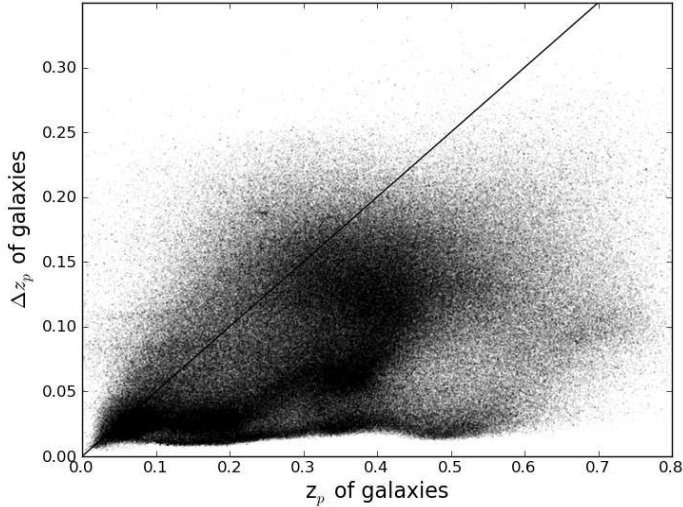}}}
  \caption[The photometric redshift error, $ \bigtriangleup z_p $, is plotted 
against the photometric redshift, $z_p$]{The photometric redshift error, 
$ \bigtriangleup z_p $, is plotted 
against the photometric redshift, $z_p$, of the galaxy sample with 
$ m_{r} \leq 22$ mag and $\bigtriangleup m_{r} < 0.5 $ mag. The solid line 
indicates the 50 percent relative error of the photometric redshift of the 
galaxy sample.} 
  \label{f:ezp_zp}
\end{figure}

To clean the galaxy sample from faint objects or from galaxies with poor 
photometric measurements, we only use 
galaxies with $ m_{r} \leq 22$ mag and $\bigtriangleup m_{r} < 0.5 $ mag. 
The resulting galaxy sample still includes galaxies with large photometric 
redshift errors, which reach a 100 percent in many cases. The photometric 
redshift errors of the galaxy sample with the applied magnitude cut   
are plotted against the photometric redshifts in Figure~\ref{f:ezp_zp}.
To exclude low redshift galaxies with significantly large relative 
photometric redshift errors as well as to keep high redshift galaxies 
with slightly large relative errors, but could be acceptable, we decided 
to apply a relative photometric redshift error cut ($< 50$ percent) and 
not to use a fixed absolute error. The 50 percent relative error line is 
plotted in Figure~\ref{f:ezp_zp}.

\begin{figure}
\centering{
  \resizebox{100mm}{!}{\includegraphics{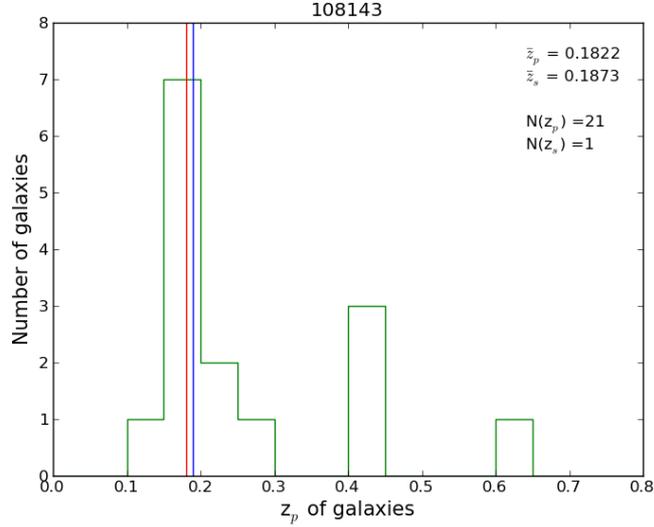}}}
  \caption[The photometric redshift distribution of all galaxies within 
one arcmin from the X-ray centroid]{The photometric redshift distribution of 
all galaxies within one arcmin from the X-ray centroid with $ m_{r} \leq 22$ mag, 
$\bigtriangleup m_{r} < 0.5 $ mag, and $ \bigtriangleup z_{p} / z_{p} < 0.5$. 
The cluster photometric redshift (red line), $\bar{z}_p$, spectroscopic redshift 
(blue line), $\bar{z}_s$, and the cluster member candidates within 560 kpc with 
photometric redshift, N($z_p$), and spectroscopic redshift, N($z_s$), are 
written in upper right, respectively.} 
  \label{f:108143_hist1}
\end{figure}

\begin{figure}
\centering{
  \resizebox{100mm}{!}{\includegraphics{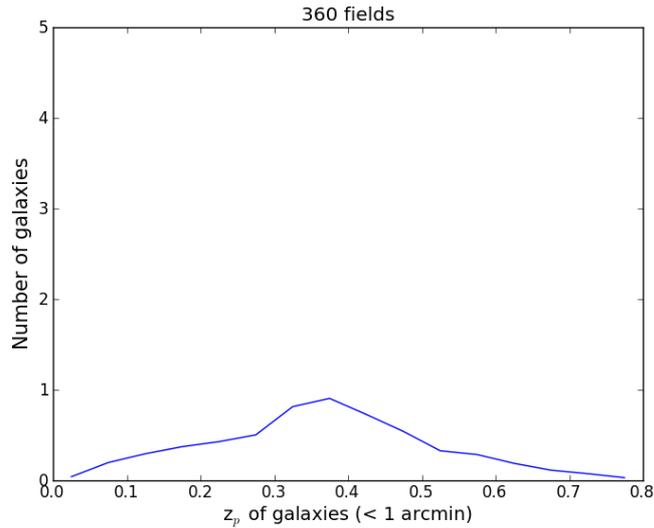}}}
  \caption[The distribution of the mean galaxy counts within one arcmin from 
the positions of 360 random fields in the SDSS footprint]{The distribution of 
the mean galaxy counts, same distribution as 
Figure~\ref{f:108143_hist1}, within one arcmin from the positions 
of 360 random fields in the SDSS footprint with $ m_{r} \leq 22$ mag, 
$\bigtriangleup m_{r} < 0.5 $ mag, and $ \bigtriangleup z_{p} / z_{p} < 0.5$. 
} 
  \label{f:360_fields}
\end{figure}

\begin{figure}
\centering{
  \resizebox{100mm}{!}{\includegraphics{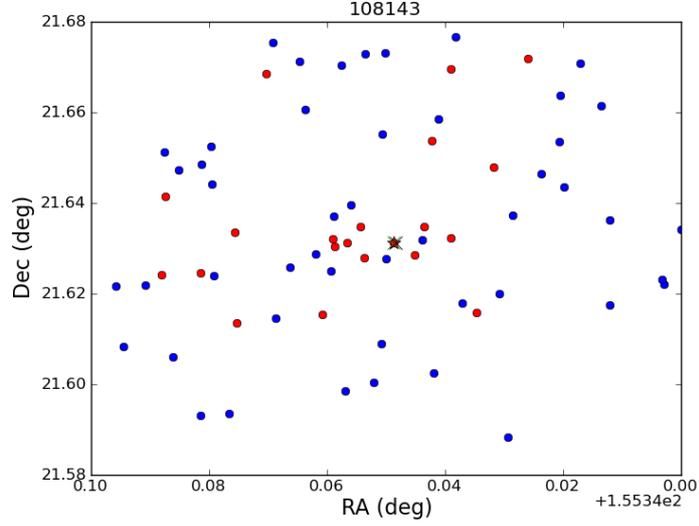}}}
  \caption[The sky distribution of cluster galaxies (red dots) and the field 
galaxies (blue dots) within 560 kpc from the X-ray position (black X marker)]{The 
sky distribution of cluster galaxies (red dots) and the field 
galaxies (blue dots) within 560 kpc from the X-ray position (black X marker).
The BCG with an available spectroscopic redshift (marked by star) is identical 
located (green x marker) with the X-ray cluster position.} 
  \label{f:108143_dist}
\end{figure}

\begin{figure}
\centering{
  \resizebox{90mm}{!}{\includegraphics{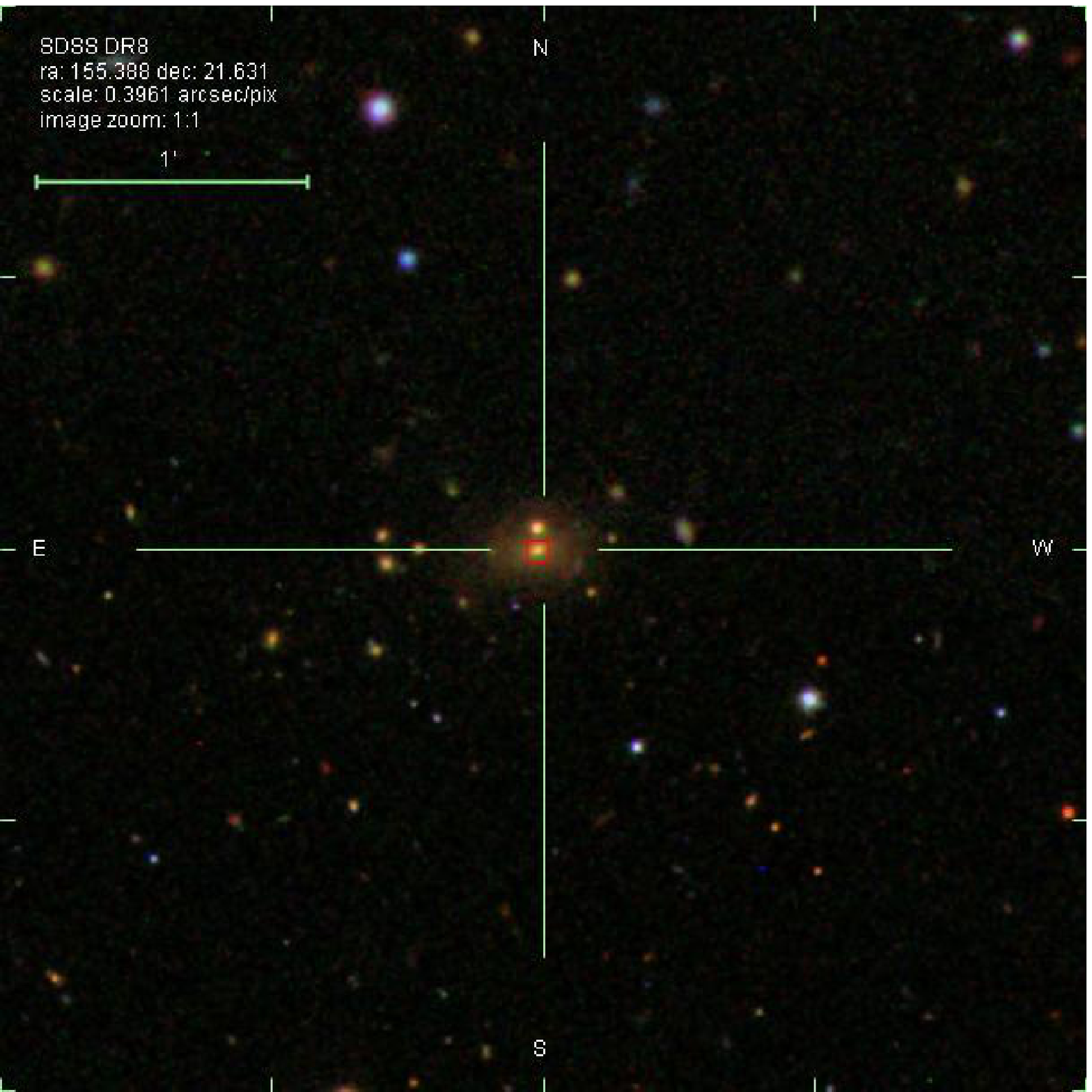}}}
  \caption[The SDSS colour image of the 2XMM J102133.2+213752]{The SDSS colour 
image of the 2XMM J102133.2+213752 with 
four arcmins a side field of view centred on the X-ray peak position 
as indicated by the cross hair. The BCG with a spectroscopic redshift is marked 
by a square and it is coincident with the X-ray position.} 
  \label{f:108143_color}
\end{figure}

The main idea of the finding algorithm is to identify the likely brightest cluster 
galaxy (BCG) among those galaxies with similar redshift within one arcmin from 
the X-ray centroid position and then search for an overdensity of surrounding 
member candidates. To confirm the X-ray cluster candidates optically and to 
measure their redshifts, we do the following steps:

\begin{enumerate}

\item plot the photometric redshift histogram of all galaxies within one arcmin 
from the X-ray position with $ m_{r} \leq 22$ mag, $ \bigtriangleup m_{r} < 0.5 $ mag, 
and fractional error of the photometric redshift, 
$ \bigtriangleup z_{p} / z_{p} < 0.5 $ as shown in Figure~\ref{f:108143_hist1}.

\item compute a tentative photometric redshift of the cluster as the centre
of the redshift bin in the main peak,  $ z_{p,\,M} $.
To make sure that the distributions of the photometric redshifts of background 
galaxies can not produce such a peak in the histogram, we selected 360 random 
positions in the SDSS sky coverage and count the galaxies with identical 
magnitude and photometric redshift criteria used in the previous step within 
one arcmin from the field positions. We have chosen that large number 
of fields in order to obtain the average redshift distribution of background 
galaxies.
Figure~\ref{f:360_fields} shows the average distribution of the galaxy counts 
within those fields as a function of redshift. The distribution does not 
exceed 0.91 per redshift bin. It is unlikely that the background 
galaxies have a significant influence on the redshift determination.    
Therefore, we can neglect subtraction of the background galaxies in the current 
step to compute a tentative cluster redshift.

\item the BCG is identified as the brightest galaxy among those galaxies within 
one arcmin around the X-ray position with a photometric redshift in the
interval  $ z_{p,\,M} \pm 0.04(1+z_{p,\,M}) $. 
If the algorithm finds multiple peaks in the redshift histogram, we select the 
closest BCG candidate to the X-ray position.
\citet{Wen09} have shown that a redshift interval of $\pm 0.04(1+z_{p,\,M})$
comprises 80 percent of the clusters members. We assume that our tentative redshift 
gives a less reliable but still a robust estimate of cluster membership.
The redshift of the likely BCG does not 
necessarily lie in the peak bin of the redshift histogram but may be within  
one of the adjacent bins. Therefore, we initially allow that the BCG
candidate lies either in the central or in one of the adjacent redshift bins.
We then chose as BCG the brightest galaxy in those bins nearest to the X-ray
position.

\item 
to detect an overdensity of galaxies in 3D space, all galaxies within  
a radius of 560 kpc from the X-ray emission peak within the redshift interval 
$ z_{p,\,BCG} \pm 0.04(1+z_{p,\,BCG})$ are considered as cluster member 
candidates, N($<$560 kpc).  
The redshift range used here is the same as that used by \citep{Wen09}.
Since the physical size of the cluster is not a priori known, 
we choose a radius of 560 kpc as the average of $R_{500}$ from Paper I. 
The radius used is similar to the radius used  
by \citep{Wen09} to detect an overdensity of galaxies. 
They showed that 
a radius of 500 kpc gives a high overdensity level and a low false 
detection rate using  Monte Carlo simulation tests. Since we are not 
computing the cluster richness in the current work, we did not subtract 
the background galaxies. The identified cluster member candidates are 
only used to compute the cluster redshift.  

\item  
The cluster photometric redshift, $\bar{z}_p$, is finally determined as the weighted
average of the photometric redshift of N($<$560 kpc)  with weights given as
$w_i = 1/(\bigtriangleup z_{p,\,i})^2$. The redshift value for our example cluster is
marked by the  vertical red line in Figure~\ref{f:108143_hist1}.
If there are available spectroscopic redshifts of N($<$560 kpc), the cluster 
spectroscopic redshift, $\bar{z}_s$, is the weighted average of 
those available spectroscopic redshifts as indicated by a blue line in 
Figure~\ref{f:108143_hist1}. For the example cluster, only the BCG has 
a spectroscopic redshift. 
Figure~\ref{f:108143_dist} shows the sky distribution of the cluster member 
candidates within 560 kpc from the X-ray centroid that are represented by 
red dots and the field galaxies that are represented by blue dots.

\item A cluster is considered as detected if there are at least 8 cluster member 
galaxies within 560 kpc and 2 members within one arcmin. If
N($<$560 kpc) $<8$ but the estimated redshift is consistent with either an
available redshift from the literature or an spectroscopic redshift from the
current algorithm, we also consider it as a detected cluster. 
The final decision to confirm the optical cluster detection is done through 
a visual inspection of the SDSS colour image of the cluster field, which led  
to the exclusion of misidentified optical counterparts in a few cases. 
Figure~\ref{f:108143_color} shows the SDSS colour image of the example 
cluster with a field of view 4$'$ $\times$ 4$'$ centred at the X-ray position.

\end{enumerate}

\begin{figure}
\centering{
  \resizebox{100mm}{!}{\includegraphics{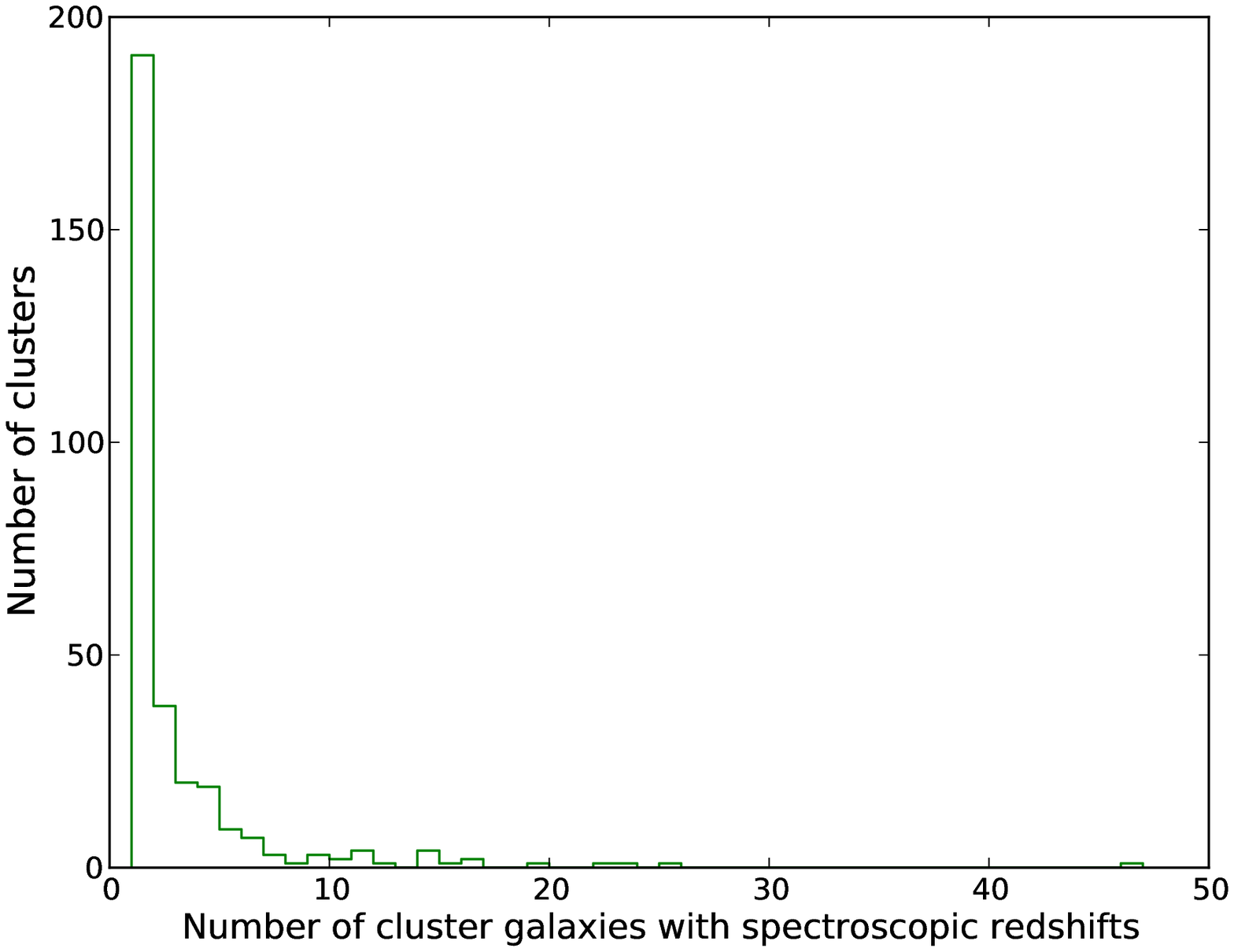}}}
  \caption[The distribution of the number of cluster members with spectra of the 
spectroscopically confirmed clusters]{The distribution of the number of cluster 
members with spectra of the spectroscopically confirmed clusters. The bin size 
of the histogram is one.} 
  \label{f:Hist_Nzs}
\end{figure}

\begin{figure}
\centering{
  \resizebox{100mm}{!}{\includegraphics{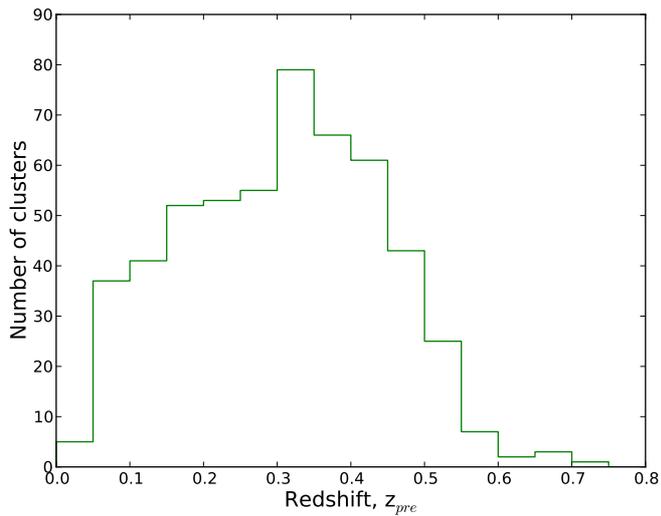}}}
  \caption{The distribution of estimated photometric redshifts and, 
if available, the spectroscopic ones of the optically confirmed 
cluster sample.}
  \label{f:Hist_zc}
\end{figure}

\begin{figure}
\centering{
  \resizebox{100mm}{!}{\includegraphics{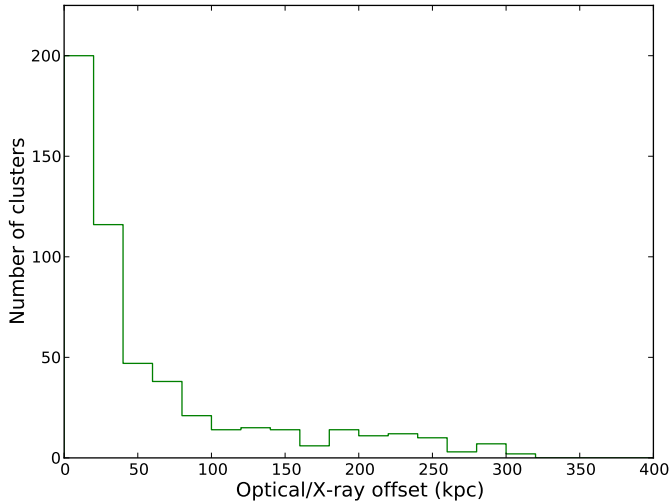}}}
  \caption{The distribution of the linear separation between the likely BCG 
and the X-ray emission peak of the optically confirmed cluster sample.} 
  \label{f:Hist_offset}
\end{figure}

Our procedure yielded 530 optically confirmed galaxy clusters with measured
redshifts. We refer to this sample as the optically confirmed cluster sample,
which spans  a wide redshift range from 0.03 to 0.70. About 60 percent of this
sample have spectroscopic confirmation. Figure~\ref{f:Hist_Nzs} shows the
distribution of the number of cluster galaxies per cluster with spectroscopic
redshifts.  
Figure~\ref{f:Hist_zc} shows the distribution of the estimated photometric 
redshifts and, if available, spectroscopic ones of the optically confirmed 
cluster sample.  
The projected separation between the X-ray centres and the optical 
centres (chosen to be the BCGs positions) of the cluster sample is 
shown in Figure~\ref{f:Hist_offset}. The distribution has a median offset of
29\,kpc, 86 percent of the BCGs are found within 150\,kpc. 
The maximum projected separation between the BCGs and X-ray peaks is about 320
kpc. The reason for the small observed offset lies in the way of sample
construction, on the other hand the offset distribution seems to be in 
agreement with the corresponding one derived for the maxBCG survey and 
ROSAT clusters \citep{Rykoff08}.

\subsubsection{Redshift uncertainty and the comparison with published redshifts}

\begin{figure}
\centering{
  \resizebox{100mm}{!}{\includegraphics{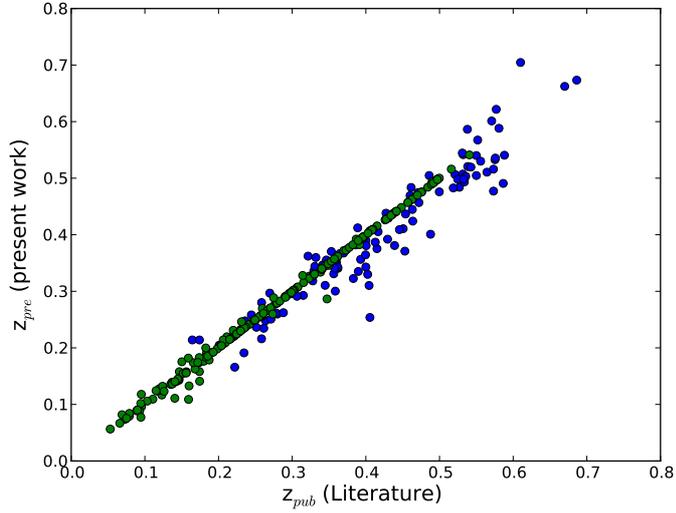}}}
  \caption[The comparison between the estimated redshifts, $z_{pre}$,  and 
the published ones, $z_{pub}$, of the optically confirmed cluster sample]{The 
comparison between the estimated redshifts, $z_{pre}$,  and 
the published ones, $z_{pub}$, of the optically confirmed cluster sample. 
The green dots represent the clusters with spectroscopic redshifts while 
blue dots represent the clusters with only photometric redshifts.  } 
  \label{f:compz}
\end{figure}

To assess the optical detection algorithm and the estimation of the cluster 
redshift, we queried the NASA Extragalactic Data base (NED) to search for 
published optical redshifts. 
 The NED lists 301 objects 
including those from our Paper I. 
Figure~\ref{f:compz} shows the relation between the present estimation of 
the redshifts, $z_{pre}$, and the published ones, $z_{pub}$. The clusters 
with available spectroscopic redshifts are represented by the green dots 
while those clusters with only photometric redshifts are represented by 
the blue dots. In general, the newly estimated redshifts are in a very good 
agreement with the published ones.

For those clusters with redshift difference $| z_{pre} - z_{pub}| > 0.05$, 
about 5 percent of the sample, we visually re-investigated their colour 
image (as in Figure~\ref{f:108143_color}) and the distribution on sky 
of the identified cluster members (as in Figure~\ref{f:108143_dist}). 
This leads in all cases to a revision of the published redshifts and we thus 
regard the newly determined redshifts more reliable than the published ones
which were based only on optical search methods. We note, that redshifts used 
in Paper I needed to be revised also for about 5 percent of the objects for 
the same reason.

\begin{figure}
\centering{
  \resizebox{100mm}{!}{\includegraphics{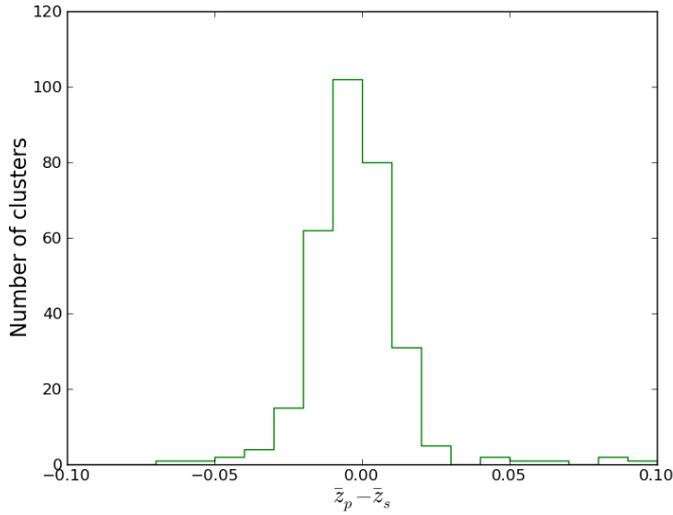}}}
  \caption{The distribution of the differences between the photometric, $\bar{z}_p$,
and spectroscopic, $\bar{z}_s$, redshifts of the optically confirmed cluster sample.} 
  \label{f:Hist_zp_zs}
\end{figure}

Among the optically confirmed cluster sample, 310 galaxy clusters are 
spectroscopically confirmed with at least one member galaxy with 
spectroscopic redshift from the existing SDSS data (SDSS-DR8). To assess 
the accuracy of our weighted average photometric redshift, $\bar{z}_p$,
we compared it with the weighted average spectroscopic redshift, $\bar{z}_s$. 
Figure~\ref{f:Hist_zp_zs} shows the distribution of the redshift differences, 
$\bar{z}_p$ - $\bar{z}_s$, of the sample. 
The standard deviation of these redshift differences is 0.02, which roughly  
indicates the accuracy of the estimated photometric redshifts.  
Therefore, we are confident about the reliability of the photometric 
redshift measurements.

\section{X-ray data analysis}

We use a similar procedure as in Paper I to reduce and analyse the X-ray 
data of the optically confirmed cluster sample. The raw XMM-Newton data were 
downloaded using the Archive InterOperability System (AIO), which provides 
access to the XMM-Newton Science Arcchive \citep[XSA:][]{Arviset02}. 
These data were reprocessed to generate the calibrated and filtered event 
lists for EPIC (MOS1, MOS1, and PN) cameras with a recent version of the 
XMM-Newton Science Analysis Software (SAS11.0.1). 
To determine the source extraction radii with the maximum signal-to-noise
ratio (SNR), we created the radial profiles in the energy band [0.5-2.0] keV
of each  camera as well as for EPIC. Then the SNR was calculated as a function
of radius taking into account the background values as given in the 2XMMi  
catalogue.

The X-ray spectra of each cluster were generated from a region with the 
determined optimal extraction radius, which was corresponding to the 
highest EPIC SNR. 
The spectral extraction from the optimal aperture was chosen in order to 
reduce the statistical uncertainty in the derived temperatures and 
luminosities from the spectral fits.
The background 
spectra were extracted from a circular annulus around the cluster with inner 
and outer radii equalling two and three times the optimum radius, respectively. 
Other field sources embedded in the source and background regions of the 
cluster were removed.

The extracted spectra were binned to one count per bin. Spectra for each 
cluster were simultaneously fitted in XSPEC \citep[version 12.7.0]{Arnaud96} 
with a single-temperature optically thin thermal plasma model modified by 
galactic absorption of neutral matter, $\tt TBabs*MEKAL$ in XSPEC 
terminology \citep{Mewe86,Wilms00}. The temperature and the normalization 
of the plasma model were allowed to vary while the abundance was fixed at 
0.4 $Z_\odot$. The Hydrogen absorbing column density, $N_{\rm H}$,  was 
derived from the Leiden/Argentine/Bonn (LAB) survey \citep{Kalberla05} 
and fixed to this value. The spectral fit was performed using the Cash 
statistics with one count per bin, a recommended strategy for sources with 
low photon counts \citep[e.g.][]{Krumpe08}. 

To avoid a conversion of the fit to a local minimum of the fitting
statistics, we ran series of fits stepping the temperature from 0.1
to 15 keV  with a step size = 0.05 using the $\tt{steppar}$ command within
XSPEC. We note that when the model spectrum is interpolated from 
a pre-calculated table, the cluster temperatures in some cases tend to 
converge exactly at the temperature grid points of the model table.  
Therefore, we run the $\tt MEKAL$ code with the option of calculating 
the model spectrum for each temperature during the fitting and stepping process.   

The spectral fitting provided us with the X-ray temperatures, fluxes in 
[0.5-2.0] keV and luminosities in the rest frame energy band [0.5-2.0] keV 
and their errors. 
The errors of the X-ray temperatures, fluxes, and luminosities represent the 68  
percent confidence range. The bolometric X-ray luminosity over the rest frame 
energy range (0.1 to 50.0) keV was determined from the dummy response matrices 
based on the best fitting model parameters. The fractional error in the 
bolometric luminosity was assumed to be the same as the fractional 
error of the luminosity in the given energy band. 
To make sure this assumption is valid, we varied the temperatures by 
$\pm$ 1 $\sigma$ in a few cases. We found the measured band luminosities 
are within their errors.   

We accepted the X-ray parameters (temperature, flux, luminosity) of a 
cluster if the relative errors of both  the temperature and luminosity were
smaller than 50 percent. A final check was made to ensure that neither the
source nor the background region were affected by detector artefacts and/or
astronomical objects. We also visually screened spectral fits applied to
the data and rejected poor spectral fits. 

The finally derived bolometric luminosities were used to estimate the cluster 
luminosities and masses at $R_{500}$ through an iterative method as
briefly described below and in more detail in Paper I.



\section{Results and discussion}

 We could derive reliable X-ray parameters from spectral fits for 345 systems 
of the optically confirmed cluster sample. In the next subsections, we compare 
our new results with the common clusters from (a) the XMM Cluster Survey 
\citep{Mehrtens12}, (b) the MCXC catalogue \citep{Piffaretti11}, and 
(c) the Paper I sample.  
We then go ahead to derive an updated \ltr relation based on this new sample.
For the remaining 185 clusters of the optically confirmed sample
without proper spectral fit, we used the X-ray flux as given in the 2XMMi-DR3
catalogue to estimate the luminosity and the mass. We finally present the 
X-ray luminosity-redshift distribution of the whole optically confirmed 
cluster sample.



\subsection{The cluster sample with reliable X-ray parameters from the 
spectral fits}

\begin{figure}
\centering{
  \resizebox{100mm}{!}{\includegraphics{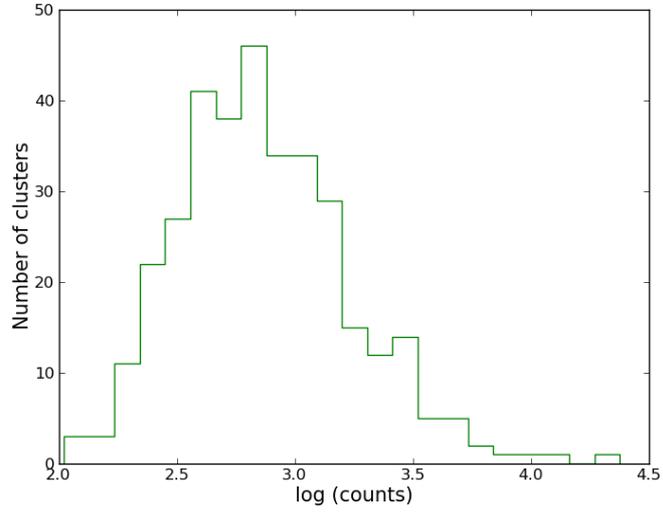}}}
  \caption{The distribution of the aperture net EPIC photon counts in 
[0.5-2.0] keV derived from the spectral fitting for the cluster sample 
with X-ray spectroscopic parameters.}
  \label{f:counts-his}
\end{figure}

\begin{figure}
\centering{
  \resizebox{100mm}{!}{\includegraphics{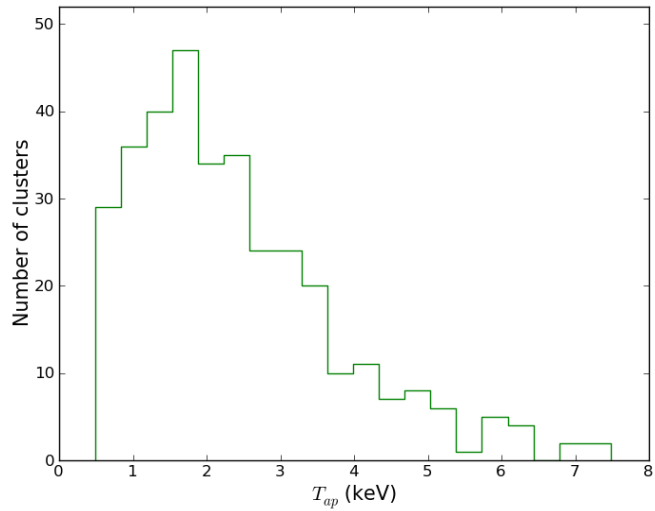}}}
  \caption{X-ray spectroscopic temperature distribution of the cluster sample
with reliable X-ray parameters from the spectral fit.} 
  \label{f:kT-his}
\end{figure}

Figure~\ref{f:counts-his} shows the distribution of the net EPIC photon counts 
in the energy interval [0.5-2.0] keV for those clusters that could be fitted successfully.
It shows that 87 percent of our
clusters have more than 300 source photons. In some cases a successful
spectral fit could be achieved with just a few more than 100 photons due to
the combination of clean X-ray data and previous knowledge of the cluster
redshift. Our new sample has a wide range of temperatures from 0.5 to 7.5 keV,
which is shown in Figure~\ref{f:kT-his}. The average relative errors of the 
temperatures and luminosities are 0.20 and 0.06, respectively.

We follow an iterative method (see Paper I) to compute physical parameters for
each cluster. The estimated aperture X-ray bolometric luminosity and its
error,  optimal extraction radius, and the redshift were used as input 
to determine the radius $R_{500}$, the X-ray bolometric 
luminosity within $R_{500}$, $L_{500}$, and the cluster mass at $R_{500}$,
$M_{500}$.
The main idea of the iterative method is to extrapolate the aperture bolometric 
flux to the bolometric flux at $R_{500}$ based on a $\beta$ model of the form: 
\begin{equation}
 S(r) = S(0)\,\Bigg[1 + \Big(\frac{r}{r_{\rm c}} \Big)^2 \Bigg]^{-3\beta + 1/2}.
\end{equation}
where $\beta$ and core radius, r$_{\rm c}$, depend on temperature (see Eq.~4 
and 5 in paper I). The correction factor of the flux is used to extrapolate 
the aperture bolometric luminosity to the bolometric $L_{500}$. Finally, 
$M_{500}$ is computed based on the $L_{500}-M_{500}$ relation from 
\citep{Pratt09}. 
The error budget of the estimated $L_{500}$ and $M_{500}$ includes the 
errors of the input parameters, the intrinsic scatter in the utilized \ltr and 
$L_{\rm X}-M$ relations, and the propagated errors of their slopes  and the
intercepts. The median correction factor between the extrapolated bolometric
luminosity to  $R_{500}$ and the aperture bolometric luminosity,
$L_{500}/L_{bol}$, was 1.7.

Table~\ref{tbl:SF_sample}, available entirely at the CDS, represents the first 
10 entries of the X-ray selected cluster sample with a total of 345 rows.
For each cluster the catalogue lists the cluster 
identification number (detection Id, detid) and its name (IAUNAME) in 
(cols.~[1] and [2]), the right ascension and declination of X-ray emission in 
equinox J2000.0 (cols.~[3] and [4]), the XMM-Newton observation Id (obsid) 
(col.~[5]), the optical redshift (col.~[6]), the scale at the cluster redshift 
in kpc/$''$ (col.~[7]), 
the aperture and $R_{500}$ radii in kpc (cols.~[8] and [9]), the cluster aperture  
X-ray temperature $T_{ap}$ and its positive and negative errors in keV 
(cols.~[10], [11] and [12], respectively), the aperture X-ray flux 
$F_{ap}$ [0.5-2.0] keV and its positive and negative errors in units of 
$10^{-14}$\ erg\ cm$^{-2}$\ s$^{-1}$ (cols.~[13], [14] and [15], respectively), 
the aperture X-ray luminosity $L_{ap}$ [0.5-2.0] keV and its positive and 
negative errors in units of $10^{42}$\ erg\ s$^{-1}$ (cols.~[16], [17] and 
[18], respectively), the cluster bolometric luminosity $L_{500}$ and its error 
in units of $10^{42}$\ erg\ s$^{-1}$ (cols.~[19] and [20]), the cluster mass 
$M_{500}$ and its error in units of $10^{13}$\ M$_\odot$ (cols.~[21] and [22]), 
the Galactic HI column in units $10^{22}$\ cm$^{-2}$ (col.~[23]), the 
{\tt objid} of the likely BCG in the SDSS-DR8  (col.~[24]), the BCG right 
ascension and declination in equinox J2000.0 (cols.~[25] and [26]), the 
estimated photometric and, if available, spectroscopic redshift of the cluster 
(col.~[27] and col.~[28]), the number of cluster members within 560 kpc with
available spectroscopic redshifts, which were used to compute the cluster
spectroscopic redshift, (col.~[29]), the redshift  type (col.~[30]), the
linear offset between the cluster X-ray position and the BCG position
(col.~[31]), and the NED name and its references (col.~[32] and  col.~[33]).



\subsection{The cluster sample with X-ray flux from the 2XMMi-DR3 catalogue}

For clusters with insufficient X-ray data to perform a proper spectral fit, we  
estimated their X-ray parameters based on the EPIC flux and its error in 
0.5-2.0 keV as given in the 2XMMi-DR3 catalogue. 
The catalogue provides aperture corrected fluxes, which are  
calculated by the SAS tasks {\tt emldetect}. For the individual cameras, 
individual-band fluxes are calculated from the respective band count rate 
using the filter- and camera-dependent energy conversion factors and corrected 
for the dead time due to the read-out phase. The EPIC flux in each band is
estimated as the mean of the band-specific detections in all cameras weighted 
by their errors \citep{Watson09}. Here we use the combined EPIC flux in band 2 
(0.5 - 1.0 keV) and band 3 (1.0 - 2.0 keV) and its propagated error, 
$F_{\rm cat}$ in [0.5-2.0] keV.

\begin{figure}
\centering{
  \resizebox{100mm}{!}{\includegraphics{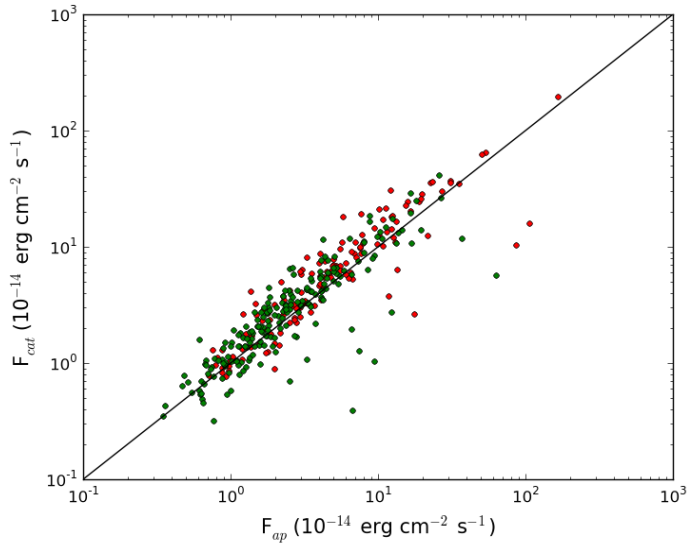}}}
  \caption[The cluster flux, $F_{cat}$, in 0.5-2.0 keV from the 2XMMI-DR3 
catalogue is plotted against the cluster flux, $F_{ap}$, in 0.5-2.0 keV from 
the the spectral fitting of the cluster sample with spectroscopic 
parameters]{The cluster flux, $F_{cat}$, in 0.5-2.0 keV from the 2XMMI-DR3 
catalogue is plotted against the cluster flux, $F_{ap}$, in 0.5-2.0 keV from 
the best-fitting model parameters for the cluster sample with spectroscopic 
parameters. The red dots represent the first cluster sample, Paper I, while the 
green ones represent the extended sample with reliable X-ray parameters.} 
  \label{f:Fxcat-Fxsf}
\end{figure}

Figure~\ref{f:Fxcat-Fxsf} shows the relation between the flux given in the 
2XMMi-DR3 catalogue and the aperture flux determined by us for the 
345 clusters with reliable X-ray parameters from the spectral fits. 
It shows a linear relation between the two flux measurements apart from 
some outliers (of order 5 percent) which we found to be contaminated by point 
sources in the 2XMMi-DR3 catalogue. In general terms the catalogued flux 
is larger than the aperture flux, since the former was computed for the 
integrated $\beta$-model.

\begin{figure}
\centering{
  \resizebox{100mm}{!}{\includegraphics{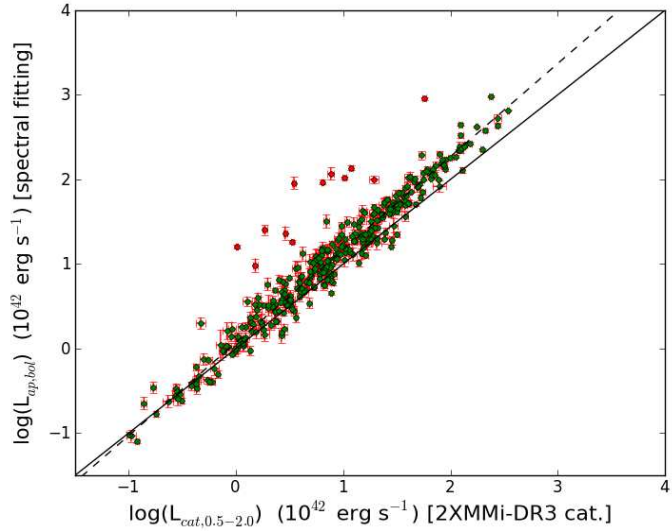}}}
  \caption[The aperture bolometric luminosities, $L_{\rm ap,\,bol}$, are 
plotted against the 2XMMi-DR3 catalogue luminosities, $L_{\rm cat,\,0.5-2.0}$, 
of the cluster sample with reliable parameters from the spectral fits.]{The 
aperture bolometric luminosities, $L_{\rm ap,\,bol}$, are 
plotted against the 2XMMi-DR3 catalogue luminosities, $L_{\rm cat,\,0.5-2.0}$, 
of the cluster sample with reliable parameters from the spectral fits. 
The one-to-one relationship is represented by solid line. The dashed line 
represents the best-fit using the BCES orthogonal regression method after 
excluding 12 outliers that are represented by red dots.} 
  \label{f:Lcatband-Lapbol}
\end{figure}

Figure~\ref{f:Lcatband-Lapbol} shows the relation between the aperture 
bolometric luminosities, $L_{\rm ap,\,bol}$, and $L_{\rm cat,\,0.5-2.0}$
of the cluster sample with X-ray spectroscopic parameters, where 
$L_{\rm cat,\,0.5-2.0}$ is based on $F_{\rm cat}$ in [0.5-2.0] keV.
 Generally, there is a linear 
relation between the two luminosities except for 12 outliers with 
$L_{\rm ap,\,bol} / L_{\rm cat,\,0.5-2.0} > 2$.   
Ignoring these outliers we performed a linear regression between their logarithms 
to convert $L_{\rm cat,\,0.5-2.0}$ to $L_{\rm ap,\,bol}$ for the 185 clusters 
without proper spectral fit. 
The best-fit linear relation derived using the BCES orthogonal regression 
method \citep{Akritas96} is represented by the dashed line in 
Figure~\ref{f:Lcatband-Lapbol} and is given by Eq.~2 :

\begin{equation}
  \log\ (L_{\rm ap,\,bol}) = 0.07 + 1.10\ \log\ (L_{\rm cat,\,0.5-2.0}).  
\end{equation}

\begin{figure}
\centering{
 \resizebox{100mm}{!}{\includegraphics{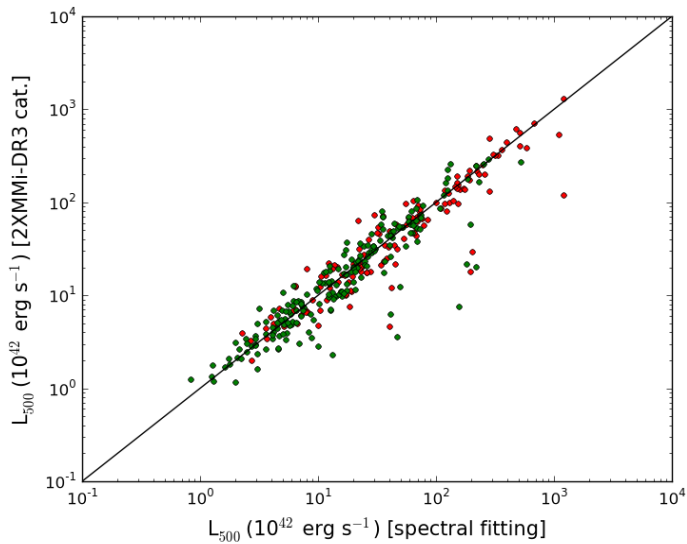}}}
 \caption[Comparison between the measured bolometric luminosity $L_{500}$ based 
on the flux given in the 2XMMi-DR3 catalogue and the bolometric $L_{500}$ based 
on the spectral fits flux for the first (red dots) and extended (green dots) 
cluster sample with X-ray spectroscopic parameters from the survey]{Comparison 
between the measured bolometric luminosity $L_{500}$ based 
on the flux given in the 2XMMi-DR3 catalogue and the bolometric $L_{500}$ based 
on the spectral fits flux for the first (red dots) and extended (green dots) 
cluster sample with X-ray spectroscopic parameters from the survey. The solid 
line shows the one-to-one relationship.} 
 \label{f:L500cat-L500sf}
\end{figure}

Using the iterative method as described above we computed bolometric $L_{500}$ 
per cluster with the redshift, aperture radius $R_{ap}$ and aperture bolometric 
luminosity $L_{\rm ap,\,bol}$ as input. The aperture radius used here is still 
the radius that is corresponding to the maximum EPIC SNR, see Section 3. 
We finally determined for each cluster $R_{500}$, $M_{500}$, and $T_{500}$ 
and the corresponding errors using the extrapolated values for $L_{500}$. 
From the comparison between the bolometric $L_{500}$ based on the catalogue 
flux and the bolometric $L_{500}$ based on the spectroscopic flux, there is 
no obvious systematic differences between the two luminosities as shown in 
Figure~\ref{f:L500cat-L500sf}. Therefore, the conversion from 
$L_{\rm cat,\,0.5-2.0}$ to $L_{\rm ap,\,bol}$ and the iterative 
procedure are acceptable.

Table~\ref{tbl:cat_sample}, a full version of this table is provided at the CDS,  
represents the first 10 entries of the X-ray selected cluster sample comprising 
185 clusters with X-ray parameters based on the flux given in the 2XMMi-DR3 
catalogue. For each cluster, the catalogue 
provides the cluster identification number (detection Id, detid) and its name 
(IAUNAME) in (cols.~[1] and [2]), the right ascension and declination of 
X-ray emission in equinox J2000.0 (cols.~[3] and [4]), the XMM-Newton 
observation Id (obsid) (col.~[5]), the optical redshift (col.~[6]), the scale 
at the cluster redshift in kpc/$''$ (col.~[7]), the $R_{500}$ in kpc 
(col.~[8]), the 2XMMi-DR3 X-ray flux $F_{cat}$ [0.5-2.0] keV and its error in 
units of $10^{-14}$\ erg\ cm$^{-2}$\ s$^{-1}$ (cols.~[9], and [10]), 
the estimated X-ray luminosity $L_{cat}$ [0.5-2.0] keV and its error 
in units of $10^{42}$\ erg\ s$^{-1}$ (cols.~[11], and [12]), 
the cluster bolometric luminosity $L_{500}$ and its error 
in units of $10^{42}$\ erg\ s$^{-1}$ (cols.~[13] and [14]), the cluster mass 
$M_{500}$ and its error in units of $10^{13}$\ M$_\odot$ (cols.~[15] and [16]), 
the $T_{500}$ and its error in units of keV (cols.~[17] and [18]),
the {\tt objid} of the likely BCG in SDSS-DR8 (col.~[19]), the BCG right 
ascension and declination in equinox J2000.0 (cols.~[20] and [21]), the 
estimated photometric and, if available,  spectroscopic redshift of the 
cluster (col.~[22] and col.~[23]), the number of 
cluster members within 560 kpc with available spectroscopic redshifts, which were
used to compute the cluster spectroscopic redshift, (col.~[24]), the redshift 
type (col.~[25]), the linear offset between the cluster X-ray position and the 
BCG position (col.~[26]), and the NED name and its references (col.~[27] and 
col.~[28]).



\subsection{Analysis of the cluster sample with reliable X-ray parameters }

We present a comparison of the measured parameters (temperatures, luminosities, 
and masses) of the cluster sample (345 systems) that have reliable X-ray 
parameters from the spectral fits with the available values in the literature.  

\subsubsection{Comparison with the XCS sample}

\begin{figure}
\centering{
  \resizebox{100mm}{!}{\includegraphics{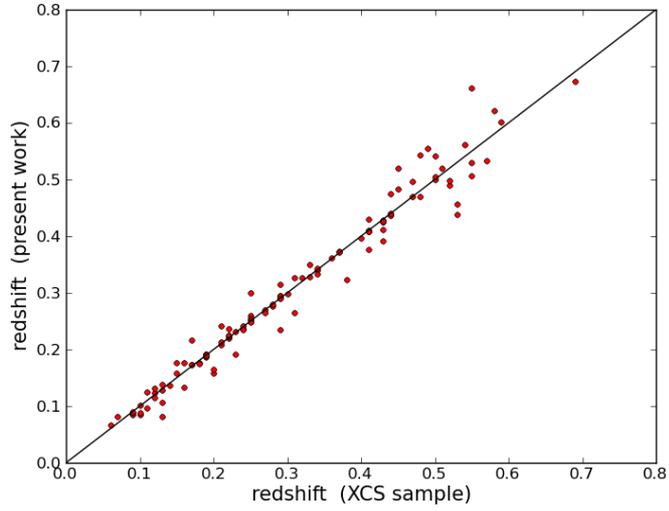}}}
  \caption[The comparison of the estimated redshifts of the common sample 
between XCS catalogue and our sample with X-ray spectroscopic parameters]{The 
comparison of the estimated redshifts of the common sample 
between XCS catalogue and our sample with X-ray spectroscopic parameters. 
The solid line in the figure indicates the unity line.} 
  \label{f:zs-zxcs}
\end{figure}

The so far largest published catalogue of X-ray clusters based on the entire 
XMM-Newton archive was compiled by the XMM Cluster Survey team
\citep[XCS,][]{Romer01, Lloyd-Davies11, Mehrtens12}. The catalogue consists 
of 503 optically confirmed clusters. Of these, 463 systems have redshifts in 
the range 0.05 to 1.46. The X-ray temperatures are measured for 401 clusters.
We cross-matched our cluster sample and the XCS sample with available 
temperature measurements within a matching 
radius of 30 arcsec which yielded 114 common clusters. About half of the 
common sample was previously published by us in Paper I. The standard 
deviation of the redshift difference $(z_{\rm XCS} - z_{\rm pre})$ is 0.027 
and thus of order of the photometric redshift accuracy. There is no systematic 
deviation as e.g.~a function of redshift present as shown in 
Figure~\ref{f:zs-zxcs}.

\begin{figure}
\centering{
  \resizebox{100mm}{!}{\includegraphics{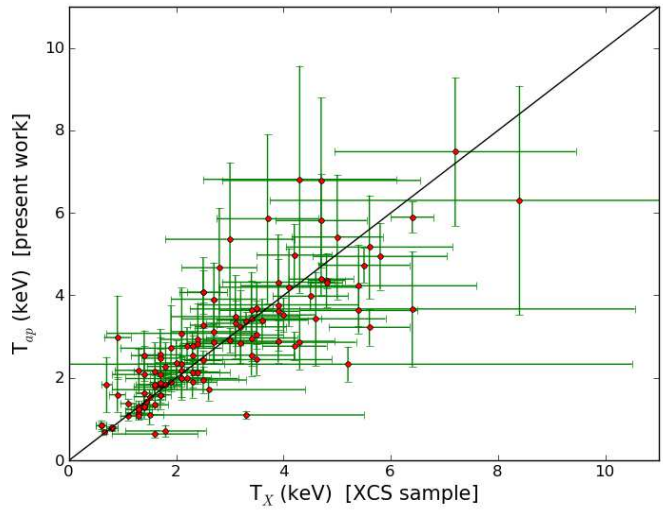}}}
  \caption[Comparison of measured temperatures between $T_{ap}$ in our sample 
and $T_{X}$ in XSC sample]{Comparison of measured temperatures between $T_{ap}$ in our sample 
and $T_{X}$ in XSC sample. The solid line shows the one-to-one relationship. 
The errors are the average errors of positive and negative errors that are 
provided by the spectral fitting procedure.} 
  \label{f:Ts-Txcs}
\end{figure}

Regarding the temperature measurements, even though we extracted the cluster 
spectra from a different aperture than the aperture used in the XCS project and 
using a different spectral fitting procedure, in general there is agreement
between the two measurements. Figure~\ref{f:Ts-Txcs} shows the comparison of
the measured temperatures from  the two projects. We plot the symmetric errors
of each temperature as the average of the positive and negative errors. Our
procedure reveals a slightly smaller temperature uncertainty than derived in
the XCS with a median 
$\bigtriangleup T_{\rm pre}/\bigtriangleup T_{\rm XCS} = 0.84$. 
The differences between the two temperature measurements have a mean of 
0.02 keV and a standard deviation of 0.93 keV that is comparable with the 
standard deviation, 0.82 keV, of the error measurements in temperatures of 
the XCS sample.

\begin{figure}
\centering{
  \resizebox{100mm}{!}{\includegraphics{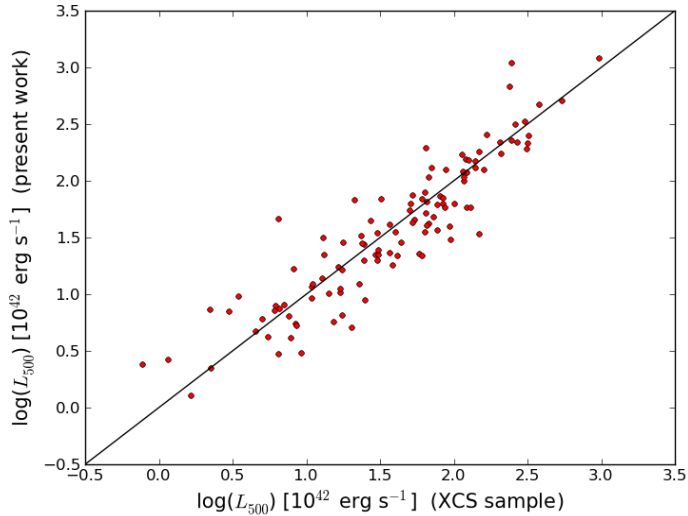}}}
  \caption[Comparison between the bolometric luminosities, $L_{500}$, from the 
present work and the corresponding ones from the XCS sample]{Comparison between 
the bolometric luminosities, $L_{500}$, from the 
present work and the corresponding ones from the XCS sample. The solid line 
shows the one-to-one relationship. } 
  \label{f:Ls-Lxcs}
\end{figure}

In the XCS project, the cluster luminosity $L_{500}$ was calculated by using an 
analytical function of $\beta$ model fitted to the surface brightness profile. 
The same profile was used to determine a scaling factor between the aperture 
luminosity and $L_{500}$ \citep{ Lloyd-Davies11}.          
Our procedure of the extrapolation was described above and in more detail in
Paper I. We find a good agreement between both determinations of $L_{500}$
as shown in Figure~\ref{f:Ls-Lxcs}. The ratio between the current luminosity 
measurements to that of the XCS has a median of 0.93.

\subsubsection{Comparison with the MCXC sample }

The MCXC catalogue, a meta-catalogue of X-ray detected clusters of galaxies, 
is compiled from published ROSAT All Sky Survey-based and serendipitous cluster 
catalogues \citep{Piffaretti11}. The catalogue comprises 1743 clusters that
span  a wide redshift range up to 1.3. For each cluster the catalogue lists
redshift, luminosity $L_{500}$ in the 0.1-2.4 keV band, total mass $M_{500}$,  
and radius $R_{500}$. 
Within a cross-matching radius of the cluster centres of 30 arcsec there are
only 23 common clusters. The small overlap is mainly due to our small survey
area and our strategy to investigate serendipitous clusters only, not cluster
targets.

\begin{figure}
\centering{
  \resizebox{100mm}{!}{\includegraphics{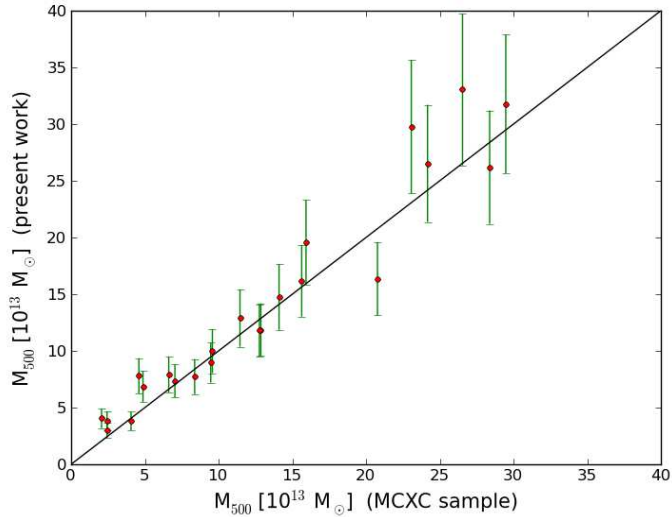}}}
  \caption[Comparison of our sample mass estimations within $R_{500}$ with the 
estimated values from the MCXC catalogue]{Comparison of our sample mass 
estimations within $R_{500}$ with the 
estimated values from the MCXC catalogue. The solid line represents the 
one-to-one relationship.} 
  \label{f:Ms-Mmcxc}
\end{figure}

We compared the masses of the common sample in  Figure~\ref{f:Ms-Mmcxc} and
found consisting results. This comparison made sure that our mass measurements 
are reliable and not affected by any systematic bias. We also found consistency 
between the redshifts used in both catalogues.

\subsubsection{Comparison with the sample from Paper I}

Since we have developed an algorithm to estimate the redshifts of the X-ray 
cluster candidates, the redshifts of the first cluster sample from the 
survey, Paper I, were revised, as discussed in Section 2. We also revised 
the X-ray spectroscopic parameters for the first cluster sample. The common 
sample between the current sample with reliable X-ray parameters and the first 
cluster sample consists of 141 systems. The remaining 34 clusters from Paper I 
did not pass the quality criterion applied in the present work. 
Those missed clusters are nevertheless included in the published cluster
catalogue from this paper with less reliable parameters (see above).

\begin{figure}
\centering{
  \resizebox{100mm}{!}{\includegraphics{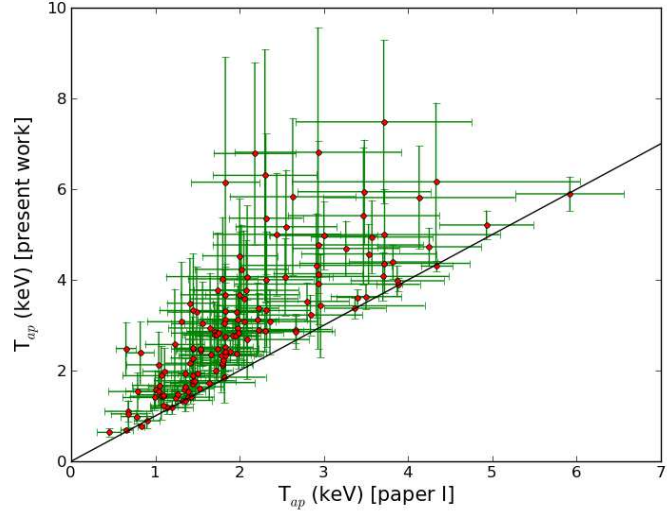}}}
  \caption[The current X-ray aperture temperature estimations are plotted 
against the corresponding ones of the first cluster sample from Paper I]{The 
current X-ray aperture temperature estimations are plotted 
against the corresponding ones of the first cluster sample from Paper I. 
For ease comparison we plot the unity line.} 
  \label{f:kT1-kT2}
\end{figure}

We found a systematic bias of the temperature measurements of the sample in
Paper I and the current sample as shown in Figure~\ref{f:kT1-kT2}. When investigating 
possible reasons for the discrepancy we realized that the X-ray data in Paper
I were analysed in a non-appropriate manner. The X-ray spectra were not 
grouped and  binned before a spectral model was applied. This led to 
a systematic shift of many of the derived temperatures determined in 
Paper I towards too low values.

\begin{figure}
\centering{
  \resizebox{100mm}{!}{\includegraphics{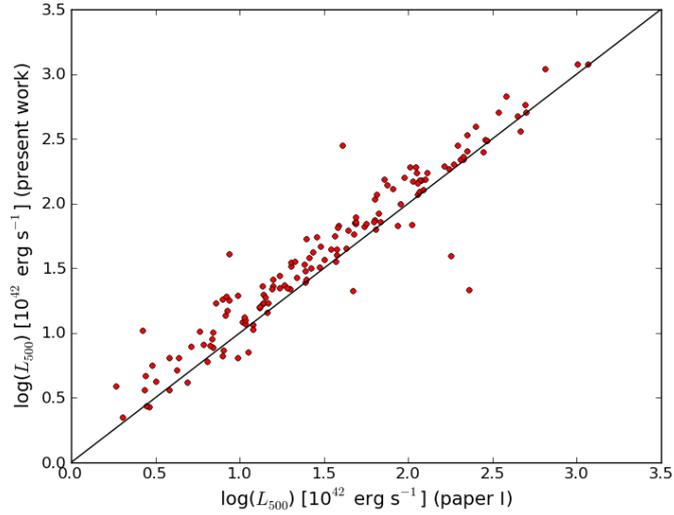}}}
  \caption[Comparison between the bolometric luminosities, $L_{500}$, from the 
present work and the corresponding ones from Paper I]{Comparison between the
 bolometric luminosities, $L_{500}$, from the 
present work and the corresponding ones from Paper I. The solid line shows the 
one-to-one relationship.} 
  \label{f:Lp1-Lp2}
\end{figure}

As a consequence the luminosities were biased towards smaller values
(Figure~\ref{f:Lp1-Lp2}) by a factor of 20 percent. Revised redshifts 
and in some cases revised spectral extraction regions led to a few 
outliers in that figure.

We also presented in Paper I the \ltr of the first cluster sample, which we
now regard as affected by the underestimated X-ray temperatures. We are
confident through several sanity checks that our updated temperatures and 
luminosities are reliable and we re-determine the \ltr relation based on 
the much enlarged sample in the next subsection.

\subsection{The \ltr relation of the cluster sample with reliable X-ray 
parameters}

Based on the cluster sample with X-ray spectroscopic parameters, we investigate 
the \ltr relation as well as the evolution of its slope and intrinsic scatter 
as presented in the following subsections.

\subsubsection{The \ltr relation of the full sample}

The bolometric luminosities $L_{500}$ and the aperture temperatures $T_{ap}$ based on 
X-ray spectral fits were used to investigate the $L_{500}-T_{ap}$ relation 
for the cluster sample with reliable X-ray parameters. 
We note that we could not make an attempt to excise the cores in most cases 
of the cluster sample which is due to the rather low resolution of the X-ray 
optics of the XMM-Newton telescopes, not too long exposure times and the 
rather large distance of most of our clusters. This caveat needs to be 
made when comparing our results with those in the literature which are 
partly based on nearby clusters with Chandra follow-up.

Figure~\ref{f:L-T} shows the relation between the measured X-ray bolometric
luminosity, $L_{500}$, modified with the evolution parameter for self-similar
evolution and the X-ray aperture temperature, $T_{ap}$. We used the BCES
orthogonal regression method \citep{Akritas96} to derive the best-fit linear
relation between the logarithms of $L_{500}$ and $T_{ap}$ taking into account
their errors as well as the intrinsic scatter. The best fit is shown in 
Figure~\ref{f:L-T} and is given by Eq.~3:  

\begin{equation}
\log\ (h(z)^{-1}\ L_{500}) = (44.39 \pm 0.06)  + (2.80 \pm 0.12)\ \log\ (T_{ap}/5).
\end{equation} 

where $h(z)$  is the Hubble constant normalised to its present-day value, $
h(z) = \bigl[\Omega_{\rm M} (1+z)^{3} + \Omega_{\Lambda}\Bigr]^{1/2}$, $L_{500}$
in erg s$^{-1}$, and $T_{ap}$ in keV.
By an analysis of common objects between our list and that of the XCS 
we have shown that our $T_{ap}$ and $L_{500}$ compare well with $T_{\rm X}$ 
and $L_{500}$ of the XCS sample. Not unexpectedly, the slopes and intercepts 
of the corresponding \ltr relations in three redshift bins agree with each 
other within 1-2 $\sigma$ (see subsection 4.4.2).

\begin{figure}[t]
\centering{
  \resizebox{\hsize}{!}{\includegraphics[viewport=10  5  525 400, clip]{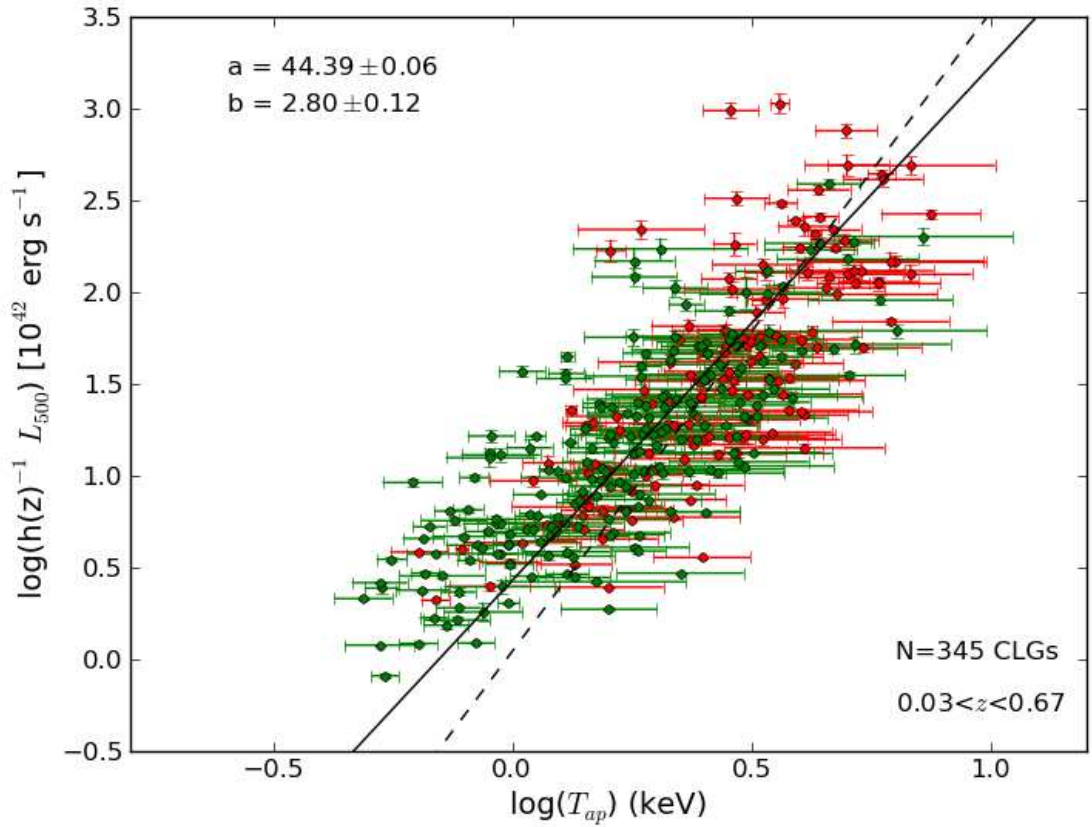}}}
  \caption[The X-ray bolometric luminosity, $L_{500}$ , is plotted against 
aperture X-ray spectroscopic temperature, $T_{ap}$, for the first
(red dots) and expanding (green dots) cluster sample with reliable X-ray 
parameters]{The X-ray bolometric luminosity, $L_{500}$ , is plotted against 
aperture X-ray spectroscopic temperature, $T_{ap}$, for the first
(red dots) and expanding (green dots) cluster sample with reliable X-ray parameters. 
The solid line is the fit to both samples using the BCES orthogonal regression.
The intercept, a, and the slope, b, of the fitted line are written in the upper 
left legend of the figure while the sample number, N, and its redshift range are 
written in the lower right side. The dashed line represents the relation fit 
of the first cluster sample using the current parameter measurements.}
  \label{f:L-T}
\end{figure}

Relations between the luminosity and the temperature, $L_{500}-T_{500}$, 
were published for the REXCESS and HIFLUGCS sample \citep{Pratt09,Mittal11}. 
The REXCESS sample 
comprises 31 nearby ($ z < 0.2 $) galaxy clusters with temperature range 
from 2 to 9 keV which have been observed with the XMM-Newton. The HIFLUGCS
sample comprises the 64 brightest galaxy clusters in the sky with $kT \ge
1$\,keV and $ z \le 0.2 $, with high quality {\tt Chandra} data. In both
samples the core emission could be excised but \ltr were published for the
non-excised data as well. The REXCESS team used a fitting procedure similar to
us, the HIFLUGCS sample was fitted with a BCES-bisector routine.

The present slope of the relation in Eq.~(3), 2.80 $\pm$ 0.12, is slightly
lower than that from the REXCESS sample, 3.35 $\pm$ 0.32, but still within
1.8$\sigma$. We also found the present slope in agreement with the slope given
by \citet{Mittal11}, $2.94 \pm 0.16$. 

The current cluster sample includes groups with much lower luminosity than
REXCESS and HIFLUGCS, which might influence the slope of the \ltr relation.
If we exclude systems with luminosities $ L_{500} < 5 \times 10^{42}$ erg
s$^{-1}$, the slope of the relation becomes $3.07 \pm 0.19$, in much closer
agreement with the published ones for REXCESS, HIFLUGCS, and XCS samples.  
The normalization of the relation, 
$44.46 \pm 0.07$, is still much lower than the one, $44.85 \pm 0.06$, 
of the REXCESS sample. This is due to the much wider temperature range 
covered by the current large sample. In addition to establishing the current 
relation based on aperture temperatures that are in general slightly higher 
than the temperatures at $R_{500}$. We found that the median scaling factor 
of $T_{ap}$ and $T_{500}$ of the full sample, $T_{ap}/T_{500}$,  was 1.2, where  
$T_{500}$ is the the predicted temperature based on the $L_{500}-T_{500}$ 
relation by \cite{Pratt09} using our value for $L_{500}$.  
%

\cite{Eckmiller11} found the slope of \ltr relation of galaxy groups 
(26 systems, $ L_x \sim 1 - 26 \times 10^{42}$\, erg s$^{-1}$ ) is slightly 
shallower than the one derived for clusters (HIFLUGCS), but they are still 
consistent within the errors. 
They also found no significant change of the slope derived from a sample 
combining groups and clusters than clusters only, which is consistent with 
the results by \cite{Osmond04}. We found the slope 
derived from the current sample (including groups and clusters) is in 
good agreement with the slope obtained from clusters only (HIFLUGCS sample) 
but it is lower than the slope of REXCESS sample.

The current slope of the \ltr relation is significantly lower than the one 
published in Paper I, $3.41\pm 0.15$. The redshifts, temperatures, and 
luminosities of the previous sample were revised. Using the updated values 
we still find a rather steep slope of $3.48 \pm 0.30$ thus confirming
the initial result (the new fit is shown with a dashed line in
Figure~\ref{f:L-T}). 
The much shallower slope found here based on 
the full sample is clearly due to the inclusion of new objects that 
have a wider temperature and luminosity range. As discussed above, when 
excluding the low luminosity systems from the full sample the slope 
becomes steep, $3.07 \pm 0.19$, which is consistent within 1.4$\sigma$ 
with the updated slope, $3.48 \pm 0.30$, for the Paper I sample.

To estimate the intrinsic scatter in the \ltr relation, we followed the method 
used by \citet{Pratt09}. First we estimated the raw scatter using the 
error-weighted orthogonal distances to the regression line  \citep[Eqs.~3 and
  4 in][]{Pratt09}. Then we computed the intrinsic scatter as the mean value
of the quadratic differences between the raw scatters and the statistical
errors.  The error of the intrinsic scatter was computed as the standard error
of its  value. The computed intrinsic scatter value of the relation, 
$0.48 \pm 0.03$, is slightly higher than the value of REXCESS sample, 
$0.32 \pm 0.06$.

The updated \ltr relation is derived for the first time from  a sample comprising 
345 clusters drawn from a single survey and spans a wide redshift range 
($0.03 < z < 0.67$). Of these, 210 clusters have spectroscopic redshifts for at 
least one cluster member galaxy. The redshifts and the X-ray parameters of the 
sample are measured in a consistent way. The sample has X-ray spectroscopic 
temperature measurements from 0.5 to 7.5 keV and bolometric luminosity range 
$L_{500} \sim 1.0 \times 10^{42} - 1.0 \times 10^{45}$\ erg s$^{-1}$.

Based on the SDSS we could identify only about half of our X-ray cluster
candidates. The other 50 percent probably represents a more luminous 
population. The omission of that subsample may have a yet to be quantified 
influence on the \ltr relation. However including luminous distant clusters 
does not have a significant effect on the slope of the \ltr relation 
\citep{Hilton12} as described in the subsequent section.  
Also the current sample does not include distant clusters beyond $z=0.7$, 
therefore we defer the measurement of the evolution of the normalisation of 
the \ltr relation to a future study.



\subsubsection{Evolution of the slope and intrinsic scatter}

Based on the first data release of the XCS, \citet{Hilton12} investigated a
possible evolution of the slope and intrinsic scatter of the \ltr relation
in three redshift bins. A sample of 211 clusters with spectroscopic redshift
up to 1.5 was used for this exercise. No evidence for evolution in
either the slope or intrinsic scatter as a function of redshift was found.

Using our much larger sample of clusters with measured X-ray spectroscopic 
parameters we further investigate a possible evolution of the mentioned 
parameters of the \ltr relation.
We divided our sample into three subsamples with similar redshift
bins as used by \citet{Hilton12}, $0.03 \le z < 0.25, 0.25 \le z < 0.5$, and
$0.5 \le z < 0.7$. The numbers of clusters per redshift bin are listed in
Table~\ref{tbl:LTrelation}. Our two low redshift subsamples are about twice 
as large as the XCS corresponding subsamples. The number of clusters in the 
high redshift bin are comparable, however the XCS comprises clusters up to 
redshift 1.5. In general, there are about 75 common clusters between our 
sample and the XCS-DR1 sample which were used to derive the \ltr relation. 
Of these common clusters, 44 systems were published from our survey in 
Paper I.

\begin{figure}[t]
\centering{ 
  \resizebox{\hsize}{!}{\includegraphics[viewport=15  95  570 320, clip]{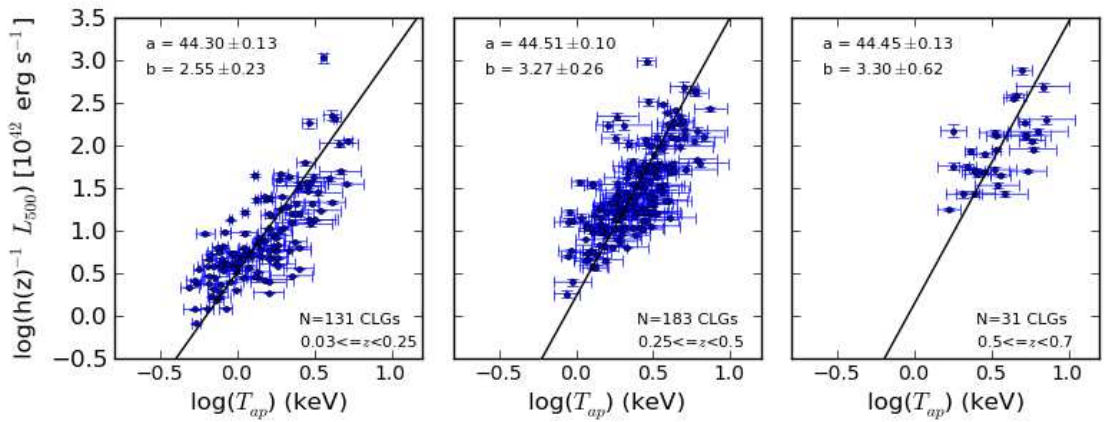}}}
  \caption[$L_{500}-T_{ap}$ relations for the three subsamples in redshift 
bins]{$L_{500}-T_{ap}$ relations for the three subsamples in redshift bins. 
The redshift bin and the cluster number of these subsamples are written in the 
lower right of the figure. The best fit line of the subsamples is presented 
by the solid line, which their parameters (intercepts, a, and slopes, b) are 
written in the upper left in the figure.} 
  \label{f:L-T-zbins}
\end{figure}

The \ltr relations of our subsamples are shown in Figure~\ref{f:L-T-zbins}. 
When fitting these susamples using the BCES orthogonal regression method, 
we find that the relation slope of the subsamples in the intermediate and 
high redshift bins are consistent while the subsample in the lowest redshift 
bin has a shallower slope. The reason is that the low redshift subsample 
includes groups/clusters with low temperature and luminosity, which produces a
shallower slope. Also the present slope of the low redshift subsample is lower
than the published one of the corresponding XCS-DR1 subsample and those 
of the REXCESS and HIFLUGCS samples. On the other
hand, the slopes of the intermediate and high redshift subsamples are in
agreement with the slopes of the corresponding XCS-DR1 subsamples.
The intrinsic scatter of all subsamples agree with each other.
Table~\ref{tbl:LTrelation} also lists the fitted parameters (intercept and
slope) of the \ltr relations, 
their intrinsic scatter, together with published values (slope, sample size,
reference). 

If we fit the \ltr for the low redshift subsample after excluding the groups 
with low luminosity (i.e. $ L_{500} < 5 \times 10^{42}$ erg s$^{-1}$), the 
slope of the relation becomes $2.86 \pm 0.41$, which is in agreement with the 
intermediate and high redshift subsamples as well as the corresponding 
published slopes given in Table~\ref{tbl:LTrelation}.
We thus confirm the finding by \cite{Hilton12} that the \ltr relation 
does not show a significant change of its slope and its intrinsic scatter 
as a function of redshift.

\begin{table}
\caption[The fit parameters of the $L_{500}-T_{ap}$ relation, derived from the 
BCES orthogonal regression method, for the three subsamples in redshift bins]{The 
fit parameters of the $L_{500}-T_{ap}$ relation, derived from the 
BCES orthogonal regression method, for the three subsamples in redshift bins. 
The fitted model is $\log\ (h(z)^{-1}\ L_{500}) = a  + b\ \log\ (T_{ap}/5)$, 
and the fit parameters (a and b) are also shown in the legend of 
Figure~\ref{f:L-T-zbins}.} 
\label{tbl:LTrelation}     
\centering
\scriptsize{                        
\begin{tabular}{c c c c c c c c c}       
\hline\hline                        
     redshift range & N$_{CLGs}$ & intercept & current slope &  $\sigma_{log}L_{500}$  &published slope &  N$_{CLGs, pub.}$ & ref. \\   
\hline  
 0.03 $\le$ z $<$ 0.25  & 131 &44.30 $\pm$ 0.13 & 2.55 $\pm$ 0.23 & 0.45 $\pm$ 0.04 &  3.18 $\pm$ 0.22 & 96 & 1 \\ 
                        &     &                 &                 &                 &  3.35 $\pm$ 0.32 & 31 & 2 \\ 
                        &     &                 &                 &                 &  2.94 $\pm$ 0.16 & 64 & 3 \\                            
 0.25 $\le$ z $<$ 0.50  & 183 &44.51 $\pm$ 0.10 & 3.27 $\pm$ 0.26 & 0.49 $\pm$ 0.04 &  2.82 $\pm$ 0.25 & 77 & 1 \\ 
 0.50 $\le$ z $<$ 0.70  & 31  &44.45 $\pm$ 0.13 & 3.30 $\pm$ 0.62 & 0.41 $\pm$ 0.07 &  2.89 $\pm$ 0.45 & 38 & 1 \\ 
\hline 
\end{tabular}
}
\tablebib{1- \cite{Hilton12}; 2- \cite{Pratt09}; 3- \cite{Mittal11}.}
\end{table}

\subsection{The distribution of the luminosity with redshift}

\begin{figure}[t]
\centering{
  \resizebox{\hsize}{!}{\includegraphics{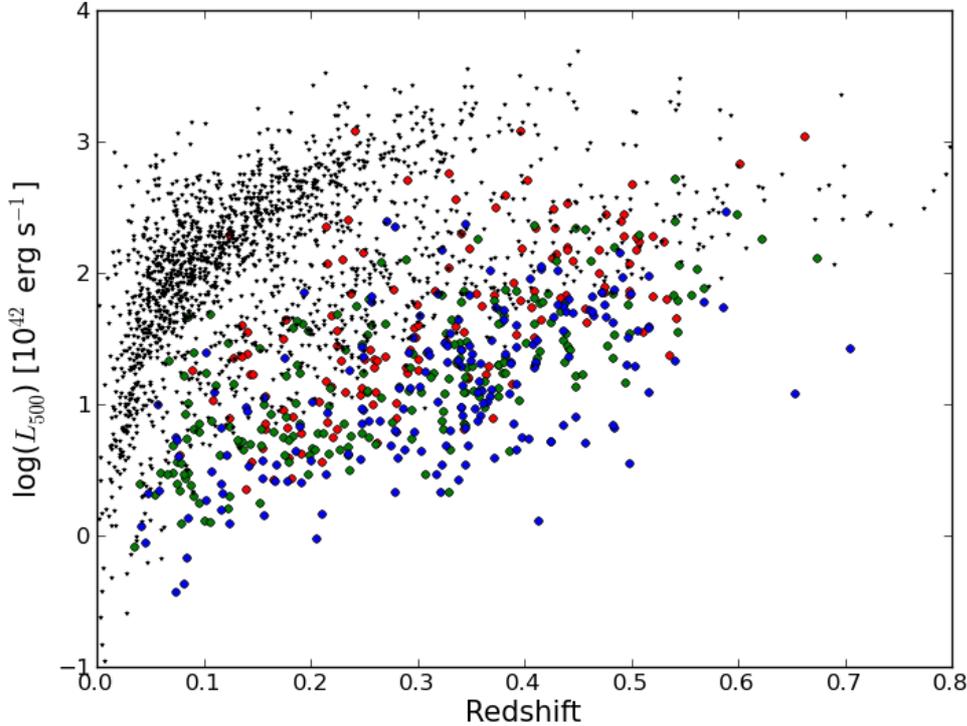}}}
  \caption{The distribution of the estimated bolometric luminosity, $L_{500}$,  
as a function of the redshift for the first (red dots) and extended (green dots) 
cluster sample with X-ray spectroscopic parameters, the cluster sample 
(blue dots) with X-ray parameters based on the given flux in the 2XMMi-DR3 
catalogue, and the MCXC cluster sample (black stars) \citep{Piffaretti11}.} 
  \label{f:L-z-all}
\end{figure}

Figure~\ref{f:L-z-all} shows the distribution of the bolometric luminosity
$L_{500}$ as a function of the redshift for all 530 clusters with redshifts
that were determined in the present work. 
Included are also the 1730 systems from the MCXC catalogue below redshift 0.8.
The X-ray luminosity $L_{500}$ in $0.1-2.4$ keV of the MCXC sample was 
converted to the bolometric luminosity $L_{500}$ by assuming the factor 
$L_{\rm bol,\,500} / L_{0.1-2.4,\,500} = 1.3$. This factor was derived as 
a median of $L_{\rm bol,\,500} / L_{0.1-2.4,\,500}$ for the 23 common systems 
between the cluster sample with reliable parameters from the spectral fitting 
and MCXC cataloge.

It is clearly obvious that our X-ray selected samples extend to include 
groups and clusters with low luminosity. The sensitivity of XMM-Newton 
and deeper exposures for some fields allow us to detect less luminous 
clusters over the redshift range as shown in Figure~\ref{f:L-z-all}.

\section{Summary and outlook}

We have presented the optically confirmed cluster sample of 530 galaxy 
groups and clusters from the 2XMMi/SDSS Galaxy Cluster Survey. The survey 
consists of 1180 X-ray cluster candidates with at least 80 net photon counts 
selected from the second XMM-Newton serendipitous source catalogue (2XMMi-DR3), 
which are in the footprint of the SDSS-DR7. The survey area is 210 deg$^2$ 
considering the XMM-Newton field of view has a radius of 15 arcmin.
We developed a finding algorithm to detect the optical counterparts of 
the X-ray cluster candidates and to constrain their redshifts using the 
photometric and, if available, the spectroscopic redshifts of surrounding 
galaxies from the SDSS-DR8 data. The cluster is recognized if there are at 
least 8 member galaxies within a radius of 560 kpc from the X-ray emission 
peak with photometric redshift in the redshift interval of the redshift of 
the likely identified BCG, $z_{\rm p,\,BCG} \pm 0.04(1+z_{\rm p,\,BCG})$.
 
The cluster photometric and spectroscopic redshift is measured as the 
weighted average of the photometric and the available spectroscopic 
redshifts, respectively, of the cluster galaxies within 560 kpc from 
the X-ray position. The measured redshifts are in a good 
agreement with available redshifts in the literature, to date 301 
clusters are known as optically selected clusters with redshift measurements.
Also, 310 clusters of the optically confirmed cluster sample have spectroscopic 
redshifts for at least one cluster member. The measured photometric redshifts 
are in a good agreement with the measured spectroscopic ones from the survey. 
The cluster redshifts of the optically confirmed cluster sample span  
a wide redshift range from 0.03 to 0.70. We reduced and analysed the X-ray 
data of this sample in an automated way to compute their X-ray properties. 

We present a cluster catalogue from the survey comprising 345 X-ray 
selected groups and clusters with their X-ray parameters derived from the 
spectral fits including the published sample in Paper I. 
In addition to the best fitting parameters, we estimated 
the physical properties ($R_{500}$, $L_{500}$ and $M_{500}$) of this sample
from an iterative procedure based on published scaling relations.
We investigated the \ltr relations for the first time based on a large 
cluster sample with X-ray spectroscopic parameters drawn from a single survey.
The current sample includes groups and clusters with wide ranges of 
temperatures and luminosities. The slope of the relation is consistent with 
the published ones of clusters with high temperatures and luminosities.
After excluding the low luminosity  groups, we find no significant change 
of the slope and the intrinsic scatter of the relation with redshift when 
dividing the sample into three redshift bins. When including the low luminosity  
groups in the low redshift subsample, the slope is no longer consistent with the 
intermediate and high redshift subsamples.

In addition to the cluster sample with X-ray spectroscopic data, we present the 
remainder of the optically confirmed cluster sample with their X-ray parameters 
based on the given flux in the 2XMMi-DR3 catalogue. We used 
the 2XMMi-DR3 flux because of their low quality X-ray data, which is not 
sufficient to perform the spectral fitting. This sample comprises 185 groups 
and clusters with their fluxes and luminosity in the energy band 0.5-2.0 keV 
and their physical parameters ($R_{500}$, $L_{500}$, $M_{500}$, and $T_{500}$).

This is the largest X-ray selected cluster catalogue to date based on  
XMM-Newton observations. It comprises 530 clusters with their optical and 
X-ray properties, spanning the redshift range $0.03 < z < 0.70$. 
More than 75 percent of the cluster sample are newly discovered clusters 
in X-ray wavelengths. About 40 percent of the sample are new systems to the 
literature according to current entries in the NED.      

In the future we plan to study the remainder of the X-ray cluster candidates, which 
were not detected by the current detection algorithm. They are either poor or 
at high redshifts. For the distant clusters, we plan follow-up by imaging and 
spectroscopy. For those X-ray cluster candidates that have galaxy members detected 
in SDSS imaging and not be identified by the current algorithm, we plan to improve  
the current finding algorithm to constrain their redshifts. The new sample from the 
survey especially the distant ones will allow us to investigate 
the evolution of \ltr relation and X-ray-optical relations.

\clearpage
\begin{landscape}
\begin{table}
\caption{The first 10 entries of the X-ray selected 
group/cluster sample (345 objects) from the 2XMMi/SDSS Galaxy Cluster Survey 
with X-ray parameters from the spectral fitting.}
\label{tbl:SF_sample}
{\tiny
\begin{tabular}{c c c c c c c c c c c c c c c c c c}
\hline \hline
  \multicolumn{1}{c}{detid\tablefootmark{a}} &
  \multicolumn{1}{c}{Name\tablefootmark{a}} &
  \multicolumn{1}{c}{ra\tablefootmark{a}} &
  \multicolumn{1}{c}{dec\tablefootmark{a}} &
  \multicolumn{1}{c}{obsid\tablefootmark{a}} &
  \multicolumn{1}{c}{z\tablefootmark{b}} &
  \multicolumn{1}{c}{scale} &
  \multicolumn{1}{c}{$R_{\rm ap}$} &
  \multicolumn{1}{c}{$R_{\rm 500}$} &
  \multicolumn{1}{c}{$T_{\rm ap}$} &
  \multicolumn{1}{c}{$+eT_{\rm ap}$} &
  \multicolumn{1}{c}{$-eT_{\rm ap}$} &
  \multicolumn{1}{c}{$F_{\rm ap}$\tablefootmark{c}} &
  \multicolumn{1}{c}{$+eF_{\rm ap}$} &
  \multicolumn{1}{c}{$-eF_{\rm ap}$} &
  \multicolumn{1}{c}{$L_{\rm ap}$\tablefootmark{d}} &
  \multicolumn{1}{c}{$+eL_{\rm ap}$} &
  \multicolumn{1}{c}{$-eL_{\rm ap}$}  \\

  \multicolumn{1}{c}{} &
  \multicolumn{1}{c}{IAUNAME} &
  \multicolumn{1}{c}{(deg)} &
  \multicolumn{1}{c}{(deg)} &
  \multicolumn{1}{c}{} &
  \multicolumn{1}{c}{} &
  \multicolumn{1}{c}{kpc/$''$} &
  \multicolumn{1}{c}{(kpc)} &
  \multicolumn{1}{c}{(kpc)} &
  \multicolumn{1}{c}{(keV)} &
  \multicolumn{1}{c}{(keV)} &
  \multicolumn{1}{c}{(keV)} &
  \multicolumn{1}{c}{} &
  \multicolumn{1}{c}{} &
  \multicolumn{1}{c}{} &
  \multicolumn{1}{c}{} &
  \multicolumn{1}{c}{} &
  \multicolumn{1}{c}{} \\

(1)  &  (2)  &  (3)  & (4)  &   (5)   &  (6)  &   (7)   &   (8)   &  (9)   &  (10)    &  (11) &  (12) & (13) & (14) & (15) &  (16)  & (17)  &  (18)   \\
\hline 

002294 &     2XMM J001817.2+161740 &   4.57190 &  16.29470 & 0111000101 & 0.54 & 6.35 & 476.50 &  810.94 & 4.57 & 0.78 & 0.60 &  16.74 & 0.58 & 0.79 & 144.72 &  5.58 &  5.52 \\
004444 &     2XMM J003318.4-212447 &   8.32687 & -21.41319 & 0044350101 & 0.19 & 3.17 & 161.44 &  579.22 & 2.25 & 0.66 & 0.40 &   3.75 & 0.28 & 0.27 &   3.66 &  0.19 &  0.21 \\
005825 &     2XMM J003917.9+004200 &   9.82489 &   0.70013 & 0203690101 & 0.28 & 4.24 & 152.81 &  483.64 & 1.43 & 0.77 & 0.29 &   0.88 & 0.06 & 0.06 &   2.10 &  0.13 &  0.17 \\
005842 &     2XMM J003922.4+004809 &   9.84343 &   0.80269 & 0203690101 & 0.41 & 5.49 & 395.23 &  618.80 & 4.02 & 0.64 & 0.52 &   3.55 & 0.08 & 0.08 &  18.33 &  0.44 &  0.48 \\
005901 &     2XMM J003942.2+004533 &   9.92584 &   0.75919 & 0203690101 & 0.42 & 5.50 & 247.44 &  589.18 & 2.35 & 0.43 & 0.33 &   2.51 & 0.12 & 0.08 &  14.02 &  0.67 &  0.55 \\
006070 &     2XMM J004039.2+253106 &  10.16344 &  25.51840 & 0153030101 & 0.15 & 2.64 & 142.48 &  632.85 & 1.51 & 0.13 & 0.10 &  10.55 & 0.49 & 0.41 &   6.37 &  0.33 &  0.22 \\
006469 &     2XMM J004156.8+253151 &  10.48690 &  25.53105 & 0153030101 & 0.13 & 2.28 & 150.73 &  579.30 & 3.18 & 1.09 & 0.77 &   5.21 & 0.19 & 0.34 &   2.10 &  0.10 &  0.09 \\
006920 &     2XMM J004231.2+005114 &  10.63008 &   0.85401 & 0090070201 & 0.16 & 2.73 & 114.55 &  501.99 & 2.16 & 0.92 & 0.47 &   1.37 & 0.10 & 0.09 &   0.89 &  0.07 &  0.04 \\
007340 &     2XMM J004252.6+004259 &  10.71952 &   0.71650 & 0090070201 & 0.27 & 4.13 & 421.41 &  579.12 & 3.12 & 0.90 & 0.61 &   4.14 & 0.19 & 0.15 &   8.45 &  0.44 &  0.34 \\
007362 &     2XMM J004253.7-093423 &  10.72397 &  -9.57311 & 0065140201 & 0.41 & 5.43 & 260.60 &  613.30 & 3.29 & 1.25 & 0.74 &   3.03 & 0.20 & 0.24 &  15.14 &  1.16 &  0.94 \\

\hline
\end{tabular}
}
\end{table}


\addtocounter{table}{-1}
\begin{table}
\caption{\label{} continued.}
{\tiny
\begin{tabular}{c c c c c c c c c c c c c c c c}
\hline
\hline
  \multicolumn{1}{c}{detid\tablefootmark{a}} &
  \multicolumn{1}{c}{$L_{500}$\tablefootmark{e}} &
  \multicolumn{1}{c}{$\pm eL_{500}$} &
  \multicolumn{1}{c}{$M_{500}$\tablefootmark{f}} &
  \multicolumn{1}{c}{$\pm eM_{500}$} &
  \multicolumn{1}{c}{nH\tablefootmark{g}} &
  \multicolumn{1}{c}{objid\tablefootmark{h}} &
  \multicolumn{1}{c}{RA\tablefootmark{h}} &
  \multicolumn{1}{c}{Dec\tablefootmark{h}} &
  \multicolumn{1}{c}{$\bar{z}_{\rm p}$\tablefootmark{h}} &
  \multicolumn{1}{c}{$\bar{z}_{\rm s}$\tablefootmark{h}} &
  \multicolumn{1}{c}{$N_{z_{\rm s}}$\tablefootmark{h}} &
  \multicolumn{1}{c}{$z_{\rm type}$\tablefootmark{h}} &
  \multicolumn{1}{c}{offset\tablefootmark{h}} &
  \multicolumn{1}{c}{NED-Name} &
  \multicolumn{1}{c}{ref.}  \\

  \multicolumn{1}{c}{} &
  \multicolumn{1}{c}{} &
  \multicolumn{1}{c}{} &
  \multicolumn{1}{c}{} &
  \multicolumn{1}{c}{} &
  \multicolumn{1}{c}{} &
  \multicolumn{1}{c}{(BCG)} &
  \multicolumn{1}{c}{(deg)} &
  \multicolumn{1}{c}{(deg)} &
  \multicolumn{1}{c}{} &
  \multicolumn{1}{c}{} &
  \multicolumn{1}{c}{} &
  \multicolumn{1}{c}{} &
  \multicolumn{1}{c}{(kpc)} &
  \multicolumn{1}{c}{} &
  \multicolumn{1}{c}{}  \\

  (1)  &  (19)  & (20)  &  (21)  &  (22)   &  (23) &  (24)  &   (25)   &  (26) & (27) & (28)  & (29) & (30) & (31) & (32) & (33)   \\
\hline

002294 &  521.28 &  31.45 & 27.28 &  5.24 & 0.0393 & 1237679454926995783 &   4.57107 &  16.29433 & 0.54 & 0.00 &  0 & photo-z &  20.49 &                     RX J0018.2+1617 &                  1,2 \\
004444 &   17.49 &   0.71 &  6.67 &  1.35 & 0.0153 & 1237673016766496932 &   8.32630 & -21.41445 & 0.19 & 0.00 &  0 & photo-z &  17.22 &                                   - &                    - \\
005825 &    7.80 &   0.28 &  4.28 &  0.91 & 0.0198 & 1237663204918493337 &   9.82501 &   0.69981 & 0.27 & 0.28 &  1 &  spec-z &   5.14 &       
SDSS CE J009.833157+00.701518 &                3,4,5 \\
005842 &   59.67 &   2.50 & 10.46 &  2.03 & 0.0197 & 1237663204918493446 &   9.84605 &   0.79222 & 0.39 & 0.41 &  3 &  spec-z & 213.01 &                                   - &                    - \\
005901 &   44.13 &   2.10 &  9.04 &  1.78 & 0.0195 & 1237663204918493223 &   9.92730 &   0.76163 & 0.40 & 0.42 &  2 &  spec-z &  56.24 &                WHL J003942.5+004541 &                  5,6 \\
006070 &   26.71 &   1.08 &  8.36 &  1.65 & 0.0368 & 1237678580906524886 &  10.16314 &  25.51779 & 0.15 & 0.00 &  0 & photo-z &   6.62 &                                   - &                    - \\
006469 &   14.21 &   0.07 &  6.26 &  1.27 & 0.0384 & 1237680071245365404 &  10.48821 &  25.52932 & 0.13 & 0.00 &  0 & photo-z &  18.10 &                                   - &                    - \\
006920 &    6.43 &   0.06 &  4.20 &  0.89 & 0.0179 & 1237663716555882709 &  10.63094 &   0.85020 & 0.15 & 0.16 &  4 &  spec-z &  36.87 &           
GMBCG J010.63096+00.85021 &                  4,6 \\
007340 &   23.13 &   1.33 &  7.27 &  1.46 & 0.0178 & 1237663204918886547 &  10.71962 &   0.71844 & 0.26 & 0.27 &  4 &  spec-z &  28.79 &       
SDSS CE J010.717058+00.725393 &                3,5,7 \\
007362 &   54.86 &   4.07 & 10.09 &  1.99 & 0.0270 & 1237652947993428384 &  10.72131 &  -9.57365 & 0.41 & 0.00 &  0 & photo-z &  52.83 &           
GMBCG J010.72131-09.57365 &                  5,6 \\
\hline
\end{tabular}
}
\tablefoot{ The full catalogue is available at CDS.
\tablefoottext{a}{All these parameters are extracted from the 2XMMi-DR3 catalogue.} 
\tablefoottext{b}{The cluster redshift from col. (28) or col. (27).}     
\tablefoottext{c}{Aperture X-ray flux $F_{\rm ap}$ [0.5-2.0] keV and its positive and negative errors in units of $10^{-14}$\ erg\ cm$^{-2}$\ s$^{-1}$.}
\tablefoottext{d}{Aperture X-ray luminosity $L_{\rm ap}$ [0.5-2.0] keV and its positive and negative errors in units of $10^{42}$\ erg\ s$^{-1}$.} 
\tablefoottext{e}{X-ray bolometric luminosity $L_{500}$ and its error in units of $10^{42}$\ erg\ s$^{-1}$.} 
\tablefoottext{f}{The cluster mass $M_{500}$ and its error  in units of $10^{13}$\ M$_\odot$.}
\tablefoottext{g}{The Galactic HI column in units $10^{22}$\ cm$^{-2}$.}
\tablefoottext{h}{These parameters are obtained from the developed optical detection algorithm.}    }   
\tablebib{
{\scriptsize
1-  \cite{Romer00};      2-  \cite{Kolokotronis06}; 3- \cite{Goto02}; 
4-  \cite{Koester07};    5-  \cite{Wen09};          6- \cite{Hao10}; 
7-  \cite{Plionis05};    8-  \cite{Lopes04};        9- \cite{Bahcall03}; 
10- \cite{Vikhlinin98};  11- \cite{Mullis03};      12- \cite{Burenin07};
13- \cite{Miller05};     14- \cite{Gal03};         15- \cite{Horner08};
16- \cite{Finoguenov07}; 17- \cite{Merchan05};     18- \cite{Olsen07};
19- \cite{Grove09};      20- \cite{Falco99};       21- \cite{Ramella01};
22- \cite{Boschin02};    23- \cite{Zwicky61};      24- \cite{dellAntonio94}; 
25- \cite{Berlind06};    26- \cite{McConnachie09}; 27- \cite{Dietrich07};
28- \cite{Gunn86};       29- \cite{Gladders05};    30- \cite{Yoon08}; 
31- \cite{Barkhouse06};  32- \cite{McDowell03};    33- \cite{Schuecker04};
34- \cite{Wittman06};    35- \cite{Carlberg01};    36- \cite{Finoguenov09};
37- \cite{Hughes98};     38- \cite{Sehgal08};      39- \cite{Postman02}. }
}
\end{table}
\end{landscape}



\clearpage
\begin{landscape}
\begin{table}
\caption{The first 10 entries of the X-ray selected 
group/cluster sample (185 systems) from the 2XMMi/SDSS Galaxy Cluster Survey 
with X-ray parameters based on the given flux in the 2XMMi-DR3 catalogue.}
\label{tbl:cat_sample}
{\tiny
\begin{tabular}{c c c c c c c c c c c c c c c c }
\hline
\hline
  \multicolumn{1}{c}{detid\tablefootmark{a}} &
  \multicolumn{1}{c}{Name\tablefootmark{a}} &
  \multicolumn{1}{c}{ra\tablefootmark{a}} &
  \multicolumn{1}{c}{dec\tablefootmark{a}} &
  \multicolumn{1}{c}{obsid\tablefootmark{a}} &
  \multicolumn{1}{c}{z\tablefootmark{b}} &
  \multicolumn{1}{c}{scale} &
  \multicolumn{1}{c}{$R_{500}$} &
  \multicolumn{1}{c}{$F_{\rm cat}$\tablefootmark{a,c}} &
  \multicolumn{1}{c}{$\pm eF_{\rm cat}$} &
  \multicolumn{1}{c}{$L_{\rm cat}$\tablefootmark{d}} &
  \multicolumn{1}{c}{$\pm eL_{\rm cat}$} &
  \multicolumn{1}{c}{$L_{500}$\tablefootmark{e}} &
  \multicolumn{1}{c}{$\pm eL_{500}$} &
  \multicolumn{1}{c}{$M_{500}$\tablefootmark{f}} &
  \multicolumn{1}{c}{$\pm eM_{500}$} \\

  \multicolumn{1}{c}{} &
  \multicolumn{1}{c}{IAUNAME} &
  \multicolumn{1}{c}{(deg)} &
  \multicolumn{1}{c}{(deg)} &
  \multicolumn{1}{c}{} &
  \multicolumn{1}{c}{} &
  \multicolumn{1}{c}{kpc/$''$} &
  \multicolumn{1}{c}{(kpc)} &
  \multicolumn{1}{c}{(keV)} &
  \multicolumn{1}{c}{(keV)} &
  \multicolumn{1}{c}{(keV)} &
  \multicolumn{1}{c}{} &
  \multicolumn{1}{c}{} &
  \multicolumn{1}{c}{} &
  \multicolumn{1}{c}{} &
  \multicolumn{1}{c}{} \\

(1) &  (2)  &  (3)  & (4)  &   (5)   &  (6)  &   (7)  &   (8)  &  (9)  &  (10)  &  (11) &  (12) & (13) & (14) & (15) &  (16)   \\
\hline 

006511 &   "2XMM J004205.5-093613" &  10.52296 &  -9.60375 & 0065140201 & 0.33 & 4.71 &  582.39 &   3.30 & 0.47 &  11.51 &  1.63 &   29.35 &   4.66 &  7.87 &  1.67  \\
007481 &   "2XMM J004259.7-092634" &  10.74900 &  -9.44286 & 0065140201 & 0.42 & 5.49 &  678.59 &   7.97 & 1.21 &  49.11 &  7.44 &  106.34 &  20.98 & 13.80 &  2.93 \\
011071 &   "2XMM J005608.0+004103" &  14.03365 &   0.68427 & 0303110401 & 0.46 & 5.84 &  588.27 &   2.71 & 0.60 &  21.40 &  4.74 &   51.85 &  13.25 &  9.48 &  2.18 \\
014038 &   "2XMM J010606.8+004926" &  16.52863 &   0.82407 & 0150870201 & 0.26 & 3.98 &  680.44 &  13.69 & 1.54 &  27.55 &  3.11 &   60.33 &   7.82 & 11.62 &  2.34  \\
014050 &   "2XMM J010610.0+005110" &  16.54201 &   0.85302 & 0150870201 & 0.26 & 3.99 &  689.89 &  15.44 & 1.42 &  31.12 &  2.86 &   65.80 &   7.60 & 12.12 &  2.41 \\
021043 &   "2XMM J015558.5+053159" &  28.99394 &   5.53329 & 0153030701 & 0.43 & 5.62 &  671.20 &   5.82 & 0.85 &  39.27 &  5.76 &  105.58 &  20.88 & 13.61 &  2.90  \\
021688 &   "2XMM J020056.5-092119" &  30.23615 &  -9.35526 & 0203840201 & 0.34 & 4.83 &  549.44 &   2.29 & 0.24 &   8.72 &  0.92 &   21.37 &   2.51 &  6.70 &  1.40 \\
023255 &   "2XMM J021447.5-005425" &  33.69817 &  -0.90720 & 0201020201 & 0.27 & 4.08 &  484.82 &   0.88 & 0.12 &   1.91 &  0.27 &    7.50 &   0.78 &  4.24 &  0.93  \\
033092 &   "2XMM J024810.2+311511" &  42.04268 &  31.25311 & 0111490401 & 0.39 & 5.27 &  532.12 &   2.17 & 0.39 &  11.31 &  2.03 &   20.99 &   4.49 &  6.44 &  1.46  \\
034341 &   "2XMM J030212.0+001108" &  45.55036 &   0.18583 & 0041170101 & 0.65 & 6.94 &  413.91 &   0.43 & 0.08 &   7.91 &  1.49 &   12.12 &   2.98 &  4.15 &  1.01 \\

\hline
\end{tabular}
}

\end{table}


\addtocounter{table}{-1}
\begin{table}
\caption{\label{} continued.}
{\tiny
\begin{tabular}{c c c c c c c c c c c c c}
\hline
\hline
  \multicolumn{1}{c}{detid\tablefootmark{a}} &
  \multicolumn{1}{c}{$T_{500}$} &
  \multicolumn{1}{c}{$\pm eT_{500}$} &
  \multicolumn{1}{c}{objid\tablefootmark{g}} &
  \multicolumn{1}{c}{RA\tablefootmark{g}} &
  \multicolumn{1}{c}{DEC\tablefootmark{g}} &
  \multicolumn{1}{c}{$\bar{z}_{\rm p}$\tablefootmark{g}} &
  \multicolumn{1}{c}{$\bar{z}_{\rm s}$\tablefootmark{g}} &
  \multicolumn{1}{c}{$N_{z_{\rm s}}$\tablefootmark{g}} &
  \multicolumn{1}{c}{$z_{\rm type}$\tablefootmark{g}} &
  \multicolumn{1}{c}{offset\tablefootmark{g}} &
  \multicolumn{1}{c}{NED-Name} &
  \multicolumn{1}{c}{ref.}  \\

  \multicolumn{1}{c}{}&
  \multicolumn{1}{c}{(keV)}&
  \multicolumn{1}{c}{(keV)} &
  \multicolumn{1}{c}{(BCG)} &
  \multicolumn{1}{c}{(deg)} &
  \multicolumn{1}{c}{(deg)} &
  \multicolumn{1}{c}{} &
  \multicolumn{1}{c}{} &
  \multicolumn{1}{c}{} &
  \multicolumn{1}{c}{} &
  \multicolumn{1}{c}{(kpc)} &
  \multicolumn{1}{c}{} &
  \multicolumn{1}{c}{}  \\

  (1)  & (17)  &  (18) &  (19)  & (20)  &  (21)  &  (22)   &  (23) &  (24)  &   (25)   &  (26) & (27) & (28)  \\
\hline

006511 & 1.84 & 0.45 & 1237652947993297563 &  10.51514 &  -9.60060 & 0.33 & 0.00 &  0 & photo-z & 138.56 &                                   - &
                    - \\
007481 & 2.65 & 0.63 & 1237652630713795354 &  10.75138 &  -9.43350 & 0.42 & 0.00 &  0 & photo-z & 191.29 &                                   - &
                    - \\
011071 & 2.12 & 0.53 & 1237663204920328298 &  14.04250 &   0.68188 & 0.46 & 0.00 &  0 & photo-z & 192.67 &                                   - &
                    - \\
014038 & 2.30 & 0.55 & 1237663204921376994 &  16.52926 &   0.81949 & 0.25 & 0.26 &  4 &  spec-z &  68.21 &       SDSS CE J016.528793+00.817471 &
            2,3,4,5,6 \\
014050 & 2.36 & 0.56 & 1237663785278374092 &  16.54324 &   0.85569 & 0.25 & 0.26 &  5 &  spec-z &  42.90 &          MaxBCG J016.54324+00.85569 &
                4,7,8 \\
021043 & 2.64 & 0.63 & 1237678663047250389 &  28.98754 &   5.53073 & 0.43 & 0.00 &  0 & photo-z & 143.57 &                                   - &
                    - \\
021688 & 1.67 & 0.41 & 1237652900224303421 &  30.23274 &  -9.35660 & 0.35 & 0.34 &  1 &  spec-z &  62.93 &                                   - &
                    - \\
023255 & 1.23 & 0.32 & 1237680000377684204 &  33.69949 &  -0.90894 & 0.26 & 0.26 &  1 &  spec-z &  32.01 &                                   - &
                    - \\
033092 & 1.64 & 0.42 & 1237670458043073373 &  42.04490 &  31.25411 & 0.39 & 0.00 &  0 & photo-z &  39.90 &                                   - &
                    - \\
034341 & 1.33 & 0.35 & 1237663784217346252 &  45.54822 &   0.18751 & 0.65 & 0.65 &  1 &  spec-z &  67.99 &                 BLOX J0302.2+0010.5 &
                    9 \\

%

\hline
\end{tabular}
}

\tablefoot{ The full catalogue is available at CDS and contains the information given in columns (1)-(28) in Table~\ref{tbl:cat_sample}.
\tablefoottext{a}{All these parameters are extracted from the 2XMMi-DR3 catalogue.} 
\tablefoottext{b}{The cluster redshift from col. (23) or col. (22).}     
\tablefoottext{c}{The given flux in the 2XMMi-DR3 $F_{\rm cat}$ [0.5-2.0] keV and its errors in units of $10^{-14}$\ erg\ cm$^{-2}$\ s$^{-1}$.}
\tablefoottext{d}{The computed X-ray luminosity $L_{\rm cat}$ [0.5-2.0] keV and its errors in units of $10^{42}$\ erg\ s$^{-1}$.} 
\tablefoottext{e}{X-ray bolometric luminosity $L_{500}$ and its error in units of $10^{42}$\ erg\ s$^{-1}$.} 
\tablefoottext{f}{The cluster mass $M_{500}$ and its error  in units of $10^{13}$\ M$_\odot$.}
\tablefoottext{g}{These parameters are obtained from our detection algorithm in the optical band.}
}

\tablebib{
1-  \cite{Merchan05};      2-  \cite{Goto02};      3-  \cite{Lopes04};
4-  \cite{Barkhouse06};    5-  \cite{Wen09};       6-  \cite{Hao10};
7-  \cite{Koester07};      8-  \cite{Bahcall03};   9-  \cite{Dietrich07};
10- \cite{Gunn86};         11- \cite{Knobel09};    12- \cite{Olsen07};
13- \cite{McConnachie09};  14- \cite{Gal03};       15- \cite{Kolokotronis06};
16- \cite{Horner08};       17- \cite{Zwicky61};    18- \cite{Falco99}; 
19- \cite{Abell58};        20- \cite{Abell89};     21- \cite{Romer00};
22- \cite{Wittman06};      23- \cite{Burenin07};   24- \cite{Yoon08};
25- \cite{White99}.
}

\end{table}
\end{landscape}

%
%







\chapter[III. Clusters associated with spectroscopically targeted LRGs in SDSS-DR9]{III. Clusters associated with spectroscopically targeted LRGs in SDSS-DR9\footnote{This chapter will be submitted to {\it Astronomy \& Astrophysics} } }

 
\section*{Abstract} 
We present a sample of 324 X-ray selected galaxy groups and clusters with 
spectroscopic redshift measurements (up to $z \sim 0.77$) from the 
2XMMi/SDSS Galaxy Cluster Survey. 
The X-ray cluster candidates were selected as serendipitous extended 
sources from the 2XMMi-DR3 catalogue in the foot print of the Sloan Digital 
Sky survey (SDSS-DR7). The cluster galaxies with available spectroscopic 
redshifts are selected from the SDSS-DR9.  
We developed an algorithm to identify the cluster candidates associated with 
spectroscopically targeted Luminous Red Galaxies (LRGs) and to measure 
the cluster spectroscopic redshift. A cross correlation of the identified 
cluster sample with published optically selected cluster catalogues shows 
that 241/324 sources were previously identified with available redshifts. 
The present redshift measurements are consistent with the published ones. 
The current cluster sample extends the optically confirmed cluster sample from 
our cluster survey by 44 objects and provides spectroscopic confirmation for  
49 clusters among the published sample with only photometric redshifts. 
Among the extended cluster sample, about 80 percent are newly X-ray discovered
systems and 55 percent are newly discovered as galaxy clusters in optical and 
X-ray wavelengths. Based on the measured redshifts and the fluxes given in
the 2XMMi-DR3 catalogue we estimated the X-ray luminosities and masses of the 
cluster sample. 
%


\section{Introduction}

Galaxy clusters are the largest gravitationally bound objects in the universe. 
They have been formed from the densest regions in the large-scale matter 
distribution of the universe and have collapsed to form their own proper 
equilibrium structure. Their form can be well assessed by observations 
and well described by theoretical modelings
\citep[e.g.][]{Sarazin88, Bahcall88, Voit05, Boehringer06}.
X-ray and optical observations show that clusters of galaxies are well defined 
connected structural entities where the diffuse X-ray emission from the hot 
intracluster medeium (ICM) trace contiguously the whole structure of the cluster. 
They are excellent giant laboratories sites for several astrophysical studies, 
for example investigation of galaxy evolution in their dense environments 
\citep[e.g.][]{Dressler80, Goto03}, evolution of the dynamical and thermal 
structure \citep[e.g.][]{Balestra07, Maughan08, Anderson09},  chemical 
enrichment of the intracluster medium \citep[e.g.][]{Cora06, Heath07}, 
to study lensed high redshift background galaxies 
\citep[e.g.][]{Metcalfe03, Santos04, Bartelmann10}, and to investigate the 
evolution of the universe in order to test the cosmological models 
\citep[e.g.][]{Rosati02, Reiprich02, Voit05, Vikhlinin09a, Allen11}.

Due to the multi-component nature of galaxy clusters, they can be observed 
and identified through multiple observable signals across the electromagnetic 
spectrum. Tens of thousands of galaxy clusters have been identified through 
detecting their galaxies in optical and NIR band 
\citep[e.g.][]{Abell58, Abell89, Zwicky61, Gladders05, Merchan05, Koester07, 
Wen09, Hao10, Szabo11, Geach11, Durret11, Wen12, Gettings12}. Recently, 
several galaxy cluster surveys have been conducted at mm wavelength through 
the Sunyaev-Zeldovich (SZ) effect \citep[e.g.][]{Vanderlinde10, Marriage11, 
Planck11, Reichardt13, Planck13}, which provided hundreds of SZ selected clusters.     
 
X-ray cluster surveys provide pure and complete cluster catalogues, in addition 
the tight correlations between X-ray observables and masses of clusters 
\citep[e.g.][]{Allen11}. Hundreds of galaxy clusters were detected in X-rays 
based on previous X-ray missions mainly from ROSAT data 
\citep[e.g.][]{Ebeling98, Boehringer04, Reiprich02, Ebeling10, Rosati98, 
Burenin07}. The current X-ray telescopes (XMM-Newton, Chandra, Swift/X-ray) 
provide contiguous surveys for small areas \citep[e.g.][]{Finoguenov07, 
Finoguenov10, Adami11,Suhada12}, in addition to serendipitous cluster surveys  
\citep[e.g.][]{Barkhouse06, Kolokotronis06, Fassbender11, Takey11, Mehrtens12, 
Clerc12, Tundo12, de-Hoon13, Takey13a}. So far these surveys provided 
a substantial cluster sample of few hundreds up to redshift of 1.57.   

We have conducted a systematic search for X-ray detected galaxy clusters based 
on XMM-Newton fields that are in the footprint of the SDSS-DR7. 
The catalogue of XMM-Newton serendipitous extended sources detected in EPIC 
images was the basic database to select the X-ray cluster candidates, which 
comprises 1180 objects.
The redshifts of about half of the cluster candidates were measured based on 
the galaxy redshifts given in the SDSS-DR8. By having an optically confirmed 
groups/clusters with redshift measurements, we derived their X-ray luminosities 
and temperatures and investigated the X-ray luminosity-temperature relation.
The selection criteria, redshift measurements, and the X-ray properties of 
the optically confirmed sample were described in more detail by 
\citet[][Paper I, Paper II, hereafter]{Takey11, Takey13a}.         

In this work we compile a new sample of X-ray detected galaxy clusters 
that are associated with LRGs, which have spectroscopic redshifts in the 
SDSS-DR9 data. We present the procedure used to construct a cluster sample 
among the X-ray cluster candidates that have spectroscopic confirmation based 
on the spectroscopic redshifts of LRGs and then to measure their redshifts. 
We also present the measurements of X-ray luminosity and luminosity-based mass 
of the constructed cluster sample. 

The paper is organized as follows: in Section 2 we describe briefly the 
selection procedure of the X-ray cluster candidates. In Section 3 we describe 
the cluster sample associated with LRGs and their redshift measurements. 
The X-ray parameters of the cluster sample are presented in Section 4. 
The summary of the paper is presented in Section 5. Throughout this paper, 
we used the cosmological parameters  $\Omega_{\rm M}=0.3$, 
$\Omega_{\Lambda}=0.7$ and $H_0=70$\ km\ s$^{-1}$\ Mpc$^{-1}$.


\section{Description of the X-ray cluster candidates}

Galaxy clusters are simply identified among the X-ray sources as X-ray 
luminous, spatially extended, extragalactic sources \citep{Allen11}. The so 
far largest X-ray sources catalogue \cite[2XMMi-DR3,][]{Watson09} was comprised 
based on all XMM-Newton observations (till October 2009) taken by the EPIC 
(PN, MOS1, MOS2) cameras, which comprises 353191 detections corresponding to 
262902 unique sources. Among these detections, 30470 extended detections that 
are including both real and spurious extended sources as well as multiple 
detections of same sources.
 
We selected the X-ray cluster candidates from the reliable extended sources 
(with no warning about being spurious) in the 2XMMi-DR3 catalogue at high 
galactic latitudes, $|b| > 20^{\circ}$. 
The survey was constrained to those XMM-Newton fields that are in the 
footprint of the SDSS-DR7 in order to be able to measure the optical redshifts 
of the possible optical counterparts. The overlap area of XMM-Newton fields and 
imaging area of the SDSS-DR7 is 210 deg$^2$. After excluding possible spurious 
X-ray detections and low redshift galaxies that appear resolved at X-ray 
wavelengths through visual inspections of X-ray images and X-ray-optical 
overlays, the X-ray cluster candidates list comprised 1180 objects. The 
selection procedure was described in more detail in Papers I and II.         

The X-ray cluster candidates have a wide range of the net photon counts from 
80 up to few thousands. For the X-ray bright candidates in the list 
(about 4 percent with 2000 counts or more), the X-ray spectroscopy provides 
a tool to measure the X-ray redshift \citep[][]{Lamer08, Yu11}. We defer 
measuring X-ray redshifts for those clusters to a future work. 
The alternative and the main way to obtain the cluster redshifts is based 
on the optical data. This can by done by either cross-matching the X-ray 
cluster candidates with the available optically selected galaxy clusters 
catalogues in the literature (see Paper I) or by measuring the cluster 
photometric redshifts based on galaxy redshifts given in the SDSS catalogues 
(see paper II). Using those methods we could establish an optically 
confirmed cluster sample comprising 530 groups/clusters with redshift 
measurements. 

\begin{figure}[t]
  \centering{
  \resizebox{100mm}{!}{\includegraphics{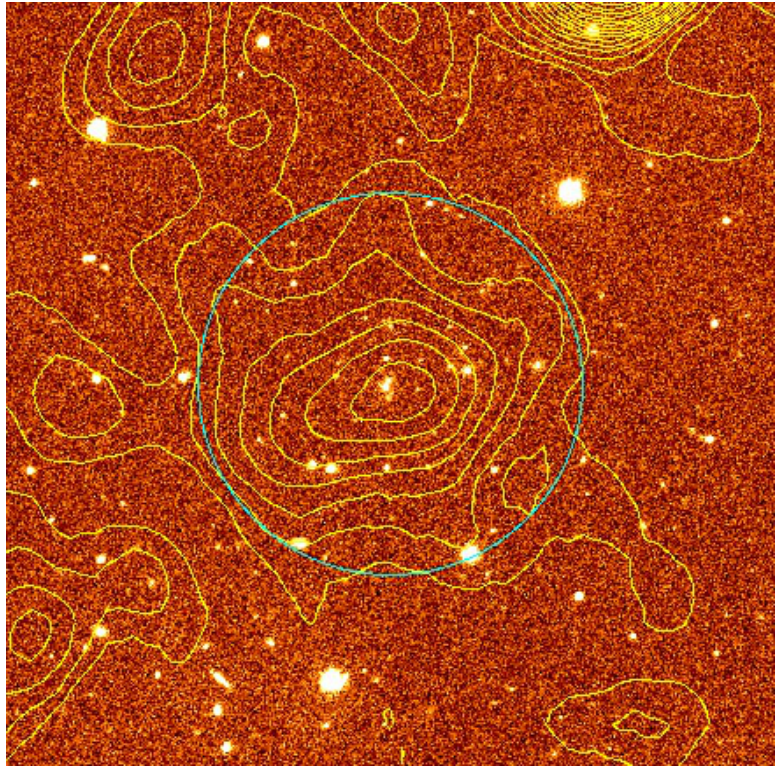}}}
  \caption[SDSS image of the example cluster 2XMMi J143742.9+340810 at 
redshift = 0.5446 ]{SDSS image of the example cluster at 
redshift = 0.5446,
 2XMMi J143742.9+340810, with X-ray surface brightness contours 
(0.2 - 4.5 keV) overlaid  in yellow. The plotted cyan circle has a radius of 
one arcmin around the X-ray emission peak position. The field of view is 
$4'\times 4'$ centred on the X-ray cluster position.} 
  \label{f:275341_overlay}
  \end{figure}

In the current work we are going to identify a subsample of the X-ray 
cluster candidates associated with LRGs that have spectroscopic redshifts 
in the SDSS-DR9 in order to construct a sample with spectroscopic confirmations. 
As an example, Figure~\ref{f:275341_overlay} shows a newly X-ray discovered 
galaxy cluster associated with two LRGS as cluster galaxies with available 
spectroscopic redshift of 0.5446. We use this cluster to show the procedure 
of the redshift measurements in the next section.

\section{Clusters associated with spectroscopically targeted LRGs in SDSS-DR9}

Generally, the brightest cluster galaxies (BCGs) are elliptical massive galaxies 
and reside near the cluster center of mass. The BCGs tend to be very luminous 
and red galaxies (LRGs) \citep[e.g.][]{Postman95, Eisenstein01, Wen12}, therefore 
LRGs with available spectroscopic redshifts give spectroscopic confirmations 
for clusters that are associated with them \citep[e.g.][]{Goto02, Mehrtens12}. 

We identify LRGs from SDSS-DR9 data that are cluster members of any of the 
X-ray cluster candidates. The spectroscopic redshifts of these cluster members 
are used to measure the cluster redshifts. In the next subsection we describe 
how to select the LRGs from the recent data of the SDSS and the procedure 
used to measure the cluster redshifts. We also present some statistical 
properties of the constructed cluster sample, e.g. the redshift distribution 
and the linear offsets from the X-ray positions.
Then a comparison of the current redshift measurements with the published 
ones is presented. 

\subsection{Luminous Red Galaxy sample}

The so far latest data release from the SDSS is Data Release 9 
\citep[DR9,][]{Ahn12}, which provides the first spectroscopic data from the 
SDSS-III's Baryon Oscillation Spectroscopic Survey (BOSS) as well as imaging 
and spectroscopic data from the previous SDSS data releases. BOSS is an 
ongoing project and its  first data release includes more than 800,000 spectra 
of galaxies, in addition to thousands of quasar and stellar spectra over 
3,300 deg$^2$. One aim of BOSS is to  obtain spectra of 1.5 million galaxies 
with $0.15 < z < 0.8$ over 10,000 deg$^2$, therefore it will be a valuable 
resource to obtain spectroscopic confirmation for luminous cluster galaxies.

For each X-ray cluster candidate, we created a sample of galaxies that 
are located within 10 arcmins from the X-ray source position. The search 
radius corresponds to a physical radius of 500\,kpc at a redshift 
0.04, which is about our low redshift limit.
The galaxies were selected from the {\tt galaxy} view table in the SDSS-DR9, 
which contains the photometric parameters measured for resolved primary objects, 
classified as galaxies.  Also, the photometric redshifts ($z_{\rm p}$) and, 
if available, the spectroscopic redshifts ($z_{\rm s}$) of the galaxy sample 
were selected from the {\tt Photoz} and {\tt SpecObj} tables, respectively. 
The {\tt SpecObj} table includes spectroscopic redshifts that were measured 
from clean galaxy spectra taken by the new and old spectrographs in the SDSS 
projects. The extracted parameters of the galaxy sample include the coordinates, 
the (model and composite model) magnitudes in $r-$ and $i-$band, the photometric 
redshifts, and, if available, the spectroscopic redshifts.
We used the magnitudes in the {\tt galaxy} table that are corrected for 
Galactic extinction following \citep{Schlegel98}. To clean the galaxy sample 
from faint objects beyond the detection limits of SDSS, we only deal with  
galaxies that have $ m_{r} \leq 22.2$ mag, $\bigtriangleup m_{r} < 0.5 $ mag, 
and $\bigtriangleup z_{p}/z_{p} < 0.5$.  

BOSS data includes two main target galaxy samples; first the BOSS $``$ LOWZ $``$
galaxy sample with $z \le 0.4$; second the BOSS constant-mass, 
$``$ CMASS $``$, galaxy sample with $0.4 < z < 0.8$. The target 
selection algorithms for galaxies in BOSS are significantly different 
from those used in the previous SDSS projects due to the different scientific 
goals \citep{Ahn12}. BOSS targets significantly fainter galaxies than galaxy 
targets in the previous SDSS projects with the aim of measuring large-scale 
clustering of galaxies at higher redshifts.    
To select a homogeneous luminous red galaxy sample from BOSS and previous
SDSS data releases, we apply the same selection criteria on both data. 
We selected the LRGs with available spectroscopic redshifts from the 
constructed galaxy sample within 10 arcmin from the X-ray positions.  
The applied selection criteria of LRGs are based on the colour and 
magnitude cuts that are described by Padmanabhan et al. (2013, in preparation) 
and given in Appendix B.
We also made sure that the selected objects are confirmed galaxies using 
the spectroscopic class parameter given in {\tt SpecObj} table in order to 
exclude those objects targeted as galaxies but turned out to be stars or 
quasars.  
The selected LRG sample is used to identify the BCGs of the X-ray cluster 
candidates as described in the next subsection.


\subsection{Optical identifications and redshift measurements}

\begin{figure}[h]
\centering{
  \resizebox{100mm}{!}{\includegraphics[viewport=35  10  525 410, clip]{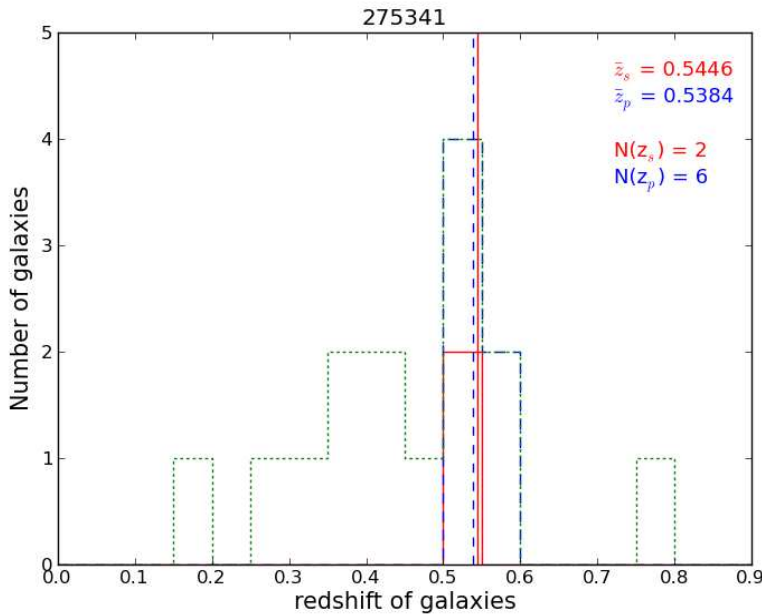}}}
  \caption[The histogram of the spectroscopic ($z_{\rm s}$, solid red) 
and photometric ($z_{\rm p}$, dashed blue) redshifts of the cluster member 
candidates ($N_{z_{\rm s}}$ and $N_{z_{\rm p}}$) within 500 kpc of the example 
cluster, 2XMMi J143742.9+340810]{The histogram of the spectroscopic ($z_{\rm s}$, solid red) 
and photometric ($z_{\rm p}$, dashed blue) redshifts of the cluster member 
candidates ($N_{z_{\rm s}}$ and $N_{z_{\rm p}}$) within 500 kpc of the example 
cluster, 2XMMi J143742.9+340810. 
The green doted histogram represents the distribution of $z_{\rm p}$ of 
field galaxies with 500 kpc. The cluster spectroscopic, $\bar z_{\rm s}$, 
and photometric, $\bar z_{\rm p}$, redshifts are represented by solid red 
and dashed blue vertical lines, respectively, and written in the upper 
right legend.} 
\label{f:275341_hist_DR9}
\end{figure}

We identify the optical counterparts of the X-ray cluster candidates based 
on the spectroscopic and photometric redshifts of galaxies from SDSS-DR9.   
To measure the redshifts of cluster candidates, we firstly identify the 
BCG candidates, then we select cluster member candidates with available similar 
$z_{\rm s}$ of the BCG's spectroscopic redshift. The procedure is described as 
follows: 

\begin{enumerate}

\item identify a BCG candidate as a LRG within 200 kpc (computed based on the 
$z_{\rm s}$ of the LRG) from the X-ray position of the X-ray cluster candidate. 
If there is only one LRG, we consider it as the BCG candidate. If there are many 
LRGs we create groups of galaxies with similar redshift. For each group, 
we select the brightest galaxy as a BCG candidate. Then we select the nearest 
BCG candidate to the X-ray position. At low redshifts the search radius 200 kpc 
subtends a large angle on the sky and might cause a wrong association of LRGs
with the X-ray cluster candidates. Therefore, we put a maximum angular 
separation limit of the BCGs offset from the X-ray emission peak of 90 arcsec.
The search radius of 200 kpc is used since we found 90 percent of the BCGs in 
Paper II are located within 200 kpc from the X-ray positions.      

\item identify the cluster member candidates within 500 kpc from the X-ray peak
based on the spectroscopic redshift of the identified BCG candidate. 
The cluster galaxies with available $z_{\rm s}$ are selected within a small 
redshift interval of $z_{\rm s,\rm BCG} \pm 0.01$. While the cluster member 
candidates with only $z_{\rm p}$ are selected within a slightly larger
redshift interval of $z_{\rm s,\rm BCG} \pm 0.04(1+z_{\rm s,\rm BCG})$.     
The distribution of the redshifts of the cluster member candidates and field 
galaxies for the example cluster is shown in Figure~\ref{f:275341_hist_DR9}. 
The redshift interval used to identify the cluster members with $z_{\rm p}$ 
gives 80 percent of the cluster members \citep{Wen09}. They also showed 
that a radius of 500 kpc gives a high overdensity level and a low false 
detection rate.
The identified BCG candidate could be the second or third brightest cluster 
galaxy, thus we re-identify the likely BCG as the brightest galaxy among the 
cluster member candidates within 500 kpc.

\item compute the spectroscopic, $\bar z_{\rm s}$, and photometric, 
$\bar z_{\rm p}$, redshift of a cluster as a weighted average of the 
spectroscopic and photometric redshifts of the cluster member candidates 
within 500 kpc, respectively. The weighted redshift errors are also computed. 
If there is only one cluster galaxy with available spectroscopic redshift, we 
consider its redshift as the cluster redshift.    

\item We consider the optical counterpart and the redshift measurement of an 
X-ray cluster candidate if the optical detection passed the quality assessment 
that are done through the following visual inspection process. 

\end{enumerate}


\begin{figure}[h]
\centering{
  \resizebox{100mm}{!}{\includegraphics[viewport=10  10  525 410, clip]{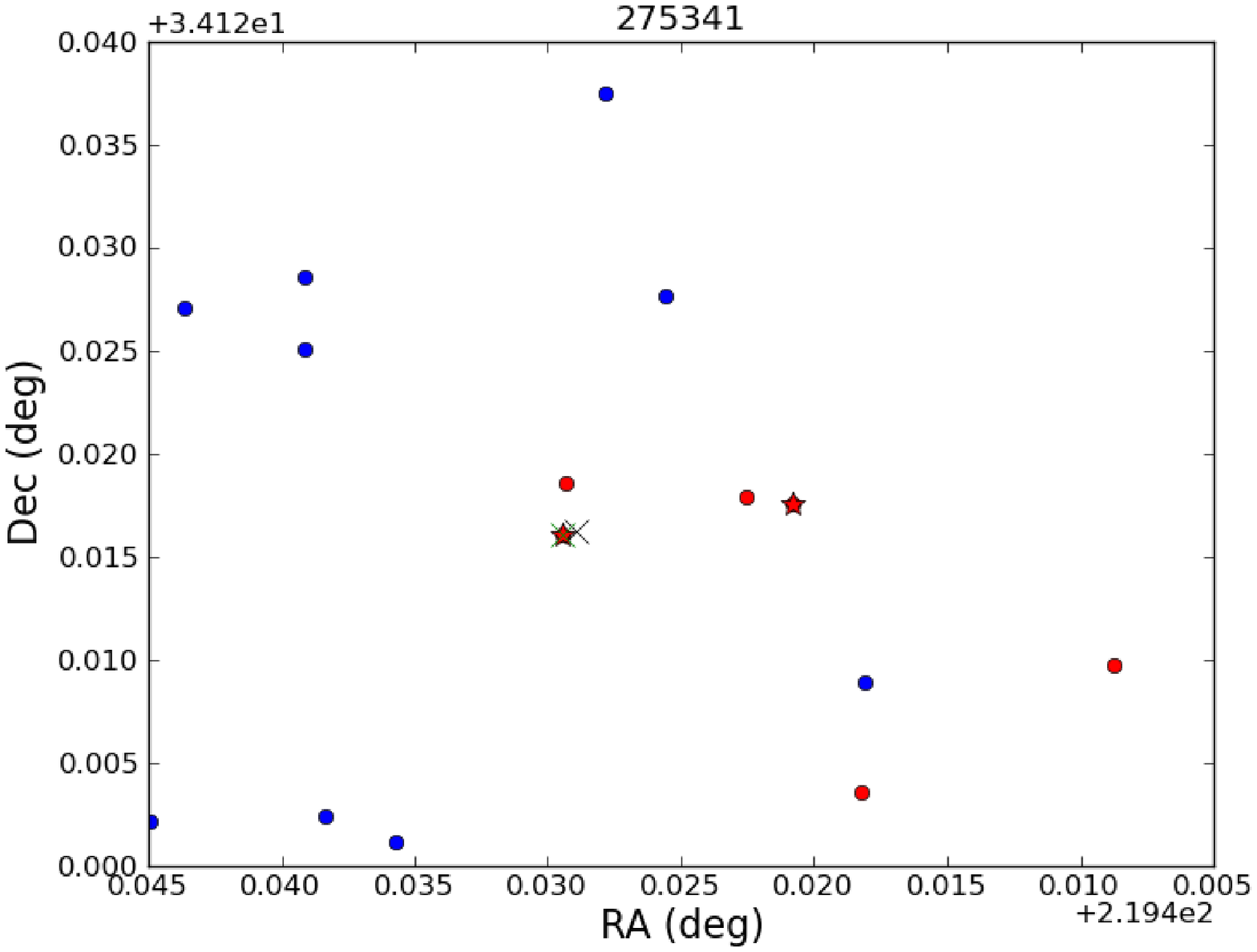}}}
  \caption[The distribution on the sky of the cluster member candidates 
(red dots) and field galaxies (blue dots) within 500 kpc ($\sim$ 1.3 arcmin) 
from the X-ray position (marked by black cross marker) of the example cluster, 
2XMMi J143742.9+340810]{The distribution on the sky of the cluster member candidates 
(red dots) and field galaxies (blue dots) within 500 kpc ($\sim$ 1.3 arcmin) 
from the X-ray position (marked by black cross marker) of the example cluster, 
2XMMi J143742.9+340810. Note the different 
scale in Figure~\ref{f:275341_SDSS}. The cluster galaxies with 
available $z_{\rm s}$ are marked by stars. The BCG is marked by green 
cross marker, which has a linear separation from the X-ray position of 
$\sim$ 11 kpc. We only presented galaxies that have $ m_{r} \leq 22.2$ mag, 
$\bigtriangleup m_{r} < 0.5$ mag, and $\bigtriangleup z_{p}/z_{p} < 0.5$. } 
  \label{f:275341_dist_DR9}
\end{figure}

\begin{figure}
\centering{
  \resizebox{85mm}{!}{\includegraphics{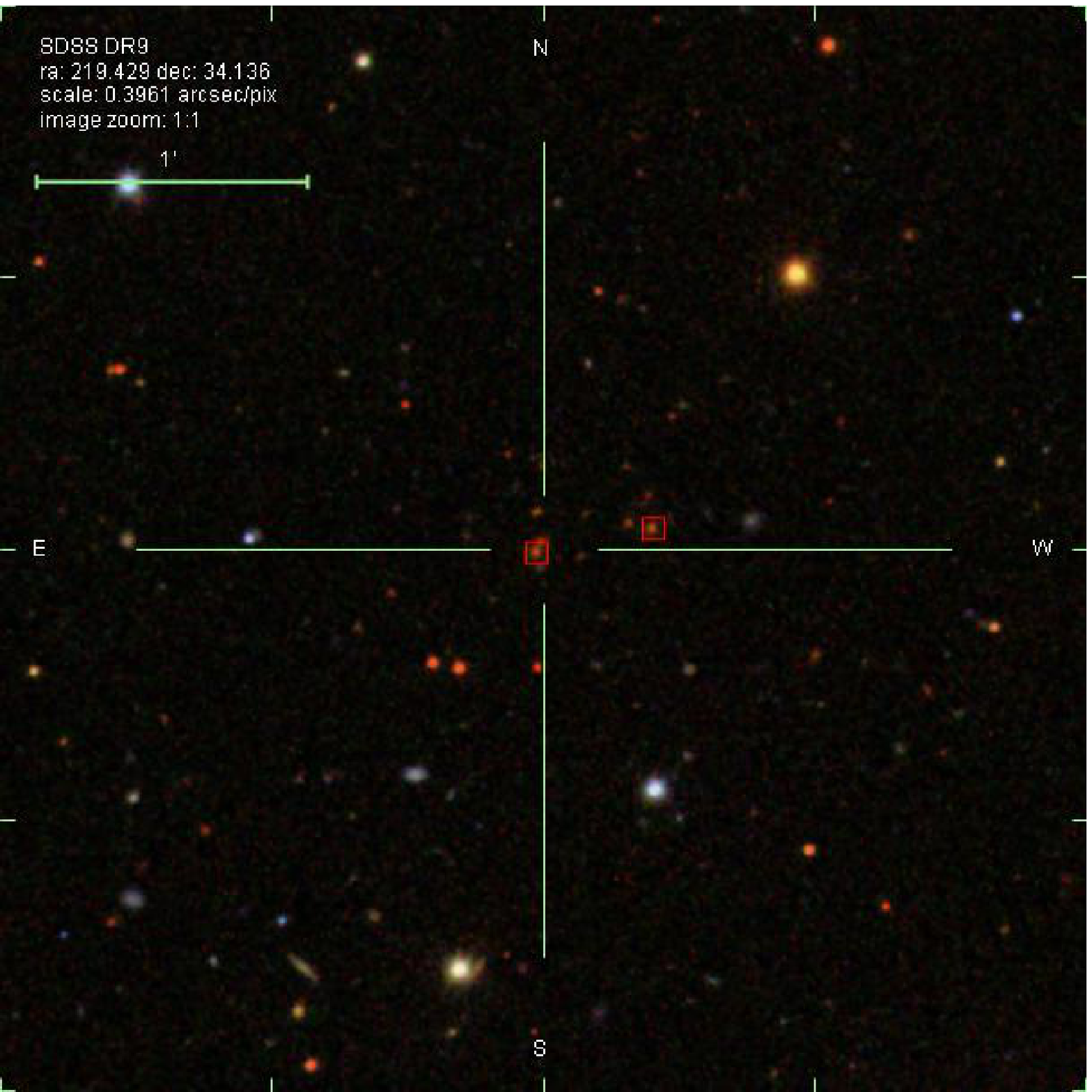}}}
  \caption[The SDSS colour image of the example cluster, 2XMMi J143742.9+340810]{The 
SDSS colour image of the example cluster, 2XMMi J143742.9+340810, 
with 4 arcmin a side centred on the X-ray position that is marked by cross-hair. 
Galaxies with spectra are marked by red squares. The measured spectroscopic 
redshift for this system is 0.5446.} 
  \label{f:275341_SDSS}
\end{figure}

The current procedure yields an initial list of optical counterparts that 
comprises 350 systems.     
To accept the optical detection we compare the identified BCG and cluster 
member candidates with the corresponding SDSS colour image of the same field. 
The distribution of cluster members on the sky of the example cluster is 
shown in Figure~\ref{f:275341_dist_DR9} while Figure~\ref{f:275341_SDSS} 
shows the corresponding SDSS colour image. From both images, it is clearly 
obvious that the algorithm picked the right associated LRGs (and thus the 
BCG too) and the cluster member candidates. Since SDSS provides a shallow 
survey the fainter cluster galaxies are not detected in SDSS imaging.

For a few cases about 7 percent of the initial resulting optical counterparts 
sample, we found miss-matched association of LRGs and consequently lead to 
wrong  redshift estimations. These wrong cases resulted due to the overlapping 
clusters along the line of sight or due to identifying a field galaxy as BCG 
candidate.

\begin{figure}
\centering{
  \resizebox{85mm}{!}{\includegraphics{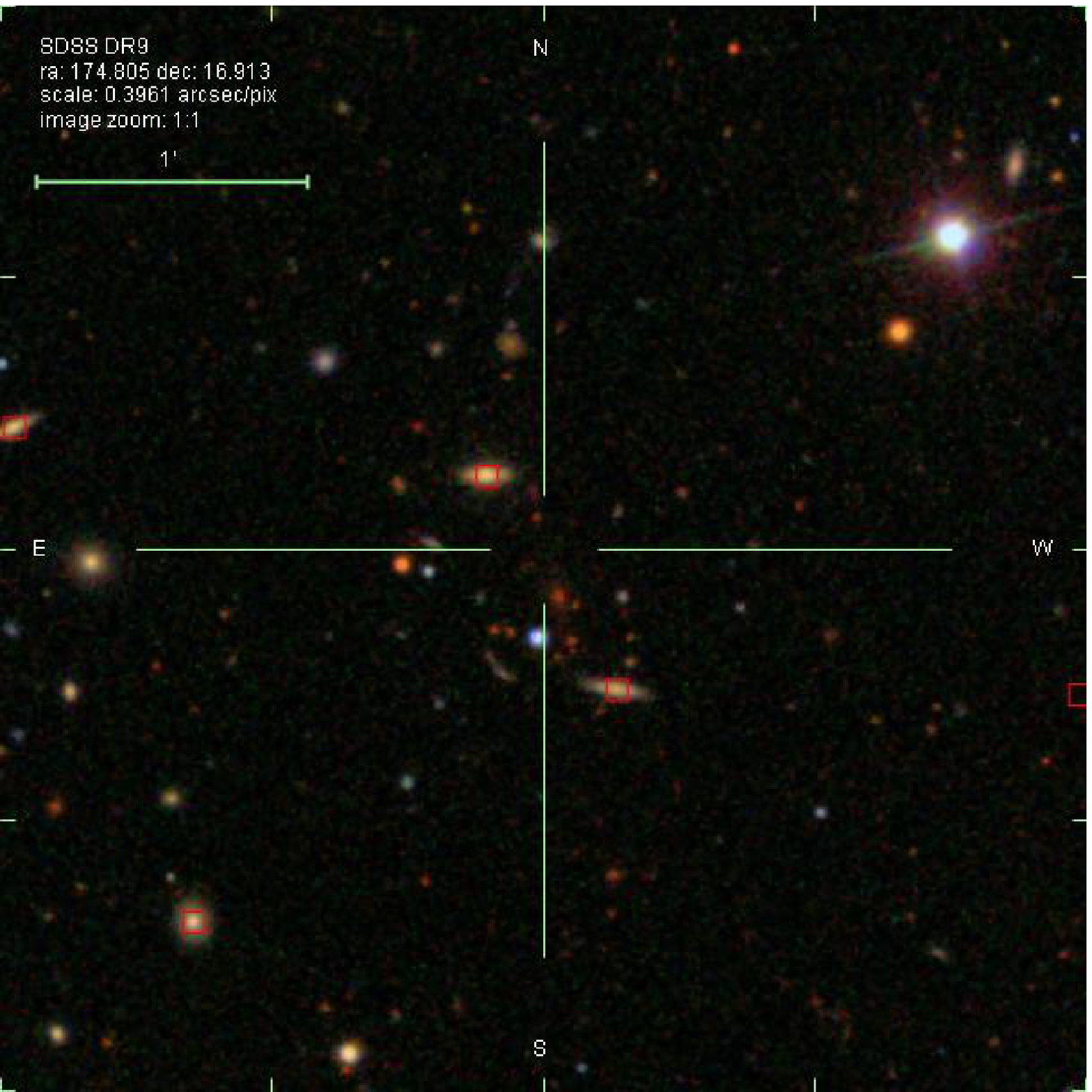}}}
  \caption{A similar image to Figure~\ref{f:275341_SDSS} but for another 
X-ray cluster candidate, 2XMM J113913.0+165446, as an example of confusing 
cases due to two overlapping clusters along the line of sight.} 
  \label{f:129017_SDSS}
\end{figure}

Figure~\ref{f:129017_SDSS} shows a case of overlap of two clusters along the 
line of sight of the X-ray peak.
There is a slight overdensity of relatively distant galaxies with 
similar colour around the X-ray emission peak, which was detected as a cluster 
with photometric redshift = 0.58 by \citet{Szabo11}. For this system, there 
is no cluster galaxy (as LRG) with available $z_{\rm s}$ in the current 
SDSS release.  In addition to that distant cluster there is another nearby 
galaxy group at redshift = 0.1.  
The current procedure picked the nearby galaxy group with a BCG candidate (LRG) 
at $z_{\rm s} = 0.1059$ with angular and linear separation from the X-ray 
position equal 21 arcsec and 40 kpc, respectively. 
In such difficult cases, it is likely impossible to associate the X-ray emission 
to one of the two optical detected clusters following the current procedure. 
Therefore we excluded similar cases from the initial compiled cluster list 
using the present algorithm described above. The final list of the optically 
validated cluster sample includes 324 systems with spectroscopic 
confirmations. 


\subsection{The optically validated cluster sample}

\begin{figure}
\centering{
  \resizebox{100mm}{!}{\includegraphics[viewport=25  10  530 400, clip]{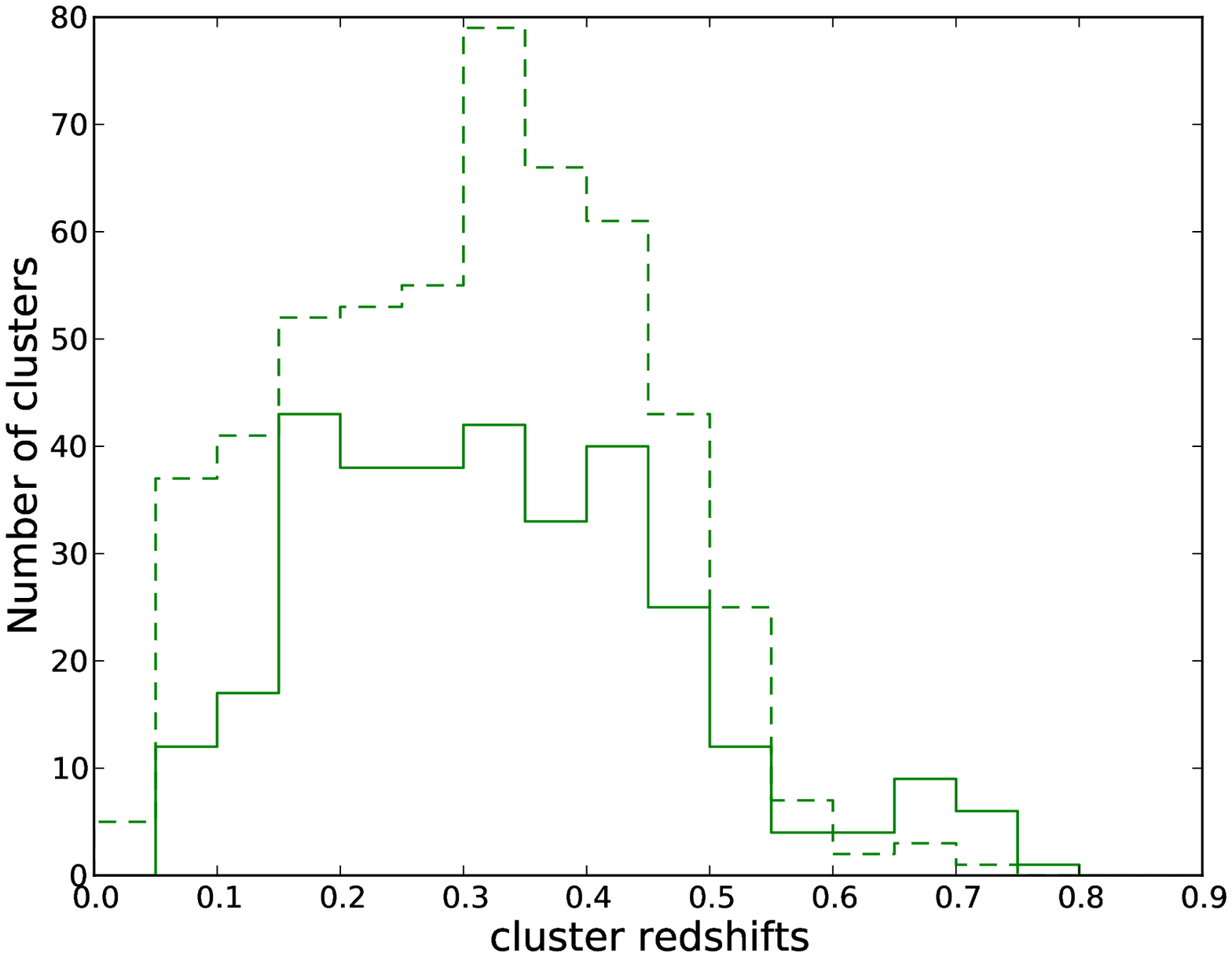}}}
  \caption{ The distribution of measured spectroscopic redshifts of the 
cluster sample associated with LRGs that have spectra is presented by the 
solid line, while the redshift distribution of the optically confirmed 
cluster sample in Paper II is presented by the dashed line.}
  \label{f:Hist_zc_Dr9}
\end{figure}

\begin{figure}
\centering{
  \resizebox{100mm}{!}{\includegraphics[viewport=25  10  530 400, clip]{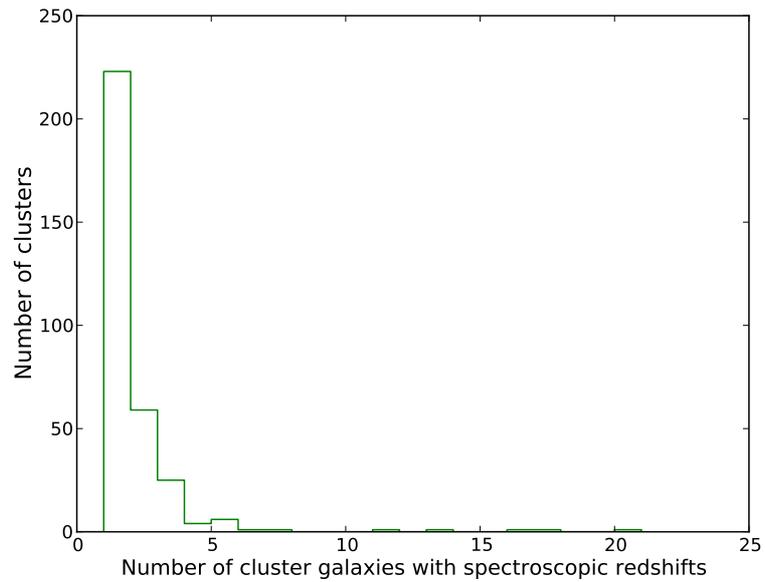}}}
  \caption{The distribution of the cluster galaxies with spectroscopic 
redshifts within 500 kpc from the X-ray positions for the optically validated 
cluster sample.} 
  \label{f:Hist_Nzs_DR9}
\end{figure}

The current procedure yielded 324 galaxy groups and clusters with spectroscopic 
redshifts based on at least one LRG with available $z_{\rm s}$  from the 
SDSS-DR9. The redshift of the sample spans a wide range from 0.05 to 0.77
with a median of 0.31. The redshift distribution of the current cluster sample 
as well as the optically confirmed cluster sample (530 systems) in Paper II is 
shown in Figure~\ref{f:Hist_zc_Dr9}. The common objects between the two 
samples are 280 systems, see the next subsection for the redshift comparison.
It is clearly shown that the current cluster sample includes a handful of 
clusters beyond $z = 0.6$ thanks to the first data release of BOSS in the 
SDSS-DR9. Additionally, the current sample extends the sample in paper II by 
44 systems, of which about 55 percent are newly discovered systems as 
clusters of galaxies.     
The majority of the clusters in the sample have one or two LRGs with 
$z_{\rm s}$ while few cluster have three LRGs or more with available 
$z_{\rm s}$. The distribution of the cluster galaxies with $z_{\rm s}$ per 
a cluster of the cluster sample is shown in Figure~\ref{f:Hist_Nzs_DR9}.

\begin{figure}
\centering{
  \resizebox{100mm}{!}{\includegraphics[viewport=25  10  530 400, clip]{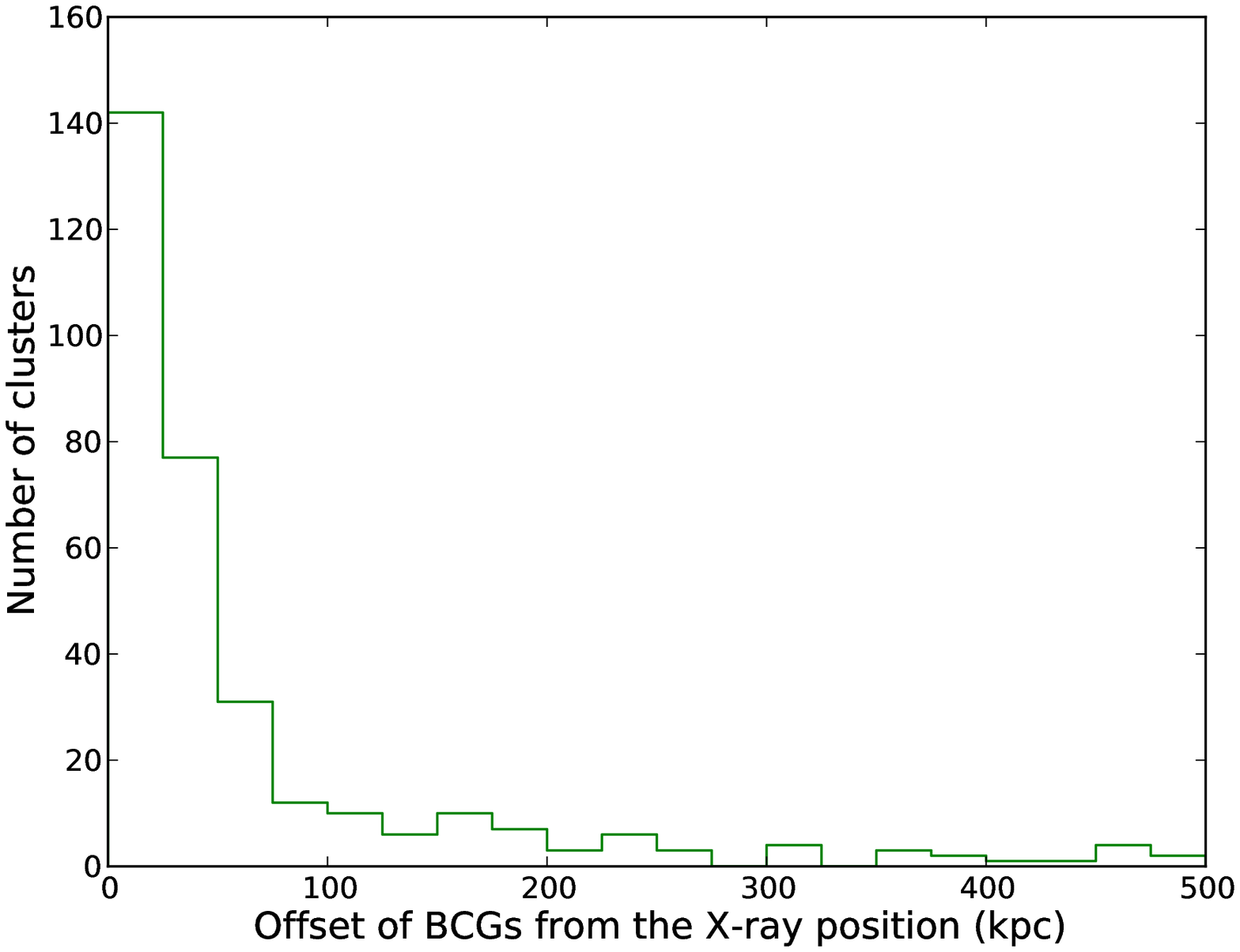}}}
  \caption{The distribution of the linear separations between the likely BCGs 
and the X-ray emission peaks of the cluster sample.} 
  \label{f:Hist_offset_DR9}
\end{figure}

Based on the cluster redshift and the angular separation of the BCGs to the 
X-ray peaks, we computed their linear offsets. The distribution of the linear 
separations between the likely BCGs and the X-ray emission peaks is shown in 
Figure~\ref{f:Hist_offset_DR9}. We found the majority of the BCGs (about  
90 percent) have offsets smaller than 200 kpc, which is in agreement with the 
offsets of BCGs sample in paper II. By using the current selection procedure of 
the BCGs, the maximum offset is of about 500 kpc. The large offset of the 
BCGs from the X-ray centroids might appear in systems with an ongoing 
merger or in dynamically active clusters \citep{Rykoff08}.


\subsection{Comparison with published redshifts}

The so far largest optically selected galaxy cluster sample was compiled 
by \citet[][WHL12 hereafter]{Wen12}, based on overdensities of galaxies in 
photometric redshift space from the SDSS-DR8 data. It comprises 132,684 
clusters with photometric redshift measurements in the range of 
$0.05 \le z_p < 0.8$. Cross-matching our sample with the WHL12 catalogue 
yielded 174 common clusters. 
We also queried the NASA Extragalactic Database (NED) for available redshift 
measurements for the remainder of the cluster sample. As a result, 67 clusters
with redshift estimations from different projects were found. In total, 241 
clusters are previously known in the literature mostly as optically confirmed 
galaxy clusters.

\begin{figure}
\centering{
  \resizebox{100mm}{!}{\includegraphics[viewport=25  10  530 400, clip]{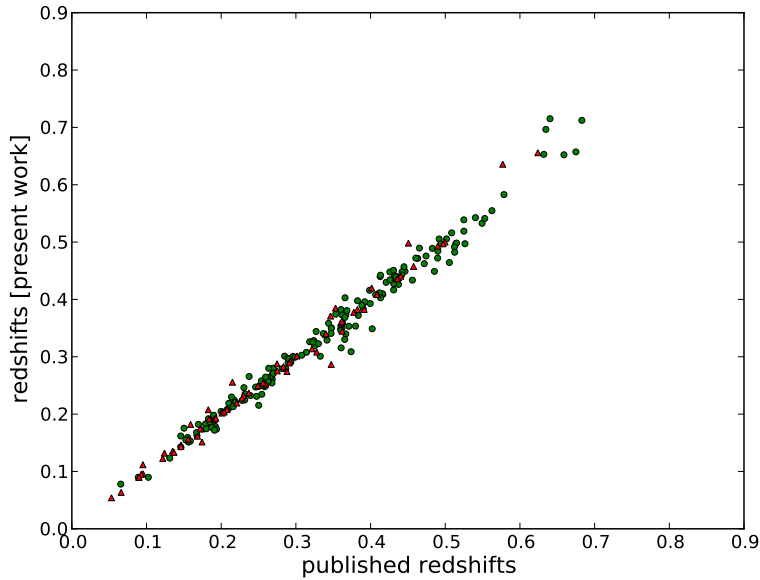}}}
  \caption{Comparison of the present spectroscopic redshift measurements with 
the published photometric ones (green dots) by \citet{Wen12} and the available 
(either spectroscopic or photometric) redshifts from the NED (red triangle).} 
\label{f:zpre_zpub}
\end{figure}

Figure~\ref{f:zpre_zpub} shows the comparison between the present redshift 
measurements and the WHL12 ones as well as the available redshifts from the NED.
The good agreement between the current redshift measurements and the published 
ones is clearly obvious. The differences between the two measurements, 
$z_{pre} - z_{pub}$, have a mean and standard deviation of 0.0001 and 0.0166, 
respectively.  

We also compared the current redshift measurements with the published ones of 
the optically confirmed cluster sample from our ongoing survey (Paper II). 
There are 280 common clusters between the two samples, of these 231 have 
spectroscopic redshifts for at least one cluster galaxy in Paper II.
The current procedure provided spectroscopic confirmation for the remainder 
of the common sample with only photometric redshifts (49 systems). 
We noted that the current procedure did not identify the whole sample 
in Paper II with spectroscopic confirmations (310 clusters). This is due 
to using the criterion of having a LRG with $z_{\rm s}$ within 200 kpc. 
In addition we used in Paper II the spectroscopic data  from the SDSSI/II 
projects, which comprises a galaxy sample with $z_{\rm s}$ including LRGs 
as well as a magnitude-limited galaxy sample that are not necessarily LRGs.      

\begin{figure}
\centering{
  \resizebox{100mm}{!}{\includegraphics[viewport=25  10  530 400, clip]{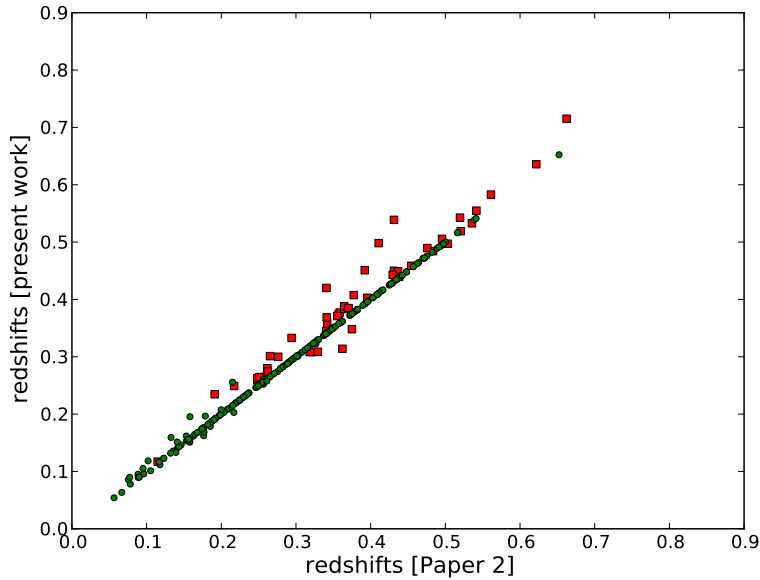}}}
  \caption{ Comparison of the present spectroscopic redshift measurements with 
the measured spectroscopic redshifts (green dots) and photometric ones 
(red squares) from Paper II.} 
\label{f:zp3_zp2}
\end{figure}

Figure~\ref{f:zp3_zp2} shows the comparison of the present redshift 
measurements with the ones from Paper II of the common sample. It shows 
a good agreement between the two measurements. The mean and standard 
deviation of the differences between the measured values, 
$\bigtriangleup z = z_{pre} - z_{pII}$, are 0.0035 and 0.0136, respectively. 
There is only 3 percent with redshift differences of 
$|\bigtriangleup z| > 2\sigma$, where $\sigma = 0.02$ the uncertainty of the 
measured photometric redshifts in Paper II. These redshift differences were  
found for the subsample with only photometric redshifts in our previous work.  
 
We also noted that the present spectroscopic redshifts are not identical with 
the spectroscopic redshifts in Paper II for a few cases.  There are  11 
clusters with  redshift differences of $0.01 < |\bigtriangleup z| < 0.04$.
This is due to the selection of cluster galaxies in Paper II was based on their 
photometric redshifts within a redshift interval of 
$z_{\rm p,\rm BCG} \pm 0.04(1+z_{\rm p,\rm BCG})$, then the cluster 
spectroscopic redshift was measured as the weighted average of the available 
spectroscopic redshifts of the identified cluster galaxies. This led to 
selecting galaxies with available $z_{\rm s}$ that have redshift outside 
the redshift interval used in this work ($z_{\rm s,\rm BCG} \pm 0.01$, 
see Section 3.2).              

%


\section{X-ray parameters}

Among the current optically validated cluster sample there are 280 clusters 
were published in Paper II together with their X-ray properties 
($F_{\rm X}, L_{\rm X}, T_{\rm X},$ and $M_{500}$). 
The current sample extends the optically confirmed cluster sample from our 
survey by 44 objects and adds a handful of clusters at higher redshift up to 
$z = 0.77$. Among the extended sample, about 80 percent are new X-ray 
detected galaxy groups/clusters. Since we found a good agreement 
between the measured redshifts of the common sample (see Sec. 3.4), 
there are no expected significant changes of their X-ray parameters.

In paper II we provided two subsamples of clusters; the first one with X-ray 
spectroscopic parameters since their X-ray data is sufficient to measure the 
parameters from X-ray spectral fits; the second subsample with X-ray parameters 
based on the X-ray flux given in the 2XMMi-DR3 catalogue since they have low 
quality X-ray data. We found a good agreement between the X-ray parameters 
measured from the two procedures.

In the current work, we measured the X-ray parameters for the present optically 
validated cluster sample (324 systems) based on the 2XMMi-DR3 flux. We computed 
the X-ray luminosities in (0.5-2 keV) using the catalogue fluxes in the same 
energy band and the measured redshifts. Then we extrapolated the computed 
luminosities to the bolometric luminosities, $L_{500}$, based on well 
established scaling relations in the literature through an iterative 
procedure that was described in detail in Paper I and Paper II. The 
derived $L_{500}$ was used to compute $M_{500}$. 

\begin{figure}
\centering{
  \resizebox{100mm}{!}{\includegraphics[viewport=25  5  530 400, clip]{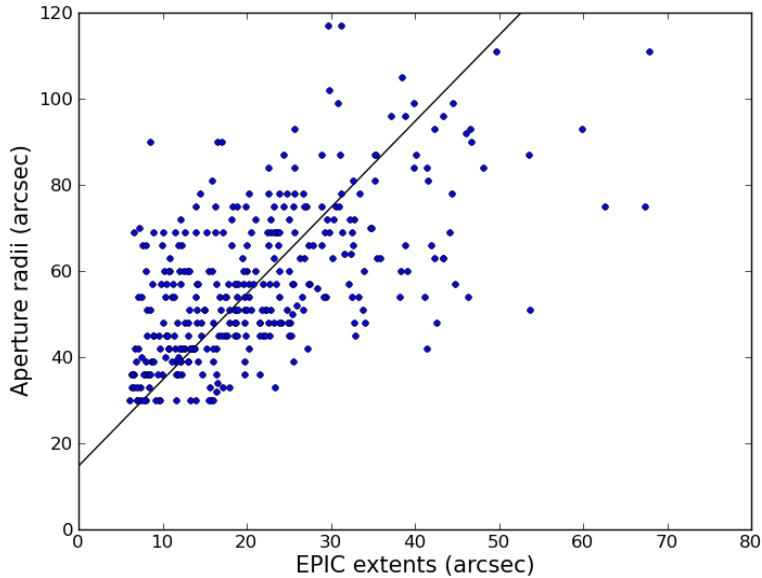}}}
  \caption[The aperture radii are plotted against the core radii (EPIC extents) 
obtained from the 2XMMi-DR3 catalogue for the cluster sample with X-ray 
spectroscopic parameters in Paper II]{The aperture radii are plotted against 
the core radii (EPIC extents) 
obtained from the 2XMMi-DR3 catalogue for the cluster sample with X-ray 
spectroscopic parameters in Paper II. The solid line represents the best fit  
using the BCES orthogonal regression. The slope and intercept of the best fit 
line are 2.00 and 14.71, respectively.} 
  \label{f:Rap-Rcat}
\end{figure}

The inputs of this method are redshift, an optimal aperture, and the enclosed 
bolometric luminosity within this aperture. The first input, redshift, is 
measured based on the current redshift procedure as described above. The 
other inputs are not available for the extended cluster sample. Therefore, 
we use the properties of the cluster sample with reliable X-ray parameters 
from spectral fits in Paper II in order to estimate the values of these 
input parameters. The optimal aperture radius is the cluster radius that 
represents the maximum signal-to-noise ratio. 
We used the linear relation between the optimal aperture radii and the core 
radii given in the 2XMMi-DR3 catalogue in order to estimate an optimal aperture 
radius based on the core radii available in the 2XMMi-DR3 catalogue.
Figure~\ref{f:Rap-Rcat} shows the relation between the optimal aperture radii 
and the core radii.The slope and the intercept of the best fit line obtained 
using the BCES orthogonal regression methods \citep{Akritas96} are 2.00 and 
14.71, respectively.

The third input is the aperture bolometric luminosity. For the extended cluster 
sample (44 systems), we only have the X-ray luminosity in [0.5-2.0] keV 
computed based on the integrated $\beta$ model flux in [0.5-2.0] keV.         
We also used the linear relation between the aperture bolometric luminosities 
and luminosities in [0.5-2.0] keV in order to convert the computed band 
luminosities to aperture bolometric luminosities, see Section 4.2 in Paper II 
for more information about this conversion. Based on these inputs, the 
iterative method provided the measurements of $R_{500}$, $L_{500}$, 
and $M_{500}$ for each cluster.
 
\begin{figure}
\centering{
  \resizebox{100mm}{!}{\includegraphics[viewport=15  5  530 420, clip]{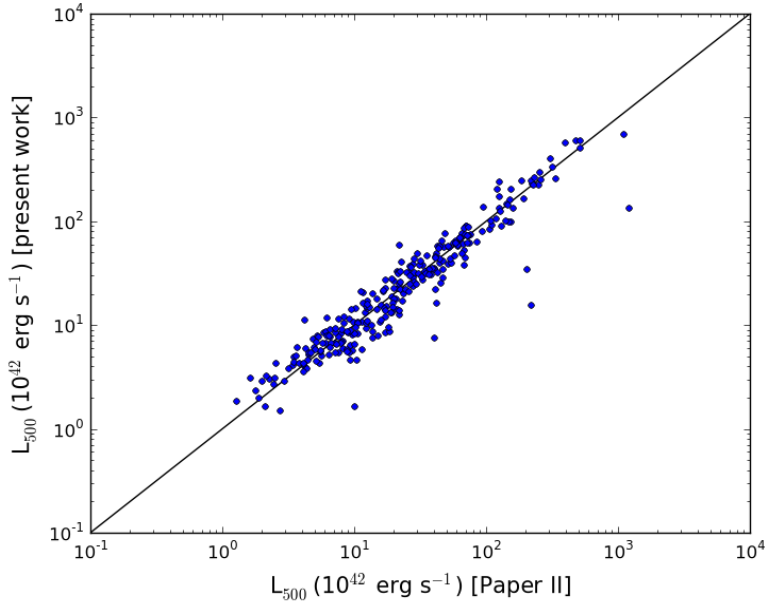}}}
  \caption[Comparison between the measured bolometric luminosity $L_{500}$ based 
on the given flux in the 2XMMi-DR3 catalogue and the bolometric $L_{500}$ from 
Paper II of the common cluster sample]{Comparison between the measured bolometric 
luminosity $L_{500}$ based 
on the given flux in the 2XMMi-DR3 catalogue and the bolometric $L_{500}$ from 
Paper II of the common cluster sample. The solid line shows the one-to-one 
relationship.} 
  \label{f:L500p2-L500cat}
\end{figure}

We compared the current measurements of $L_{500}$ based on the flux given 
in the 2XMMi-DR3 catalogue and the corresponding luminosities of the common 
cluster sample (280 systems) in Paper II. 
Figure~\ref{f:L500p2-L500cat} shows the comparison between the two measurements of 
$L_{500}$. It shows good agreement between the measured values except a few 
cases (about 4 percent) that are contaminated by point sources. 
The median value of the ratios  of the present $L_{500}$ measurements to the 
ones from Paper II is 1.03. Since there is such a good agreement between the 
measured luminosities from the current work and the previous one, we trust
the current procedure to measure the X-ray parameters for the extended cluster 
sample.

\begin{figure}
\centering{
  \resizebox{100mm}{!}{\includegraphics[viewport=15  5  530 420, clip]{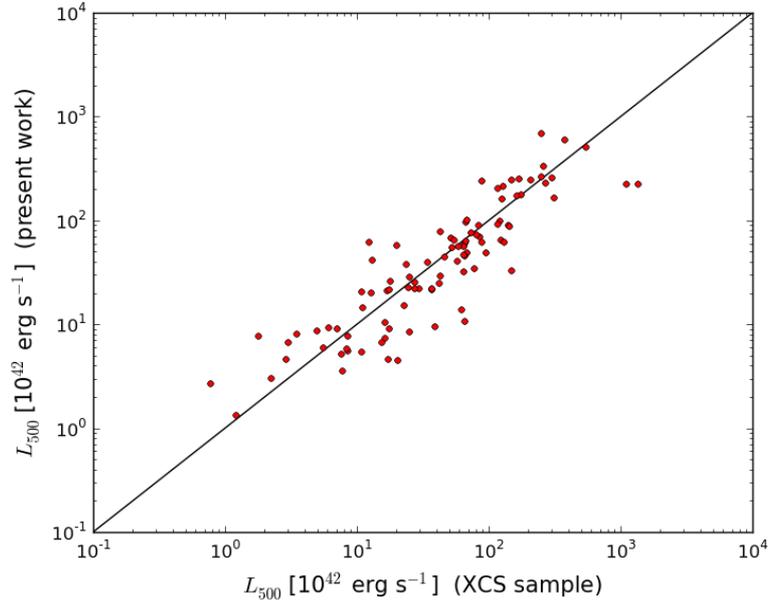}}}
  \caption[Comparison of the present measurements of $L_{500}$ and the ones 
from the XCS project]{Comparison of the present measurements of $L_{500}$ and the ones 
from the XCS project. The solid line shows the one-to-one relationship.} 
  \label{f:L500-Lxcs}
\end{figure}

The first data release of the XMM-Newton Cluster Survey 
\cite[XCS,][]{Mehrtens12} comprises about 500 clusters. Cross-matching the 
cluster sample with the XCS sample that have redshift and X-ray luminosity 
measurements within 30 arcsec yields 97 common objects. Among these, 89 
clusters are included in the optically confirmed cluster sample from our 
survey (Paper II).    
Figure~\ref{f:L500-Lxcs} shows the comparison of the current $L_{500}$ 
measurements and the corresponding ones from the XCS project. The median ratio 
between the two measurements is 0.94. In Paper II we showed a better agreement 
of $L_{500}$ values for the common sample if we constrain the comparison to  
the parameters derived from the spectral fits. 
Among the common sample, 29/97 have only photometric redshifts in the XCS sample, 
therefore our cluster sample provides spectroscopic confirmation for these 
systems. 
 
\begin{figure}
\centering{
  \resizebox{100mm}{!}{\includegraphics[viewport=15  5  530 420, clip]{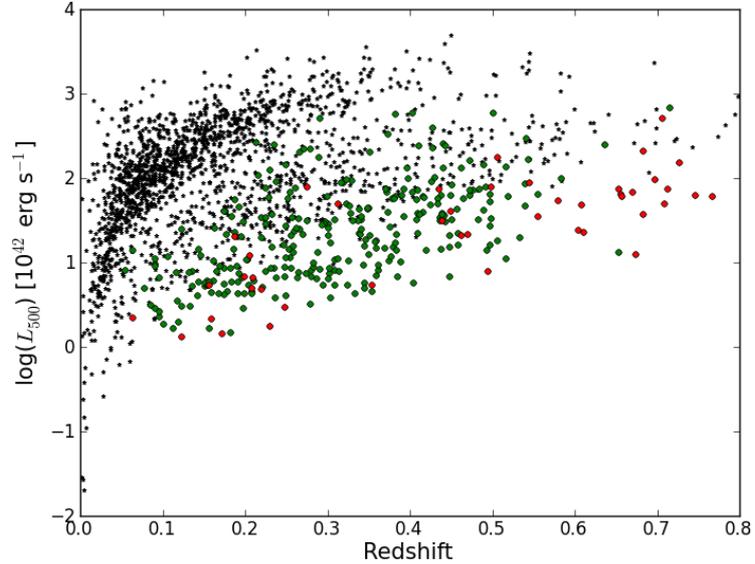}}}
  \caption{The distribution of the X-ray bolometric luminosities, $L_{500}$, 
with redshifts of the common cluster sample in Paper II (green dots), the 
extended cluster sample (red dots) from the current procedure, and a sample 
of 1730 clusters (black stars) below redshift 0.8 detected from ROSAT data 
\citep{Piffaretti11}.} 
  \label{f:L500-z-DR9}
\end{figure}

The mass range of the cluster sample is $\sim 2 - 33 \times 10^{13}$\ M$_\odot$ 
and luminosity range is $\sim 1 - 700 \times 10^{42}$\ erg\ s$^{-1}$.
Figure~\ref{f:L500-z-DR9} shows the distributions of $L_{500}$ as a function 
of redshift for the extended cluster sample and the common objects in Paper II 
as well as for 1730 clusters ($z < 0.8$) from MCXC catalogue that was comprised 
based on published X-ray selected cluster catalogues from ROSAT data 
\citep{Piffaretti11}.  Due to the sensitivity of XMM-Newton and deeper 
exposures for some fields the current cluster sample includes low luminosity 
groups and clusters at each redshift as shown in Figure~\ref{f:L500-z-DR9}.

Since the current X-ray luminosities are comparable with the previous 
measurements in Paper II, we only present a catalogue of the extended 
cluster sample (44 systems) listing their optical and X-ray parameters.
In addition to this sample we also provide the spectroscopic redshifts 
and the X-ray properties for 49 clusters among the common sample 
in paper II that had only photometric redshifts. These two subsample 
are compiled together in Table~\ref{tbl:catalog_sample} with a note 
refers to each subsample.

Table~\ref{tbl:catalog_sample}, available at the Appendix C, lists the first 10 
entries of the extended cluster sample (44 objects) from the current work 
in addition to a subsample (49 systems) that have only photometric redshifts 
in Paper II and have spectroscopic confirmations in the current work. The 
X-ray parameters are measured based on the flux given in the 2XMMi-DR3 
catalogue. 
Cols.~[1] and [2] report the cluster identification number (detection Id, detid) 
and its name (IAUNAME), cols.~[3] and [4] provide the coordinates of X-ray 
emission in equinox J2000.0, col.~[5] the XMM-Newton observation Id (obsid), 
col.~[6] the cluster spectroscopic redshift, col.~[7] the scale at the cluster 
redshift in kpc/$''$, col.~[8] the $R_{500}$ in kpc, cols.~[9], and [10] the 
2XMMi-DR3 X-ray flux $F_{\rm cat}$ [0.5-2.0] keV and its error in units of 
$10^{-14}$\ erg\ cm$^{-2}$\ s$^{-1}$, cols.~[11], and [12] the estimated 
X-ray luminosity $L_{\rm cat}$ [0.5-2.0] keV and its error in units of 
$10^{42}$\ erg\ s$^{-1}$, cols.~[13] and [14] the cluster bolometric 
luminosity $L_{500}$ and its error in units of $10^{42}$\ erg\ s$^{-1}$, 
cols.~[15] and [16] the cluster mass $M_{500}$ and its error in units of 
$10^{13}$\ M$_\odot$, 
col.~[17] the {\tt objid} of the likely BCG in SDSS-DR9, cols.~[18] and [19] 
the BCG coordinates in equinox J2000.0, col.~[20] the apparent magnitude 
m$_{\rm r}$ of the BCG, col.~[21] and [22] the weighted average 
spectroscopic redshift and the number of cluster members within 500 kpc 
with available spectroscopic redshifts that were used to compute the average 
redshift, col.~[23] and [24] the weighted average photometric redshift and 
the number of identified cluster member candidates within 500 kpc based on 
their photometric redshifts, col.~[25] the linear separation between the 
cluster X-ray position and the BCG position, 
and col.~[26] a note indicating the object status, 
$``$extended$``$ refers to a new system detected using the current procedure, 
and $``$Paper II$``$ refers to a previously known system from Paper II with 
photometric redshift and confirmed spectroscopically in the current work.


\section{Summary}

We present a sample of 324 X-ray selected galaxy groups and clusters 
associated with at least one Luminous Red Galaxy (LRG) that has a spectroscopic 
redshift in the SDSS-DR9. The redshifts of the associated LRGs are used to 
identify the BCGs and the other cluster galaxies with spectroscopic redshifts. 
The cluster spectroscopic redshift is computed as the weighted average of the 
available spectroscopic redshifts of the cluster galaxies within 500 kpc. 
The cluster sample spans a wide redshift range $0.05 < z < 0.77$ with a 
median of $z = 0.31$.  Among the cluster sample, 241 are previously known 
as optically selected galaxy clusters. 
The measured redshifts are consistent with the available redshifts in the 
literature. In addition to re-identify and confirm the redshift measurements 
of 280 clusters among the published cluster sample from our survey, we extend 
the optically confirmed cluster sample by 44 systems. Among the extended sample, 
55 percent are newly discovered groups and clusters and 80 percent are new 
X-ray detected clusters. 
The measured redshifts and the X-ray flux given in the 2XMMi-DR3 catalogue are 
used to determine the X-ray luminosity of the cluster sample. We also derived  
X-ray-luminosity-based mass of the sample. 
Comparing the current estimation of the X-ray bolometric luminosity, $L_{500}$, 
with the available ones from Paper II and the XCS project, we found a good 
agreement between the two measurements. The distribution of X-ray luminosities
of our cluster sample and ROSAT clusters with redshifts showed that we detected 
less luminous groups and clusters at each redshift. 
\clearpage
\begin{landscape}
\begin{table}
\caption{\label{tbl:catalog_sample} The first 10 entries of the extended 
cluster sample (44 objects) from the current work in addition to a subsample 
(49 systems) that had only photometric redshifts in Paper II and have 
spectroscopic confirmations in the current work.}
{\tiny
\begin{tabular}{c c c c c c c c c c c c c c c c }
\hline
\hline
  \multicolumn{1}{c}{detid\tablefootmark{a}} &
  \multicolumn{1}{c}{Name\tablefootmark{a}} &
  \multicolumn{1}{c}{ra\tablefootmark{a}} &
  \multicolumn{1}{c}{dec\tablefootmark{a}} &
  \multicolumn{1}{c}{obsid\tablefootmark{a}} &
  \multicolumn{1}{c}{z\tablefootmark{b}} &
  \multicolumn{1}{c}{scale} &
  \multicolumn{1}{c}{$R_{500}$} &
  \multicolumn{1}{c}{$F_{cat}$\tablefootmark{a,c}} &
  \multicolumn{1}{c}{$\pm eF_{cat}$} &
  \multicolumn{1}{c}{$L_{cat}$\tablefootmark{d}} &
  \multicolumn{1}{c}{$\pm eL_{cat}$} &
  \multicolumn{1}{c}{$L_{500}$\tablefootmark{e}} &
  \multicolumn{1}{c}{$\pm eL_{500}$} &
  \multicolumn{1}{c}{$M_{500}$\tablefootmark{f}} &
  \multicolumn{1}{c}{$\pm eM_{500}$} \\

  \multicolumn{1}{c}{} &
  \multicolumn{1}{c}{IAUNAME} &
  \multicolumn{1}{c}{(deg)} &
  \multicolumn{1}{c}{(deg)} &
  \multicolumn{1}{c}{} &
  \multicolumn{1}{c}{} &
  \multicolumn{1}{c}{kpc/$''$} &
  \multicolumn{1}{c}{(kpc)} &
  \multicolumn{1}{c}{(keV)} &
  \multicolumn{1}{c}{(keV)} &
  \multicolumn{1}{c}{(keV)} &
  \multicolumn{1}{c}{} &
  \multicolumn{1}{c}{} &
  \multicolumn{1}{c}{} &
  \multicolumn{1}{c}{} &
  \multicolumn{1}{c}{}  \\

(1) &  (2)  &  (3)  & (4)  &   (5)   &  (6)  &   (7)  &   (8)  &  (9)  &  (10)  &  (11) &  (12) & (13) & (14) & (15) &  (16)   \\
\hline 
005735 &     2XMM J003840.4+004746 &   9.66841 &   0.79636 & 0203690101 & 0.5549 & 6.44 &  522.21 &   1.44 & 0.18 &  17.83 &  2.19 &   35.42 &   5.05 &  7.42 &  1.56 \\
007554 &     2XMM J004304.2-092801 &  10.76751 &  -9.46695 & 0065140201 & 0.1866 & 3.12 &  594.22 &   6.84 & 1.06 &   6.74 &  1.05 &   20.30 &   3.72 &  7.18 &  1.56 \\
010986 &     2XMM J005556.9+003806 &  13.98720 &   0.63507 & 0303110401 & 0.2047 & 3.36 &  541.55 &   5.01 & 0.95 &   6.06 &  1.15 &   12.11 &   2.64 &  5.54 &  1.28 \\
021043 &     2XMM J015558.5+053159 &  28.99394 &   5.53329 & 0153030701 & 0.4499 & 5.76 &  640.92 &   5.82 & 0.85 &  43.44 &  6.37 &   84.97 &  15.25 & 12.11 &  2.54 \\
021597 &     2XMM J020019.2+001931 &  30.08012 &   0.32553 & 0101640201 & 0.6825 & 7.07 &  643.12 &   4.17 & 0.52 &  85.10 & 10.59 &  213.07 &  37.70 & 16.13 &  3.35 \\
030746 &     2XMM J023346.9-085054 &  38.44543 &  -8.84844 & 0150470601 & 0.2653 & 4.08 &  587.32 &   5.95 & 1.09 &  12.93 &  2.36 &   24.86 &   5.10 &  7.55 &  1.67 \\
030889 &     2XMM J023458.7-085055 &  38.74463 &  -8.84868 & 0150470601 & 0.2590 & 4.01 &  586.22 &   4.15 & 0.53 &   8.54 &  1.09 &   24.02 &   3.31 &  7.45 &  1.56 \\
089821 &     2XMM J083114.4+523447 & 127.81014 &  52.57993 & 0092800201 & 0.6107 & 6.74 &  470.06 &   0.79 & 0.11 &  12.25 &  1.64 &   22.80 &   3.72 &  5.78 &  1.26 \\
089885 &     2XMM J083146.1+525056 & 127.94516 &  52.84719 & 0092800201 & 0.5190 & 6.23 &  582.48 &   3.50 & 0.21 &  36.78 &  2.24 &   60.96 &   5.82 &  9.86 &  1.97 \\
091280 &     2XMM J083926.4+193658 & 129.86017 &  19.61622 & 0101440401 & 0.3742 & 5.16 &  481.60 &   0.93 & 0.13 &   4.50 &  0.62 &   10.74 &   1.48 &  4.71 &  1.03 \\
\hline
\end{tabular}
}
\end{table}


\addtocounter{table}{-1}
\begin{table}
\caption{\label{} continued.}
{\tiny
\begin{tabular}{c c c c c c c c c c c }
\hline
\hline
  \multicolumn{1}{c}{detid\tablefootmark{a}} &
  \multicolumn{1}{c}{objid\tablefootmark{g}} &
  \multicolumn{1}{c}{RA\tablefootmark{g}} &
  \multicolumn{1}{c}{DEC\tablefootmark{g}} &
  \multicolumn{1}{c}{$m_{r}$\tablefootmark{g}} &
  \multicolumn{1}{c}{$z_{s}$\tablefootmark{g}} &
  \multicolumn{1}{c}{$N_{z_{s}}$\tablefootmark{g}} &
  \multicolumn{1}{c}{$z_{p}$\tablefootmark{g}} &
  \multicolumn{1}{c}{$N_{z_{p}}$\tablefootmark{g}} &
  \multicolumn{1}{c}{offset\tablefootmark{g}} &
  \multicolumn{1}{c}{note\tablefootmark{h}}  \\
  
  \multicolumn{1}{c}{}&
  \multicolumn{1}{c}{(BCG)} &
  \multicolumn{1}{c}{(deg)} &
  \multicolumn{1}{c}{(deg)} &
  \multicolumn{1}{c}{(BCG)} &
  \multicolumn{1}{c}{} &
  \multicolumn{1}{c}{} &
  \multicolumn{1}{c}{} &
  \multicolumn{1}{c}{} &
  \multicolumn{1}{c}{(kpc)} &
  \multicolumn{1}{c}{}  \\
  (1)  &  (17)  & (18)  &  (19)  &  (20)   &  (21) &  (22)  &   (23)   &  (24) & (25) & (26)  \\
\hline

005735 & 1237663204918428144 &   9.68054 &   0.78241 & 20.047 & 0.5549 &  3 & 0.5127 &  7 & 429.63 & Extended \\
007554 & 1237652630713860232 &  10.80832 &  -9.47863 & 17.213 & 0.1866 &  2 & 0.1794 & 18 & 473.45 & Extended \\
010986 & 1237663784740388918 &  14.02537 &   0.62659 & 17.537 & 0.2047 &  3 & 0.1951 & 20 & 473.61 & Extended \\
021043 & 1237678663047250389 &  28.98754 &   5.53072 & 19.553 & 0.4499 &  1 & 0.4258 & 12 & 142.33 & Paper-II \\
021597 & 1237657071160263439 &  30.08100 &   0.32491 & 20.448 & 0.6825 &  1 & 0.6555 &  3 &  27.36 & Extended \\
030746 & 1237653500970139807 &  38.44673 &  -8.84925 & 17.540 & 0.2653 &  1 & 0.2547 & 17 &  22.30 & Paper-II \\
030889 & 1237653500970270877 &  38.74547 &  -8.84926 & 17.762 & 0.2590 &  2 & 0.2528 & 17 &  14.70 & Paper-II \\
089821 & 1237651701914141241 & 127.80965 &  52.57912 & 20.467 & 0.6107 &  1 & 0.6465 &  4 &  20.97 & Extended \\
089885 & 1237651272960967114 & 127.94343 &  52.84937 & 19.251 & 0.5190 &  2 & 0.5165 & 13 &  54.28 & Paper-II \\
091280 & 1237667107965108712 & 129.86275 &  19.61566 & 18.114 & 0.3742 &  3 & 0.3542 & 15 &  46.34 & Paper-II \\

\hline
\end{tabular}
}
\tablefoot{ The entire cluster catalogue is available in the Appendix C.
\tablefoottext{a}{Parameters extracted from the 2XMMi-DR3 catalogue.} 
\tablefoottext{b}{Spectroscopic redshift as given in col.~(21).}     
\tablefoottext{c}{2XMMi-DR3 flux, $F_{cat}$ [0.5-2.0] keV, and its errors in units of $10^{-14}$\ erg\ cm$^{-2}$\ s$^{-1}$.}
\tablefoottext{d}{Computed X-ray luminosity, $L_{cat}$ [0.5-2.0] keV, and its errors in units of $10^{42}$\ erg\ s$^{-1}$.} 
\tablefoottext{e}{X-ray bolometric luminosity, $L_{500}$, and its error in units of $10^{42}$\ erg\ s$^{-1}$.} 
\tablefoottext{f}{X-ray-luminosity-based mass $M_{500}$ and its error  in units of $10^{13}$\ M$_\odot$.}
\tablefoottext{g}{Parameters obtained from the current detection algorithm in the optical band.}
\tablefoottext{h}{A note about each system as $``$extended$``$: new cluster from the current algorithm, and $``$paper-II$``$: a cluster in paper II and spectroscopically confirmed from the present procedure.}
}

%
\end{table}
\end{landscape}

\chapter[IV. The correlation of X-ray and optical properties]{IV. The correlation of X-ray and optical properties\footnote{This chapter will be submitted to {\it Astronomy \& Astrophysics} }  }

\section*{Abstract}

We present the absolute magnitude and the optical luminosity in the $r-$band 
of the BCGs of the optically identified cluster sample (574 systems) from the 
2XMMi/SDSS Galaxy Cluster Survey. The cluster sample spans a wide redshift 
range 0.03-0.77 (median $z$=0.33). The correlation between the absolute 
magnitudes of the BCGs and the cluster redshifts is investigated. We also 
present the relation between the optical luminosity of the BCG and the 
cluster mass. For a subsample of 214 systems 
below a redshift of 0.42 with measured X-ray spectroscopic temperatures, we 
determined the optical richness and luminosity within $R_{500}$. We found 
a tight relation between the cluster richness and optical luminosity.
This subsample is used to perform a comparison of the appearance of galaxy 
clusters in the X-ray and optical bands. The X-ray parameters and the masses 
at the radius $R_{500}$ are obtained from the published cluster catalogue 
from our survey.     
We investigate the correlation between the X-ray properties (temperature 
$T_{\rm ap}$, luminosity $L_{500}$, and X-ray-luminosity-based mass $M_{500}$) 
and optical properties (richness and luminosity). The relation between the 
cluster richness and $T_{\rm ap}$, $L_{500}$, and $M_{500}$ has an orthogonal 
scatter of 41$\%$, 51$\%$, and 41$\%$, respectively while the relation 
between the optical luminosity and the same properties has an orthogonal 
scatter of 38$\%$, 48$\%$, 38$\%$, respectively. 
%
%



\section{Introduction}

Galaxy clusters are the largest clearly defined objects in the universe. 
Their baryonic matter has two components: first, the individual galaxies 
that can be studied through optical or infrared imaging and spectroscopy; 
the second, a hot diffuse intracluster medium (ICM) that can be observed 
with X-ray and microwave instruments. Clusters of galaxies are ideal 
laboratories to study the cosmic evolution of the  ICM and the cluster 
galaxy population. Precise observations of large cluster samples provide 
a powerful tool to constrain the cosmological parameters 
\citep[e.g.][]{Sarazin88, Rosati02, Voit05, Boehringer10, Allen11}. 

Solid observational evidence indicates a strong interaction between the two 
baryonic components of galaxy clusters. The evolution of galaxies in clusters 
is influenced by the hot diffuse gas in the ICM. The observed metal abundance in 
the ICM is produced by the pollution metals expelled from galaxies via galactic 
winds \citep[][]{Finoguenov01, DeGrandi02}. To understand the complex physics 
of galaxy cluster baryon components, it is required to combine X-ray and optical 
observations of a large sample of these systems \citep{Popesso04}.        

Several cluster surveys have been conducted in optical/IR wavelengths 
\citep[e.g.][]{Abell58, Abell89, Zwicky61, Gladders05, Merchan05, Koester07, 
Wen09, Hao10, Szabo11, Geach11, Durret11, Wen12, Gettings12} or in X-rays based 
on data from previous missions mainly from ROSAT \citep[e.g.][]{Ebeling98, 
Boehringer04, Reiprich02, Ebeling10, Rosati98, Burenin07} or using the 
current X-ray telescopes (XMM-Newton, Chandra, Swift/X-ray) 
\citep[e.g.][]{Finoguenov07, Finoguenov10, Adami11,Suhada12, Barkhouse06, 
Kolokotronis06, Fassbender11, Takey11, Mehrtens12, Clerc12, Tundo12, de-Hoon13, 
Takey13a, Takey13b}. So far, tens of thousands clusters have been optically 
detected while only a few thousands of clusters have been identified in X-rays.  

We have initiated a serendipitous search for galaxy clusters in XMM-Newton 
observation fields that are located in the footprint of the Sloan Digital Sky 
Survey, SDSS-DR7. The main aim of the survey is to construct a large catalogue 
of newly discovered X-ray selected groups and clusters up to high redshifts 
in order to investigate the X-ray scaling relations as well as X-ray-optical 
relations. Recently, we have published the optically confirmed cluster sample 
that comprises 530 galaxy groups and clusters from our survey based on the 
SDSS-DR8 \citep{Takey13a}. 
In addition, we have extended the optically confirmed cluster
sample by 44 objects based on the recent SDSS-DR9 \citep{Takey13b}. 
The published catalogues provide the optical redshift as well as the X-ray 
luminosity and X-ray luminosity-based mass for all systems. The X-ray data 
allowed to measure the X-ray temperatures for 60 percent of the sample. 
The published cluster sample with X-ray temperature measurements, which  
spans a wide redshift range up to 0.7, was used to investigate the 
X-ray luminosity-temperature relation \citep{Takey13a}.

In the present paper, we investigate the correlation between the cluster X-ray 
and the optical parameters that have been determined in a homogenous way based 
on the current optically identified cluster sample from our survey. Firstly, we 
investigate the correlations between the BCG properties (absolute magnitude 
and optical luminosity) and the cluster global properties (redshift and mass). 
Secondly, we  
compute the richness and the optical luminosity within $R_{500}$ (the radius at 
which the cluster mean density is 500 times the critical density of the 
universe) of a nearby subsample ($z \le 0.42$, with a complete 
membership detection from the SDSS data) with measured X-ray temperatures from 
our published catalogue. The relation between the estimated optical luminosity 
and richness is also presented. Finally, the correlation between the cluster 
optical properties (richness and luminosity) and the cluster global properties 
(X-ray temperature, luminosity, mass) are investigated.


Several studies have presented the correlations between the X-ray 
observables such as luminosity and temperature as well as the cluster mass  
with both optical richness and luminosity \citep[e.g.][]{Popesso04, Popesso05, 
Lopes06, Rykoff08, Lopes09, Wen09,Szabo11} based on known massive X-ray galaxy 
clusters in the literature apart from RASS-SDSS sample that comprised galaxy 
groups and clusters compiled by \cite{Popesso04}. 
These correlations show the ability 
to predict the X-ray properties of galaxy clusters, the most expensive to 
observe, and the cluster mass, the most important parameter for the cosmological 
studies, from the optical properties, and vice versa within a certain accuracy.   
These predictions are important for future galaxy cluster surveys conducted 
only in only one band of the electromagnetic spectrum.     
Here we investigate these correlations based on our large cluster sample 
including the newly discovered X-ray groups and clusters from the survey, 
which are in low and intermediate mass regime.

The presentation of the paper is as the follows. In Section 2, we describe  
the cluster sample and their X-ray properties. In Section 3, we present the 
properties of the BCGs against the cluster properties. The strategy to compute 
the richness and optical luminosity as well as their correlations with X-ray 
properties and masses are presented in Section 4. We summarise our results in 
Section 5. 
The cosmological parameters  $\Omega_{\rm M}=0.3$, $\Omega_{\Lambda}=0.7$ and 
$H_0=70$\ km\ s$^{-1}$\ Mpc$^{-1}$ were used throughout this paper. 



\section{The study sample and their X-ray properties}

The 2XMMi/SDSS Galaxy Cluster Survey as well as the so far optically confirmed 
cluster sample are presented in \citet[][]{Takey11, Takey13a, Takey13b}. Here 
we present a brief summary about the full optically confirmed cluster sample 
from the survey, which is used in the present work.

The survey comprises 1180 X-ray cluster candidates selected from the second 
XMM-Newton serendipitous source catalogue (2XMMi-DR3), which are in the 
footprint of the SDSS-DR7 that was available at the initiating of our survey. 
The total overlap area of the XMM-Newton fields
at high galactic latitudes, $|b| > 20^{\circ}$, in the sky coverage by the SDSS-DR7 
is 210 deg$^2$. To measure the X-ray properties of these candidates, their 
redshifts need to be determined. The X-ray data quality of these candidates 
do not permit to measure their redshifts from the emission lines in the X-ray 
spectrum apart from very bright X-ray sources (about 4 percent with more than 
2000 photon counts). Therefore, the main resource to measure their redshifts 
is the optical data. The so far 
largest optical survey is the SDSS, which covers a sky area of 14,555 
deg$^2$ and provides positions, magnitudes, and photometric redshifts of 
galaxies, in addition to spectroscopic redshifts for about one and half 
millions galaxies \citep{Ahn12}.      

Based on the photometric redshifts of galaxies in the SDSS-DR8, we have optically 
identified the counterparts of 530 galaxy groups/clusters \citep{Takey13a}. 
Then we constructed a cluster sample of 324 systems with spectroscopic 
confirmations based on the available spectroscopic redshifts of the Luminous 
Red Galaxies (LRGs) in the SDSS-DR9 \citep{Takey13b}. 
The spectroscopic confirmed cluster sample extends 
the optically confirmed cluster sample by 44 objects and provides spectroscopic 
confirmations for 49 systems. Therefore, the full cluster list with optical 
counterparts and redshift measurements comprises 574 galaxy groups/clusters. 
This is the so far largest X-ray selected galaxy clusters catalogue based on 
XMM-Newton observations.

\begin{figure}
\centering{
  \resizebox{100mm}{!}{\includegraphics[viewport=10  10  525 420, clip]{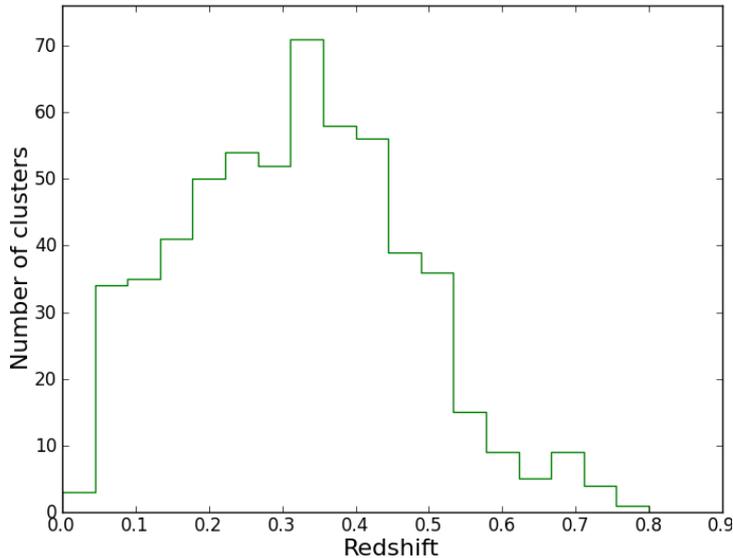}}}
  \caption[The distribution of the cluster sample redshifts]{The distribution of 
the cluster sample redshifts. The sample includes spectroscopic redshifts for 
403 objects based on at least a spectroscopic redshift of one cluster galaxy 
as well as the photometric redshifts for the remainder of the sample 
(171 systems).} 
  \label{f:zc-hist}
\end{figure}

The redshifts of the cluster sample span a wide range from 
0.03 to 0.77 with a median of 0.33. The distribution of the redshifts of the 
cluster sample is shown in Figure~\ref{f:zc-hist}. Among the confirmed cluster 
sample, 70 percent have spectroscopic redshifts for at least one cluster galaxy. 
About 40 percent of the cluster sample are newly discovered systems as galaxy 
clusters, while more than 70 percent are new X-ray detections of clusters.

\begin{figure}[t]
\centering{
  \resizebox{100mm}{!}{\includegraphics[viewport=10 10  525 420, clip]{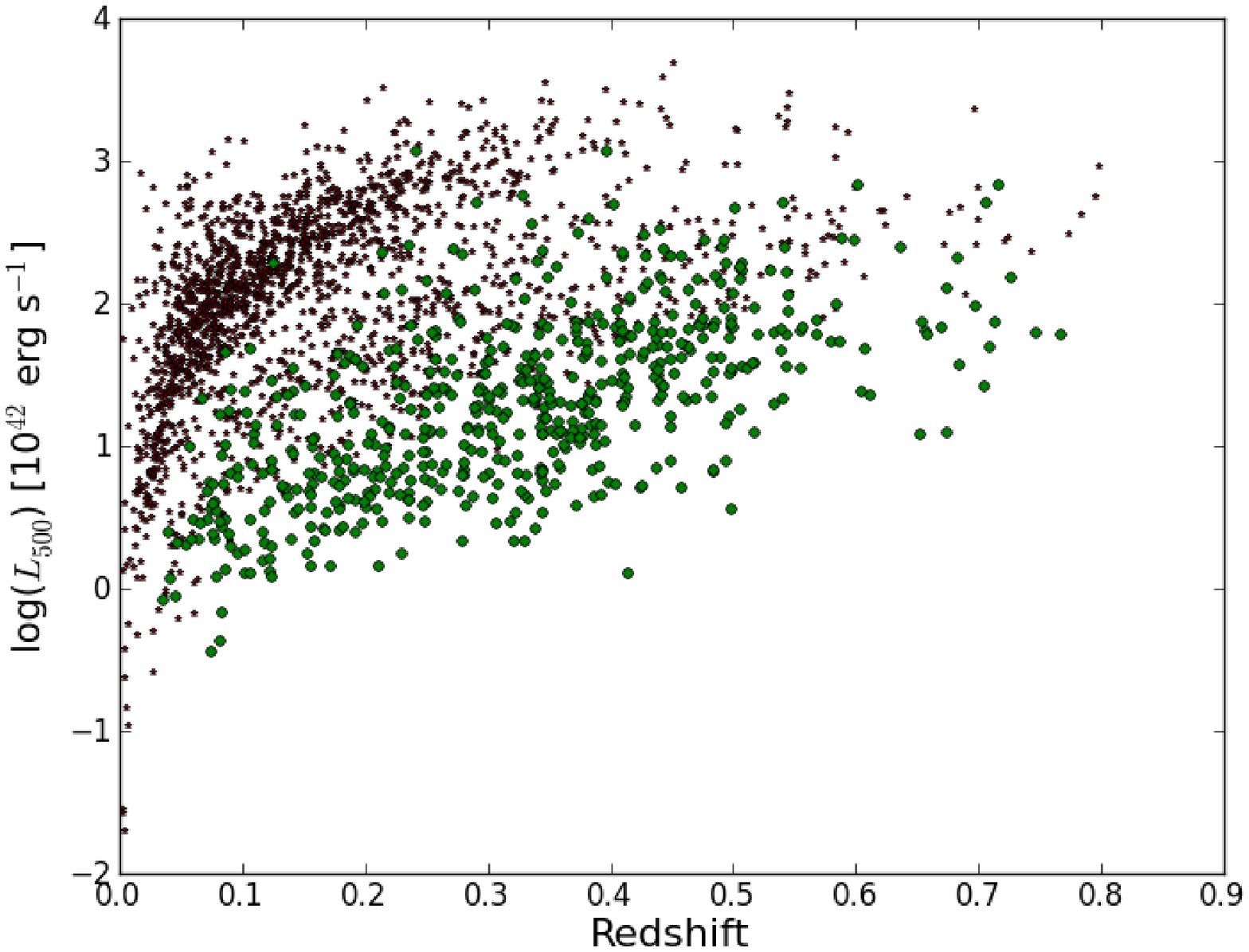}}}
  \caption[The distribution of the X-ray bolometric luminosity, $L_{500}$, 
with the redshift of the full optically confirmed cluster sample]{The 
distribution of the X-ray bolometric luminosity, $L_{500}$, 
with the redshift of the full optically confirmed cluster sample (green dots) 
and a sample of 1730 clusters (black stars) below redshift = 0.8 from the MCXC 
catalogue detected based on ROSAT data \citep{Piffaretti11}.} 
  \label{f:L500-z}
\end{figure}

For each system in the sample, we have derived the X-ray bolometric luminosity, 
$L_{500}$, based on the measured luminosity from the spectral fits or based on 
the flux given in the 2XMMi-DR3 catalogue and the measured redshift. The 
distribution of the measured luminosities of the cluster sample as well as for 
identified clusters from ROSAT data \citep{Piffaretti11} as a function of 
redshift is shown in Figure~\ref{f:L500-z}. The distribution shows that our 
X-ray detected groups and clusters are in the low and intermediate luminosity 
regime apart from few luminous systems, thanks to the XMM-Newton sensitivity 
and the available XMM-Newton deep fields. The measured luminosities are used 
to measure the cluster mass, $M_{500}$, based on the published $L - M$ relation 
by \cite{Pratt09}. The mass range of the cluster sample is 
$1.1 \times 10^{13} - 4.9 \times 10^{14}$\ M$_{\odot}$. 
For a sub-sample of 345 systems with good X-ray data quality, we 
measured their X-ray spectroscopic temperatures within an aperture with 
a radius chosen to optimize the signal-to-noise ratio. The measured 
temperatures ranges from 0.5 keV to 7.5 keV \citep{Takey13a, Takey13b}.

In the current study, we use the whole optically confirmed sample to investigate 
the relation between the absolute magnitude of the BCGs and the cluster redshift 
as well as the relation between the BCG luminosity in $r-$band and the cluster 
mass, see the following section. The systems at redshift below 0.42 with X-ray 
spectroscopic temperatures are used to investigate the correlation between the 
optical properties (richness and total luminosity) and the X-ray properties 
(temperature, luminosity, mass), see Section 4.

\section{Correlations of the BCG and cluster properties}

To determine the optical properties of the cluster sample, we created  
a galaxy sample from the SDSS-DR9 for each object by selecting the 
surrounding galaxies within 4 Mpc from the X-ray source position. 
The galaxies were selected from the {\tt Galaxy} view table, which contains 
the photometric parameters measured for resolved primary objects, classified 
as galaxies.  Also, the photometric redshifts and, if available, the 
spectroscopic redshifts of the galaxy sample were obtained from the 
{\tt Photoz} and {\tt SpecObj} tables, respectively. The extracted 
parameters of the galaxy sample include the coordinates, the apparent 
dereddened (model and composite model) magnitudes, K-correction, the photometric 
redshifts, and, if available, the spectroscopic redshifts. When the 
spectroscopic redshifts of galaxies are available, we use them instead of the 
photometric ones. To clean the galaxy sample from faint objects and from 
galaxies with large photometric redshift errors, we apply a magnitude cut of               
$ m_{r} \leq 22.2$ mag, $ \bigtriangleup m_{r} < 0.5 $ mag, and a fractional 
error cut of the photometric redshift, $ \bigtriangleup z_{p} / z_{p} < 0.5 $.

For the whole optically confirmed cluster sample we identify the BCG as the 
brightest galaxy among the cluster luminous member candidates that are 
selected within the measured $R_{500}$ and redshift interval from the 
cluster redshift of $z \pm 0.04(1+z)$. The luminous member galaxies are 
selected based on their evolution-corrected absolute magnitude that are 
brighter than $M_{r, ev} = -20.5$. The strategy for selecting the cluster 
luminous member galaxies is similar to that used by \cite{Wen12}.
For each cluster galaxy, we computed 
its absolute magnitude in the $r-$band, M$_{\rm r}$, as the follows :

\begin{equation}
M_{\rm r} = m_{\rm r} - 25 - 5\ log_{10}(D_{\rm L}/ 1 Mpc) - K_{\rm (r)}.  
\end{equation}
where m$_{\rm r}$ is the dereddened model magnitude of the cluster galaxy. 
$K_{\rm (r)}$ is the K-correction in the $r-$band derived from the templates 
and given in the {\tt Photoz} table in the SDSS-DR9 catalogue data 
\citep{Csabai03}. 
D$_{\rm L}$ is the luminosity distance that was calculated based on the cluster 
redshift. Then, we corrected the absolute magnitude for a passive evolution as : 

\begin{equation}
M_{r, ev} = M_{\rm r} + Q\, z.   
\end{equation}
where Q = 1.62 represent the evolution of the luminosity in units of magnitude 
per unit redshift, which means that the galaxy was more luminous in the past 
\citep{Blanton03}.

\begin{figure}
\centering{
  \resizebox{100mm}{!}{\includegraphics[viewport=10  10  530 420, clip]{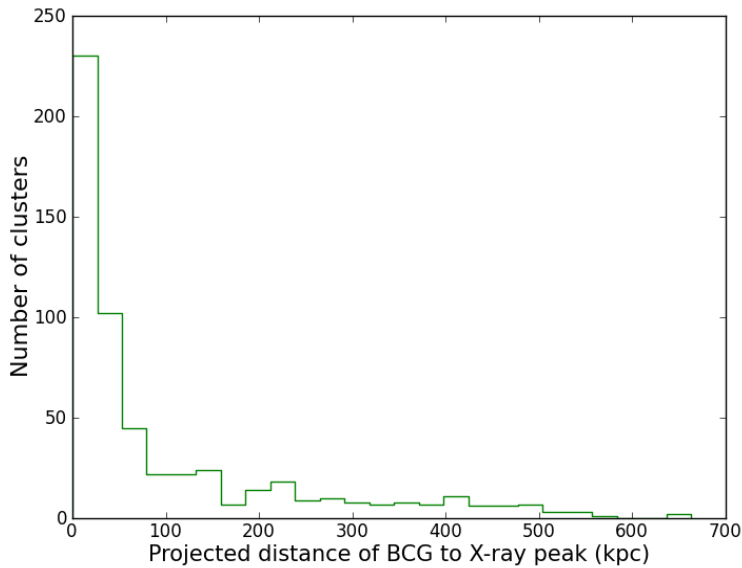}}}
  \caption{The distribution of the linear separations between the BCG positions 
and the X-ray emission peaks of the cluster sample.} 
  \label{f:BCG-lsep}
\end{figure}

Based on the cluster redshift and the angular separation of the BCG from the 
X-ray peak, we computed its linear separation. The distribution of the linear 
separations of the BCGs from the X-ray centroids of the sample is shown in 
Figure~\ref{f:BCG-lsep}. About 80 percent of the BCGs reside within 
200 kpc from the X-ray positions. This percentage is lower than what we found 
in our published catalogue (90 percent). This is due to selecting the 
BCG in the previous work as brightest galaxy among the cluster member 
candidates within one arcmin from the X-ray peak. The systems with large 
offset of the BCGs might have an ongoing merger or are dynamically active 
clusters \citep{Rykoff08}.

The BCGs are in general elliptical massive galaxies and tend to be luminous 
red galaxies (LRGs) \citep[e.g.][]{Postman95, Eisenstein01, Wen12}. We 
compared the BCGs magnitudes and colours with the magnitude and colour 
cuts of the spectroscopically targeted LRGs in the SDSS-III project, BOSS 
survey\footnote{\url{http://www.sdss3.org/dr9/algorithms/boss_galaxy_ts.php}}, 
(Padmanabhan et al. 2013, in preparation). 
We found 74 percent of the BCGs satisfy the LRGs criteria in the BOSS survey.
\citet{Wen12} constructed a catalogue of 132,684 clusters based on the 
photometric redshift of galaxies in the SDSS-DR8. They found 66 percent 
of the BCGs in their cluster sample fulfill the colour cuts of the SDSS LRG 
selection criteria.

\begin{figure}
\centering{
  \resizebox{100mm}{!}{\includegraphics[viewport=10  10  525 420, clip]{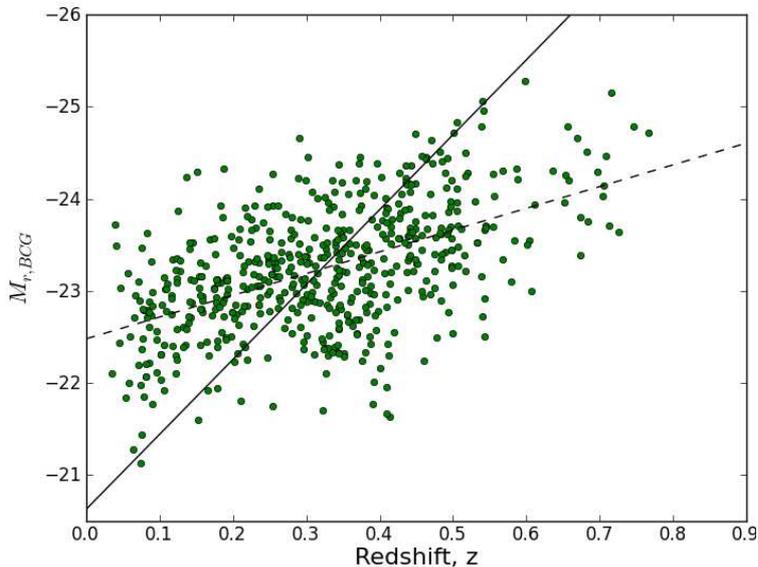}}}
  \caption[The absolute magnitude of the BCG, M$_{\rm r,\, BCG}$, are plotted 
against the cluster redshift, $z$, for the full cluster sample]{The absolute 
magnitude of the BCG, M$_{\rm r,\, BCG}$, are plotted 
against the cluster redshift, $z$, for the full cluster sample. The dashed and solid 
lines are the best-fit relations derived from the BCES(Y$|$X) and orthogonal 
regression methods, respectively.} 
  \label{f:MrBCG-z}
\end{figure}

Several studies \cite[e.g.][]{DeLucia07, Stott08} showed that the BCGs were 
formed at high redshift, $z \ge 2$, and have been passively evolving to the 
present day. The stellar population of the BCGs becomes old with cosmic time 
and consequently the BCGs become fainter at low redshifts. For each BCG we  
computed its absolute magnitude in the $r-$band, M$_{\rm r,\, BCG}$ using Eq.~1. 
Figure~\ref{f:MrBCG-z} shows the relation between M$_{\rm r,\, BCG}$ and the 
cluster redshift, $z$. It shows a weak relation indicating the higher redshift 
of the cluster, the brighter the BCG is.  

We derived the best-fit-linear relation between M$_{\rm r,\, BCG}$ and $z$ 
using the BCES regression method \citep{Akritas96}. 
The BCES algorithm provides several kinds of linear regression. Here we only 
show the results of BCES(Y$|$X) fitting method, which minimises the residuals 
in Y, and the BCES orthogonal fitting method, which minimises the squared 
orthogonal distances. The linear relation between  M$_{\rm r,\, BCG}$   
and $z$ is shown in Figure~\ref{f:MrBCG-z} and given by :

\begin{equation}
M_{\rm r,\, BCG} = a + b\, z.  
\end{equation}

The fitted parameters (intercept and slope) derived from both BCES methods are 
listed in the first row of the upper part of Table~\ref{tbl:relations}. 
The best-fit-lines 
are also plotted in Figure~\ref{f:MrBCG-z}. The slope derived by the BCES(Y$|$X) 
algorithm, -2.36 $\pm$ 0.16, is much steeper than the published one, 
-1.74 $\pm$ 0.03, by \cite{Wen12} based an optically cluster sample at 
$z \leq 0.42$. This disagreement might be due to our cluster sample 
includes less massive clusters and extends to higher redshifts than the sample 
used by \cite{Wen12}. In addition to using a different fitting method in both 
projects.     

To estimate the orthogonal scatter in the relation, we compute the differences 
between the measured absolute magnitudes and the predicted ones from the 
best-fit relation (the fit residuals) based on the cluster redshift from 
Eq.~3. The residuals of the fit and the slope of the relation are used to 
compute the orthogonal distances from the best-fit line. The standard 
deviation of these orthogonal distances was considered as the orthogonal 
scatter in the relation. The scatter error was computed as the standard 
error of the measured scatter value. The orthogonal scatters derived from 
the BCES(Y$|$X) and the orthogonal routines of the $M_{\rm r,\, BCG} - z$ 
relation are 0.22 $\pm$ 0.01 and 0.13 $\pm$ 0.01, respectively and are 
listed in Table~\ref{tbl:relations}.  


Several groups \cite[e.g.][]{Lin04, Brough08, Popesso07, Hansen09, Mittal09} 
investigated the relation between the optical luminosity of the 
BCGs and the cluster masses. They found a relation with different exponents as 
$L_{\rm BCG}\, \propto \, M^{0.1...0.6 }$. We measured the optical luminosity of 
the BCGs in $r-$band, $L_{\rm r, BCG}$, of the cluster sample. The absolute 
magnitude of each BCG was transformed to absolute luminosity in units of 
solar luminosities as :

\begin{equation} 
L_{\rm r,\, BCG}(L_{\rm r,\odot}) = 10^{(M_{\rm r,\odot} - M_{\rm r,\, BCG})/2.5}. 
\end{equation}

\begin{figure}
\centering{
  \resizebox{100mm}{!}{\includegraphics[viewport=10  0  525 420, clip]{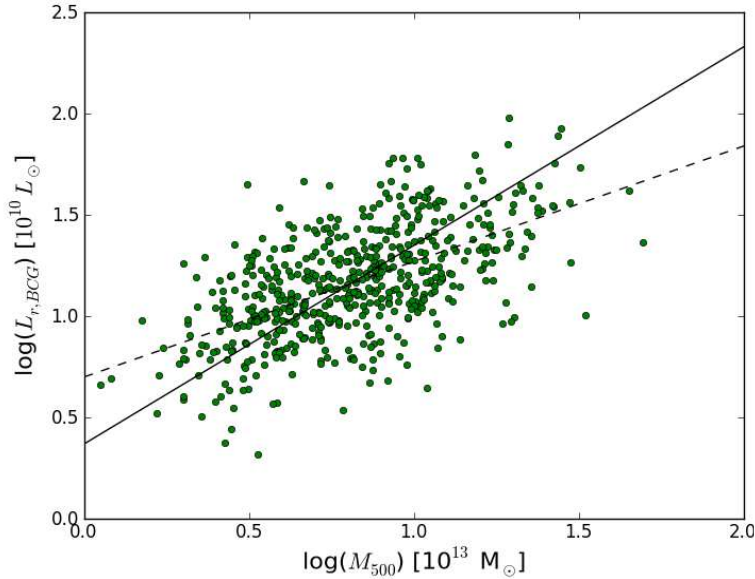}}}
  \caption[The optical luminosity of the BCGs in $r-$band, L$_{\rm r,\, BCG}$, 
versus the total cluster mass, M$_{500}$]{The optical luminosity of the BCGs 
in $r-$band, L$_{\rm r,\, BCG}$, versus the total cluster mass, M$_{500}$. The 
best-fit-lines are the same in Figure~\ref{f:MrBCG-z}.} 
  \label{f:LrBCG-M500}
\end{figure}

The solar absolute magnitude, $M_{\rm r,\odot}$, was transformed from the 
Johnson-Morgan-Cousins system to the SDSS system using the transformation 
equations by \cite{Jester05}. The best fit relation between the logarithms of 
$L_{\rm r, BCG}$ and the mass of the host cluster, $M_{500}$, is shown in 
Figure~\ref{f:LrBCG-M500} and given by :

\begin{equation}
\log\ (L_{\rm r,\, BCG}) = a + b\, \log\ (M_{500}).  
\end{equation}

The slope, intercept, and the orthogonal scatter of the relations derived from 
both BCES fitting algorithms are listed in Table~\ref{tbl:relations}. 
The scatter is computed in a similar way as in $M_{\rm r,\, BCG} - z$ relation 
but in a logarithmic scale since the fit of the $L_{\rm r,\, BCG}\ -\ M_{500}$ 
relation is performed in log-log scale. \cite{Mittal09} investigated the 
$L_{\rm K,\, BCG}\ -\ M_{500}$ relation based on the HIFLUGCS sample that 
comprises 64 bright clusters ($z \leq 0.2$) observed with {\it Chandra}.
They found a slope derived from the BCES bisector linear regression routine 
of 0.62 $\pm$ 0.05, which is consistent with the present slope 0.57 $\pm$ 0.03 
obtained using BCES(Y$|$X). The present slope is much steeper than the ones 
(around $\sim 0.3$) published by \citet[e.g.][]{Lin04, Brough08, Popesso07, 
Hansen09}.  
%

\section{Correlations of the cluster X-ray and optical properties}

We present the performance of the cluster richness and optical luminosity 
as predictors of the global X-ray properties (temperature and luminosity) as 
well as the cluster mass.     
\citet{Popesso07} reported that there is no significant difference among the 
results of the relation between the cluster global properties and the cluster 
optical luminosity in the different SDSS bands, therefore we will only present 
the relations for the optical luminosity in the $r-$band. \citet{Lopes09} 
found that the relations based on parameters extracted within $R_{500}$ had 
smaller scatter than the relations based on parameters derived within 
a fixed radius or within $R_{200}$. Therefore we only show here the relations 
between the X-ray and optical properties that were measured within $R_{500}$ 
except the X-ray temperature that was measured within an optimal aperture, 
which maximises the signal-to-noise ratio.

\subsection{The cluster richness and optical luminosity}

We use a method similar to that described by \cite{Wen12} to estimate the 
richness and optical luminosity of the study sample. To estimate the 
cluster richness, we identify the luminous member candidates within $R_{500}$ 
and a redshift interval from the cluster redshift of $z \pm 0.04 (1+z)$ and 
with evolution-corrected absolute magnitude $M_{\rm r, ev} \leq -20.5$. 
To compute $M_{\rm r, ev}$, we first determine the absolute magnitude of 
each galaxy using Eq.~1, then we correct it for a passive evolution using 
Eq.~2. The bright end of the cluster member candidates is the absolute 
magnitude of the BCG, M$_{\rm r,\, BCG}$.  The BCG was identified as the 
brightest galaxy among the cluster luminous galaxies within $R_{500}$.

The cluster luminous member galaxies with $M_{\rm r, ev} \leq -20.5$ are 
complete for clusters up to a redshift of $z = 0.42$. This is due to 
the magnitude limit of the SDSS photometric data and the faint limit absolute 
magnitude used for identifying the luminous member galaxies \citep{Wen12}.
Therefore, we only consider clusters below $z = 0.42$ in investigating the 
correlation between both the richness and the optical luminosity with the X-ray 
properties (temperatures, luminosity, and mass). We also include only those 
clusters with good quality X-ray data that permitted to measure the X-ray 
temperatures. The subsample includes 214 systems with a temperature $kT$ range 
of 0.5-6.8 keV and mass (X-ray-luminosity-based mass) range
of $M_{500} = 1.7 \times 10^{13} - 4.9 \times 10^{14}$\ M$_{\odot}$.

The net cluster member galaxies in the region with a radius of $R_{500}$ 
after subtraction the expecting foreground and background galaxies is defined 
as the cluster richness, $R$. 
To estimate the number of local foreground and background galaxies, we 
need to count the number of galaxies within a nearby region of the cluster 
with same redshift and absolute magnitude range of the luminous cluster member 
candidates. 
To do this, we divide the annuals between 2-4 Mpc from the X-ray position 
into 4 sectors. The number of foreground and background galaxies is determined 
as the mean value of the normalized number of galaxies to $R_{500}$ region in 
the 4 sectors within the used redshift and absolute magnitude intervals.   
We also compute the standard deviations $\sigma$ of the galaxy counts in 
these sectors. If a sector has a number of background galaxies larger than 
3$\sigma$ deviation from the mean galaxy count of the 4 sectors, we discard 
it and re-calculate the mean of galaxy counts in the remaining sectors. This 
was done in order to avoid any filaments or substructures belonging to the 
system.

The optical luminosity of a cluster $L_{r}$ is determined in a similar way as 
the estimation of the richness. It is computed as the sum of the 
cluster member luminosities within $R_{500}$ after subtraction the contamination 
of the background galaxies luminosity. The luminosity of background galaxies is 
computed as the mean of the summed optical luminosity of the galaxies with same 
redshift and absolute magnitude intervals of member galaxies in the 4 sectors. 
The optical luminosity of the cluster member candidates and the background 
galaxies are determined for the final galaxy counts resulting from the richness 
measurements. The absolute magnitude of a galaxy was transformed to absolute 
luminosity in units of solar luminosities using Eq.~4.

\begin{figure}
\centering{
  \resizebox{100mm}{!}{\includegraphics[viewport=10  10  525 420, clip]{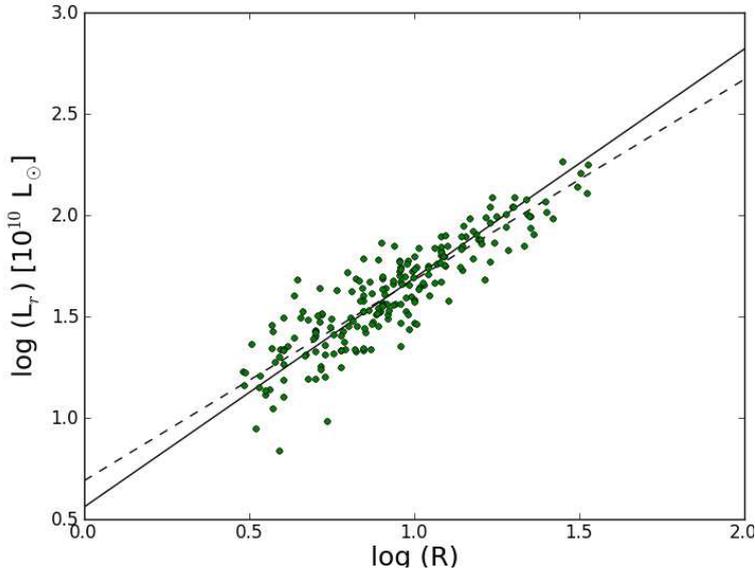}}}
  \caption[Correlation between the summed optical luminosity in $r-$band, 
$L_{\rm r}$, and the richness, $R$]{Correlation between the summed optical luminosity in $r-$band, 
$L_{\rm r}$, and the richness, $R$, which are computed within $R_{500}$, 
for the cluster subsample below $z=0.42$ with available X-ray spectroscopic 
temperature measurements from our survey. The best-fit-lines are the same 
in Figure~\ref{f:MrBCG-z}.} 
  \label{f:R_Lr_R500}
\end{figure}

Figure~\ref{f:R_Lr_R500} shows the relation between the summed $r-$band 
luminosities and richness that were computed within $R_{500}$ of the 
subsample below $z = 0.42$. We found a strong correlation between the 
measured optical luminosity and the richness. The best linear fits using 
the BCES regression methods  between their logarithms are shown in 
Figure~\ref{f:R_Lr_R500} and are expressed by :

\begin{equation}
\log\ (L_{\rm r}) = a  + b\, \log\ (R).
\end{equation}     

The slope of the relation is 1.13 $\pm$ 0.04 and 0.99 $\pm$ 0.03 obtained from 
the BCES orthogonal and the BCEX(Y$|$X) algorithms, respectively. The slopes, 
intercepts, and orthogonal scatters are given in first raw of the lower part of 
Table~\ref{tbl:relations}. The scatter is computed in a similar way as 
described above and given in a logarithmic scale. 

\citet{Popesso07} investigated the relation between the cluster $r-$band 
optical luminosity within $R_{200}$ and the net count of cluster galaxies 
(richness) based on an optically and X-ray selected group and cluster sample 
of 217 systems at $z \leq 0.25$. 
They found the relation slope is 1.00 $\pm$ 0.03 derived from an orthogonal 
regression method, which is shallower than the present slope 1.13 $\pm$ 0.04 
(orthogonal) but it is consistent with the slope 0.99 $\pm$ 0.03 derived using  
the BCEX(Y$|$X) algorithm. The linear fit derived from the BCES 
orthogonal method is affected by the two points with large scatter from the 
best-fit line.  
\citet{Wen09} found a slope of 0.97 $\pm$ 0.01 between $L_{\rm r}$ and $R$ that 
were estimated within 1 Mpc from the BCG position for a large optically 
selected cluster sample at $z \leq 0.42$, which is consistent with the present 
slope 0.99 $\pm$ 0.03 derived from the BCEX(Y$|$X) algorithm.

\subsection{The correlation of the cluster richness and optical luminosity with 
the cluster global properties}

\begin{figure*}
\centering{
  \resizebox{\hsize}{!}{\includegraphics[viewport=10  65  577 330, clip]{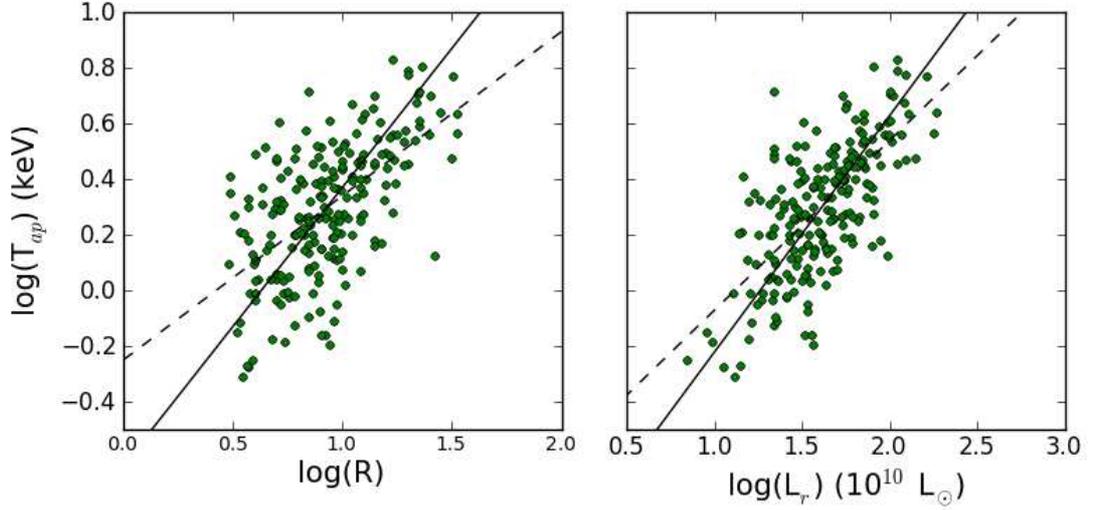}}}
  \caption[Correlation between the X-ray spectroscopic aperture temperature, 
$T_{ap}$, and both richness, $R$ (left), and optical luminosity, $L_{\rm r}$ 
(right), of the subsample with redshift of $\leq 0.42$]{Correlation between 
the X-ray spectroscopic aperture temperature, 
$T_{ap}$, and both richness, $R$ (left), and optical luminosity, $L_{\rm r}$ 
(right), of the subsample with redshift of $\leq 0.42$. Same best-fit-lines in 
Figure~\ref{f:MrBCG-z} are plotted.} 
  \label{f:Tap_RLr_R500}
\end{figure*}

\begin{figure*}
\centering{
  \resizebox{\hsize}{!}{\includegraphics[viewport=10  65  577 330, clip]{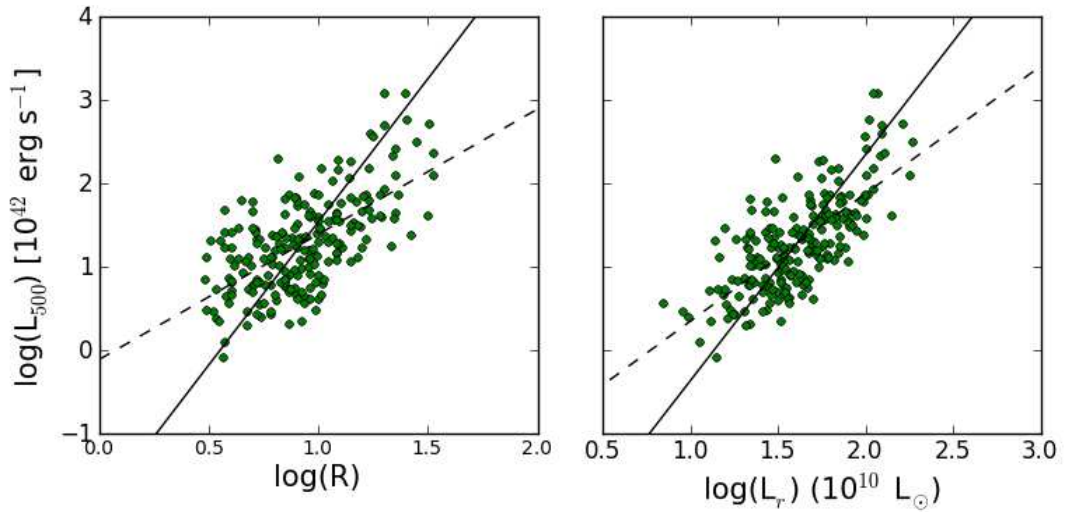}}}
  \caption[The X-ray bolometric luminosity, $L_{500}$, is plotted against 
the cluster richness, $R$ (left), and the optical luminosity, $L_{\rm r}$ 
(right)]{The X-ray bolometric luminosity, $L_{500}$, is plotted against 
the cluster richness, $R$ (left), and the optical luminosity, $L_{\rm r}$ 
(right). The best-fit-lines are the same in Figure~\ref{f:MrBCG-z}.} 
  \label{f:L500_RLr_R500}
\end{figure*}

\begin{figure*}
\centering{
  \resizebox{\hsize}{!}{\includegraphics[viewport=10  65  577 330, clip]{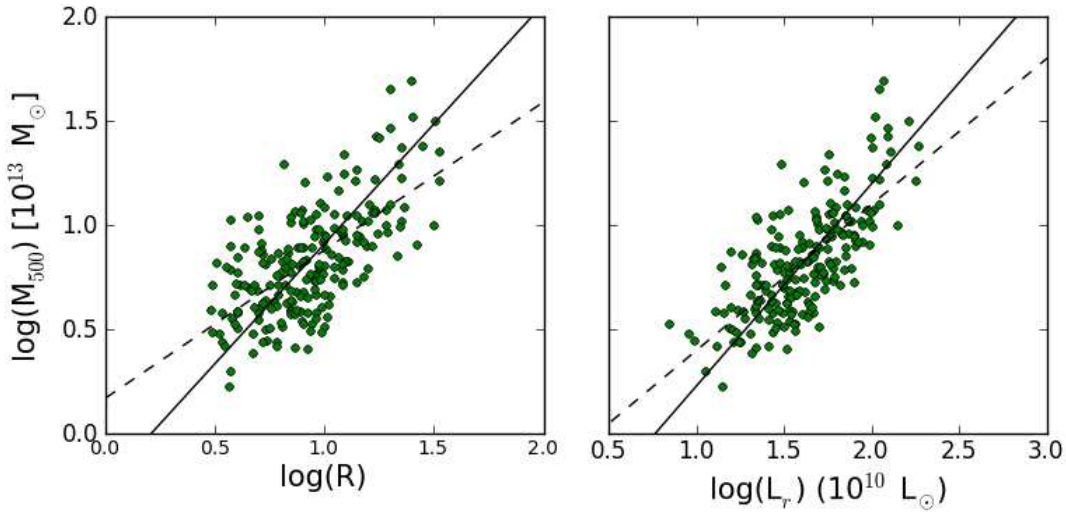}}}
  \caption[The cluster mass, $M_{500}$, is plotted versus the cluster richness, 
$R$ (left), and the optical luminosity, $L_{\rm r}$ (right)]{The cluster mass, 
$M_{500}$, is plotted versus the cluster richness, 
$R$ (left), and the optical luminosity, $L_{\rm r}$ (right). The best-fit-lines 
are the same in Figure~\ref{f:MrBCG-z}.} 
  \label{f:M500_RLr_R500}
\end{figure*}

The linear relation between the logarithms of the X-ray and optical properties 
are fitted using the BCES(Y$|$X) and orthogonal algorithms and are expressed by :

\begin{equation}
\log\ (P_{\rm X}) = a  +  b\, \log\ (P_{\rm opt}).
\end{equation}    
where $P_{\rm X}$ is the X-ray property (aperture temperature $T_{\rm ap}$, 
X-ray luminosity $L_{500}$, and X-ray-luminosity-based mass $M_{500}$) and 
$P_{\rm opt}$ is the optical property (richness $R$ and total optical 
luminosity $L_{\rm r}$). Table~\ref{tbl:relations} lists in the lower 
part the fitting parameters as well as the scatter of each relation. 
These correlations are shown in Figure~\ref{f:Tap_RLr_R500}, 
Figure~\ref{f:L500_RLr_R500}, and Figure~\ref{f:M500_RLr_R500}. The figures 
also show the best-fit-lines derived from both BCES(Y$|$X) and BCES orthogonal 
fitting methods.

If we consider the observed relations $L_{\rm X}\ \propto \ T^{3.35}$ and 
$L_{\rm X}\ \propto \ M^{2.08}$ (BCES orthogonal) derived from the 
REXCESS sample that comprises 31 nearby ($ z < 0.2 $) galaxy clusters 
\citep{Pratt09} and a mass-to-light ratio, $ M/L\ \propto \ L^{0.33}$ 
\citep{Girardi02}, we expect a relation between the X-ray temperature 
and optical luminosity as $T\ \propto \ L_{\rm r}^{0.83}$, a relation 
between the X-ray and optical luminosity as 
$L_{\rm X}\ \propto \ L_{\rm r}^{2.77}$, and a relation between the 
cluster mass and optical luminosity as $M\ \propto \ L_{\rm r}^{1.33}$. 
Since there is a tight relation between the measured optical luminosity and 
richness with an exponent deviated slightly from the unity as 
$L_{\rm r}\ \propto \ R^{1.13}$ (orthogonal), we expect a slight variation of 
the exponents of the correlations between the cluster richness and the other 
cluster properties. The expected relations are 
$T_{\rm X}\ \propto \ R^{0.94}$, $L_{\rm X}\ \propto\ R^{3.13}$, 
$M_{500}\ \propto\ R^{1.50}$ (orthogonal). Regarding the relations derived from  
the BCES(Y$|$X) algorithm, we expect that the relations between each cluster 
global parameter with both $R$ and $L_{\rm r}$ have consistent slopes 
since the slope of $L_{\rm r} - R$ relation is approximately the unity, 
0.99 $\pm$ 0.3, as shown in Table~\ref{tbl:relations}.

The current slope of the $T_{\rm ap} - L_{\rm r}$ relation, 0.85 $\pm$ 0.05 
derived from the BCES orthogonal method is consistent with the expected one, 
0.83. \citet{Popesso05} investigated the same relation based on a sample of 
49 groups and clusters ($z \leq 0.25$) with available ASCA temperature.  
Their optical luminosities were estimated within $R_{500}$ in the $r-$band. 
They found the slope of the $T_{\rm X} - L_{\rm r}$ relation obtained using 
a numerical orthogonal distance regression method is 0.59 $\pm$ 0.03, which 
is much shallower than the present slope 0.85 $\pm$ 0.05. This difference in 
the slope might be introduced from the cluster sample since their sample is 
small and contained systems with high temperatures apart from few objects 
(12 systems) with $T_{\rm X} < 2$ keV, in addition to the different 
procedures used in estimating the optical luminosities and different 
aperture used in measuring  the temperatures.  
It is the same case when comparing our relation with the relation published by 
\citet{Lopes09} who investigated it based on a small sample of 
21 clusters at $z \leq 0.1$ with $T_{\rm X} > 1.5$ keV obtained from the 
X-Ray Galaxy Clusters Database (BAX). 

\citet{Wen09} found a slope of 0.83 $\pm$ 0.08 
derived from a cluster sample of 67 systems with ASCA temperature. However, 
they estimated the optical luminosity within a fixed radius of 1 Mpc and 
used another fitting method, there is good agreement between the two slopes.
The BCES(Y$|$X) gives a shallower slope 0.61 $\pm$ 0.04 than the one  
derived from the orthogonal algorithm.            
The expected relation $T_{\rm X}\ \propto \ R^{0.94}$ (orthogonal) is 
consistent with the derived one based on our cluster sample 
$T_{\rm ap}\ \propto \ R^{1.00 \pm 0.09}$ (orthogonal). Again we found 
disagreement between the present slope and the published one 0.5 $\pm$ 0.07 
(orthogonal) by \citet{Lopes09} due to the differences in the size and mass regime 
between the two samples. The present slope 0.59  $\pm$ 0.06 (BCES(Y$|$X)) agrees 
with the one, 0.61 $\pm$ 0.05, published by \cite{Wen12} who estimated the 
richness within $R_{200}$.            

%

The observed relation, $L_{500}\ \propto \ L_{\rm r}^{2.71 \pm 0.17}$ 
(orthogonal), is in good agreement with the expected one, 
$L_{\rm X}\ \propto \ L_{\rm r}^{2.77}$ (orthogonal), from the scaling relations.
\cite{Popesso05} investigated the relation based on a sample 
of 69 clusters with optical mass, a sample of 49 clusters with mass estimated 
from the temperature with an overlap of 16 clusters, and the combined sample 
of 102 systems at $z \leq 0.25$, they found the slopes derived from the 
orthogonal algorithm are 1.59 $\pm$ 0.10, 1.89 $\pm$ 0.11, and 
1.79 $\pm$ 0.10, respectively.      
\cite{Lopes09} found a slope of 1.62 $\pm$ 0.11 derived from a sample of 104 
clusters at $z \leq 0.1$. The X-ray luminosities in their works were obtained 
from ROSAT data and were not estimated within a fixed aperture but were 
calculated from the X-ray luminosity profile. Their optical luminosities were 
estimated within $R_{500}$ based on the SDSS data.
Once again the present slope is not consistent with the ones derived in those 
studies due to using different procedure to estimate the X-ray and optical 
luminosities in addition to the different luminosity regime of the samples.

The present slope 2.71 $\pm$ 0.17 of $L_{500}\ - \ L_{\rm r}$ relation 
agrees well with the published one 2.67 $\pm$ 0.12 by \cite{Wen09} based on 
a sample of 146 clusters at $z \leq 0.42$. The BCES(Y$|$X) slope, 
1.53 $\pm$ 0.11, is shallower than the slope obtained by the 
BCES orthogonal algorithm. We note that the BCES(Y$|$X) slope is consistent 
with the slopes (orthogonal) published by \cite[][]{Popesso05, Lopes09}.      
Regarding the X-ray luminosity-richness relation, the present exponent 
$L_{500}\ \propto\ R^{3.43 \pm 0.25}$ is consistent within 1.2$\sigma$ with 
the expected one $L_{\rm X}\ \propto\ R^{3.13}$ (orthogonal) and it is higher 
than the observed one, 2.79 $\pm$ 0.13, by \cite{Wen09}. \cite{Lopes09} found 
much shallower slope 1.64 $\pm$ 0.10 (orthogonal) than the present one and 
this might be due to the same reasons mentioned above. The BCES(Y$|$X) slope, 
1.50 $\pm$ 0.14, is consistent with the published one, 1.64 $\pm$ 0.10 
(orthogonal), by \cite{Lopes09}.   


The current slope of $M_{500} - L_{\rm r}$ relation, 0.97 $\pm$ 0.07 
(orthogonal), is shallower than the expected one, 1.33 $\pm$ 0.12, 
from the scaling relations. \citet{Popesso05} investigated the same relation 
based on the three samples mentioned above and they found slopes (orthogonal) 
of 1.25 $\pm$ 0.06, 0.93 $\pm$ 0.06, 1.09 $\pm$ 0.04, respectively. Similarly, 
\cite{Lopes09} found slopes of 1.09 $\pm$ 0.07, 0.99 $\pm$ 0.10, and 
1.06 $\pm$ 0.04 derived from an optically selected sample of 127 systems, 
X-ray selected sample of 53 clusters, and the combined sample of 180 clusters 
at $z \leq 0.1$, respectively. The two parameters (mass and the optical 
luminosity in the $r-$band) were estimated within $R_{500}$ in these 
investigations as our measurements. The current slope is consistent with 
the slope values from those studies. The BCES(Y$|$X) provides a shallower 
slope of 0.70 $\pm$ 0.05.      

Regarding $M - R$ relation, we found a relation of 
$M_{500}\ \propto\ R^{1.15 \pm 0.09}$ (orthogonal) that is shallower than the 
expected one, $M_{500}\ \propto\ R^{1.50}$ (orthogonal). 
\cite{Lopes09} found the slope (orthogonal) of the relation 
derived from the three samples mentioned above are 1.01 $\pm$ 0.07, 
0.95 $\pm$ 0.09, 1.04 $\pm$ 0.04, respectively. Their richness and mass were 
estimated within $R_{500}$. The later value is consistent within 
1.2$\sigma$ from the present slope. The slope derived from the BCES(Y$|$X) 
method 0.71 $\pm$ 0.06 is shallower than the one derived from the BCES orthogonal 
regression method.   

%

%

The orthogonal scatter (in a logarithmic scale) of these relations was computed 
in a similar way as mentioned in the previous section. We note that the BCES 
orthogonal algorithm provides smaller scatter than the one derived using the 
BCES(Y$|$X) routine. Using the same fitting algorithm, we found the relation 
between the cluster properties and the optical luminosity have slightly 
smaller scatter than the relations with the cluster richness. This means 
that the cluster optical luminosity correlates slightly better than the 
richness with the cluster global properties ($T_{\rm ap}$, $L_{500}$, and 
$M_{500}$).

The cluster richness and optical luminosity show a correlation with the cluster 
X-ray temperature, luminosity and mass. The relation between the cluster 
richness and $T_{\rm ap}$, $L_{500}$, and $M_{500}$ has an orthogonal 
scatter of 41$\%$, 51$\%$, and 41$\%$, respectively while the relation 
between the optical luminosity and the same properties has an orthogonal 
scatter of 38$\%$, 48$\%$, 38$\%$, respectively. 
%

The most important parameter of galaxy clusters to be used in cosmological 
studies is the mass. We showed that the cluster mass can be estimated using 
a very cheap estimator (richness or optical luminosity) that can be determined 
from ground-based photometric data. The optical luminosity correlates with the 
cluster mass with an orthogonal scatter of 38$\%$. \citet{Popesso05} showed 
that the optical luminosity correlates with the cluster mass much better than 
the X-ray luminosity. They found an orthogonal scatter in the range 
20-30$\%$ of $L_{\rm optical} - M$ relation and an orthogonal scatter 
38-50$\%$ of $L_{\rm X} - M$ relation. Here we found a slightly higher scatter 
38$\%$ of $M_{500} - L_{\rm r}$ relation and this is due to including many galaxy 
groups among our sample, which are the main source of the scatter \citep{Popesso04}.

\section{Summary}

We re-identified the BCGs of the optically confirmed cluster sample (574 
systems) from the 2XMMi/SDSS galaxy cluster survey as the brightest galaxy 
among the luminous cluster member galaxies within $R_{500}$. We found that 
about 74 percent of the identified BCGs are fulfill the magnitude and colour 
cuts of the spectroscopically targeted LRGs in the SDSS-III'BOSS. For each BCG, 
we determined its absolute magnitude and optical luminosity in the $r-$band 
based on the cluster redshift and the SDSS photometric data. 
We presented the relation between the absolute magnitude of the BCGs with the 
cluster redshift. The relation between the optical luminosity of the BCGs 
with the cluster mass was investigated, which shows that the massive clusters 
have luminous BCGs. 
We determined the net count of the luminous cluster member galaxies, richness, 
and their summed luminosity in  the $r-$band, cluster optical luminosity, 
within $R_{500}$, of 214 galaxy groups and clusters in the redshift range of 
0.03-0.42 with available X-ray temperatures from our survey. This subsample 
comprises systems in low and intermediate mass regime.  The estimated optical 
luminosity is tightly correlated with the cluster richness. We investigated the 
correlations of the X-ray temperature, luminosity, and the cluster mass with 
both the optical richness and luminosity of this subsample. 
The orthogonal scatter in these relations was measured, which indicates that 
the optical luminosity correlates slightly better than the richness with the 
cluster global properties. The relation between the cluster 
richness and $T_{\rm ap}$, $L_{500}$, and $M_{500}$ has an orthogonal 
scatter of 41$\%$, 51$\%$, and 41$\%$, respectively while the relation 
between the optical luminosity and the same properties has an orthogonal 
scatter of 38$\%$, 48$\%$, 38$\%$, respectively. 
%
%
%

\clearpage
\begin{landscape}

\begin{table}[h]
\caption[The relations between the cluster X-ray and optical properties]{The 
relations between the optical properties (M$_{\rm r, BCG}$, 
L$_{\rm r, BCG}$) of the BCGs and the cluster properties ($z$, M$_{500}$)
of the whole optically confirmed cluster sample are presented in the upper 
part of the table. The lower part lists the relations between the X-ray 
properties ($T_{ap}$, $L_{500}$, $M_{500}$) and optical properties 
($R$ and $L_{\rm r}$) as well as the relation between $L_{\rm r}$ and $R$
of a subsample of 214 systems with $z \leq 0.42$ and temperature measurements. 
Col.~1 indicates the types of the objects used in the relations (BCGs or ClGs: 
clusters of galaxies). Col.~2 lists the relation between the investigated 
parameters. Cols.~3, 4, and 5 represent the intercept, slope, orthogonal scatter, 
respectively, derived from BCES(Y$|$X), while Cols.~6, 7, and 8 list the 
same fitted parameters from the BCES orthogonal method. Cols.~9 and 10 
represent the published slope of the corresponding relation in the literature 
and its reference.\\}
\label{tbl:relations}     

 {\small                          
\begin{tabular}{c | c | c c c | c c c | c c}       
\hline\hline 
  &           &  & BCES(Y$|$X)    &   &  & Orthogonal    &   & published & \\                    
  &  relation & a  & b & scatter\tablefootmark{\star} & a  & b & scatter\tablefootmark{\star} & b & ref. \\   
\hline
BCGs  &M$_{\rm r, BCG}$ - z&-22.48 $\pm$ 0.05&-2.36 $\pm$ 0.16&0.22 $\pm$ 0.01&-20.63 $\pm$ 0.18&-8.13 $\pm$ 0.55&0.13 $\pm$ 0.01&-1.74 $\pm$ 0.03&1 \\ 
      &L$_{\rm r, BCG}$ - M$_{500}$&0.70 $\pm$ 0.03&0.57 $\pm$ 0.03&0.19 $\pm$ 0.01&0.37 $\pm$ 0.05&0.98 $\pm$ 0.06&0.18 $\pm$ 0.01&0.62 $\pm$ 0.05 &2\\
\hline
ClGs&L$_{r}$ - R&0.69 $\pm$ 0.04&0.99 $\pm$ 0.03&0.09 $\pm$ 0.01&0.56 $\pm$ 0.04&1.13 $\pm$ 0.04&0.08 $\pm$ 0.01&1.28 $\pm $ 0.31&3 \\ 
&T$_{ap}$ - R&-0.25 $\pm$ 0.06&0.59 $\pm$ 0.06&0.16 $\pm$ 0.01&-0.63 $\pm$ 0.08&1.00 $\pm$ 0.09&0.15 $\pm$ 0.01&0.61 $\pm$ 0.05&1 \\ 
&T$_{ap}$ - L$_{r}$&-0.68 $\pm$ 0.08&0.61 $\pm$ 0.04&0.15 $\pm$ 0.01&-1.07 $\pm$ 0.09&0.85 $\pm$ 0.05&0.14 $\pm$ 0.01&0.59 $\pm$ 0.03&4 \\ 
&L$_{500}$ - R&-0.11 $\pm$ 0.13&1.50 $\pm$ 0.14&0.24 $\pm$ 0.02&-1.89 $\pm$ 0.23&3.43 $\pm$ 0.25&0.18 $\pm$ 0.01&1.59 $\pm$ 0.09&1 \\ 
&L$_{500}$ - L$_{r}$&-1.18 $\pm$ 0.18&1.53 $\pm$ 0.11&0.22 $\pm$ 0.01&-3.07 $\pm$ 0.28&2.71 $\pm$ 0.17&0.17 $\pm$ 0.01&2.67 $\pm$ 0.12&5 \\ 
&M$_{500}$ - R&0.17 $\pm$ 0.06&0.71 $\pm$ 0.06&0.16 $\pm$ 0.01&-0.24 $\pm$ 0.08&1.15 $\pm$ 0.09&0.15 $\pm$ 0.01&1.04 $\pm$ 0.04&6 \\ 
&M$_{500}$ - L$_{r}$&-0.30 $\pm$ 0.08&0.70 $\pm$ 0.05&0.15 $\pm$ 0.01&-0.74 $\pm$ 0.11&0.97 $\pm$ 0.07&0.14 $\pm$ 0.01&0.99 $\pm$ 0.10&6 \\ 
%
\hline
\end{tabular}
 }
\\
\tablefoottext{\star}{The scatter in all relations is the orthogonal scatter in logarithmic scale except the $M_{\rm r, BCG} - z$ relation that has the  scatter in linear scale.} 
\tablebib{
1- \cite{Wen12} ; 2- \cite{Mittal09} ; 3- \cite{Popesso07}; 4- \cite{Popesso05}; 5- \cite{Wen09}; 6- \cite{Lopes09}. 
}
\end{table}

\end{landscape}

\chapter{Summary, Conclusions, and Outlook}



\section{Summary and Conclusions}

In this thesis I have conducted a serendipitous search for galaxy clusters 
based on the largest X-ray serendipitous source catalogue, 2XMMi-DR3, and 
the largest optical survey, SDSS. The aims of my survey are to identify new 
X-ray detected clusters, to trace the evolution of their X-ray scaling 
relations, and to investigate the correlations between their X-ray and the 
optical properties. The X-ray cluster 
candidates are selected among the extended X-ray sources in the 2XMMi-DR3 
catalogue. To study the properties of these cluster candidates, the redshifts 
need to be determined. Therefore, the survey is constrained to those candidates 
that are in the footprint of the SDSS-DR7 in order to measure their optical 
redshifts. However, the X-ray spectroscopy provides a tool to measure the X-ray 
redshift, unfortunately it is only possible for very bright X-ray sources (about 
4 percent of the cluster candidates list). Optical redshifts can be obtained from 
the literature \citep{Takey11}, using the photometric redshift of the cluster 
galaxies \citep{Takey13a}, or using the spectroscopic redshift of the cluster 
luminous red galaxies (LRGs) from the SDSS data base \citep{Takey13b}. 
The optically confirmed cluster sample with redshift measurements can be used 
to investigates various scaling relations e.g. the X-ray luminosity-temperature 
relation \citep{Takey11, Takey13a}, relations between the BCG and the cluster 
properties as well as the correlations between the X-ray and optical properties 
of the clusters \citep{Takey13c}. The summary  of the main contributions from 
the thesis work is as the follows:       
     
\begin{enumerate}

\item  In Chapter 2 I described the 2XMMi/SDSS Galaxy Cluster Survey. 
The X-ray cluster candidates are selected from extended sources classified 
as reliable detections (with no warning about being spurious) in the 2XMMi-DR3 
catalogue \citep{Watson09} at high galactic latitudes, $|b| > 20^{\circ}$. 
The overlap area between the XMM-Newton 
fields and SDSS imaging is 210 deg$^2$. After excluding possible spurious 
X-ray detections and low redshift galaxies that appear resolved at X-ray 
wavelengths through visual inspections of X-ray images and X-ray-optical 
overlays, the X-ray cluster candidates list comprised 1180 objects with at 
least 80 net photon counts, of which more than 75 percent are new 
X-ray discoveries. We have demonstrated that the 2XMMi-DR3 catalogue that 
was based on the archival XMM-Newton data is a rich resource for identifying 
new X-ray detected clusters \citep{Takey11}.

\item A quarter of the X-ray cluster candidates had been previously published 
in optical cluster catalogues based on the SDSS data. Cross-correlations 
of the X-ray cluster cluster candidates with the recent and largest optical 
cluster catalogues  constructed by \citet[][]{Hao10, Wen09, Koester07, Szabo11} 
within a matching radius of one arcmin confirmed 275 clusters and provided us 
with the photometric redshifts for all of them and the spectroscopic redshifts 
for 120 BCGs. Based on the cluster redshifts given in those catalogues, we 
extracted all available spectroscopic redshifts for the cluster members from 
recent SDSS data (SDSS-DR8). Among the confirmed cluster sample, 182 clusters 
have spectroscopic redshifts for at least one cluster galaxy. More than 80 
percent of the confirmed sample are newly identified X-ray clusters and the 
others had been previously identified using ROSAT, Chandra, or XMM-Newton 
data \citep{Takey11}.

\item I have reduced and analysed the X-ray data of the optically confirmed 
cluster sample with redshifts from the literature in an automated way following 
the standard pipelines of processing the XMM-Newton data. In this pipeline, I 
extracted the cluster spectra from EPIC(PN, MOS1, MOS2) images within an 
optimal aperture with a radius representing the highest signal-to-noise ratio. 
The spectral fitting procedure provided the X-ray temperature for 175 systems, 
which was published as the first cluster sample from our survey. In addition, 
I derived the physical properties ($R_{500}$, $L_{500}$ and $M_{500}$) of this 
sample from an iterative procedure using the published scaling relations 
\citep{Takey11}.

\item In Chapter 3 I described a finding algorithm developed to detect the 
optical counterparts of the X-ray cluster candidates and to measure their 
redshifts using the photometric and, if available, the spectroscopic redshifts 
of surrounding galaxies from the SDSS-DR8 data. The cluster is recognized 
if there are at least 8 member galaxies within a radius of 560 kpc from 
the X-ray emission peak with photometric redshift in the redshift interval 
of the redshift of the likely identified BCG, 
$z_{\rm p,\,BCG} \pm 0.04(1+z_{\rm p,\,BCG})$. The BCG was identified as 
the brightest galaxy among those galaxies within one arcmin from the X-ray 
position that show a peak in the histogram of their photometric redshifts.

The cluster photometric and spectroscopic redshift is measured as the 
weighted average of the photometric and the available spectroscopic 
redshifts, respectively, of the cluster galaxies within 560 kpc from 
the X-ray position. The measured redshifts are in a good 
agreement with available redshifts in the literature, to date 301 
clusters are known as optically selected clusters with redshift measurements.
Also, 310 clusters of the optically confirmed cluster sample have spectroscopic 
redshifts for at least one cluster member galaxy. The measured photometric redshifts 
are in a good agreement with the measured spectroscopic ones from the survey. 
The cluster redshifts of the optically confirmed cluster sample span  
a wide redshift range from 0.03 to 0.70 \citep{Takey13a}.  

\item I reduced and analysed the X-ray data of this sample in a similar way 
as described above. We presented a cluster catalogue from the survey comprising 
345 X-ray selected groups and clusters with their X-ray parameters derived 
from the spectral fits including the published sample by \cite{Takey11}. 
In addition to the cluster sample with X-ray spectroscopic data, we presented the 
remainder of the optically confirmed cluster sample with their X-ray parameters 
based on the flux given in the 2XMMi-DR3 catalogue. We used the 2XMMi-DR3 flux 
because of their low quality X-ray data, which is not sufficient to perform  
spectral fitting. This sample comprises 185 groups and clusters with their 
fluxes and luminosities in the energy band 0.5-2.0 keV. 

For both subsamples, 
we estimated the physical properties ($R_{500}$, $L_{500}$ and $M_{500}$) 
from an iterative procedure based on published scaling relations. The measured 
temperatures and luminosities are in good agreement with published values for 
the sample overlap (114 systems) with the XMM Cluster Survey 
\citep[XCS,][]{Mehrtens12}. Comparison of the cluster luminosities of the sample 
with the luminosities of the identified clusters from ROSAT data 
\citep{Piffaretti11} shows that our X-ray detected groups and clusters are in 
the low and intermediate luminosity regimes apart from few luminous systems, 
thanks to the XMM-Newton sensitivity and the available XMM-Newton deep fields 
\citep{Takey13a}.

\item As a first application of the confirmed cluster sample with measured 
X-ray temperatures, we investigated the $L_{\rm X}-T$ relation for the first 
time based on a large cluster sample of 345 systems with X-ray spectroscopic 
parameters drawn from a single survey. The current sample includes groups 
and clusters with wide ranges of redshifts, temperatures, and luminosities. 
The slope of the relation is consistent with the published ones of nearby 
clusters with higher temperatures and luminosities  \citep{Pratt09,Mittal11}. 
The derived relation is still much steeper than that predicted by self-similar 
evolution (see Chapter 3, Section 3.4.4). 

We also investigated the evolution of the slope and the scatter of the \ltr 
relation with the cluster redshift.
After excluding the low luminosity groups, we find no significant changes 
of the slope and the intrinsic scatter of the relation with redshift when 
dividing the sample into three redshift bins.  In addition the slopes of the 
relations of the three subsamples are in agreement with the corresponding 
published slopes of the XCS subsamples \citep{Hilton12}. When including the 
low luminosity groups in the low redshift subsample, we found its \ltr relation 
becomes flatter than  the relation of the intermediate and high redshift 
subsamples \citep{Takey13a}.

\item In Chapter 4 I presented a sample of 324 X-ray selected galaxy groups and 
clusters associated with at least one Luminous Red Galaxy (LRG) that has 
a spectroscopic redshift in the SDSS-DR9 \citep{Ahn12}. The redshifts of the 
associated LRGs are used to identify the BCGs and the other cluster galaxies 
with spectroscopic redshifts. The cluster spectroscopic redshift is computed 
as the weighted average of the available spectroscopic redshifts of the cluster 
galaxies within 500 kpc. The cluster sample spans a wide redshift range from 
0.05 to 0.77 with a median redshift of $z = 0.31$.  In addition to re-identify 
and confirm the redshift measurements of 280 clusters among the published 
cluster sample from our survey by \cite{Takey13a}, we extended the optically 
confirmed cluster sample by 44 systems. Among the newly constructed sample, 55 
percent are newly discovered systems and 80 percent are new X-ray detected 
galaxy groups and clusters. The measured redshifts and the X-ray flux given 
in the 2XMMi-DR3 catalogue were used to determine the X-ray luminosity and 
mass of the cluster sample \citep{Takey13b}. 

\item  In Chapter 5 we investigated the correlations between the optical 
properties of the BCGs with the cluster properties, as a second application 
of the optically confirmed cluster sample from our ongoing survey. To check 
those correlations, I re-identified the BCGs of the whole optically confirmed 
cluster sample (574 systems) as the brightest galaxy among the luminous cluster 
member galaxies within $R_{500}$. We found that 74 percent of the re-identified 
BCGs fulfill the magnitude and colour cuts of the spectroscopically targeted 
LRGs in the BOSS (SDSS-III). For each BCG, we determined its absolute 
magnitude and optical luminosity in the $r-$band.  We investigated the relation 
between the absolute magnitude of the BCGs with the cluster redshift.  It shows 
a linear relation indicating the BCGs are brighter in the high redshift clusters, 
which is consistent with the result obtained by \cite{Wen12}. The relation 
between the optical luminosity of the BCGs with the cluster mass was also 
investigated, which indicated that the more massive the cluster, the more 
luminous the BCG is. The slope of this relation is consistent with the 
published one by \cite{Mittal09} \citep{Takey13c}.

\item As a third application of the optically confirmed cluster sample, we 
investigated the correlation between the optical properties (richness and 
luminosity) and the X-ray properties (temperature, luminosity, mass). To 
investigate these relations, we determined the net luminous cluster member 
galaxies within $R_{500}$, richness, and their summed luminosity in  the 
$r-$band, cluster optical luminosity, of 214 galaxy groups and clusters in 
the redshift range of 0.03-0.42 with available X-ray temperatures from our 
survey. This subsample comprises systems in low and intermediate mass regime.  
The estimated optical luminosity is tightly correlated with the cluster richness. 

We investigated the correlations of the X-ray temperature, luminosity, and 
the cluster mass with both the optical richness and luminosity of this 
subsample. The orthogonal scatter in these relations was measured. 
According th the measured scatters, the optical luminosity correlates slightly 
better than the richness with the cluster global properties. The relation between 
the cluster richness and $T_{\rm ap}$, $L_{500}$, and $M_{500}$ has an orthogonal 
scatter of 41$\%$, 51$\%$, and 41$\%$, respectively, while the relation 
between the optical luminosity and the same properties has an orthogonal 
scatter of 38$\%$, 48$\%$, 38$\%$, respectively \citep{Takey13c}.  

%
%
%

\item Finally, the constructed sample from our ongoing survey is the largest 
X-ray selected cluster catalogue to date based on X-ray data from the current 
X-ray observatories (XMM-Newton, Chandra, Suzaku, and Swift/XRT). It 
comprises 574 groups and clusters with their optical and X-ray properties, 
spanning the redshift range $0.03 \leq z \leq 0.77$. More than 75 percent of the 
cluster sample are newly discovered clusters at X-ray wavelengths. About 
40 percent of the sample are new systems to the literature according to 
current entries in the NED \citep{Takey13c}.

\end{enumerate} 



\section{Future perspectives} 

I have presented the optically confirmed cluster sample (574 systems) from the 
2XMMi/SDSS Galaxy Cluster Survey, which is about half of the X-ray cluster 
candidates list (1180 objects). In the future I plan to study the remainder 
of the X-ray cluster candidates, which were not identified by the current 
detection algorithms since they are either poor or at high redshifts. For the 
distant clusters, we plan follow-up by imaging and spectroscopy. For those 
X-ray cluster candidates that have member galaxies detected in the SDSS 
imaging and not be identified as groups or clusters by the algorithms 
used in this thesis, I plan to improve these algorithms to constrain their 
redshifts. By confirming the remainder of the X-ray cluster candidates, 
especially distant ones, and measuring their redshifts, we will be able to 
trace the evolution of $L_{\rm X}-T$ relation and X-ray-optical relations 
as well as to asses the selection effects on those relations. 

Since the current survey was successful, a natural extension is to continue 
to enlarge the catalogue of X-ray selected galaxy groups and clusters 
based on the upcoming version (not yet published) of the XMM-Newton 
serendipitous source catalogue, 3XMM. 
The 3XMM catalogue is about 30 percent larger than the 2XMMi-DR3 
catalogue. The cluster redshifts will be measured based on the available large  
optical surveys e.g. the Sloan Digital Sky Survey (SDSS), the Canada-France-Hawaii 
Telescope Legacy Survey (CHFTLS), the European Southern Observatory (ESO) archive 
and the Hubble Space Telescope (HST) archive, and Infrared surveys by UKIDSS,  
the NASA's Spitzer Space Telescope (SST) and the Wide-field Infrared Survey 
Explorer (WISE). This extension will be in the frame work of the ARCHES  
(Astronomical Resource Cross-matching for High Energy Studies) project, 
which is funded by the 7th Framework of the European Union.        
  






\backmatter 
\appendix
\chapter{Appendix A}

{\bf Gallery}

We present a gallery of four galaxy clusters from the first cluster sample 
with different X-ray fluxes and data quality at different redshifts covering 
the whole redshift range of the sample. For each cluster, 
X-ray flux contours (0.2-4.5 keV) are overlaid on combined image from 
$r$, $i$, and $z$-SDSS images. The upper panel in each figure shows the 
X-ray-optical overlays. The field of view is $4'\times 4'$ centred on the 
X-ray cluster position. In each overlay, the cross-hair indicates the position 
of the brightest cluster galaxy (BCG), while in Figure A.4 the cross-hair 
indicates the cluster stellar mass centre although it is obvious that the BCG is 
located at the X-ray emission peak. In each figure, the bottom panel shows 
the X-ray spectra (EPIC PN (black), MOS1 (green), MOS2 (red)) and the best 
fitting MEKAL model. The full gallery of the first cluster sample is available at
\url{http://www.aip.de/groups/xray/XMM_SDSS_CLUSTERS/17498.html}.

\begin{figure}
\centering{
    \resizebox{120mm}{!}{\includegraphics{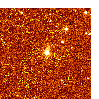}}
    \resizebox{120mm}{!}{\includegraphics[angle=-90, viewport=30  0  520 660 ,clip]{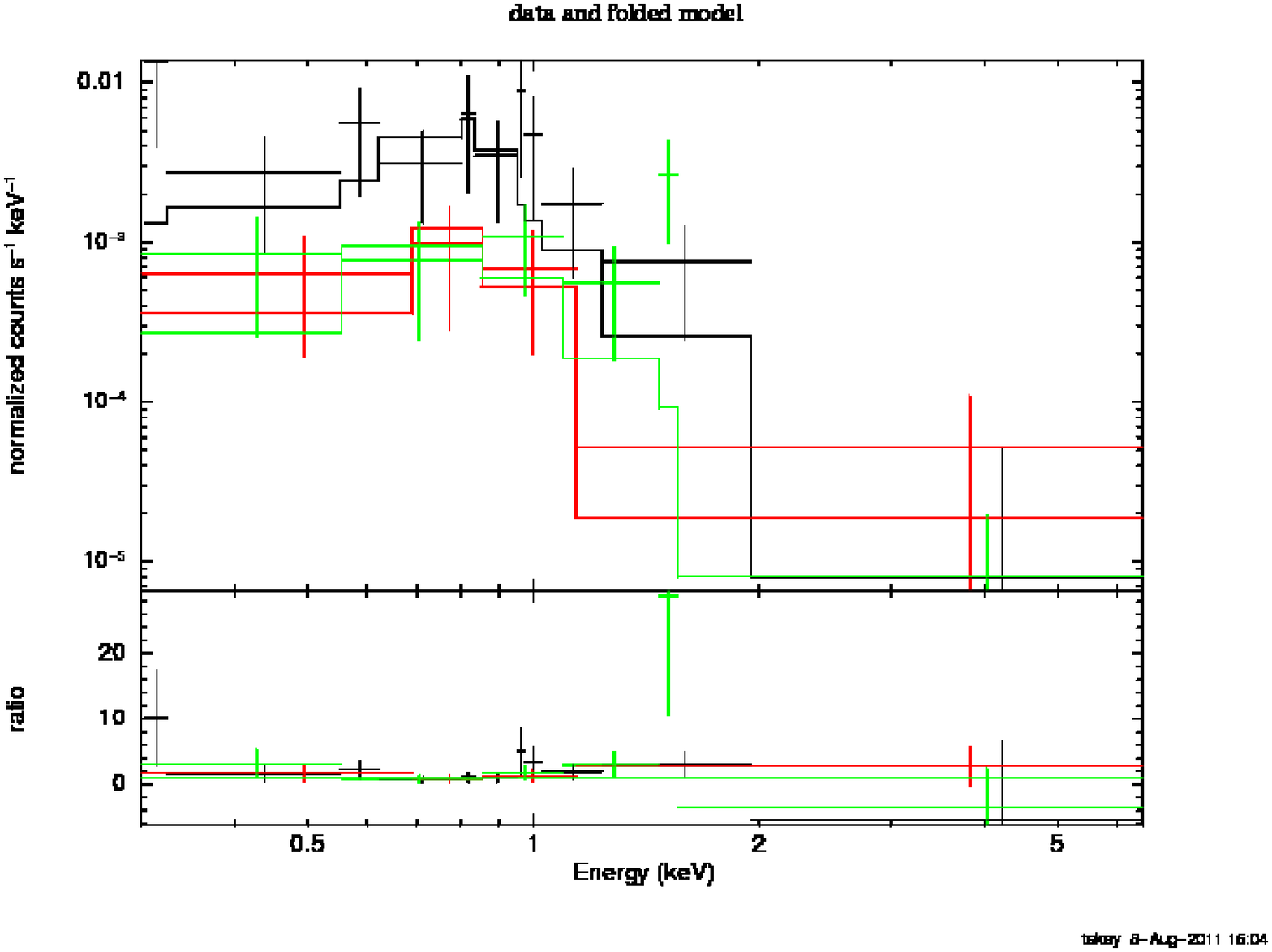}}}
    \caption{ detid = 275409: 2XMMI J143929.0+024605 at $z_{s}$ = 0.1447
   ($F_{ap}\ [0.5-2]\ keV = 0.63 \times 10^{-14}$ \ erg\ cm$^{-2}$\ s$^{-1}$).}
  \label{Fig.A.1}
\end{figure}

\begin{figure}
\centering{
  \resizebox{120mm}{!}{\includegraphics{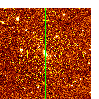}}
  \resizebox{120mm}{!}{\includegraphics[angle=-90, viewport=30  0  520 660 ,clip]{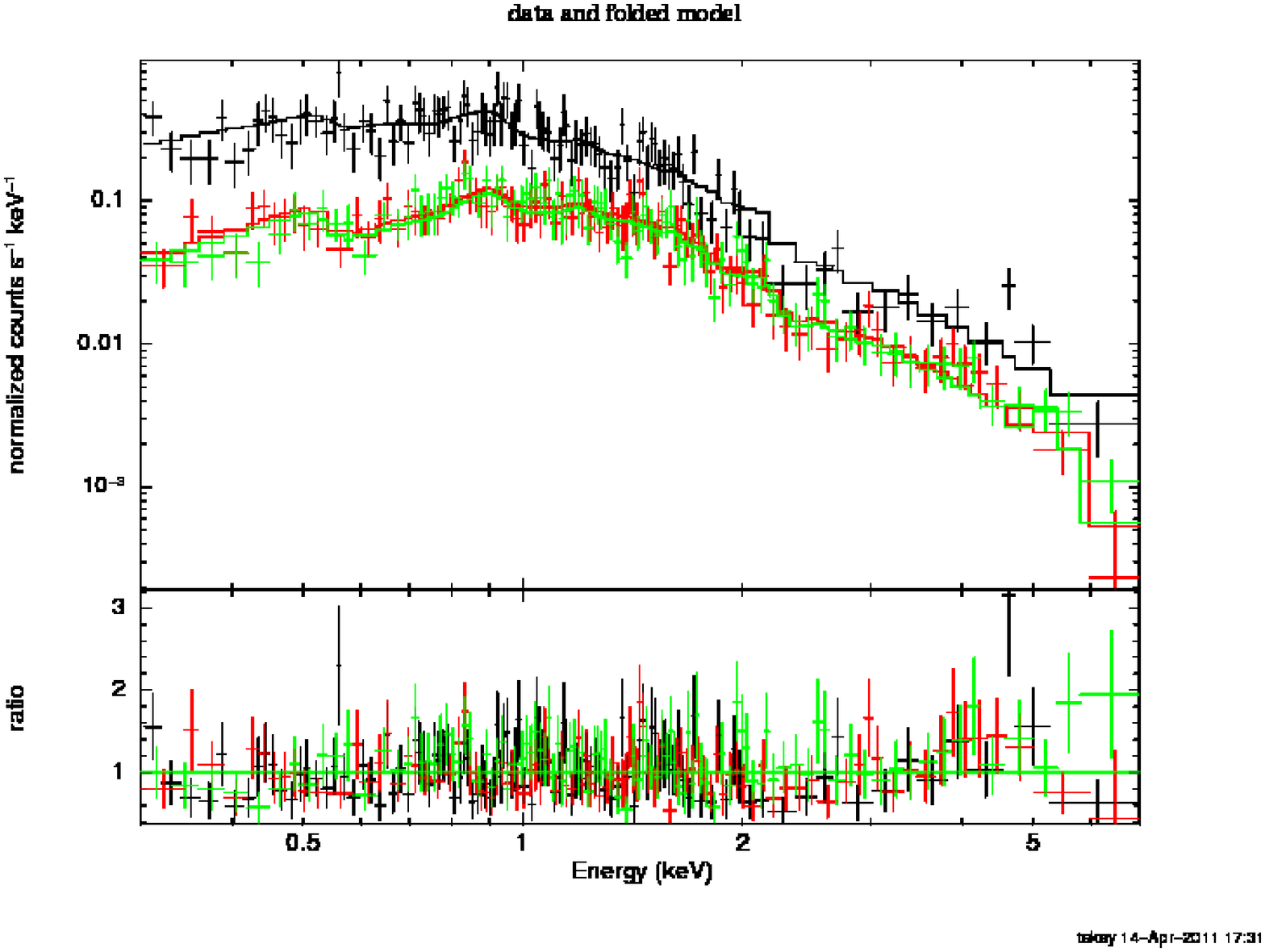}}}
  \caption{detid = 090256: 2XMM J083454.8+553422 at $z_{s}$ = 0.2421 
($F_{ap}\ [0.5-2]\ keV = 165.21 \times 10^{-14}$ \ erg\ cm$^{-2}$\ s$^{-1}$).  }
  \label{Fig.A.2}
\end{figure}

\begin{figure}
\centering{
  \resizebox{120mm}{!}{\includegraphics{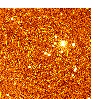}}
  \resizebox{120mm}{!}{\includegraphics[angle=-90, viewport=30  0  520 660 ,clip]{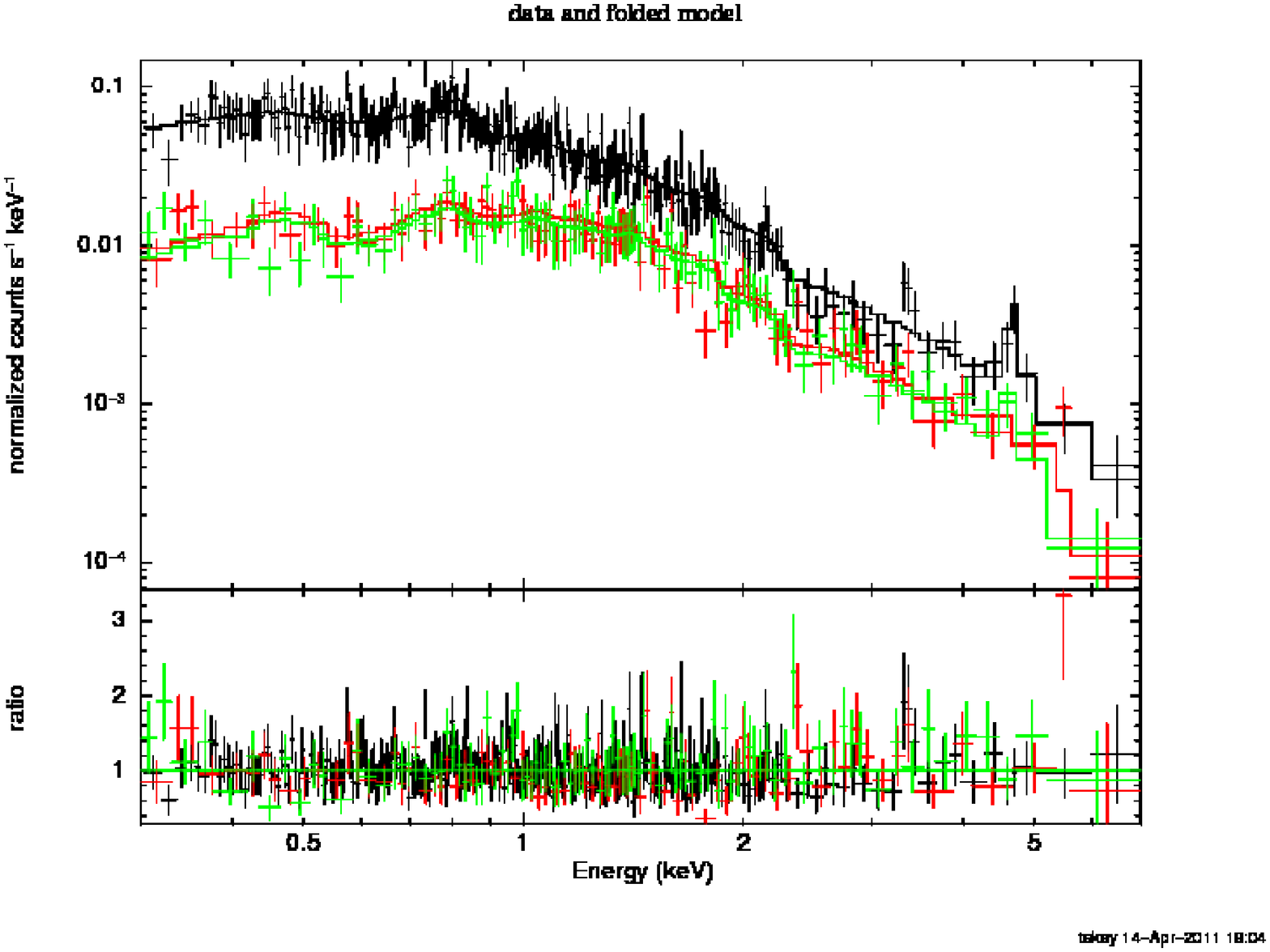}}}
  \caption{detid = 312615: 2XMM J091935.0+303157 at $z_{s}$ = 0.4273
($F_{ap}\ [0.5-2]\ keV = 16.03 \times 10^{-14}$ \ erg\ cm$^{-2}$\ s$^{-1}$).  }
  \label{Fig.A.3}
\end{figure}

\begin{figure}
\centering{
  \resizebox{120mm}{!}{\includegraphics{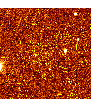}}
  \resizebox{120mm}{!}{\includegraphics[angle=-90, viewport=30  0  520 660 ,clip]{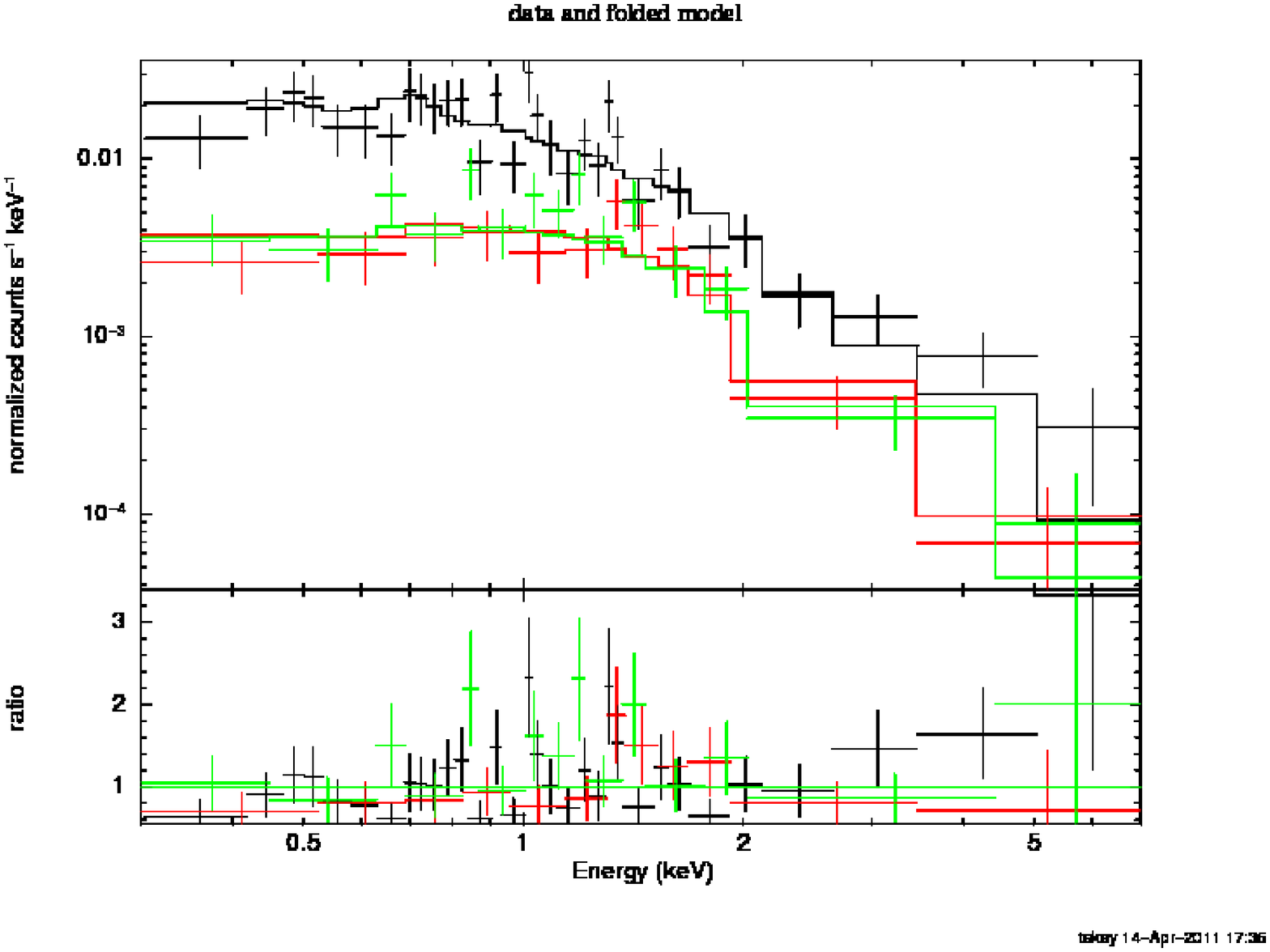}}}
  \caption{detid = 097911: 2XMM J092545.5+305858  at $z_{p}$ = 0.5865
($F_{ap}\ [0.5-2]\ keV = 7.59 \times 10^{-14}$ \ erg\ cm$^{-2}$\ s$^{-1}$). }
  \label{Fig.A.4}
\end{figure}

\chapter{Appendix B}
 
{\bf Selection criteria of LRGs}

We used the colour and magnitude cuts that were used to select the spectroscopic  
targets to construct BOSS galaxy sample in the SDSS-III project. The  
selection criteria of galaxies targeted in BOSS were given in 
(Padmanabhan et al. 2013) and were provided in the BOSS homepage
\footnote{\url{http://www.sdss3.org/dr9/algorithms/boss_galaxy_ts.php}}.

The BOSS includes two samples of galaxies, 

{\it (i) The BOSS $``$LOWZ$``$ Galaxy Sample, $z \le 0.4$}\\
The selection cuts are as follows:
\begin{enumerate}
\item $|c_{\perp}| < 0.2$, to define the colour boundaries of the sample around 
a passive stellar population, where $c_{\perp} = (r - i) - (g - r)/4.0 - 0.18$ 

\item $r < 13.5 + c_{||}/0.3$, to select the brightest galaxies at each redshift, 
where $c_{||} = 0.7(g - r) + 1.2[(r - i) - 0.18)]$ 

\item $16 < r < 19.6$, to define the faint and bright limits
 
\end{enumerate}

{\it (ii) The BOSS $``$CMASS$``$ Galaxy Sample, $0.4 < z < 0.8$}\\
The colour and magnitude cuts are as follows:

\begin{enumerate}

\item  $d_{\perp} > 0.55$, to isolate high-redshift objects, where 
       $d_{\perp} = (r-i) - (g-r)/8.0$

\item  $i < 19.86+ 1.6(d_{\perp} - 0.8)$, to select the brightest or more 
       massive galaxies with redshift
   
\item  $17.5 < i < 19.9 $, to define the faint and bright limits

\item  $r - i < 2 $, to protect from some outliers

\end{enumerate}

Note that we did not apply the criteria that were used to perform a star-galaxy
separation since we dealt only with objects that were classified as galaxies 
indicated by spectroscopic class parameters given in the {\tt SpecObj} table.
Note also all colours are computed using model magnitudes while the magnitude 
cuts are applied on composite model (cmodel) magnitudes. All magnitudes are 
corrected for Galactic extinction following \cite{Schlegel98}.



\begin{landscape}

\chapter{Appendix C}

\begin{table}[h]
\caption[The full table of Table~\ref{tbl:catalog_sample}]{The entire cluster 
catalogue (Table~\ref{tbl:catalog_sample}) of the extended 
cluster sample (44 objects) from the current work in addition to a subsample 
(49 systems) that had only photometric redshifts in Paper II and have 
spectroscopic confirmations in the current work.}
\label{tbl:PIII }
\end{table}

{\tiny
\begin{longtable}{c c c c c c c c c c c c c c c c }
\hline
\hline
  \multicolumn{1}{c}{detid\tablefootmark{a}} &
  \multicolumn{1}{c}{Name\tablefootmark{a}} &
  \multicolumn{1}{c}{ra\tablefootmark{a}} &
  \multicolumn{1}{c}{dec\tablefootmark{a}} &
  \multicolumn{1}{c}{obsid\tablefootmark{a}} &
  \multicolumn{1}{c}{z\tablefootmark{b}} &
  \multicolumn{1}{c}{scale} &
  \multicolumn{1}{c}{$R_{500}$} &
  \multicolumn{1}{c}{$F_{cat}$\tablefootmark{a,c}} &
  \multicolumn{1}{c}{$\pm eF_{cat}$} &
  \multicolumn{1}{c}{$L_{cat}$\tablefootmark{d}} &
  \multicolumn{1}{c}{$\pm eL_{cat}$} &
  \multicolumn{1}{c}{$L_{500}$\tablefootmark{e}} &
  \multicolumn{1}{c}{$\pm eL_{500}$} &
  \multicolumn{1}{c}{$M_{500}$\tablefootmark{f}} &
  \multicolumn{1}{c}{$\pm eM_{500}$} \\

  \multicolumn{1}{c}{} &
  \multicolumn{1}{c}{IAUNAME} &
  \multicolumn{1}{c}{(deg)} &
  \multicolumn{1}{c}{(deg)} &
  \multicolumn{1}{c}{} &
  \multicolumn{1}{c}{} &
  \multicolumn{1}{c}{kpc/$''$} &
  \multicolumn{1}{c}{(kpc)} &
  \multicolumn{1}{c}{(keV)} &
  \multicolumn{1}{c}{(keV)} &
  \multicolumn{1}{c}{(keV)} &
  \multicolumn{1}{c}{} &
  \multicolumn{1}{c}{} &
  \multicolumn{1}{c}{} &
  \multicolumn{1}{c}{} &
  \multicolumn{1}{c}{}  \\

(1) &  (2)  &  (3)  & (4)  &   (5)   &  (6)  &   (7)  &   (8)  &  (9)  &  (10)  &  (11) &  (12) & (13) & (14) & (15) &  (16)   \\
\hline 

005735 &     2XMM J003840.4+004746 &   9.66841 &   0.79636 & 0203690101 & 0.5549 & 6.44 &  522.21 &   1.44 & 0.18 &  17.83 &  2.19 &   35.42 &   5.05 &  7.42 &  1.56 \\
007554 &     2XMM J004304.2-092801 &  10.76751 &  -9.46695 & 0065140201 & 0.1866 & 3.12 &  594.22 &   6.84 & 1.06 &   6.74 &  1.05 &   20.30 &   3.72 &  7.18 &  1.56 \\
010986 &     2XMM J005556.9+003806 &  13.98720 &   0.63507 & 0303110401 & 0.2047 & 3.36 &  541.55 &   5.01 & 0.95 &   6.06 &  1.15 &   12.11 &   2.64 &  5.54 &  1.28 \\
021043 &     2XMM J015558.5+053159 &  28.99394 &   5.53329 & 0153030701 & 0.4499 & 5.76 &  640.92 &   5.82 & 0.85 &  43.44 &  6.37 &   84.97 &  15.25 & 12.11 &  2.54 \\
021597 &     2XMM J020019.2+001931 &  30.08012 &   0.32553 & 0101640201 & 0.6825 & 7.07 &  643.12 &   4.17 & 0.52 &  85.10 & 10.59 &  213.07 &  37.70 & 16.13 &  3.35 \\
030746 &     2XMM J023346.9-085054 &  38.44543 &  -8.84844 & 0150470601 & 0.2653 & 4.08 &  587.32 &   5.95 & 1.09 &  12.93 &  2.36 &   24.86 &   5.10 &  7.55 &  1.67 \\
030889 &     2XMM J023458.7-085055 &  38.74463 &  -8.84868 & 0150470601 & 0.2590 & 4.01 &  586.22 &   4.15 & 0.53 &   8.54 &  1.09 &   24.02 &   3.31 &  7.45 &  1.56 \\
089821 &     2XMM J083114.4+523447 & 127.81014 &  52.57993 & 0092800201 & 0.6107 & 6.74 &  470.06 &   0.79 & 0.11 &  12.25 &  1.64 &   22.80 &   3.72 &  5.78 &  1.26 \\
089885 &     2XMM J083146.1+525056 & 127.94516 &  52.84719 & 0092800201 & 0.5190 & 6.23 &  582.48 &   3.50 & 0.21 &  36.78 &  2.24 &   60.96 &   5.82 &  9.86 &  1.97 \\
091280 &     2XMM J083926.4+193658 & 129.86017 &  19.61622 & 0101440401 & 0.3742 & 5.16 &  481.60 &   0.93 & 0.13 &   4.50 &  0.62 &   10.74 &   1.48 &  4.71 &  1.03 \\
091629 &     2XMM J084124.2+004640 & 130.35101 &   0.77799 & 0202940201 & 0.4075 & 5.43 &  615.91 &   4.28 & 0.29 &  25.27 &  1.69 &   56.45 &   5.47 & 10.23 &  2.03 \\
099426 &     2XMM J094034.6+355945 & 145.14429 &  35.99594 & 0021740101 & 0.3011 & 4.47 &  560.28 &   3.01 & 0.41 &   8.75 &  1.19 &   21.08 &   3.39 &  6.82 &  1.46 \\
099550 &     2XMM J094107.7+032912 & 145.28229 &   3.48674 & 0306050201 & 0.2490 & 3.90 &  443.09 &   0.77 & 0.14 &   1.45 &  0.27 &    4.04 &   0.63 &  3.18 &  0.74 \\
103306 &     2XMM J095945.4+023634 & 149.93957 &   2.60961 & 0302351601 & 0.3445 & 4.89 &  443.47 &   0.82 & 0.16 &   3.26 &  0.65 &    5.75 &   1.33 &  3.55 &  0.87 \\
104390 &     2XMM J100056.7+023343 & 150.23611 &   2.56332 & 0203360601 & 0.2194 & 3.54 &  463.19 &   1.53 & 0.26 &   2.16 &  0.37 &    4.80 &   0.77 &  3.52 &  0.81 \\
104705 &     2XMM J100117.0+021658 & 150.32104 &   2.28282 & 0302350801 & 0.1225 & 2.20 &  398.12 &   0.75 & 0.14 &   0.30 &  0.06 &    1.35 &   0.22 &  2.02 &  0.50 \\
105537 &     2XMM J100217.2+015625 & 150.57174 &   1.94037 & 0203360401 & 0.3082 & 4.54 &  448.94 &   1.41 & 0.39 &   4.33 &  1.19 &    5.43 &   2.02 &  3.54 &  1.00 \\
106241 &     2XMM J100423.5+410008 & 151.09813 &  41.00230 & 0207130201 & 0.3583 & 5.02 &  454.11 &   1.02 & 0.28 &   4.46 &  1.24 &    7.01 &   2.42 &  3.87 &  1.05 \\
106515 &     2XMM J100713.3+124718 & 151.80582 &  12.78839 & 0140550601 & 0.2481 & 3.89 &  422.58 &   0.49 & 0.12 &   0.91 &  0.22 &    3.00 &   0.58 &  2.76 &  0.67 \\
109295 &     2XMM J102748.7+000336 & 156.95333 &   0.06013 & 0305980401 & 0.7060 & 7.17 &  730.83 &   7.85 & 0.62 & 173.96 & 13.79 &  518.41 &  57.07 & 24.35 &  4.78 \\
109850 &     2XMM J103003.5+052023 & 157.51499 &   5.33982 & 0148560501 & 0.3082 & 4.54 &  480.19 &   1.49 & 0.14 &   4.56 &  0.42 &    8.26 &   0.95 &  4.33 &  0.95 \\
109997 &     2XMM J103027.8+310802 & 157.61623 &  31.13392 & 0102040301 & 0.4622 & 5.85 &  510.06 &   1.51 & 0.22 &  12.00 &  1.71 &   21.41 &   3.59 &  6.19 &  1.35 \\
110111 &     2XMM J103046.8+052751 & 157.69516 &   5.46432 & 0148560501 & 0.3772 & 5.18 &  489.68 &   0.88 & 0.07 &   4.31 &  0.35 &   12.04 &   0.99 &  4.96 &  1.05 \\
112052 &     2XMM J104104.0+060130 & 160.26679 &   6.02522 & 0151390101 & 0.5056 & 6.14 &  483.60 &   0.85 & 0.11 &   8.34 &  1.12 &   18.13 &   2.86 &  5.55 &  1.21 \\
118189 &     2XMM J105215.5+441759 & 163.06461 &  44.29999 & 0146990201 & 0.4982 & 6.09 &  535.68 &   1.16 & 0.22 &  11.07 &  2.05 &   33.37 &   6.71 &  7.48 &  1.65 \\
118353 &     2XMM J105226.8+441859 & 163.11217 &  44.31609 & 0146990201 & 0.4983 & 6.09 &  615.15 &   3.50 & 0.44 &  33.38 &  4.20 &   79.15 &  12.35 & 11.33 &  2.34 \\
121351 &     2XMM J110344.0+360513 & 165.93374 &  36.08715 & 0205370101 & 0.4972 & 6.09 &  478.28 &   0.84 & 0.13 &   7.93 &  1.24 &   16.38 &   3.11 &  5.32 &  1.20 \\
121575 &     2XMM J110355.9+360025 & 165.98326 &  36.00699 & 0205370101 & 0.3328 & 4.78 &  427.50 &   0.49 & 0.08 &   1.81 &  0.31 &    4.38 &   0.71 &  3.14 &  0.74 \\
125233 &     2XMM J111659.0+174953 & 169.24600 &  17.83141 & 0099030101 & 0.4386 & 5.68 &  550.28 &   2.28 & 0.41 &  15.99 &  2.89 &   31.44 &   6.65 &  7.56 &  1.69 \\
128668 &     2XMM J113804.0+031525 & 174.51687 &   3.25708 & 0111970701 & 0.4491 & 5.75 &  610.45 &   3.59 & 0.52 &  26.66 &  3.84 &   62.51 &  10.34 & 10.45 &  2.18 \\
132492 &     2XMM J120544.4+352307 & 181.43514 &  35.38554 & 0148742401 & 0.6530 & 6.94 &  553.21 &   1.89 & 0.41 &  34.66 &  7.50 &   74.25 &  19.42 &  9.91 &  2.29 \\
133193 &     2XMM J120933.9+392234 & 182.39078 &  39.37648 & 0112830201 & 0.5548 & 6.44 &  575.26 &   2.72 & 0.20 &  33.53 &  2.48 &   64.74 &   6.35 &  9.91 &  1.98 \\
134697 &     2XMM J121205.3+131739 & 183.02238 &  13.29430 & 0112550501 & 0.4489 & 5.75 &  569.61 &   2.88 & 0.31 &  21.41 &  2.34 &   40.55 &   5.22 &  8.49 &  1.75 \\
135421 &     2XMM J121658.3+065826 & 184.24317 &   6.97416 & 0204650201 & 0.4842 & 6.00 &  615.82 &   3.26 & 0.70 &  28.96 &  6.24 &   75.49 &  19.08 & 11.18 &  2.54 \\
135555 &     2XMM J121733.6+071035 & 184.39023 &   7.17641 & 0204650201 & 0.5796 & 6.58 &  552.31 &   2.13 & 0.38 &  29.18 &  5.15 &   55.28 &  11.54 &  9.03 &  1.98 \\
139177 &     2XMM J122643.5+334555 & 186.68155 &  33.76528 & 0200340101 & 0.7664 & 7.40 &  499.09 &   0.91 & 0.09 &  24.59 &  2.36 &   60.66 &   7.14 &  8.34 &  1.71 \\
140028 &     2XMM J122903.6+015012 & 187.26514 &   1.83680 & 0159960101 & 0.3139 & 4.59 &  475.65 &   1.26 & 0.35 &   4.03 &  1.12 &    7.95 &   2.51 &  4.23 &  1.10 \\
142993 &     2XMM J123232.4+200350 & 188.13529 &  20.06410 & 0301450201 & 0.0635 & 1.22 &  444.98 &   1.91 & 0.39 &   0.19 &  0.04 &    2.23 &   0.33 &  2.66 &  0.63 \\
143008 &     2XMM J123235.9+000233 & 188.14990 &   0.04251 & 0203170301 & 0.6832 & 7.08 &  486.42 &   0.86 & 0.07 &  17.68 &  1.44 &   37.40 &   3.26 &  6.99 &  1.43 \\
147020 &     2XMM J124425.3+164759 & 191.10564 &  16.79993 & 0302581501 & 0.2346 & 3.73 &  505.70 &   1.66 & 0.17 &   2.73 &  0.28 &    8.76 &   0.43 &  4.66 &  0.98 \\
147115 &     2XMM J124448.8-001949 & 191.20335 &  -0.33035 & 0110980201 & 0.4693 & 5.90 &  510.22 &   1.24 & 0.12 &  10.26 &  0.96 &   22.05 &   2.51 &  6.25 &  1.31 \\
149238 &     2XMM J125313.4+155612 & 193.30610 &  15.93687 & 0082990101 & 0.2754 & 4.19 &  702.73 &  12.17 & 1.19 &  28.83 &  2.82 &   78.97 &  12.24 & 13.07 &  2.68 \\
158679 &     2XMM J132505.0+302445 & 201.27118 &  30.41268 & 0025740201 & 0.4348 & 5.65 &  635.41 &   4.24 & 0.60 &  29.14 &  4.15 &   76.03 &  12.89 & 11.59 &  2.42 \\
159400 &     2XMM J132631.5+074503 & 201.63141 &   7.75093 & 0200730201 & 0.5327 & 6.31 &  484.00 &   0.86 & 0.12 &   9.67 &  1.39 &   20.23 &   3.24 &  5.75 &  1.26 \\
160031 &     2XMM J132905.8+582655 & 202.27296 &  58.44784 & 0142770301 & 0.1566 & 2.71 &  488.76 &   1.48 & 0.27 &   0.99 &  0.18 &    5.42 &   0.63 &  3.87 &  0.86 \\
160695 &     2XMM J133046.2+334620 & 202.69291 &  33.77228 & 0305361601 & 0.4587 & 5.82 &  515.54 &   1.94 & 0.29 &  15.17 &  2.23 &   22.58 &   4.50 &  6.37 &  1.42 \\
162382 &     2XMM J133458.4+375019 & 203.74351 &  37.83850 & 0109661001 & 0.3834 & 5.23 &  499.71 &   1.38 & 0.11 &   7.03 &  0.56 &   13.99 &   1.36 &  5.31 &  1.12 \\
162545 &     2XMM J133514.1+374908 & 203.80896 &  37.81876 & 0109661001 & 0.6965 & 7.13 &  561.72 &   2.07 & 0.12 &  44.47 &  2.67 &   96.69 &   8.08 & 10.93 &  2.15 \\
163928 &     2XMM J133852.5+044309 & 204.71904 &   4.71926 & 0152940101 & 0.7264 & 7.25 &  593.32 &   3.00 & 0.35 &  71.26 &  8.29 &  152.84 &  23.61 & 13.35 &  2.73 \\
164535 &     2XMM J134138.7+001721 & 205.41160 &   0.28936 & 0111281001 & 0.5054 & 6.14 &  698.23 &   9.49 & 1.17 &  93.56 & 11.54 &  179.29 &  26.54 & 16.71 &  3.38 \\
164909 &     2XMM J134304.9-000053 & 205.77057 &  -0.01498 & 0202460101 & 0.7151 & 7.21 &  761.31 &  12.50 & 0.44 & 285.99 & 10.01 &  693.11 &  48.62 & 27.83 &  5.37 \\
165043 &     2XMM J134330.3+403051 & 205.87636 &  40.51441 & 0070340701 & 0.7084 & 7.18 &  501.33 &   0.92 & 0.11 &  20.58 &  2.46 &   49.80 &   6.53 &  7.88 &  1.63 \\
171595 &     2XMM J141256.9-030918 & 213.23753 &  -3.15574 & 0013140101 & 0.6697 & 7.02 &  541.50 &   1.62 & 0.27 &  31.57 &  5.30 &   69.32 &  13.66 &  9.48 &  2.05 \\
171668 &     2XMM J141313.0-031732 & 213.30419 &  -3.29242 & 0013140101 & 0.3088 & 4.54 &  500.87 &   1.54 & 0.31 &   4.74 &  0.96 &   10.77 &   2.20 &  4.91 &  1.13 \\
173756 &     2XMM J142305.5+382807 & 215.77325 &  38.46869 & 0147570601 & 0.4508 & 5.77 &  550.27 &   1.79 & 0.27 &  13.39 &  2.03 &   32.92 &   4.90 &  7.67 &  1.62 \\
174085 &     2XMM J142732.7+264122 & 216.88659 &  26.68946 & 0111290601 & 0.2293 & 3.67 &  393.65 &   0.53 & 0.14 &   0.83 &  0.21 &    1.80 &   0.47 &  2.18 &  0.58 \\
178050 &     2XMM J145317.4+033446 & 223.32278 &   3.57946 & 0150350101 & 0.3710 & 5.13 &  658.33 &   5.52 & 0.28 &  26.11 &  1.33 &   74.58 &   6.87 & 11.97 &  2.35 \\
178120 &     2XMM J145351.0+032319 & 223.46267 &   3.38865 & 0150350101 & 0.3685 & 5.11 &  509.66 &   1.64 & 0.26 &   7.61 &  1.23 &   14.96 &   2.87 &  5.54 &  1.25 \\
180311 &     2XMM J150918.8-001424 & 227.32855 &  -0.24013 & 0305750201 & 0.2625 & 4.05 &  592.47 &   5.56 & 0.93 &  11.81 &  1.97 &   25.99 &   4.61 &  7.72 &  1.67 \\
184398 &     2XMM J154931.9+213259 & 237.38303 &  21.54999 & 0136040101 & 0.6357 & 6.86 &  679.01 &   5.08 & 0.31 &  87.20 &  5.29 &  249.36 &  24.50 & 17.95 &  3.50 \\
185578 &     2XMM J155930.6+350441 & 239.87761 &  35.07807 & 0112600801 & 0.4895 & 6.04 &  557.90 &   2.30 & 0.44 &  21.04 &  3.98 &   41.60 &   9.08 &  8.37 &  1.86 \\
187930 &     2XMM J161713.8+123805 & 244.30779 &  12.63495 & 0103461001 & 0.1987 & 3.28 &  497.04 &   1.27 & 0.25 &   1.44 &  0.29 &    6.94 &   1.04 &  4.26 &  0.95 \\
200392 &     2XMM J172335.5+341155 & 260.89826 &  34.19881 & 0102040101 & 0.4428 & 5.71 &  762.32 &  11.78 & 1.79 &  84.63 & 12.89 &  244.14 &  54.38 & 20.21 &  4.39 \\
262357 &    2XMMI J081129.1+481011 & 122.87145 &  48.17000 & 0402780701 & 0.6075 & 6.72 &  531.93 &   1.53 & 0.31 &  23.53 &  4.75 &   48.71 &  11.43 &  8.34 &  1.89 \\
264469 &    2XMMI J092623.1+362132 & 141.59634 &  36.35902 & 0402370101 & 0.6558 & 6.96 &  539.87 &   1.81 & 0.17 &  33.56 &  3.19 &   64.46 &   7.79 &  9.24 &  1.88 \\
264943 &    2XMMI J095610.3-002151 & 149.04328 &  -0.36440 & 0206430101 & 0.5828 & 6.59 &  607.25 &   3.60 & 0.32 &  50.02 &  4.45 &  101.16 &  11.49 & 12.06 &  2.40 \\
265123 &    2XMMI J095801.8+020147 & 149.50756 &   2.02998 & 0302352401 & 0.6740 & 7.04 &  410.88 &   0.65 & 0.65 &  12.96 &  7.78 &   12.59 &  10.50 &  4.16 &  1.89 \\
274756 &    2XMMI J134549.4+074440 & 206.45601 &   7.74465 & 0405950501 & 0.6573 & 6.96 &  536.03 &   1.48 & 0.27 &  27.48 &  4.95 &   61.99 &  13.51 &  9.06 &  2.01 \\
275328 &    2XMMI J143713.9+341519 & 219.30818 &  34.25553 & 0405200101 & 0.5426 & 6.37 &  674.95 &   6.93 & 0.77 &  80.95 &  9.05 &  167.42 &  24.46 & 15.78 &  3.19 \\
275341 &    2XMMI J143742.9+340810 & 219.42892 &  34.13626 & 0405200101 & 0.5446 & 6.38 &  609.39 &   3.44 & 0.44 &  40.58 &  5.21 &   89.20 &  15.36 & 11.64 &  2.43 \\
278534 &    2XMMI J164729.3+271241 & 251.87220 &  27.21165 & 0304071801 & 0.4937 & 6.06 &  427.28 &   0.57 & 0.22 &   5.33 &  2.09 &    8.00 &   3.85 &  3.78 &  1.19 \\
287820 &    2XMMI J220605.5-001110 & 331.52310 &  -0.18635 & 0401180101 & 0.3878 & 5.27 &  475.32 &   0.89 & 0.18 &   4.68 &  0.94 &   10.41 &   2.19 &  4.59 &  1.07 \\
309493 &    2XMMI J081058.2+500529 & 122.74276 &  50.09156 & 0401270401 & 0.4029 & 5.40 &  638.50 &   6.08 & 0.45 &  34.94 &  2.57 &   69.46 &   7.11 & 11.33 &  2.25 \\
309859 &    2XMMI J082432.7+294757 & 126.13630 &  29.79925 & 0504102001 & 0.2094 & 3.42 &  490.66 &   1.06 & 0.18 &   1.35 &  0.23 &    6.65 &   0.73 &  4.14 &  0.91 \\
310772 &    2XMMI J084052.6+383847 & 130.21920 &  38.64663 & 0502060201 & 0.1171 & 2.12 &  466.97 &   1.64 & 0.15 &   0.58 &  0.05 &    3.58 &   0.13 &  3.24 &  0.72 \\
311197 &    2XMMI J085253.3+175718 & 133.22230 &  17.95522 & 0305480301 & 0.2076 & 3.40 &  469.72 &   1.02 & 0.09 &   1.28 &  0.12 &    5.03 &   0.14 &  3.63 &  0.79 \\
312824 &    2XMMI J092209.5+063230 & 140.53973 &   6.54177 & 0502920101 & 0.1720 & 2.93 &  393.26 &   0.79 & 0.21 &   0.65 &  0.17 &    1.47 &   0.38 &  2.05 &  0.55 \\
315216 &    2XMMI J103007.0-030638 & 157.52957 &  -3.11081 & 0404840201 & 0.4394 & 5.68 &  567.31 &   2.75 & 0.18 &  19.42 &  1.26 &   38.13 &   2.85 &  8.29 &  1.66 \\
315488 &    2XMMI J103421.4+395153 & 158.58926 &  39.86479 & 0506440101 & 0.6043 & 6.71 &  476.15 &   0.75 & 0.15 &  11.37 &  2.32 &   24.10 &   5.37 &  5.96 &  1.37 \\
317881 &    2XMMI J112712.0+025956 & 171.80019 &   2.99890 & 0551021201 & 0.5388 & 6.35 &  552.50 &   2.54 & 0.59 &  29.21 &  6.81 &   47.31 &  13.61 &  8.62 &  2.07 \\
318253 &    2XMMI J114541.4+025415 & 176.42267 &   2.90425 & 0551022701 & 0.2799 & 4.24 &  474.30 &   1.05 & 0.16 &   2.58 &  0.40 &    6.90 &   0.82 &  4.04 &  0.89 \\
318315 &    2XMMI J114619.2+025609 & 176.58015 &   2.93602 & 0551022701 & 0.1584 & 2.74 &  421.56 &   1.00 & 0.27 &   0.69 &  0.18 &    2.17 &   0.46 &  2.49 &  0.62 \\
319555 &    2XMMI J120045.3+342454 & 180.18887 &  34.41505 & 0551630301 & 0.2627 & 4.05 &  475.86 &   1.22 & 0.15 &   2.60 &  0.32 &    6.62 &   0.63 &  4.00 &  0.88 \\
320719 &    2XMMI J121921.0+060134 & 184.83777 &   6.02628 & 0502120101 & 0.3535 & 4.97 &  438.54 &   0.72 & 0.17 &   3.06 &  0.71 &    5.54 &   1.45 &  3.47 &  0.87 \\
321268 &     2XMM J123050.8+413415 & 187.71095 &  41.57120 & 0556300101 & 0.7455 & 7.32 &  509.26 &   1.11 & 0.15 &  28.22 &  3.68 &   63.43 &   9.38 &  8.64 &  1.80 \\
322726 &    2XMMI J124919.4+051837 & 192.33104 &   5.31033 & 0503610101 & 0.4200 & 5.53 &  489.75 &   1.18 & 0.16 &   7.46 &  1.00 &   14.16 &   2.42 &  5.22 &  1.16 \\
322873 &    2XMMI J125003.9+052118 & 192.51654 &   5.35501 & 0503610101 & 0.3849 & 5.25 &  444.65 &   0.83 & 0.11 &   4.27 &  0.58 &    6.79 &   1.21 &  3.75 &  0.87 \\
325696 &    2XMMI J133805.9-013501 & 204.52478 &  -1.58373 & 0502060101 & 0.3481 & 4.92 &  614.91 &   3.33 & 0.44 &  13.55 &  1.79 &   44.75 &   6.65 &  9.50 &  1.97 \\
327024 &    2XMMI J142114.9+030745 & 215.31213 &   3.12938 & 0502480701 & 0.3122 & 4.58 &  638.61 &   6.90 & 0.61 &  21.82 &  1.92 &   49.66 &   5.66 & 10.22 &  2.06 \\
327387 &    2XMMI J143113.8-000618 & 217.80765 &  -0.10501 & 0501540201 & 0.7122 & 7.19 &  533.53 &   1.71 & 0.31 &  38.72 &  7.09 &   74.55 &  16.73 &  9.54 &  2.12 \\
327608 &    2XMMI J143927.3+001249 & 219.86378 &   0.21364 & 0551200101 & 0.3000 & 4.45 &  464.71 &   0.96 & 0.15 &   2.77 &  0.44 &    6.54 &   0.95 &  3.89 &  0.88 \\
329022 &    2XMMI J145709.1-010057 & 224.28792 &  -1.01585 & 0502780601 & 0.2742 & 4.18 &  606.21 &   2.98 & 0.30 &   6.99 &  0.71 &   31.27 &   4.44 &  8.38 &  1.74 \\
331167 &    2XMMI J151529.5+003943 & 228.87305 &   0.66205 & 0556210501 & 0.2645 & 4.07 &  616.16 &   3.47 & 0.39 &   7.50 &  0.85 &   33.44 &   5.15 &  8.71 &  1.82 \\
\hline
\end{longtable}
}


%

{\tiny
\begin{longtable}{c c c c c c c c c c c }
\hline
\hline
  \multicolumn{1}{c}{detid\tablefootmark{a}} &
  \multicolumn{1}{c}{objid\tablefootmark{g}} &
  \multicolumn{1}{c}{RA\tablefootmark{g}} &
  \multicolumn{1}{c}{DEC\tablefootmark{g}} &
  \multicolumn{1}{c}{$m_{r}$\tablefootmark{g}} &
  \multicolumn{1}{c}{$z_{s}$\tablefootmark{g}} &
  \multicolumn{1}{c}{$N_{z_{s}}$\tablefootmark{g}} &
  \multicolumn{1}{c}{$z_{p}$\tablefootmark{g}} &
  \multicolumn{1}{c}{$N_{z_{p}}$\tablefootmark{g}} &
  \multicolumn{1}{c}{offset\tablefootmark{g}} &
  \multicolumn{1}{c}{note\tablefootmark{h}}  \\
  
  \multicolumn{1}{c}{}&
  \multicolumn{1}{c}{(BCG)} &
  \multicolumn{1}{c}{(deg)} &
  \multicolumn{1}{c}{(deg)} &
  \multicolumn{1}{c}{(BCG)} &
  \multicolumn{1}{c}{} &
  \multicolumn{1}{c}{} &
  \multicolumn{1}{c}{} &
  \multicolumn{1}{c}{} &
  \multicolumn{1}{c}{(kpc)} &
  \multicolumn{1}{c}{}  \\
  (1)  &  (17)  & (18)  &  (19)  &  (20)   &  (21) &  (22)  &   (23)   &  (24) & (25) & (26)  \\
\hline

005735 & 1237663204918428144 &   9.68054 &   0.78241 & 20.047 & 0.5549 &  3 & 0.5127 &  7 & 429.63 & Extended \\
007554 & 1237652630713860232 &  10.80832 &  -9.47863 & 17.213 & 0.1866 &  2 & 0.1794 & 18 & 473.45 & Extended \\
010986 & 1237663784740388918 &  14.02537 &   0.62659 & 17.537 & 0.2047 &  3 & 0.1951 & 20 & 473.61 & Extended \\
021043 & 1237678663047250389 &  28.98754 &   5.53072 & 19.553 & 0.4499 &  1 & 0.4258 & 12 & 142.33 & Paper-II \\
021597 & 1237657071160263439 &  30.08100 &   0.32491 & 20.448 & 0.6825 &  1 & 0.6555 &  3 &  27.36 & Extended \\
030746 & 1237653500970139807 &  38.44673 &  -8.84925 & 17.540 & 0.2653 &  1 & 0.2547 & 17 &  22.30 & Paper-II \\
030889 & 1237653500970270877 &  38.74547 &  -8.84926 & 17.762 & 0.2590 &  2 & 0.2528 & 17 &  14.70 & Paper-II \\
089821 & 1237651701914141241 & 127.80965 &  52.57912 & 20.467 & 0.6107 &  1 & 0.6465 &  4 &  20.97 & Extended \\
089885 & 1237651272960967114 & 127.94343 &  52.84937 & 19.251 & 0.5190 &  2 & 0.5165 & 13 &  54.28 & Paper-II \\
091280 & 1237667107965108712 & 129.86275 &  19.61566 & 18.114 & 0.3742 &  3 & 0.3542 & 15 &  46.34 & Paper-II \\
091629 & 1237648722282742639 & 130.35096 &   0.77637 & 19.415 & 0.4075 &  1 & 0.3774 & 16 &  31.67 & Paper-II \\
099426 & 1237661139030638763 & 145.14560 &  35.99356 & 17.982 & 0.3011 &  1 & 0.2795 &  7 &  41.85 & Paper-II \\
099550 & 1237654601024471324 & 145.28619 &   3.50159 & 18.009 & 0.2490 &  1 & 0.2231 & 13 & 215.51 & Paper-II \\
103306 & 1237651754534044209 & 149.93796 &   2.60627 & 19.098 & 0.3445 &  1 & 0.3452 & 10 &  65.24 & Paper-II \\
104390 & 1237651754534175091 & 150.23989 &   2.56276 & 17.632 & 0.2194 &  1 & 0.2132 &  9 &  48.74 & Extended \\
104705 & 1237653665258995921 & 150.31973 &   2.28670 & 16.715 & 0.1225 &  5 & 0.1332 & 12 &  32.19 & Extended \\
105537 & 1237653664722256181 & 150.56128 &   1.94206 & 18.564 & 0.3082 &  1 & 0.3201 &  8 & 173.06 & Paper-II \\
106241 & 1237661852532277722 & 151.10591 &  41.01332 & 19.300 & 0.3583 &  1 & 0.3419 & 16 & 225.40 & Paper-II \\
106515 & 1237661069785956568 & 151.80525 &  12.78700 & 17.503 & 0.2481 &  1 & 0.2361 & 14 &  20.95 & Extended \\
109295 & 1237654670275511096 & 156.95279 &   0.06054 & 20.848 & 0.7060 &  1 & 0.7141 &  4 &  17.56 & Extended \\
109850 & 1237654602640457895 & 157.51425 &   5.33829 & 18.500 & 0.3082 &  1 & 0.3056 & 10 &  27.78 & Paper-II \\
109997 & 1237665127460045508 & 157.61411 &  31.13688 & 19.109 & 0.4622 &  1 & 0.4610 &  7 &  73.14 & Extended \\
110111 & 1237658297923076324 & 157.69698 &   5.46371 & 18.764 & 0.3772 &  1 & 0.3607 & 12 &  35.55 & Paper-II \\
112052 & 1237658298461061646 & 160.26775 &   6.02345 & 19.502 & 0.5056 &  1 & 0.4980 & 10 &  44.34 & Paper-II \\
118189 & 1237660634386727383 & 163.06443 &  44.29762 & 19.898 & 0.4982 &  1 & 0.4835 &  8 &  52.02 & Paper-II \\
118353 & 1237660634386792717 & 163.11172 &  44.31504 & 19.925 & 0.4983 &  3 & 0.5078 & 11 &  24.13 & Extended \\
121351 & 1237664338249187791 & 165.93781 &  36.08689 & 19.650 & 0.4972 &  1 & 0.4977 & 13 &  72.41 & Paper-II \\
121575 & 1237664338249187846 & 165.98502 &  36.00626 & 18.686 & 0.3328 &  1 & 0.3167 & 11 &  27.64 & Paper-II \\
125233 & 1237668496852844963 & 169.24231 &  17.83243 & 19.318 & 0.4386 &  1 & 0.4189 &  8 &  74.85 & Extended \\
128668 & 1237654030866776457 & 174.51838 &   3.25657 & 18.134 & 0.4491 &  2 & 0.4363 & 12 &  32.92 & Paper-II \\
132492 & 1237665129079570844 & 181.43720 &  35.38470 & 20.295 & 0.6530 &  1 & 0.6295 &  3 &  46.82 & Extended \\
133193 & 1237664667907850801 & 182.38888 &  39.37339 & 19.409 & 0.5548 &  1 & 0.5436 & 13 &  79.41 & Paper-II \\
134697 & 1237661950790599049 & 183.01990 &  13.29903 & 19.310 & 0.4489 &  1 & 0.4753 &  8 & 110.07 & Extended \\
135421 & 1237661974936223829 & 184.23923 &   6.97396 & 18.743 & 0.4842 &  1 & 0.4826 & 18 &  84.67 & Paper-II \\
135555 & 1237661971185271488 & 184.38379 &   7.16889 & 20.754 & 0.5796 &  1 & 0.5731 &  6 & 233.71 & Extended \\
139177 & 1237665126933922648 & 186.68128 &  33.76685 & 20.275 & 0.7664 &  1 & 0.6427 &  2 &  42.03 & Extended \\
140028 & 1237651752939749568 & 187.26808 &   1.83400 & 18.417 & 0.3139 &  1 & 0.3003 & 18 &  67.11 & Paper-II \\
142993 & 1237668298205364430 & 188.13313 &  20.06167 & 16.077 & 0.0635 &  2 & 0.0614 & 11 &  13.94 & Extended \\
143008 & 1237648704579175014 & 188.14958 &   0.04373 & 21.039 & 0.6832 &  1 & 0.7176 &  2 &  32.14 & Extended \\
147020 & 1237668623551889555 & 191.10600 &  16.79916 & 17.357 & 0.2346 &  1 & 0.2167 & 22 &  11.38 & Paper-II \\
147115 & 1237671764249543261 & 191.22085 &  -0.32448 & 19.986 & 0.4693 &  1 & 0.4536 &  7 & 391.96 & Extended \\
149238 & 1237664292079403190 & 193.30398 &  15.93415 & 18.095 & 0.2754 &  1 & 0.2673 & 18 &  51.39 & Extended \\
158679 & 1237665225700934189 & 201.27591 &  30.41477 & 18.416 & 0.4348 &  1 & 0.4170 & 10 &  93.27 & Extended \\
159400 & 1237671956442120782 & 201.63075 &   7.75169 & 20.070 & 0.5327 &  1 & 0.5208 &  9 &  22.75 & Paper-II \\
160031 & 1237655107845029998 & 202.26078 &  58.44167 & 16.451 & 0.1566 &  2 & 0.1577 &  7 &  86.45 & Extended \\
160695 & 1237665128550367803 & 202.69388 &  33.76724 & 19.998 & 0.4587 &  1 & 0.4488 & 14 & 106.97 & Paper-II \\
162382 & 1237664294223085804 & 203.76069 &  37.83613 & 18.949 & 0.3834 &  1 & 0.3735 & 25 & 259.46 & Paper-II \\
162545 & 1237664294223086790 & 203.81047 &  37.81983 & 20.737 & 0.6965 &  1 & 0.7021 &  2 &  41.24 & Extended \\
163928 & 1237671992413782929 & 204.71949 &   4.72021 & 21.535 & 0.7264 &  1 & 0.7145 &  4 &  27.58 & Extended \\
164535 & 1237651504881140622 & 205.41102 &   0.28245 & 19.338 & 0.5054 &  2 & 0.4906 & 16 & 153.82 & Extended \\
164909 & 1237651504344400779 & 205.76896 &  -0.01547 & 19.692 & 0.7151 &  1 & 0.6751 &  6 &  43.58 & Paper-II \\
165043 & 1237662194533401363 & 205.87815 &  40.51334 & 20.387 & 0.7084 &  1 & 0.6737 &  2 &  44.71 & Extended \\
171595 & 1237655497594897279 & 213.23824 &  -3.15516 & 20.370 & 0.6697 &  1 & 0.7371 &  2 &  23.08 & Extended \\
171668 & 1237655493299536387 & 213.30337 &  -3.29271 & 18.403 & 0.3088 &  1 & 0.3229 & 18 &  14.34 & Paper-II \\
173756 & 1237662195073745343 & 215.76928 &  38.46832 & 19.469 & 0.4508 &  1 & 0.4292 & 12 &  64.95 & Paper-II \\
174085 & 1237665442598092974 & 216.88254 &  26.69292 & 17.312 & 0.2293 &  1 & 0.1969 &  5 &  66.02 & Extended \\
178050 & 1237651822175715561 & 223.32185 &   3.57965 & 18.760 & 0.3710 &  2 & 0.3531 & 18 &  17.57 & Paper-II \\
178120 & 1237654879667028594 & 223.46466 &   3.39233 & 19.004 & 0.3685 &  2 & 0.3576 & 19 &  77.10 & Paper-II \\
180311 & 1237648704059474189 & 227.32956 &  -0.23877 & 17.821 & 0.2625 &  1 & 0.2517 & 29 &  24.73 & Paper-II \\
184398 & 1237665536540279868 & 237.38329 &  21.55053 & 20.089 & 0.6357 &  1 & 0.6584 &  2 &  14.54 & Paper-II \\
185578 & 1237662474765992736 & 239.87379 &  35.07410 & 19.752 & 0.4895 &  2 & 0.4835 &  8 & 109.72 & Paper-II \\
187930 & 1237668367463481778 & 244.31101 &  12.62973 & 17.493 & 0.1987 &  1 & 0.1889 & 14 &  72.10 & Extended \\
200392 & 1237662701873267520 & 260.90067 &  34.19946 & 18.321 & 0.4428 &  2 & 0.4313 & 14 &  42.99 & Paper-II \\
262357 & 1237651495759315826 & 122.87630 &  48.16215 & 21.000 & 0.6075 &  2 & 0.5786 &  2 & 205.32 & Extended \\
264469 & 1237660962404172624 & 141.58913 &  36.35895 & 19.799 & 0.6558 &  1 & 0.6501 &  4 & 145.52 & Extended \\
264943 & 1237654669735166950 & 149.04432 &  -0.36284 & 20.462 & 0.5828 &  1 & 0.5651 & 10 &  44.43 & Paper-II \\
265123 & 1237653664721797488 & 149.50358 &   2.02399 & 20.780 & 0.6740 &  1 & 0.5520 &  3 & 182.09 & Extended \\
274756 & 1237662247131349450 & 206.46481 &   7.75003 & 20.539 & 0.6573 &  1 & 0.6696 &  5 & 256.86 & Extended \\
275328 & 1237662662134661959 & 219.31932 &  34.25158 & 20.046 & 0.5426 &  4 & 0.5322 & 19 & 229.84 & Paper-II \\
275341 & 1237662662134727278 & 219.42946 &  34.13609 & 19.954 & 0.5446 &  2 & 0.5386 &  8 &  10.82 & Extended \\
278534 & 1237662500012425472 & 251.86614 &  27.21090 & 20.408 & 0.4937 &  1 & 0.5004 &  6 & 118.77 & Extended \\
287820 & 1237663543146644275 & 331.52062 &  -0.17853 & 19.208 & 0.3878 &  1 & 0.3632 &  8 & 155.73 & Paper-II \\
309493 & 1237651274032874098 & 122.74182 &  50.09320 & 18.559 & 0.4029 &  1 & 0.4049 & 16 &  34.04 & Paper-II \\
309859 & 1237660635983118556 & 126.13809 &  29.79794 & 16.667 & 0.2094 &  1 & 0.2021 & 14 &  25.08 & Extended \\
310772 & 1237657400804966579 & 130.21947 &  38.64636 & 15.502 & 0.1171 &  2 & 0.1154 & 16 &   2.67 & Paper-II \\
311197 & 1237667293186883997 & 133.22122 &  17.95514 & 16.979 & 0.2076 &  2 & 0.2016 & 21 &  12.72 & Extended \\
312824 & 1237658425155977363 & 140.55332 &   6.53523 & 17.140 & 0.1720 &  2 & 0.1884 &  9 & 157.99 & Extended \\
315216 & 1237650369398375098 & 157.53108 &  -3.10994 & 18.820 & 0.4394 &  2 & 0.4442 & 14 &  35.39 & Paper-II \\
315488 & 1237661383851049990 & 158.58657 &  39.87086 & 20.719 & 0.6043 &  1 & 0.6413 &  4 & 154.84 & Extended \\
317881 & 1237651754543612643 & 171.79315 &   3.01452 & 20.377 & 0.5388 &  1 & 0.5087 &  6 & 391.38 & Paper-II \\
318253 & 1237654030330691620 & 176.41303 &   2.88182 & 18.177 & 0.2799 &  3 & 0.2631 & 17 & 373.00 & Paper-II \\
318315 & 1237671142555975817 & 176.58184 &   2.93817 & 16.675 & 0.1584 &  1 & 0.1564 & 17 &  26.90 & Extended \\
319555 & 1237665024366477463 & 180.18868 &  34.41193 & 18.037 & 0.2627 &  2 & 0.2580 & 17 &  45.66 & Paper-II \\
320719 & 1237655126620045679 & 184.83849 &   6.02894 & 18.724 & 0.3535 &  1 & 0.3507 & 10 &  49.29 & Extended \\
321268 & 1237662193453695521 & 187.71084 &  41.57208 & 20.261 & 0.7455 &  1 & 0.7264 &  2 &  23.31 & Extended \\
322726 & 1237654880727139062 & 192.33753 &   5.30017 & 19.905 & 0.4200 &  1 & 0.3916 & 11 & 239.92 & Paper-II \\
322873 & 1237671264962150728 & 192.50153 &   5.34915 & 18.727 & 0.3849 &  2 & 0.3757 & 16 & 303.29 & Paper-II \\
325696 & 1237655499738579243 & 204.52444 &  -1.58302 & 18.794 & 0.3481 &  1 & 0.3663 & 11 &  13.86 & Paper-II \\
327024 & 1237651821098500609 & 215.30550 &   3.13251 & 18.623 & 0.3122 &  1 & 0.3260 & 10 & 120.69 & Extended \\
327387 & 1237648721247273154 & 217.80191 &  -0.10467 & 21.057 & 0.7122 &  1 & 0.6585 &  5 & 148.89 & Extended \\
327608 & 1237648704593068052 & 219.84926 &   0.21340 & 18.493 & 0.3000 &  2 & 0.2831 &  9 & 233.23 & Paper-II \\
329022 & 1237655693551796667 & 224.28785 &  -1.01563 & 18.441 & 0.2742 &  1 & 0.2813 &  9 &   3.40 & Paper-II \\
331167 & 1237655467525407096 & 228.87274 &   0.66177 & 18.122 & 0.2645 &  1 & 0.2516 & 24 &   6.16 & Paper-II \\

\hline
\end{longtable}
}
 \tablefoot{ The entire cluster catalogue of Table~\ref{tbl:catalog_sample}.
\tablefoottext{a}{Parameters extracted from the 2XMMi-DR3 catalogue.} 
\tablefoottext{b}{Spectroscopic redshift as given in col.~(21).}     
\tablefoottext{c}{2XMMi-DR3 flux, $F_{cat}$ [0.5-2.0] keV, and its errors in units of $10^{-14}$\ erg\ cm$^{-2}$\ s$^{-1}$.}
\tablefoottext{d}{Computed X-ray luminosity, $L_{cat}$ [0.5-2.0] keV, and its errors in units of $10^{42}$\ erg\ s$^{-1}$.} 
\tablefoottext{e}{X-ray bolometric luminosity, $L_{500}$, and its error in units of $10^{42}$\ erg\ s$^{-1}$.} 
\tablefoottext{f}{X-ray-luminosity-based mass $M_{500}$ and its error  in units of $10^{13}$\ M$_\odot$.}
\tablefoottext{g}{Parameters obtained from the current detection algorithm in the optical band.}
\tablefoottext{h}{A note about each system as $``$extended$``$: new cluster from the current algorithm, $``$paper-II$``$: a cluster in paper II 
and spectroscopically confirmed from the present procedure.}
}
%

\end{landscape}


\bibliographystyle{References/aa}
\renewcommand{\bibname}{References} 
\bibliography{References/refbib_thesis}

\begin{thebibliography}{40}
\expandafter\ifx\csname natexlab\endcsname\relax\def\natexlab#1{#1}\fi

\bibitem[{{Arnaud}(1996)}]{Arnaud96}
{Arnaud}, K.~A. 1996, in Astronomical Society of the Pacific Conference Series,
  Vol. 101, Astronomical Data Analysis Software and Systems V, ed.
  {G.~H.~Jacoby \& J.~Barnes}, 17

\bibitem[{{Arnaud} {et~al.}(2010){Arnaud}, {Pratt}, {Piffaretti},
  {B{\"o}hringer}, {Croston}, \& {Pointecouteau}}]{Arnaud10}
{Arnaud}, M., {Pratt}, G.~W., {Piffaretti}, R., {et~al.} 2010, \aap, 517, A92

\bibitem[{{Arviset} {et~al.}(2002){Arviset}, {Guainazzi}, {Hernandez},
  {Dowson}, {Osuna}, \& {Venet}}]{Arviset02}
{Arviset}, C., {Guainazzi}, M., {Hernandez}, J., {et~al.} 2002, ArXiv
  Astrophysics e-prints

\bibitem[{{Barkhouse} {et~al.}(2006){Barkhouse}, {Green}, {Vikhlinin}, {Kim},
  {Perley}, {Cameron}, {Silverman}, {Mossman}, {Burenin}, {Jannuzi}, {Kim},
  {Smith}, {Smith}, {Tananbaum}, \& {Wilkes}}]{Barkhouse06}
{Barkhouse}, W.~A., {Green}, P.~J., {Vikhlinin}, A., {et~al.} 2006, \apj, 645,
  955

\bibitem[{{Boggs} \& {Rogers}(1990)}]{Boggs90}
{Boggs}, P.~T. \& {Rogers}, J.~E. 1990, Contemporary Mathematics, 112, 183

\bibitem[{{B{\"o}hringer} {et~al.}(2002){B{\"o}hringer}, {Collins}, {Guzzo},
  {Schuecker}, {Voges}, {Neumann}, {Schindler}, {Chincarini}, {De Grandi},
  {Cruddace}, {Edge}, {Reiprich}, \& {Shaver}}]{Boehringer02}
{B{\"o}hringer}, H., {Collins}, C.~A., {Guzzo}, L., {et~al.} 2002, \apj, 566,
  93

\bibitem[{{Fassbender} {et~al.}(2007){Fassbender}, {B{\"o}hringer}, {Santos},
  {Schuecker}, {Lamer}, {Schwope}, {Kohnert}, {Rosati}, {Mullis}, \&
  {Quintana}}]{Fassbender07}
{Fassbender}, R., {B{\"o}hringer}, H., {Santos}, J., {et~al.} 2007, in Heating
  versus Cooling in Galaxies and Clusters of Galaxies, ed. {H.~B{\"o}hringer,
  G.~W.~Pratt, A.~Finoguenov, \& P.~Schuecker }, 54

\bibitem[{{Finoguenov} {et~al.}(2007){Finoguenov}, {Guzzo}, {Hasinger},
  {Scoville}, {Aussel}, {B{\"o}hringer}, {Brusa}, {Capak}, {Cappelluti},
  {Comastri}, {Giodini}, {Griffiths}, {Impey}, {Koekemoer}, {Kneib},
  {Leauthaud}, {Le F{\`e}vre}, {Lilly}, {Mainieri}, {Massey}, {McCracken},
  {Mobasher}, {Murayama}, {Peacock}, {Sakelliou}, {Schinnerer}, {Silverman},
  {Smol{\v c}i{\'c}}, {Taniguchi}, {Tasca}, {Taylor}, {Trump}, \&
  {Zamorani}}]{Finoguenov07}
{Finoguenov}, A., {Guzzo}, L., {Hasinger}, G., {et~al.} 2007, \apjs, 172, 182

\bibitem[{{Finoguenov} {et~al.}(2010){Finoguenov}, {Watson}, {Tanaka},
  {Simpson}, {Cirasuolo}, {Dunlop}, {Peacock}, {Farrah}, {Akiyama}, {Ueda},
  {Smol{\v c}i{\'c}}, {Stewart}, {Rawlings}, {van Breukelen}, {Almaini},
  {Clewley}, {Bonfield}, {Jarvis}, {Barr}, {Foucaud}, {McLure}, {Sekiguchi}, \&
  {Egami}}]{Finoguenov10}
{Finoguenov}, A., {Watson}, M.~G., {Tanaka}, M., {et~al.} 2010, \mnras, 403,
  2063

\bibitem[{{Hao} {et~al.}(2010){Hao}, {McKay}, {Koester}, {Rykoff}, {Rozo},
  {Annis}, {Wechsler}, {Evrard}, {Siegel}, {Becker}, {Busha}, {Gerdes},
  {Johnston}, \& {Sheldon}}]{Hao10}
{Hao}, J., {McKay}, T.~A., {Koester}, B.~P., {et~al.} 2010, \apjs, 191, 254

\bibitem[{{Kalberla} {et~al.}(2005){Kalberla}, {Burton}, {Hartmann}, {Arnal},
  {Bajaja}, {Morras}, \& {P{\"o}ppel}}]{Kalberla05}
{Kalberla}, P.~M.~W., {Burton}, W.~B., {Hartmann}, D., {et~al.} 2005, \aap,
  440, 775

\bibitem[{{Koester} {et~al.}(2007){Koester}, {McKay}, {Annis}, {Wechsler},
  {Evrard}, {Bleem}, {Becker}, {Johnston}, {Sheldon}, {Nichol}, {Miller},
  {Scranton}, {Bahcall}, {Barentine}, {Brewington}, {Brinkmann}, {Harvanek},
  {Kleinman}, {Krzesinski}, {Long}, {Nitta}, {Schneider}, {Sneddin}, {Voges},
  \& {York}}]{Koester07}
{Koester}, B.~P., {McKay}, T.~A., {Annis}, J., {et~al.} 2007, \apj, 660, 239

\bibitem[{{Krumpe} {et~al.}(2008){Krumpe}, {Lamer}, {Corral}, {Schwope},
  {Carrera}, {Barcons}, {Page}, {Mateos}, {Tedds}, \& {Watson}}]{Krumpe08}
{Krumpe}, M., {Lamer}, G., {Corral}, A., {et~al.} 2008, \aap, 483, 415

\bibitem[{{Lamer} {et~al.}(2008){Lamer}, {Hoeft}, {Kohnert}, {Schwope}, \&
  {Storm}}]{Lamer08}
{Lamer}, G., {Hoeft}, M., {Kohnert}, J., {Schwope}, A., \& {Storm}, J. 2008,
  \aap, 487, L33

\bibitem[{{Marriage} {et~al.}(2011){Marriage}, {Acquaviva}, {Ade}, {Aguirre},
  {Amiri}, {Appel}, {Barrientos}, {Battistelli}, {Bond}, {Brown}, {Burger},
  {Chervenak}, {Das}, {Devlin}, {Dicker}, {Bertrand Doriese}, {Dunkley},
  {D{\"u}nner}, {Essinger-Hileman}, {Fisher}, {Fowler}, {Hajian}, {Halpern},
  {Hasselfield}, {Hern{\'a}ndez-Monteagudo}, {Hilton}, {Hilton}, {Hincks},
  {Hlozek}, {Huffenberger}, {Handel Hughes}, {Hughes}, {Infante}, {Irwin},
  {Baptiste Juin}, {Kaul}, {Klein}, {Kosowsky}, {Lau}, {Limon}, {Lin},
  {Lupton}, {Marsden}, {Martocci}, {Mauskopf}, {Menanteau}, {Moodley},
  {Moseley}, {Netterfield}, {Niemack}, {Nolta}, {Page}, {Parker}, {Partridge},
  {Quintana}, {Reese}, {Reid}, {Sehgal}, {Sherwin}, {Sievers}, {Spergel},
  {Staggs}, {Swetz}, {Switzer}, {Thornton}, {Trac}, {Tucker}, {Warne},
  {Wilson}, {Wollack}, \& {Zhao}}]{Marriage11}
{Marriage}, T.~A., {Acquaviva}, V., {Ade}, P.~A.~R., {et~al.} 2011, \apj, 737,
  61

\bibitem[{{Maughan} {et~al.}(2011){Maughan}, {Giles}, {Randall}, {Jones}, \&
  {Forman}}]{Maughan11}
{Maughan}, B.~J., {Giles}, P.~A., {Randall}, S.~W., {Jones}, C., \& {Forman},
  W.~R. 2011, ArXiv e-prints

\bibitem[{{Maughan} {et~al.}(2008){Maughan}, {Jones}, {Forman}, \& {Van
  Speybroeck}}]{Maughan08}
{Maughan}, B.~J., {Jones}, C., {Forman}, W., \& {Van Speybroeck}, L. 2008,
  \apjs, 174, 117

\bibitem[{{Mewe} {et~al.}(1986){Mewe}, {Lemen}, \& {van den Oord}}]{Mewe86}
{Mewe}, R., {Lemen}, J.~R., \& {van den Oord}, G.~H.~J. 1986, \aaps, 65, 511

\bibitem[{{Pierre} {et~al.}(2006){Pierre}, {Pacaud}, {Duc}, {Willis},
  {Andreon}, {Valtchanov}, {Altieri}, {Galaz}, {Gueguen}, {Le F{\`e}vre},
  {F{\`e}vre}, {Ponman}, {Sprimont}, {Surdej}, {Adami}, {Alshino}, {Bremer},
  {Chiappetti}, {Detal}, {Garcet}, {Gosset}, {Jean}, {Maccagni}, {Marinoni},
  {Mazure}, {Quintana}, \& {Read}}]{Pierre06}
{Pierre}, M., {Pacaud}, F., {Duc}, P.-A., {et~al.} 2006, \mnras, 372, 591

\bibitem[{{Piffaretti} {et~al.}(2011){Piffaretti}, {Arnaud}, {Pratt},
  {Pointecouteau}, \& {Melin}}]{Piffaretti11}
{Piffaretti}, R., {Arnaud}, M., {Pratt}, G.~W., {Pointecouteau}, E., \&
  {Melin}, J.-B. 2011, \aap, 534, A109

\bibitem[{{Planck Collaboration} {et~al.}(2011){Planck Collaboration}, {Ade},
  {Aghanim}, {Arnaud}, {Ashdown}, {Aumont}, {Baccigalupi}, {Balbi}, {Banday},
  {Barreiro}, \& et~al.}]{Planck11}
{Planck Collaboration}, {Ade}, P.~A.~R., {Aghanim}, N., {et~al.} 2011, \aap,
  536, A8

\bibitem[{{Pratt} {et~al.}(2010){Pratt}, {Arnaud}, {Piffaretti},
  {B{\"o}hringer}, {Ponman}, {Croston}, {Voit}, {Borgani}, \&
  {Bower}}]{Pratt10}
{Pratt}, G.~W., {Arnaud}, M., {Piffaretti}, R., {et~al.} 2010, \aap, 511, A85

\bibitem[{{Pratt} {et~al.}(2009){Pratt}, {Croston}, {Arnaud}, \&
  {B{\"o}hringer}}]{Pratt09}
{Pratt}, G.~W., {Croston}, J.~H., {Arnaud}, M., \& {B{\"o}hringer}, H. 2009,
  \aap, 498, 361

\bibitem[{{Reiprich} \& {B{\"o}hringer}(2002)}]{Reiprich02}
{Reiprich}, T.~H. \& {B{\"o}hringer}, H. 2002, \apj, 567, 716

\bibitem[{{Romer} {et~al.}(2001){Romer}, {Viana}, {Liddle}, \&
  {Mann}}]{Romer01}
{Romer}, A.~K., {Viana}, P.~T.~P., {Liddle}, A.~R., \& {Mann}, R.~G. 2001,
  \apj, 547, 594

\bibitem[{{Rosati} {et~al.}(2002){Rosati}, {Borgani}, \& {Norman}}]{Rosati02}
{Rosati}, P., {Borgani}, S., \& {Norman}, C. 2002, \araa, 40, 539

\bibitem[{{Rykoff} {et~al.}(2008){Rykoff}, {McKay}, {Becker}, {Evrard},
  {Johnston}, {Koester}, {Rozo}, {Sheldon}, \& {Wechsler}}]{Rykoff08}
{Rykoff}, E.~S., {McKay}, T.~A., {Becker}, M.~R., {et~al.} 2008, \apj, 675,
  1106

\bibitem[{{Schuecker} {et~al.}(2003){Schuecker}, {B{\"o}hringer}, {Collins}, \&
  {Guzzo}}]{Schuecker03}
{Schuecker}, P., {B{\"o}hringer}, H., {Collins}, C.~A., \& {Guzzo}, L. 2003,
  \aap, 398, 867

\bibitem[{{Sunyaev} \& {Zeldovich}(1980)}]{Sunyaev80}
{Sunyaev}, R.~A. \& {Zeldovich}, I.~B. 1980, \araa, 18, 537

\bibitem[{{Sunyaev} \& {Zeldovich}(1972)}]{Sunyaev72}
{Sunyaev}, R.~A. \& {Zeldovich}, Y.~B. 1972, Comments on Astrophysics and Space
  Physics, 4, 173

\bibitem[{{Szabo} {et~al.}(2011){Szabo}, {Pierpaoli}, {Dong}, {Pipino}, \&
  {Gunn}}]{Szabo11}
{Szabo}, T., {Pierpaoli}, E., {Dong}, F., {Pipino}, A., \& {Gunn}, J. 2011,
  \apj, 736, 21

\bibitem[{{{\v S}uhada} {et~al.}(2010){{\v S}uhada}, {Song}, {B{\"o}hringer},
  {Benson}, {Mohr}, {Fassbender}, {Finoguenov}, {Pierini}, {Pratt},
  {Andersson}, {Armstrong}, \& {Desai}}]{Suhada10}
{{\v S}uhada}, R., {Song}, J., {B{\"o}hringer}, H., {et~al.} 2010, \aap, 514,
  L3

\bibitem[{{Vanderlinde} {et~al.}(2010){Vanderlinde}, {Crawford}, {de Haan},
  {Dudley}, {Shaw}, {Ade}, {Aird}, {Benson}, {Bleem}, {Brodwin}, {Carlstrom},
  {Chang}, {Crites}, {Desai}, {Dobbs}, {Foley}, {George}, {Gladders}, {Hall},
  {Halverson}, {High}, {Holder}, {Holzapfel}, {Hrubes}, {Joy}, {Keisler},
  {Knox}, {Lee}, {Leitch}, {Loehr}, {Lueker}, {Marrone}, {McMahon}, {Mehl},
  {Meyer}, {Mohr}, {Montroy}, {Ngeow}, {Padin}, {Plagge}, {Pryke}, {Reichardt},
  {Rest}, {Ruel}, {Ruhl}, {Schaffer}, {Shirokoff}, {Song}, {Spieler},
  {Stalder}, {Staniszewski}, {Stark}, {Stubbs}, {van Engelen}, {Vieira},
  {Williamson}, {Yang}, {Zahn}, \& {Zenteno}}]{Vanderlinde10}
{Vanderlinde}, K., {Crawford}, T.~M., {de Haan}, T., {et~al.} 2010, \apj, 722,
  1180

\bibitem[{{Vikhlinin} {et~al.}(2009){Vikhlinin}, {Burenin}, {Ebeling},
  {Forman}, {Hornstrup}, {Jones}, {Kravtsov}, {Murray}, {Nagai}, {Quintana}, \&
  {Voevodkin}}]{Vikhlinin09b}
{Vikhlinin}, A., {Burenin}, R.~A., {Ebeling}, H., {et~al.} 2009, \apj, 692,
  1033

\bibitem[{{Vikhlinin} {et~al.}(2006){Vikhlinin}, {Kravtsov}, {Forman}, {Jones},
  {Markevitch}, {Murray}, \& {Van Speybroeck}}]{Vikhlinin06}
{Vikhlinin}, A., {Kravtsov}, A., {Forman}, W., {et~al.} 2006, \apj, 640, 691

\bibitem[{{Voges} {et~al.}(1999){Voges}, {Aschenbach}, {Boller},
  {Br{\"a}uninger}, {Briel}, {Burkert}, {Dennerl}, {Englhauser}, {Gruber},
  {Haberl}, {Hartner}, {Hasinger}, {K{\"u}rster}, {Pfeffermann}, {Pietsch},
  {Predehl}, {Rosso}, {Schmitt}, {Tr{\"u}mper}, \& {Zimmermann}}]{Voges99}
{Voges}, W., {Aschenbach}, B., {Boller}, T., {et~al.} 1999, \aap, 349, 389

\bibitem[{{Watson} {et~al.}(2009){Watson}, {Schr{\"o}der}, {Fyfe}, {Page},
  {Lamer}, {Mateos}, {Pye}, {Sakano}, {Rosen}, {Ballet}, {Barcons}, {Barret},
  {Boller}, {Brunner}, {Brusa}, {Caccianiga}, {Carrera}, {Ceballos}, {Della
  Ceca}, {Denby}, {Denkinson}, {Dupuy}, {Farrell}, {Fraschetti}, {Freyberg},
  {Guillout}, {Hambaryan}, {Maccacaro}, {Mathiesen}, {McMahon}, {Michel},
  {Motch}, {Osborne}, {Page}, {Pakull}, {Pietsch}, {Saxton}, {Schwope},
  {Severgnini}, {Simpson}, {Sironi}, {Stewart}, {Stewart}, {Stobbart}, {Tedds},
  {Warwick}, {Webb}, {West}, {Worrall}, \& {Yuan}}]{Watson09}
{Watson}, M.~G., {Schr{\"o}der}, A.~C., {Fyfe}, D., {et~al.} 2009, \aap, 493,
  339

\bibitem[{{Wen} {et~al.}(2009){Wen}, {Han}, \& {Liu}}]{Wen09}
{Wen}, Z.~L., {Han}, J.~L., \& {Liu}, F.~S. 2009, \apjs, 183, 197

\bibitem[{{Wilms} {et~al.}(2000){Wilms}, {Allen}, \& {McCray}}]{Wilms00}
{Wilms}, J., {Allen}, A., \& {McCray}, R. 2000, \apj, 542, 914

\bibitem[{{Yu} {et~al.}(2011){Yu}, {Tozzi}, {Borgani}, {Rosati}, \&
  {Zhu}}]{Yu11}
{Yu}, H., {Tozzi}, P., {Borgani}, S., {Rosati}, P., \& {Zhu}, Z.-H. 2011, \aap,
  529, A65

\end{thebibliography}


\begin{thebibliography}{56}
\expandafter\ifx\csname natexlab\endcsname\relax\def\natexlab#1{#1}\fi

\bibitem[{{Adami} {et~al.}(2011){Adami}, {Mazure}, {Pierre}, {Sprimont},
  {Libbrecht}, {Pacaud}, {Clerc}, {Sadibekova}, {Surdej}, {Altieri}, {Duc},
  {Galaz}, {Gueguen}, {Guennou}, {Hertling}, {Ilbert}, {Le F{\`e}vre},
  {Quintana}, {Valtchanov}, {Willis}, {Akiyama}, {Aussel}, {Chiappetti},
  {Detal}, {Garilli}, {Lebrun}, {Lef{\`e}vre}, {Maccagni}, {Melin}, {Ponman},
  {Ricci}, \& {Tresse}}]{Adami11}
{Adami}, C., {Mazure}, A., {Pierre}, M., {et~al.} 2011, \aap, 526, A18

\bibitem[{{Akritas} \& {Bershady}(1996)}]{Akritas96}
{Akritas}, M.~G. \& {Bershady}, M.~A. 1996, \apj, 470, 706

\bibitem[{{Allen} {et~al.}(2011){Allen}, {Evrard}, \& {Mantz}}]{Allen11}
{Allen}, S.~W., {Evrard}, A.~E., \& {Mantz}, A.~B. 2011, \araa, 49, 409

\bibitem[{{Arnaud}(1996)}]{Arnaud96}
{Arnaud}, K.~A. 1996, in Astronomical Society of the Pacific Conference Series,
  Vol. 101, Astronomical Data Analysis Software and Systems V, ed.
  {G.~H.~Jacoby \& J.~Barnes}, 17

\bibitem[{{Arnaud} {et~al.}(2010){Arnaud}, {Pratt}, {Piffaretti},
  {B{\"o}hringer}, {Croston}, \& {Pointecouteau}}]{Arnaud10}
{Arnaud}, M., {Pratt}, G.~W., {Piffaretti}, R., {et~al.} 2010, \aap, 517, A92

\bibitem[{{Arviset} {et~al.}(2002){Arviset}, {Guainazzi}, {Hernandez},
  {Dowson}, {Osuna}, \& {Venet}}]{Arviset02}
{Arviset}, C., {Guainazzi}, M., {Hernandez}, J., {et~al.} 2002, ArXiv
  Astrophysics e-prints

\bibitem[{{Barkhouse} {et~al.}(2006){Barkhouse}, {Green}, {Vikhlinin}, {Kim},
  {Perley}, {Cameron}, {Silverman}, {Mossman}, {Burenin}, {Jannuzi}, {Kim},
  {Smith}, {Smith}, {Tananbaum}, \& {Wilkes}}]{Barkhouse06}
{Barkhouse}, W.~A., {Green}, P.~J., {Vikhlinin}, A., {et~al.} 2006, \apj, 645,
  955

\bibitem[{{Blanton} {et~al.}(2011){Blanton}, {Randall}, {Clarke}, {Sarazin},
  {McNamara}, {Douglass}, \& {McDonald}}]{Blanton11}
{Blanton}, E.~L., {Randall}, S.~W., {Clarke}, T.~E., {et~al.} 2011, \apj, 737,
  99

\bibitem[{{B{\"o}hringer} {et~al.}(2002){B{\"o}hringer}, {Collins}, {Guzzo},
  {Schuecker}, {Voges}, {Neumann}, {Schindler}, {Chincarini}, {De Grandi},
  {Cruddace}, {Edge}, {Reiprich}, \& {Shaver}}]{Boehringer02}
{B{\"o}hringer}, H., {Collins}, C.~A., {Guzzo}, L., {et~al.} 2002, \apj, 566,
  93

\bibitem[{{B{\"o}hringer} {et~al.}(2004){B{\"o}hringer}, {Schuecker}, {Guzzo},
  {Collins}, {Voges}, {Cruddace}, {Ortiz-Gil}, {Chincarini}, {De Grandi},
  {Edge}, {MacGillivray}, {Neumann}, {Schindler}, \& {Shaver}}]{Boehringer04}
{B{\"o}hringer}, H., {Schuecker}, P., {Guzzo}, L., {et~al.} 2004, \aap, 425,
  367

\bibitem[{{B{\"o}hringer} {et~al.}(2000){B{\"o}hringer}, {Voges}, {Huchra},
  {McLean}, {Giacconi}, {Rosati}, {Burg}, {Mader}, {Schuecker}, {Simi{\c c}},
  {Komossa}, {Reiprich}, {Retzlaff}, \& {Tr{\"u}mper}}]{Boehringer00}
{B{\"o}hringer}, H., {Voges}, W., {Huchra}, J.~P., {et~al.} 2000, \apjs, 129,
  435

\bibitem[{{Borgani} {et~al.}(2001){Borgani}, {Rosati}, {Tozzi}, {Stanford},
  {Eisenhardt}, {Lidman}, {Holden}, {Della Ceca}, {Norman}, \&
  {Squires}}]{Borgani01}
{Borgani}, S., {Rosati}, P., {Tozzi}, P., {et~al.} 2001, \apj, 561, 13

\bibitem[{{Burenin} {et~al.}(2007){Burenin}, {Vikhlinin}, {Hornstrup},
  {Ebeling}, {Quintana}, \& {Mescheryakov}}]{Burenin07}
{Burenin}, R.~A., {Vikhlinin}, A., {Hornstrup}, A., {et~al.} 2007, \apjs, 172,
  561

\bibitem[{{Clerc} {et~al.}(2012){Clerc}, {Sadibekova}, {Pierre}, {Pacaud}, {Le
  F{\`e}vre}, {Adami}, {Altieri}, \& {Valtchanov}}]{Clerc12}
{Clerc}, N., {Sadibekova}, T., {Pierre}, M., {et~al.} 2012, \mnras, 3120

\bibitem[{{Eckmiller} {et~al.}(2011){Eckmiller}, {Hudson}, \&
  {Reiprich}}]{Eckmiller11}
{Eckmiller}, H.~J., {Hudson}, D.~S., \& {Reiprich}, T.~H. 2011, \aap, 535, A105

\bibitem[{{Fassbender} {et~al.}(2011){Fassbender}, {B{\"o}hringer}, {Nastasi},
  {{\v S}uhada}, {M{\"u}hlegger}, {de Hoon}, {Kohnert}, {Lamer}, {Mohr},
  {Pierini}, {Pratt}, {Quintana}, {Rosati}, {Santos}, \&
  {Schwope}}]{Fassbender11}
{Fassbender}, R., {B{\"o}hringer}, H., {Nastasi}, A., {et~al.} 2011, New
  Journal of Physics, 13, 125014

\bibitem[{{Finoguenov} {et~al.}(2007){Finoguenov}, {Guzzo}, {Hasinger},
  {Scoville}, {Aussel}, {B{\"o}hringer}, {Brusa}, {Capak}, {Cappelluti},
  {Comastri}, {Giodini}, {Griffiths}, {Impey}, {Koekemoer}, {Kneib},
  {Leauthaud}, {Le F{\`e}vre}, {Lilly}, {Mainieri}, {Massey}, {McCracken},
  {Mobasher}, {Murayama}, {Peacock}, {Sakelliou}, {Schinnerer}, {Silverman},
  {Smol{\v c}i{\'c}}, {Taniguchi}, {Tasca}, {Taylor}, {Trump}, \&
  {Zamorani}}]{Finoguenov07}
{Finoguenov}, A., {Guzzo}, L., {Hasinger}, G., {et~al.} 2007, \apjs, 172, 182

\bibitem[{{Finoguenov} {et~al.}(2010){Finoguenov}, {Watson}, {Tanaka},
  {Simpson}, {Cirasuolo}, {Dunlop}, {Peacock}, {Farrah}, {Akiyama}, {Ueda},
  {Smol{\v c}i{\'c}}, {Stewart}, {Rawlings}, {van Breukelen}, {Almaini},
  {Clewley}, {Bonfield}, {Jarvis}, {Barr}, {Foucaud}, {McLure}, {Sekiguchi}, \&
  {Egami}}]{Finoguenov10}
{Finoguenov}, A., {Watson}, M.~G., {Tanaka}, M., {et~al.} 2010, \mnras, 403,
  2063

\bibitem[{{Forman} {et~al.}(1978){Forman}, {Jones}, {Cominsky}, {Julien},
  {Murray}, {Peters}, {Tananbaum}, \& {Giacconi}}]{Forman78}
{Forman}, W., {Jones}, C., {Cominsky}, L., {et~al.} 1978, \apjs, 38, 357

\bibitem[{{Hao} {et~al.}(2010){Hao}, {McKay}, {Koester}, {Rykoff}, {Rozo},
  {Annis}, {Wechsler}, {Evrard}, {Siegel}, {Becker}, {Busha}, {Gerdes},
  {Johnston}, \& {Sheldon}}]{Hao10}
{Hao}, J., {McKay}, T.~A., {Koester}, B.~P., {et~al.} 2010, \apjs, 191, 254

\bibitem[{{Hilton} {et~al.}(2012){Hilton}, {Romer}, {Kay}, {Mehrtens},
  {Lloyd-Davies}, {Thomas}, {Short}, {Mayers}, {Rooney}, {Stott}, {Collins},
  {Harrison}, {Hoyle}, {Liddle}, {Mann}, {Miller}, {Sahl{\'e}n}, {Viana},
  {Davidson}, {Hosmer}, {Nichol}, {Sabirli}, {Stanford}, \& {West}}]{Hilton12}
{Hilton}, M., {Romer}, A.~K., {Kay}, S.~T., {et~al.} 2012, \mnras, 3303

\bibitem[{{Kalberla} {et~al.}(2005){Kalberla}, {Burton}, {Hartmann}, {Arnal},
  {Bajaja}, {Morras}, \& {P{\"o}ppel}}]{Kalberla05}
{Kalberla}, P.~M.~W., {Burton}, W.~B., {Hartmann}, D., {et~al.} 2005, \aap,
  440, 775

\bibitem[{{Koester} {et~al.}(2007){Koester}, {McKay}, {Annis}, {Wechsler},
  {Evrard}, {Bleem}, {Becker}, {Johnston}, {Sheldon}, {Nichol}, {Miller},
  {Scranton}, {Bahcall}, {Barentine}, {Brewington}, {Brinkmann}, {Harvanek},
  {Kleinman}, {Krzesinski}, {Long}, {Nitta}, {Schneider}, {Sneddin}, {Voges},
  \& {York}}]{Koester07}
{Koester}, B.~P., {McKay}, T.~A., {Annis}, J., {et~al.} 2007, \apj, 660, 239

\bibitem[{{Kolokotronis} {et~al.}(2006){Kolokotronis}, {Georgakakis},
  {Basilakos}, {Kitsionas}, {Plionis}, {Georgantopoulos}, \&
  {Gaga}}]{Kolokotronis06}
{Kolokotronis}, V., {Georgakakis}, A., {Basilakos}, S., {et~al.} 2006, \mnras,
  366, 163

\bibitem[{{Krumpe} {et~al.}(2008){Krumpe}, {Lamer}, {Corral}, {Schwope},
  {Carrera}, {Barcons}, {Page}, {Mateos}, {Tedds}, \& {Watson}}]{Krumpe08}
{Krumpe}, M., {Lamer}, G., {Corral}, A., {et~al.} 2008, \aap, 483, 415

\bibitem[{{Lloyd-Davies} {et~al.}(2011){Lloyd-Davies}, {Romer}, {Mehrtens},
  {Hosmer}, {Davidson}, {Sabirli}, {Mann}, {Hilton}, {Liddle}, {Viana},
  {Campbell}, {Collins}, {Dubois}, {Freeman}, {Harrison}, {Hoyle}, {Kay},
  {Kuwertz}, {Miller}, {Nichol}, {Sahl{\'e}n}, {Stanford}, \&
  {Stott}}]{Lloyd-Davies11}
{Lloyd-Davies}, E.~J., {Romer}, A.~K., {Mehrtens}, N., {et~al.} 2011, \mnras,
  418, 14

\bibitem[{{Markevitch}(1998)}]{Markevitch98}
{Markevitch}, M. 1998, \apj, 504, 27

\bibitem[{{Maughan} {et~al.}(2012){Maughan}, {Giles}, {Randall}, {Jones}, \&
  {Forman}}]{Maughan12}
{Maughan}, B.~J., {Giles}, P.~A., {Randall}, S.~W., {Jones}, C., \& {Forman},
  W.~R. 2012, \mnras, 421, 1583

\bibitem[{{Mehrtens} {et~al.}(2012){Mehrtens}, {Romer}, {Hilton},
  {Lloyd-Davies}, {Miller}, {Stanford}, {Hosmer}, {Hoyle}, {Collins}, {Liddle},
  {Viana}, {Nichol}, {Stott}, {Dubois}, {Kay}, {Sahl{\'e}n}, {Young}, {Short},
  {Christodoulou}, {Watson}, {Davidson}, {Harrison}, {Baruah}, {Smith},
  {Burke}, {Mayers}, {Deadman}, {Rooney}, {Edmondson}, {West}, {Campbell},
  {Edge}, {Mann}, {Sabirli}, {Wake}, {Benoist}, {da Costa}, {Maia}, \&
  {Ogando}}]{Mehrtens12}
{Mehrtens}, N., {Romer}, A.~K., {Hilton}, M., {et~al.} 2012, \mnras, 2912

\bibitem[{{Mewe} {et~al.}(1986){Mewe}, {Lemen}, \& {van den Oord}}]{Mewe86}
{Mewe}, R., {Lemen}, J.~R., \& {van den Oord}, G.~H.~J. 1986, \aaps, 65, 511

\bibitem[{{Mittal} {et~al.}(2011){Mittal}, {Hicks}, {Reiprich}, \&
  {Jaritz}}]{Mittal11}
{Mittal}, R., {Hicks}, A., {Reiprich}, T.~H., \& {Jaritz}, V. 2011, \aap, 532,
  A133

\bibitem[{{M{\"u}hlegger}(2010)}]{Muehlegger10}
{M{\"u}hlegger}, M. 2010, PhD thesis, Technischen Universit{\"a}t M{\"u}nchen

\bibitem[{{Osmond} \& {Ponman}(2004)}]{Osmond04}
{Osmond}, J.~P.~F. \& {Ponman}, T.~J. 2004, \mnras, 350, 1511

\bibitem[{{Piffaretti} {et~al.}(2011){Piffaretti}, {Arnaud}, {Pratt},
  {Pointecouteau}, \& {Melin}}]{Piffaretti11}
{Piffaretti}, R., {Arnaud}, M., {Pratt}, G.~W., {Pointecouteau}, E., \&
  {Melin}, J.-B. 2011, \aap, 534, A109

\bibitem[{{Pratt} {et~al.}(2010){Pratt}, {Arnaud}, {Piffaretti},
  {B{\"o}hringer}, {Ponman}, {Croston}, {Voit}, {Borgani}, \&
  {Bower}}]{Pratt10}
{Pratt}, G.~W., {Arnaud}, M., {Piffaretti}, R., {et~al.} 2010, \aap, 511, A85

\bibitem[{{Pratt} {et~al.}(2009){Pratt}, {Croston}, {Arnaud}, \&
  {B{\"o}hringer}}]{Pratt09}
{Pratt}, G.~W., {Croston}, J.~H., {Arnaud}, M., \& {B{\"o}hringer}, H. 2009,
  \aap, 498, 361

\bibitem[{{Reichert} {et~al.}(2011){Reichert}, {B{\"o}hringer}, {Fassbender},
  \& {M{\"u}hlegger}}]{Reichert11}
{Reichert}, A., {B{\"o}hringer}, H., {Fassbender}, R., \& {M{\"u}hlegger}, M.
  2011, \aap, 535, A4

\bibitem[{{Reiprich} \& {B{\"o}hringer}(2002)}]{Reiprich02}
{Reiprich}, T.~H. \& {B{\"o}hringer}, H. 2002, \apj, 567, 716

\bibitem[{{Romer} {et~al.}(1994){Romer}, {Collins}, {B{\"o}hringer},
  {Cruddace}, {Ebeling}, {MacGillivray}, \& {Voges}}]{Romer94}
{Romer}, A.~K., {Collins}, C.~A., {B{\"o}hringer}, H., {et~al.} 1994, \nat,
  372, 75

\bibitem[{{Romer} {et~al.}(2001){Romer}, {Viana}, {Liddle}, \&
  {Mann}}]{Romer01}
{Romer}, A.~K., {Viana}, P.~T.~P., {Liddle}, A.~R., \& {Mann}, R.~G. 2001,
  \apj, 547, 594

\bibitem[{{Rosati} {et~al.}(2002){Rosati}, {Borgani}, \& {Norman}}]{Rosati02}
{Rosati}, P., {Borgani}, S., \& {Norman}, C. 2002, \araa, 40, 539

\bibitem[{{Rykoff} {et~al.}(2008){Rykoff}, {McKay}, {Becker}, {Evrard},
  {Johnston}, {Koester}, {Rozo}, {Sheldon}, \& {Wechsler}}]{Rykoff08}
{Rykoff}, E.~S., {McKay}, T.~A., {Becker}, M.~R., {et~al.} 2008, \apj, 675,
  1106

\bibitem[{{Sarazin}(1988)}]{Sarazin88}
{Sarazin}, C.~L. 1988, {X-ray emission from clusters of galaxies} (Cambridge,
  UK: Cambridge Univ. Press)

\bibitem[{{Scharf} {et~al.}(1997){Scharf}, {Jones}, {Ebeling}, {Perlman},
  {Malkan}, \& {Wegner}}]{Scharf97}
{Scharf}, C.~A., {Jones}, L.~R., {Ebeling}, H., {et~al.} 1997, \apj, 477, 79

\bibitem[{{Schuecker} {et~al.}(2003){Schuecker}, {B{\"o}hringer}, {Collins}, \&
  {Guzzo}}]{Schuecker03}
{Schuecker}, P., {B{\"o}hringer}, H., {Collins}, C.~A., \& {Guzzo}, L. 2003,
  \aap, 398, 867

\bibitem[{{Sunyaev} \& {Zeldovich}(1980)}]{Sunyaev80}
{Sunyaev}, R.~A. \& {Zeldovich}, I.~B. 1980, \araa, 18, 537

\bibitem[{{Sunyaev} \& {Zeldovich}(1972)}]{Sunyaev72}
{Sunyaev}, R.~A. \& {Zeldovich}, Y.~B. 1972, Comments on Astrophysics and Space
  Physics, 4, 173

\bibitem[{{Szabo} {et~al.}(2011){Szabo}, {Pierpaoli}, {Dong}, {Pipino}, \&
  {Gunn}}]{Szabo11}
{Szabo}, T., {Pierpaoli}, E., {Dong}, F., {Pipino}, A., \& {Gunn}, J. 2011,
  \apj, 736, 21

\bibitem[{{Takey} {et~al.}(2011){Takey}, {Schwope}, \& {Lamer}}]{Takey11}
{Takey}, A., {Schwope}, A., \& {Lamer}, G. 2011, \aap, 534, A120

\bibitem[{{Tundo} {et~al.}(2012){Tundo}, {Moretti}, {Tozzi}, {Teng}, {Rosati},
  {Tagliaferri}, \& {Campana}}]{Tundo12}
{Tundo}, E., {Moretti}, A., {Tozzi}, P., {et~al.} 2012, \aap, 547, A57

\bibitem[{{Vikhlinin} {et~al.}(2009){Vikhlinin}, {Burenin}, {Ebeling},
  {Forman}, {Hornstrup}, {Jones}, {Kravtsov}, {Murray}, {Nagai}, {Quintana}, \&
  {Voevodkin}}]{Vikhlinin09b}
{Vikhlinin}, A., {Burenin}, R.~A., {Ebeling}, H., {et~al.} 2009, \apj, 692,
  1033

\bibitem[{{Vikhlinin} {et~al.}(1998){Vikhlinin}, {McNamara}, {Forman}, {Jones},
  {Quintana}, \& {Hornstrup}}]{Vikhlinin98}
{Vikhlinin}, A., {McNamara}, B.~R., {Forman}, W., {et~al.} 1998, \apj, 502, 558

\bibitem[{{Voit}(2005)}]{Voit05}
{Voit}, G.~M. 2005, Reviews of Modern Physics, 77, 207

\bibitem[{{Watson} {et~al.}(2009){Watson}, {Schr{\"o}der}, {Fyfe}, {Page},
  {Lamer}, {Mateos}, {Pye}, {Sakano}, {Rosen}, {Ballet}, {Barcons}, {Barret},
  {Boller}, {Brunner}, {Brusa}, {Caccianiga}, {Carrera}, {Ceballos}, {Della
  Ceca}, {Denby}, {Denkinson}, {Dupuy}, {Farrell}, {Fraschetti}, {Freyberg},
  {Guillout}, {Hambaryan}, {Maccacaro}, {Mathiesen}, {McMahon}, {Michel},
  {Motch}, {Osborne}, {Page}, {Pakull}, {Pietsch}, {Saxton}, {Schwope},
  {Severgnini}, {Simpson}, {Sironi}, {Stewart}, {Stewart}, {Stobbart}, {Tedds},
  {Warwick}, {Webb}, {West}, {Worrall}, \& {Yuan}}]{Watson09}
{Watson}, M.~G., {Schr{\"o}der}, A.~C., {Fyfe}, D., {et~al.} 2009, \aap, 493,
  339

\bibitem[{{Wen} {et~al.}(2009){Wen}, {Han}, \& {Liu}}]{Wen09}
{Wen}, Z.~L., {Han}, J.~L., \& {Liu}, F.~S. 2009, \apjs, 183, 197

\bibitem[{{Wilms} {et~al.}(2000){Wilms}, {Allen}, \& {McCray}}]{Wilms00}
{Wilms}, J., {Allen}, A., \& {McCray}, R. 2000, \apj, 542, 914

\end{thebibliography}


\begin{thebibliography}{201}
\expandafter\ifx\csname natexlab\endcsname\relax\def\natexlab#1{#1}\fi

\bibitem[{{Abell}(1958)}]{Abell58}
{Abell}, G.~O. 1958, \apjs, 3, 211

\bibitem[{{Abell} {et~al.}(1989){Abell}, {Corwin}, \& {Olowin}}]{Abell89}
{Abell}, G.~O., {Corwin}, Jr., H.~G., \& {Olowin}, R.~P. 1989, \apjs, 70, 1

\bibitem[{{Adami} {et~al.}(2011){Adami}, {Mazure}, {Pierre}, {Sprimont},
  {Libbrecht}, {Pacaud}, {Clerc}, {Sadibekova}, {Surdej}, {Altieri}, {Duc},
  {Galaz}, {Gueguen}, {Guennou}, {Hertling}, {Ilbert}, {Le F{\`e}vre},
  {Quintana}, {Valtchanov}, {Willis}, {Akiyama}, {Aussel}, {Chiappetti},
  {Detal}, {Garilli}, {Lebrun}, {Lef{\`e}vre}, {Maccagni}, {Melin}, {Ponman},
  {Ricci}, \& {Tresse}}]{Adami11}
{Adami}, C., {Mazure}, A., {Pierre}, M., {et~al.} 2011, \aap, 526, A18

\bibitem[{{Ahn} {et~al.}(2012){Ahn}, {Alexandroff}, {Allende Prieto},
  {Anderson}, {Anderton}, {Andrews}, {Aubourg}, {Bailey}, {Balbinot}, {Barnes},
  \& et~al.}]{Ahn12}
{Ahn}, C.~P., {Alexandroff}, R., {Allende Prieto}, C., {et~al.} 2012, \apjs,
  203, 21

\bibitem[{{Akritas} \& {Bershady}(1996)}]{Akritas96}
{Akritas}, M.~G. \& {Bershady}, M.~A. 1996, \apj, 470, 706

\bibitem[{{Allen} {et~al.}(2011){Allen}, {Evrard}, \& {Mantz}}]{Allen11}
{Allen}, S.~W., {Evrard}, A.~E., \& {Mantz}, A.~B. 2011, \araa, 49, 409

\bibitem[{{Anderson} {et~al.}(2009){Anderson}, {Bregman}, {Butler}, \&
  {Mullis}}]{Anderson09}
{Anderson}, M.~E., {Bregman}, J.~N., {Butler}, S.~C., \& {Mullis}, C.~R. 2009,
  \apj, 698, 317

\bibitem[{{Arnaud}(1996)}]{Arnaud96}
{Arnaud}, K.~A. 1996, in Astronomical Society of the Pacific Conference Series,
  Vol. 101, Astronomical Data Analysis Software and Systems V, ed.
  {G.~H.~Jacoby \& J.~Barnes}, 17

\bibitem[{{Arnaud} {et~al.}(2010){Arnaud}, {Pratt}, {Piffaretti},
  {B{\"o}hringer}, {Croston}, \& {Pointecouteau}}]{Arnaud10}
{Arnaud}, M., {Pratt}, G.~W., {Piffaretti}, R., {et~al.} 2010, \aap, 517, A92

\bibitem[{{Arviset} {et~al.}(2002){Arviset}, {Guainazzi}, {Hernandez},
  {Dowson}, {Osuna}, \& {Venet}}]{Arviset02}
{Arviset}, C., {Guainazzi}, M., {Hernandez}, J., {et~al.} 2002, ArXiv
  Astrophysics e-prints

\bibitem[{{Bahcall}(1977)}]{Bahcall77}
{Bahcall}, N.~A. 1977, \araa, 15, 505

\bibitem[{{Bahcall}(1988)}]{Bahcall88}
{Bahcall}, N.~A. 1988, \araa, 26, 631

\bibitem[{{Bahcall} {et~al.}(2003){Bahcall}, {McKay}, {Annis}, {Kim}, {Dong},
  {Hansen}, {Goto}, {Gunn}, {Miller}, {Nichol}, {Postman}, {Schneider},
  {Schroeder}, {Voges}, {Brinkmann}, \& {Fukugita}}]{Bahcall03}
{Bahcall}, N.~A., {McKay}, T.~A., {Annis}, J., {et~al.} 2003, \apjs, 148, 243

\bibitem[{{Balestra} {et~al.}(2007){Balestra}, {Tozzi}, {Ettori}, {Rosati},
  {Borgani}, {Mainieri}, {Norman}, \& {Viola}}]{Balestra07}
{Balestra}, I., {Tozzi}, P., {Ettori}, S., {et~al.} 2007, \aap, 462, 429

\bibitem[{{Barbosa} {et~al.}(1996){Barbosa}, {Bartlett}, {Blanchard}, \&
  {Oukbir}}]{Barbosa96}
{Barbosa}, D., {Bartlett}, J.~G., {Blanchard}, A., \& {Oukbir}, J. 1996, \aap,
  314, 13

\bibitem[{{Barkhouse} {et~al.}(2006){Barkhouse}, {Green}, {Vikhlinin}, {Kim},
  {Perley}, {Cameron}, {Silverman}, {Mossman}, {Burenin}, {Jannuzi}, {Kim},
  {Smith}, {Smith}, {Tananbaum}, \& {Wilkes}}]{Barkhouse06}
{Barkhouse}, W.~A., {Green}, P.~J., {Vikhlinin}, A., {et~al.} 2006, \apj, 645,
  955

\bibitem[{{Bartelmann}(2010)}]{Bartelmann10}
{Bartelmann}, M. 2010, Classical and Quantum Gravity, 27, 233001

\bibitem[{{Basilakos} {et~al.}(2004){Basilakos}, {Plionis}, {Georgakakis},
  {Georgantopoulos}, {Gaga}, {Kolokotronis}, \& {Stewart}}]{Basilakos04}
{Basilakos}, S., {Plionis}, M., {Georgakakis}, A., {et~al.} 2004, \mnras, 351,
  989

\bibitem[{{Berlind} {et~al.}(2006){Berlind}, {Frieman}, {Weinberg}, {Blanton},
  {Warren}, {Abazajian}, {Scranton}, {Hogg}, {Scoccimarro}, {Bahcall},
  {Brinkmann}, {Gott}, {Kleinman}, {Krzesinski}, {Lee}, {Miller}, {Nitta},
  {Schneider}, {Tucker}, {Zehavi}, \& {SDSS Collaboration}}]{Berlind06}
{Berlind}, A.~A., {Frieman}, J., {Weinberg}, D.~H., {et~al.} 2006, \apjs, 167,
  1

\bibitem[{{Biviano}(2000)}]{Biviano00}
{Biviano}, A. 2000, in Constructing the Universe with Clusters of Galaxies

\bibitem[{{Blanton} {et~al.}(2011){Blanton}, {Randall}, {Clarke}, {Sarazin},
  {McNamara}, {Douglass}, \& {McDonald}}]{Blanton11}
{Blanton}, E.~L., {Randall}, S.~W., {Clarke}, T.~E., {et~al.} 2011, \apj, 737,
  99

\bibitem[{{Blanton} {et~al.}(2003){Blanton}, {Hogg}, {Bahcall}, {Brinkmann},
  {Britton}, {Connolly}, {Csabai}, {Fukugita}, {Loveday}, {Meiksin}, {Munn},
  {Nichol}, {Okamura}, {Quinn}, {Schneider}, {Shimasaku}, {Strauss}, {Tegmark},
  {Vogeley}, \& {Weinberg}}]{Blanton03}
{Blanton}, M.~R., {Hogg}, D.~W., {Bahcall}, N.~A., {et~al.} 2003, \apj, 592,
  819

\bibitem[{{Boggs} \& {Rogers}(1990)}]{Boggs90}
{Boggs}, P.~T. \& {Rogers}, J.~E. 1990, Contemporary Mathematics, 112, 183

\bibitem[{{B{\"o}hringer}(2006)}]{Boehringer06}
{B{\"o}hringer}, H. 2006, {X-ray Studies of Clusters of Galaxies, The Universe
  in X-rays}

\bibitem[{{B{\"o}hringer} {et~al.}(2002){B{\"o}hringer}, {Collins}, {Guzzo},
  {Schuecker}, {Voges}, {Neumann}, {Schindler}, {Chincarini}, {De Grandi},
  {Cruddace}, {Edge}, {Reiprich}, \& {Shaver}}]{Boehringer02}
{B{\"o}hringer}, H., {Collins}, C.~A., {Guzzo}, L., {et~al.} 2002, \apj, 566,
  93

\bibitem[{{B{\"o}hringer} {et~al.}(2004){B{\"o}hringer}, {Schuecker}, {Guzzo},
  {Collins}, {Voges}, {Cruddace}, {Ortiz-Gil}, {Chincarini}, {De Grandi},
  {Edge}, {MacGillivray}, {Neumann}, {Schindler}, \& {Shaver}}]{Boehringer04}
{B{\"o}hringer}, H., {Schuecker}, P., {Guzzo}, L., {et~al.} 2004, \aap, 425,
  367

\bibitem[{{B{\"o}hringer} {et~al.}(2000){B{\"o}hringer}, {Voges}, {Huchra},
  {McLean}, {Giacconi}, {Rosati}, {Burg}, {Mader}, {Schuecker}, {Simi{\c c}},
  {Komossa}, {Reiprich}, {Retzlaff}, \& {Tr{\"u}mper}}]{Boehringer00}
{B{\"o}hringer}, H., {Voges}, W., {Huchra}, J.~P., {et~al.} 2000, \apjs, 129,
  435

\bibitem[{{B{\"o}hringer} \& {Werner}(2010)}]{Boehringer10}
{B{\"o}hringer}, H. \& {Werner}, N. 2010, \aapr, 18, 127

\bibitem[{{Bolzonella} {et~al.}(2000){Bolzonella}, {Miralles}, \&
  {Pell{\'o}}}]{Bolzonella00}
{Bolzonella}, M., {Miralles}, J.-M., \& {Pell{\'o}}, R. 2000, \aap, 363, 476

\bibitem[{{Borgani} {et~al.}(2001){Borgani}, {Rosati}, {Tozzi}, {Stanford},
  {Eisenhardt}, {Lidman}, {Holden}, {Della Ceca}, {Norman}, \&
  {Squires}}]{Borgani01}
{Borgani}, S., {Rosati}, P., {Tozzi}, P., {et~al.} 2001, \apj, 561, 13

\bibitem[{{Boschin}(2002)}]{Boschin02}
{Boschin}, W. 2002, \aap, 396, 397

\bibitem[{{Bower} {et~al.}(1992){Bower}, {Lucey}, \& {Ellis}}]{Bower92}
{Bower}, R.~G., {Lucey}, J.~R., \& {Ellis}, R.~S. 1992, \mnras, 254, 601

\bibitem[{{Brough} {et~al.}(2008){Brough}, {Couch}, {Collins}, {Jarrett},
  {Burke}, \& {Mann}}]{Brough08}
{Brough}, S., {Couch}, W.~J., {Collins}, C.~A., {et~al.} 2008, \mnras, 385,
  L103

\bibitem[{{Burenin} {et~al.}(2007){Burenin}, {Vikhlinin}, {Hornstrup},
  {Ebeling}, {Quintana}, \& {Mescheryakov}}]{Burenin07}
{Burenin}, R.~A., {Vikhlinin}, A., {Hornstrup}, A., {et~al.} 2007, \apjs, 172,
  561

\bibitem[{{Carlberg} {et~al.}(2001){Carlberg}, {Yee}, {Morris}, {Lin}, {Hall},
  {Patton}, {Sawicki}, \& {Shepherd}}]{Carlberg01}
{Carlberg}, R.~G., {Yee}, H.~K.~C., {Morris}, S.~L., {et~al.} 2001, \apj, 552,
  427

\bibitem[{{Carlstrom} {et~al.}(2002){Carlstrom}, {Holder}, \&
  {Reese}}]{Carlstrom02}
{Carlstrom}, J.~E., {Holder}, G.~P., \& {Reese}, E.~D. 2002, \araa, 40, 643

\bibitem[{{Cavagnolo} {et~al.}(2008){Cavagnolo}, {Donahue}, {Voit}, \&
  {Sun}}]{Cavagnolo08}
{Cavagnolo}, K.~W., {Donahue}, M., {Voit}, G.~M., \& {Sun}, M. 2008, \apj, 682,
  821

\bibitem[{{Cavaliere} \& {Fusco-Femiano}(1976)}]{Cavaliere76}
{Cavaliere}, A. \& {Fusco-Femiano}, R. 1976, \aap, 49, 137

\bibitem[{{Cavaliere} {et~al.}(1971){Cavaliere}, {Gursky}, \&
  {Tucker}}]{Cavaliere71}
{Cavaliere}, A.~G., {Gursky}, H., \& {Tucker}, W.~H. 1971, \nat, 231, 437

\bibitem[{{Clerc} {et~al.}(2012){Clerc}, {Sadibekova}, {Pierre}, {Pacaud}, {Le
  F{\`e}vre}, {Adami}, {Altieri}, \& {Valtchanov}}]{Clerc12}
{Clerc}, N., {Sadibekova}, T., {Pierre}, M., {et~al.} 2012, \mnras, 3120

\bibitem[{{Cora}(2006)}]{Cora06}
{Cora}, S.~A. 2006, \mnras, 368, 1540

\bibitem[{{Csabai} {et~al.}(2003){Csabai}, {Budav{\'a}ri}, {Connolly},
  {Szalay}, {Gy{\H o}ry}, {Ben{\'{\i}}tez}, {Annis}, {Brinkmann}, {Eisenstein},
  {Fukugita}, {Gunn}, {Kent}, {Lupton}, {Nichol}, \& {Stoughton}}]{Csabai03}
{Csabai}, I., {Budav{\'a}ri}, T., {Connolly}, A.~J., {et~al.} 2003, \aj, 125,
  580

\bibitem[{{De Grandi} \& {Molendi}(2002)}]{DeGrandi02}
{De Grandi}, S. \& {Molendi}, S. 2002, \apj, 567, 163

\bibitem[{{de Hoon} {et~al.}(2013){de Hoon}, {Lamer}, {Schwope},
  {M{\"u}hlegger}, {Fassbender}, {B{\"o}hringer}, {Lerchster}, {Nastasi}, {{\v
  S}uhada}, {Verdugo}, {Dietrich}, {Brimioulle}, {Rosati}, {Pierini}, {Santos},
  {Quintana}, {Rabitz}, \& {Takey}}]{de-Hoon13}
{de Hoon}, A., {Lamer}, G., {Schwope}, A., {et~al.} 2013, \aap, 551, A8

\bibitem[{{De Lucia} \& {Blaizot}(2007)}]{DeLucia07}
{De Lucia}, G. \& {Blaizot}, J. 2007, \mnras, 375, 2

\bibitem[{{dell'Antonio} {et~al.}(1994){dell'Antonio}, {Geller}, \&
  {Fabricant}}]{dellAntonio94}
{dell'Antonio}, I.~P., {Geller}, M.~J., \& {Fabricant}, D.~G. 1994, \aj, 107,
  427

\bibitem[{{Dietrich} {et~al.}(2007){Dietrich}, {Erben}, {Lamer}, {Schneider},
  {Schwope}, {Hartlap}, \& {Maturi}}]{Dietrich07}
{Dietrich}, J.~P., {Erben}, T., {Lamer}, G., {et~al.} 2007, \aap, 470, 821

\bibitem[{{Dressler}(1980)}]{Dressler80}
{Dressler}, A. 1980, \apj, 236, 351

\bibitem[{{Durret} {et~al.}(2011){Durret}, {Adami}, {Cappi}, {Maurogordato},
  {M{\'a}rquez}, {Ilbert}, {Coupon}, {Arnouts}, {Benoist}, {Blaizot}, {Edorh},
  {Garilli}, {Guennou}, {Le Brun}, {Le F{\`e}vre}, {Mazure}, {McCracken},
  {Mellier}, {Mezrag}, {Slezak}, {Tresse}, \& {Ulmer}}]{Durret11}
{Durret}, F., {Adami}, C., {Cappi}, A., {et~al.} 2011, \aap, 535, A65

\bibitem[{{Ebeling} {et~al.}(1998){Ebeling}, {Edge}, {B{\"o}hringer}, {Allen},
  {Crawford}, {Fabian}, {Voges}, \& {Huchra}}]{Ebeling98}
{Ebeling}, H., {Edge}, A.~C., {B{\"o}hringer}, H., {et~al.} 1998, \mnras, 301,
  881

\bibitem[{{Ebeling} {et~al.}(2010){Ebeling}, {Edge}, {Mantz}, {Barrett},
  {Henry}, {Ma}, \& {van Speybroeck}}]{Ebeling10}
{Ebeling}, H., {Edge}, A.~C., {Mantz}, A., {et~al.} 2010, \mnras, 407, 83

\bibitem[{{Eckmiller} {et~al.}(2011){Eckmiller}, {Hudson}, \&
  {Reiprich}}]{Eckmiller11}
{Eckmiller}, H.~J., {Hudson}, D.~S., \& {Reiprich}, T.~H. 2011, \aap, 535, A105

\bibitem[{{Eisenstein} {et~al.}(2001){Eisenstein}, {Annis}, {Gunn}, {Szalay},
  {Connolly}, {Nichol}, {Bahcall}, {Bernardi}, {Burles}, {Castander},
  {Fukugita}, {Hogg}, {Ivezi{\'c}}, {Knapp}, {Lupton}, {Narayanan}, {Postman},
  {Reichart}, {Richmond}, {Schneider}, {Schlegel}, {Strauss}, {SubbaRao},
  {Tucker}, {Vanden Berk}, {Vogeley}, {Weinberg}, \& {Yanny}}]{Eisenstein01}
{Eisenstein}, D.~J., {Annis}, J., {Gunn}, J.~E., {et~al.} 2001, \aj, 122, 2267

\bibitem[{{Evans} {et~al.}(2010){Evans}, {Primini}, {Glotfelty}, {Anderson},
  {Bonaventura}, {Chen}, {Davis}, {Doe}, {Evans}, {Fabbiano}, {Galle}, {Gibbs},
  {Grier}, {Hain}, {Hall}, {Harbo}, {(Helen He}, {Houck}, {Karovska},
  {Kashyap}, {Lauer}, {McCollough}, {McDowell}, {Miller}, {Mitschang},
  {Morgan}, {Mossman}, {Nichols}, {Nowak}, {Plummer}, {Refsdal}, {Rots},
  {Siemiginowska}, {Sundheim}, {Tibbetts}, {Van Stone}, {Winkelman}, \&
  {Zografou}}]{Evans10}
{Evans}, I.~N., {Primini}, F.~A., {Glotfelty}, K.~J., {et~al.} 2010, \apjs,
  189, 37

\bibitem[{{Falco} {et~al.}(1999){Falco}, {Kurtz}, {Geller}, {Huchra}, {Peters},
  {Berlind}, {Mink}, {Tokarz}, \& {Elwell}}]{Falco99}
{Falco}, E.~E., {Kurtz}, M.~J., {Geller}, M.~J., {et~al.} 1999, \pasp, 111, 438

\bibitem[{{Fassbender} {et~al.}(2011){Fassbender}, {B{\"o}hringer}, {Nastasi},
  {{\v S}uhada}, {M{\"u}hlegger}, {de Hoon}, {Kohnert}, {Lamer}, {Mohr},
  {Pierini}, {Pratt}, {Quintana}, {Rosati}, {Santos}, \&
  {Schwope}}]{Fassbender11}
{Fassbender}, R., {B{\"o}hringer}, H., {Nastasi}, A., {et~al.} 2011, New
  Journal of Physics, 13, 125014

\bibitem[{{Fassbender} {et~al.}(2007){Fassbender}, {B{\"o}hringer}, {Santos},
  {Schuecker}, {Lamer}, {Schwope}, {Kohnert}, {Rosati}, {Mullis}, \&
  {Quintana}}]{Fassbender07}
{Fassbender}, R., {B{\"o}hringer}, H., {Santos}, J., {et~al.} 2007, in Heating
  versus Cooling in Galaxies and Clusters of Galaxies, ed. {H.~B{\"o}hringer,
  G.~W.~Pratt, A.~Finoguenov, \& P.~Schuecker }, 54

\bibitem[{{Felten} {et~al.}(1966){Felten}, {Gould}, {Stein}, \&
  {Woolf}}]{Felten66}
{Felten}, J.~E., {Gould}, R.~J., {Stein}, W.~A., \& {Woolf}, N.~J. 1966, \apj,
  146, 955

\bibitem[{{Finoguenov} {et~al.}(2001){Finoguenov}, {Arnaud}, \&
  {David}}]{Finoguenov01}
{Finoguenov}, A., {Arnaud}, M., \& {David}, L.~P. 2001, \apj, 555, 191

\bibitem[{{Finoguenov} {et~al.}(2009){Finoguenov}, {Connelly}, {Parker},
  {Wilman}, {Mulchaey}, {Saglia}, {Balogh}, {Bower}, \& {McGee}}]{Finoguenov09}
{Finoguenov}, A., {Connelly}, J.~L., {Parker}, L.~C., {et~al.} 2009, \apj, 704,
  564

\bibitem[{{Finoguenov} {et~al.}(2007){Finoguenov}, {Guzzo}, {Hasinger},
  {Scoville}, {Aussel}, {B{\"o}hringer}, {Brusa}, {Capak}, {Cappelluti},
  {Comastri}, {Giodini}, {Griffiths}, {Impey}, {Koekemoer}, {Kneib},
  {Leauthaud}, {Le F{\`e}vre}, {Lilly}, {Mainieri}, {Massey}, {McCracken},
  {Mobasher}, {Murayama}, {Peacock}, {Sakelliou}, {Schinnerer}, {Silverman},
  {Smol{\v c}i{\'c}}, {Taniguchi}, {Tasca}, {Taylor}, {Trump}, \&
  {Zamorani}}]{Finoguenov07}
{Finoguenov}, A., {Guzzo}, L., {Hasinger}, G., {et~al.} 2007, \apjs, 172, 182

\bibitem[{{Finoguenov} {et~al.}(2010){Finoguenov}, {Watson}, {Tanaka},
  {Simpson}, {Cirasuolo}, {Dunlop}, {Peacock}, {Farrah}, {Akiyama}, {Ueda},
  {Smol{\v c}i{\'c}}, {Stewart}, {Rawlings}, {van Breukelen}, {Almaini},
  {Clewley}, {Bonfield}, {Jarvis}, {Barr}, {Foucaud}, {McLure}, {Sekiguchi}, \&
  {Egami}}]{Finoguenov10}
{Finoguenov}, A., {Watson}, M.~G., {Tanaka}, M., {et~al.} 2010, \mnras, 403,
  2063

\bibitem[{{Forman} {et~al.}(1978){Forman}, {Jones}, {Cominsky}, {Julien},
  {Murray}, {Peters}, {Tananbaum}, \& {Giacconi}}]{Forman78}
{Forman}, W., {Jones}, C., {Cominsky}, L., {et~al.} 1978, \apjs, 38, 357

\bibitem[{{Forman} {et~al.}(1972){Forman}, {Kellogg}, {Gursky}, {Tananbaum}, \&
  {Giacconi}}]{Forman72}
{Forman}, W., {Kellogg}, E., {Gursky}, H., {Tananbaum}, H., \& {Giacconi}, R.
  1972, \apj, 178, 309

\bibitem[{{Gal} {et~al.}(2003){Gal}, {de Carvalho}, {Lopes}, {Djorgovski},
  {Brunner}, {Mahabal}, \& {Odewahn}}]{Gal03}
{Gal}, R.~R., {de Carvalho}, R.~R., {Lopes}, P.~A.~A., {et~al.} 2003, \aj, 125,
  2064

\bibitem[{{Gal} {et~al.}(2000){Gal}, {de Carvalho}, {Odewahn}, {Djorgovski}, \&
  {Margoniner}}]{Gal00}
{Gal}, R.~R., {de Carvalho}, R.~R., {Odewahn}, S.~C., {Djorgovski}, S.~G., \&
  {Margoniner}, V.~E. 2000, \aj, 119, 12

\bibitem[{{Geach} {et~al.}(2011){Geach}, {Murphy}, \& {Bower}}]{Geach11}
{Geach}, J.~E., {Murphy}, D.~N.~A., \& {Bower}, R.~G. 2011, \mnras, 413, 3059

\bibitem[{{Gerdes} {et~al.}(2010){Gerdes}, {Sypniewski}, {McKay}, {Hao},
  {Weis}, {Wechsler}, \& {Busha}}]{Gerdes10}
{Gerdes}, D.~W., {Sypniewski}, A.~J., {McKay}, T.~A., {et~al.} 2010, \apj, 715,
  823

\bibitem[{{Gettings} {et~al.}(2012){Gettings}, {Gonzalez}, {Stanford},
  {Eisenhardt}, {Brodwin}, {Mancone}, {Stern}, {Zeimann}, {Masci}, {Papovich},
  {Tanaka}, \& {Wright}}]{Gettings12}
{Gettings}, D.~P., {Gonzalez}, A.~H., {Stanford}, S.~A., {et~al.} 2012, \apjl,
  759, L23

\bibitem[{{Girardi} {et~al.}(2002){Girardi}, {Manzato}, {Mezzetti}, {Giuricin},
  \& {Limboz}}]{Girardi02}
{Girardi}, M., {Manzato}, P., {Mezzetti}, M., {Giuricin}, G., \& {Limboz}, F.
  2002, \apj, 569, 720

\bibitem[{{Gladders} {et~al.}(1998){Gladders}, {Lopez-Cruz}, {Yee}, \&
  {Kodama}}]{Gladders98}
{Gladders}, M.~D., {Lopez-Cruz}, O., {Yee}, H.~K.~C., \& {Kodama}, T. 1998,
  \apj, 501, 571

\bibitem[{{Gladders} \& {Yee}(2005)}]{Gladders05}
{Gladders}, M.~D. \& {Yee}, H.~K.~C. 2005, \apjs, 157, 1

\bibitem[{{Goto} {et~al.}(2002){Goto}, {Sekiguchi}, {Nichol}, {Bahcall}, {Kim},
  {Annis}, {Ivezi{\'c}}, {Brinkmann}, {Hennessy}, {Szokoly}, \&
  {Tucker}}]{Goto02}
{Goto}, T., {Sekiguchi}, M., {Nichol}, R.~C., {et~al.} 2002, \aj, 123, 1807

\bibitem[{{Goto} {et~al.}(2003){Goto}, {Yamauchi}, {Fujita}, {Okamura},
  {Sekiguchi}, {Smail}, {Bernardi}, \& {Gomez}}]{Goto03}
{Goto}, T., {Yamauchi}, C., {Fujita}, Y., {et~al.} 2003, \mnras, 346, 601

\bibitem[{{Grove} {et~al.}(2009){Grove}, {Benoist}, \& {Martel}}]{Grove09}
{Grove}, L.~F., {Benoist}, C., \& {Martel}, F. 2009, \aap, 494, 845

\bibitem[{{Gunn} {et~al.}(1986){Gunn}, {Hoessel}, \& {Oke}}]{Gunn86}
{Gunn}, J.~E., {Hoessel}, J.~G., \& {Oke}, J.~B. 1986, \apj, 306, 30

\bibitem[{{Gursky} {et~al.}(1971){Gursky}, {Kellogg}, {Murray}, {Leong},
  {Tananbaum}, \& {Giacconi}}]{Gursky71}
{Gursky}, H., {Kellogg}, E., {Murray}, S., {et~al.} 1971, \apjl, 167, L81

\bibitem[{{Hansen} {et~al.}(2009){Hansen}, {Sheldon}, {Wechsler}, \&
  {Koester}}]{Hansen09}
{Hansen}, S.~M., {Sheldon}, E.~S., {Wechsler}, R.~H., \& {Koester}, B.~P. 2009,
  \apj, 699, 1333

\bibitem[{{Hao} {et~al.}(2009){Hao}, {Koester}, {Mckay}, {Rykoff}, {Rozo},
  {Evrard}, {Annis}, {Becker}, {Busha}, {Gerdes}, {Johnston}, {Sheldon}, \&
  {Wechsler}}]{Hao09}
{Hao}, J., {Koester}, B.~P., {Mckay}, T.~A., {et~al.} 2009, \apj, 702, 745

\bibitem[{{Hao} {et~al.}(2010){Hao}, {McKay}, {Koester}, {Rykoff}, {Rozo},
  {Annis}, {Wechsler}, {Evrard}, {Siegel}, {Becker}, {Busha}, {Gerdes},
  {Johnston}, \& {Sheldon}}]{Hao10}
{Hao}, J., {McKay}, T.~A., {Koester}, B.~P., {et~al.} 2010, \apjs, 191, 254

\bibitem[{{Hasselfield} {et~al.}(2013){Hasselfield}, {Hilton}, {Marriage},
  {Addison}, {Barrientos}, {Battaglia}, {Battistelli}, {Bond}, {Crichton},
  {Das}, {Devlin}, {Dicker}, {Dunkley}, {Dunner}, {Fowler}, {Gralla}, {Hajian},
  {Halpern}, {Hincks}, {Hlozek}, {Hughes}, {Infante}, {Irwin}, {Kosowsky},
  {Marsden}, {Menanteau}, {Moodley}, {Niemack}, {Nolta}, {Page}, {Partridge},
  {Reese}, {Schmitt}, {Sehgal}, {Sherwin}, {Sievers}, {Sif{\'o}n}, {Spergel},
  {Staggs}, {Swetz}, {Switzer}, {Thornton}, {Trac}, \&
  {Wollack}}]{Hasselfield13}
{Hasselfield}, M., {Hilton}, M., {Marriage}, T.~A., {et~al.} 2013, ArXiv
  e-prints

\bibitem[{{Heath} {et~al.}(2007){Heath}, {Krause}, \& {Alexander}}]{Heath07}
{Heath}, D., {Krause}, M., \& {Alexander}, P. 2007, \mnras, 374, 787

\bibitem[{{Hilton} {et~al.}(2012){Hilton}, {Romer}, {Kay}, {Mehrtens},
  {Lloyd-Davies}, {Thomas}, {Short}, {Mayers}, {Rooney}, {Stott}, {Collins},
  {Harrison}, {Hoyle}, {Liddle}, {Mann}, {Miller}, {Sahl{\'e}n}, {Viana},
  {Davidson}, {Hosmer}, {Nichol}, {Sabirli}, {Stanford}, \& {West}}]{Hilton12}
{Hilton}, M., {Romer}, A.~K., {Kay}, S.~T., {et~al.} 2012, \mnras, 3303

\bibitem[{{Horner} {et~al.}(2008){Horner}, {Perlman}, {Ebeling}, {Jones},
  {Scharf}, {Wegner}, {Malkan}, \& {Maughan}}]{Horner08}
{Horner}, D.~J., {Perlman}, E.~S., {Ebeling}, H., {et~al.} 2008, \apjs, 176,
  374

\bibitem[{{Hubble}(1926)}]{Hubble26}
{Hubble}, E.~P. 1926, \apj, 64, 321

\bibitem[{{Huchra} \& {Geller}(1982)}]{Huchra82}
{Huchra}, J.~P. \& {Geller}, M.~J. 1982, \apj, 257, 423

\bibitem[{{Hughes} \& {Birkinshaw}(1998)}]{Hughes98}
{Hughes}, J.~P. \& {Birkinshaw}, M. 1998, \apj, 497, 645

\bibitem[{{Jester} {et~al.}(2005){Jester}, {Schneider}, {Richards}, {Green},
  {Schmidt}, {Hall}, {Strauss}, {Vanden Berk}, {Stoughton}, {Gunn},
  {Brinkmann}, {Kent}, {Smith}, {Tucker}, \& {Yanny}}]{Jester05}
{Jester}, S., {Schneider}, D.~P., {Richards}, G.~T., {et~al.} 2005, \aj, 130,
  873

\bibitem[{{Kalberla} {et~al.}(2005){Kalberla}, {Burton}, {Hartmann}, {Arnal},
  {Bajaja}, {Morras}, \& {P{\"o}ppel}}]{Kalberla05}
{Kalberla}, P.~M.~W., {Burton}, W.~B., {Hartmann}, D., {et~al.} 2005, \aap,
  440, 775

\bibitem[{{Kellogg} {et~al.}(1972){Kellogg}, {Gursky}, {Tananbaum}, {Giacconi},
  \& {Pounds}}]{Kellogg72}
{Kellogg}, E., {Gursky}, H., {Tananbaum}, H., {Giacconi}, R., \& {Pounds}, K.
  1972, \apjl, 174, L65

\bibitem[{{Kim} {et~al.}(2002){Kim}, {Kepner}, {Postman}, {Strauss}, {Bahcall},
  {Gunn}, {Lupton}, {Annis}, {Nichol}, {Castander}, {Brinkmann}, {Brunner},
  {Connolly}, {Csabai}, {Hindsley}, {Ivezi{\'c}}, {Vogeley}, \& {York}}]{Kim02}
{Kim}, R.~S.~J., {Kepner}, J.~V., {Postman}, M., {et~al.} 2002, \aj, 123, 20

\bibitem[{{King}(1962)}]{King62}
{King}, I. 1962, \aj, 67, 471

\bibitem[{{Knobel} {et~al.}(2009){Knobel}, {Lilly}, {Iovino}, {Porciani},
  {Kova{\v c}}, {Cucciati}, {Finoguenov}, {Kitzbichler}, {Carollo}, {Contini},
  {Kneib}, {Le F{\`e}vre}, {Mainieri}, {Renzini}, {Scodeggio}, {Zamorani},
  {Bardelli}, {Bolzonella}, {Bongiorno}, {Caputi}, {Coppa}, {de la Torre}, {de
  Ravel}, {Franzetti}, {Garilli}, {Kampczyk}, {Lamareille}, {Le Borgne}, {Le
  Brun}, {Maier}, {Mignoli}, {Pello}, {Peng}, {Perez Montero}, {Ricciardelli},
  {Silverman}, {Tanaka}, {Tasca}, {Tresse}, {Vergani}, {Zucca}, {Abbas},
  {Bottini}, {Cappi}, {Cassata}, {Cimatti}, {Fumana}, {Guzzo}, {Koekemoer},
  {Leauthaud}, {Maccagni}, {Marinoni}, {McCracken}, {Memeo}, {Meneux}, {Oesch},
  {Pozzetti}, \& {Scaramella}}]{Knobel09}
{Knobel}, C., {Lilly}, S.~J., {Iovino}, A., {et~al.} 2009, \apj, 697, 1842

\bibitem[{{Koester} {et~al.}(2007){Koester}, {McKay}, {Annis}, {Wechsler},
  {Evrard}, {Bleem}, {Becker}, {Johnston}, {Sheldon}, {Nichol}, {Miller},
  {Scranton}, {Bahcall}, {Barentine}, {Brewington}, {Brinkmann}, {Harvanek},
  {Kleinman}, {Krzesinski}, {Long}, {Nitta}, {Schneider}, {Sneddin}, {Voges},
  \& {York}}]{Koester07}
{Koester}, B.~P., {McKay}, T.~A., {Annis}, J., {et~al.} 2007, \apj, 660, 239

\bibitem[{{Kolokotronis} {et~al.}(2006){Kolokotronis}, {Georgakakis},
  {Basilakos}, {Kitsionas}, {Plionis}, {Georgantopoulos}, \&
  {Gaga}}]{Kolokotronis06}
{Kolokotronis}, V., {Georgakakis}, A., {Basilakos}, S., {et~al.} 2006, \mnras,
  366, 163

\bibitem[{{Kotov} \& {Vikhlinin}(2005)}]{Kotov05}
{Kotov}, O. \& {Vikhlinin}, A. 2005, \apj, 633, 781

\bibitem[{{Kravtsov} \& {Borgani}(2012)}]{Kravtsov12}
{Kravtsov}, A.~V. \& {Borgani}, S. 2012, \araa, 50, 353

\bibitem[{{Krumpe} {et~al.}(2008){Krumpe}, {Lamer}, {Corral}, {Schwope},
  {Carrera}, {Barcons}, {Page}, {Mateos}, {Tedds}, \& {Watson}}]{Krumpe08}
{Krumpe}, M., {Lamer}, G., {Corral}, A., {et~al.} 2008, \aap, 483, 415

\bibitem[{{Lamer} {et~al.}(2008){Lamer}, {Hoeft}, {Kohnert}, {Schwope}, \&
  {Storm}}]{Lamer08}
{Lamer}, G., {Hoeft}, M., {Kohnert}, J., {Schwope}, A., \& {Storm}, J. 2008,
  \aap, 487, L33

\bibitem[{{Li} \& {Yee}(2008)}]{Li08}
{Li}, I.~H. \& {Yee}, H.~K.~C. 2008, \aj, 135, 809

\bibitem[{{Limber}(1959)}]{Limber59}
{Limber}, D.~N. 1959, \apj, 130, 414

\bibitem[{{Lin} \& {Mohr}(2004)}]{Lin04}
{Lin}, Y.-T. \& {Mohr}, J.~J. 2004, \apj, 617, 879

\bibitem[{{Lloyd-Davies} {et~al.}(2011){Lloyd-Davies}, {Romer}, {Mehrtens},
  {Hosmer}, {Davidson}, {Sabirli}, {Mann}, {Hilton}, {Liddle}, {Viana},
  {Campbell}, {Collins}, {Dubois}, {Freeman}, {Harrison}, {Hoyle}, {Kay},
  {Kuwertz}, {Miller}, {Nichol}, {Sahl{\'e}n}, {Stanford}, \&
  {Stott}}]{Lloyd-Davies11}
{Lloyd-Davies}, E.~J., {Romer}, A.~K., {Mehrtens}, N., {et~al.} 2011, \mnras,
  418, 14

\bibitem[{{Lopes} {et~al.}(2006){Lopes}, {de Carvalho}, {Capelato}, {Gal},
  {Djorgovski}, {Brunner}, {Odewahn}, \& {Mahabal}}]{Lopes06}
{Lopes}, P.~A.~A., {de Carvalho}, R.~R., {Capelato}, H.~V., {et~al.} 2006,
  \apj, 648, 209

\bibitem[{{Lopes} {et~al.}(2004){Lopes}, {de Carvalho}, {Gal}, {Djorgovski},
  {Odewahn}, {Mahabal}, \& {Brunner}}]{Lopes04}
{Lopes}, P.~A.~A., {de Carvalho}, R.~R., {Gal}, R.~R., {et~al.} 2004, \aj, 128,
  1017

\bibitem[{{Lopes} {et~al.}(2009){Lopes}, {de Carvalho}, {Kohl-Moreira}, \&
  {Jones}}]{Lopes09}
{Lopes}, P.~A.~A., {de Carvalho}, R.~R., {Kohl-Moreira}, J.~L., \& {Jones}, C.
  2009, \mnras, 399, 2201

\bibitem[{{L{\'o}pez-Cruz} {et~al.}(2004){L{\'o}pez-Cruz}, {Barkhouse}, \&
  {Yee}}]{Lopez-Cruz04}
{L{\'o}pez-Cruz}, O., {Barkhouse}, W.~A., \& {Yee}, H.~K.~C. 2004, \apj, 614,
  679

\bibitem[{{Markevitch}(1998)}]{Markevitch98}
{Markevitch}, M. 1998, \apj, 504, 27

\bibitem[{{Marriage} {et~al.}(2011){Marriage}, {Acquaviva}, {Ade}, {Aguirre},
  {Amiri}, {Appel}, {Barrientos}, {Battistelli}, {Bond}, {Brown}, {Burger},
  {Chervenak}, {Das}, {Devlin}, {Dicker}, {Bertrand Doriese}, {Dunkley},
  {D{\"u}nner}, {Essinger-Hileman}, {Fisher}, {Fowler}, {Hajian}, {Halpern},
  {Hasselfield}, {Hern{\'a}ndez-Monteagudo}, {Hilton}, {Hilton}, {Hincks},
  {Hlozek}, {Huffenberger}, {Handel Hughes}, {Hughes}, {Infante}, {Irwin},
  {Baptiste Juin}, {Kaul}, {Klein}, {Kosowsky}, {Lau}, {Limon}, {Lin},
  {Lupton}, {Marsden}, {Martocci}, {Mauskopf}, {Menanteau}, {Moodley},
  {Moseley}, {Netterfield}, {Niemack}, {Nolta}, {Page}, {Parker}, {Partridge},
  {Quintana}, {Reese}, {Reid}, {Sehgal}, {Sherwin}, {Sievers}, {Spergel},
  {Staggs}, {Swetz}, {Switzer}, {Thornton}, {Trac}, {Tucker}, {Warne},
  {Wilson}, {Wollack}, \& {Zhao}}]{Marriage11}
{Marriage}, T.~A., {Acquaviva}, V., {Ade}, P.~A.~R., {et~al.} 2011, \apj, 737,
  61

\bibitem[{{Maughan} {et~al.}(2011){Maughan}, {Giles}, {Randall}, {Jones}, \&
  {Forman}}]{Maughan11}
{Maughan}, B.~J., {Giles}, P.~A., {Randall}, S.~W., {Jones}, C., \& {Forman},
  W.~R. 2011, ArXiv e-prints

\bibitem[{{Maughan} {et~al.}(2012){Maughan}, {Giles}, {Randall}, {Jones}, \&
  {Forman}}]{Maughan12}
{Maughan}, B.~J., {Giles}, P.~A., {Randall}, S.~W., {Jones}, C., \& {Forman},
  W.~R. 2012, \mnras, 421, 1583

\bibitem[{{Maughan} {et~al.}(2008){Maughan}, {Jones}, {Forman}, \& {Van
  Speybroeck}}]{Maughan08}
{Maughan}, B.~J., {Jones}, C., {Forman}, W., \& {Van Speybroeck}, L. 2008,
  \apjs, 174, 117

\bibitem[{{McConnachie} {et~al.}(2009){McConnachie}, {Patton}, {Ellison}, \&
  {Simard}}]{McConnachie09}
{McConnachie}, A.~W., {Patton}, D.~R., {Ellison}, S.~L., \& {Simard}, L. 2009,
  \mnras, 395, 255

\bibitem[{{McDowell} {et~al.}(2003){McDowell}, {Clements}, {Lamb}, {Shaked},
  {Hearn}, {Colina}, {Mundell}, {Borne}, {Baker}, \& {Arribas}}]{McDowell03}
{McDowell}, J.~C., {Clements}, D.~L., {Lamb}, S.~A., {et~al.} 2003, \apj, 591,
  154

\bibitem[{{Meekins} {et~al.}(1971){Meekins}, {Fritz}, {Chubb}, \&
  {Friedman}}]{Meekins71}
{Meekins}, J.~F., {Fritz}, G., {Chubb}, T.~A., \& {Friedman}, H. 1971, \nat,
  231, 107

\bibitem[{{Mehrtens} {et~al.}(2012){Mehrtens}, {Romer}, {Hilton},
  {Lloyd-Davies}, {Miller}, {Stanford}, {Hosmer}, {Hoyle}, {Collins}, {Liddle},
  {Viana}, {Nichol}, {Stott}, {Dubois}, {Kay}, {Sahl{\'e}n}, {Young}, {Short},
  {Christodoulou}, {Watson}, {Davidson}, {Harrison}, {Baruah}, {Smith},
  {Burke}, {Mayers}, {Deadman}, {Rooney}, {Edmondson}, {West}, {Campbell},
  {Edge}, {Mann}, {Sabirli}, {Wake}, {Benoist}, {da Costa}, {Maia}, \&
  {Ogando}}]{Mehrtens12}
{Mehrtens}, N., {Romer}, A.~K., {Hilton}, M., {et~al.} 2012, \mnras, 2912

\bibitem[{{Mei} {et~al.}(2009){Mei}, {Holden}, {Blakeslee}, {Ford}, {Franx},
  {Homeier}, {Illingworth}, {Jee}, {Overzier}, {Postman}, {Rosati}, {Van der
  Wel}, \& {Bartlett}}]{Mei09}
{Mei}, S., {Holden}, B.~P., {Blakeslee}, J.~P., {et~al.} 2009, \apj, 690, 42

\bibitem[{{Merch{\'a}n} \& {Zandivarez}(2005)}]{Merchan05}
{Merch{\'a}n}, M.~E. \& {Zandivarez}, A. 2005, \apj, 630, 759

\bibitem[{{Metcalfe} {et~al.}(2003){Metcalfe}, {Kneib}, {McBreen}, {Altieri},
  {Biviano}, {Delaney}, {Elbaz}, {Kessler}, {Leech}, {Okumura}, {Ott},
  {Perez-Martinez}, {Sanchez-Fernandez}, \& {Schulz}}]{Metcalfe03}
{Metcalfe}, L., {Kneib}, J.-P., {McBreen}, B., {et~al.} 2003, \aap, 407, 791

\bibitem[{{Mewe} {et~al.}(1986){Mewe}, {Lemen}, \& {van den Oord}}]{Mewe86}
{Mewe}, R., {Lemen}, J.~R., \& {van den Oord}, G.~H.~J. 1986, \aaps, 65, 511

\bibitem[{{Miller} {et~al.}(2005){Miller}, {Nichol}, {Reichart}, {Wechsler},
  {Evrard}, {Annis}, {McKay}, {Bahcall}, {Bernardi}, {B{\"o}hringer},
  {Connolly}, {Goto}, {Kniazev}, {Lamb}, {Postman}, {Schneider}, {Sheth}, \&
  {Voges}}]{Miller05}
{Miller}, C.~J., {Nichol}, R.~C., {Reichart}, D., {et~al.} 2005, \aj, 130, 968

\bibitem[{{Mittal} {et~al.}(2011){Mittal}, {Hicks}, {Reiprich}, \&
  {Jaritz}}]{Mittal11}
{Mittal}, R., {Hicks}, A., {Reiprich}, T.~H., \& {Jaritz}, V. 2011, \aap, 532,
  A133

\bibitem[{{Mittal} {et~al.}(2009){Mittal}, {Hudson}, {Reiprich}, \&
  {Clarke}}]{Mittal09}
{Mittal}, R., {Hudson}, D.~S., {Reiprich}, T.~H., \& {Clarke}, T. 2009, \aap,
  501, 835

\bibitem[{{M{\"u}hlegger}(2010)}]{Muehlegger10}
{M{\"u}hlegger}, M. 2010, PhD thesis, Technischen Universit{\"a}t M{\"u}nchen

\bibitem[{{Mullis} {et~al.}(2003){Mullis}, {McNamara}, {Quintana}, {Vikhlinin},
  {Henry}, {Gioia}, {Hornstrup}, {Forman}, \& {Jones}}]{Mullis03}
{Mullis}, C.~R., {McNamara}, B.~R., {Quintana}, H., {et~al.} 2003, \apj, 594,
  154

\bibitem[{{Mullis} {et~al.}(2005){Mullis}, {Rosati}, {Lamer}, {B{\"o}hringer},
  {Schwope}, {Schuecker}, \& {Fassbender}}]{Mullis05}
{Mullis}, C.~R., {Rosati}, P., {Lamer}, G., {et~al.} 2005, \apjl, 623, L85

\bibitem[{{Navarro} {et~al.}(1997){Navarro}, {Frenk}, \& {White}}]{Navarro97}
{Navarro}, J.~F., {Frenk}, C.~S., \& {White}, S.~D.~M. 1997, \apj, 490, 493

\bibitem[{{Olsen} {et~al.}(2007){Olsen}, {Benoist}, {Cappi}, {Maurogordato},
  {Mazure}, {Slezak}, {Adami}, {Ferrari}, \& {Martel}}]{Olsen07}
{Olsen}, L.~F., {Benoist}, C., {Cappi}, A., {et~al.} 2007, \aap, 461, 81

\bibitem[{{Osmond} \& {Ponman}(2004)}]{Osmond04}
{Osmond}, J.~P.~F. \& {Ponman}, T.~J. 2004, \mnras, 350, 1511

\bibitem[{{Oyaizu} {et~al.}(2008){Oyaizu}, {Lima}, {Cunha}, {Lin}, {Frieman},
  \& {Sheldon}}]{Oyaizu08}
{Oyaizu}, H., {Lima}, M., {Cunha}, C.~E., {et~al.} 2008, \apj, 674, 768

\bibitem[{{Peterson} {et~al.}(2009){Peterson}, {Jernigan}, {Gupta}, {Bankert},
  \& {Kahn}}]{Peterson09}
{Peterson}, J.~R., {Jernigan}, J.~G., {Gupta}, R.~R., {Bankert}, J., \& {Kahn},
  S.~M. 2009, \apj, 707, 878

\bibitem[{{Piconcelli}(2012)}]{Piconcelli12}
{Piconcelli}, E. 2012, ESA: XMM-Newton SOC

\bibitem[{{Pierre} {et~al.}(2006){Pierre}, {Pacaud}, {Duc}, {Willis},
  {Andreon}, {Valtchanov}, {Altieri}, {Galaz}, {Gueguen}, {Le F{\`e}vre},
  {F{\`e}vre}, {Ponman}, {Sprimont}, {Surdej}, {Adami}, {Alshino}, {Bremer},
  {Chiappetti}, {Detal}, {Garcet}, {Gosset}, {Jean}, {Maccagni}, {Marinoni},
  {Mazure}, {Quintana}, \& {Read}}]{Pierre06}
{Pierre}, M., {Pacaud}, F., {Duc}, P.-A., {et~al.} 2006, \mnras, 372, 591

\bibitem[{{Piffaretti} {et~al.}(2011){Piffaretti}, {Arnaud}, {Pratt},
  {Pointecouteau}, \& {Melin}}]{Piffaretti11}
{Piffaretti}, R., {Arnaud}, M., {Pratt}, G.~W., {Pointecouteau}, E., \&
  {Melin}, J.-B. 2011, \aap, 534, A109

\bibitem[{{Planck Collaboration} {et~al.}(2013){Planck Collaboration}, {Ade},
  {Aghanim}, {Armitage-Caplan}, {Arnaud}, {Ashdown}, {Atrio-Barandela},
  {Aumont}, {Aussel}, {Baccigalupi}, \& et~al.}]{Planck13}
{Planck Collaboration}, {Ade}, P.~A.~R., {Aghanim}, N., {et~al.} 2013, ArXiv
  e-prints

\bibitem[{{Planck Collaboration} {et~al.}(2011){Planck Collaboration}, {Ade},
  {Aghanim}, {Arnaud}, {Ashdown}, {Aumont}, {Baccigalupi}, {Balbi}, {Banday},
  {Barreiro}, \& et~al.}]{Planck11}
{Planck Collaboration}, {Ade}, P.~A.~R., {Aghanim}, N., {et~al.} 2011, \aap,
  536, A8

\bibitem[{{Plionis} {et~al.}(2005){Plionis}, {Basilakos}, {Georgantopoulos}, \&
  {Georgakakis}}]{Plionis05}
{Plionis}, M., {Basilakos}, S., {Georgantopoulos}, I., \& {Georgakakis}, A.
  2005, \apjl, 622, L17

\bibitem[{{Popesso} {et~al.}(2007){Popesso}, {Biviano}, {B{\"o}hringer}, \&
  {Romaniello}}]{Popesso07}
{Popesso}, P., {Biviano}, A., {B{\"o}hringer}, H., \& {Romaniello}, M. 2007,
  \aap, 464, 451

\bibitem[{{Popesso} {et~al.}(2005){Popesso}, {Biviano}, {B{\"o}hringer},
  {Romaniello}, \& {Voges}}]{Popesso05}
{Popesso}, P., {Biviano}, A., {B{\"o}hringer}, H., {Romaniello}, M., \&
  {Voges}, W. 2005, \aap, 433, 431

\bibitem[{{Popesso} {et~al.}(2004){Popesso}, {B{\"o}hringer}, {Brinkmann},
  {Voges}, \& {York}}]{Popesso04}
{Popesso}, P., {B{\"o}hringer}, H., {Brinkmann}, J., {Voges}, W., \& {York},
  D.~G. 2004, \aap, 423, 449

\bibitem[{{Postman} \& {Lauer}(1995)}]{Postman95}
{Postman}, M. \& {Lauer}, T.~R. 1995, \apj, 440, 28

\bibitem[{{Postman} {et~al.}(2002){Postman}, {Lauer}, {Oegerle}, \&
  {Donahue}}]{Postman02}
{Postman}, M., {Lauer}, T.~R., {Oegerle}, W., \& {Donahue}, M. 2002, \apj, 579,
  93

\bibitem[{{Postman} {et~al.}(1996){Postman}, {Lubin}, {Gunn}, {Oke}, {Hoessel},
  {Schneider}, \& {Christensen}}]{Postman96}
{Postman}, M., {Lubin}, L.~M., {Gunn}, J.~E., {et~al.} 1996, \aj, 111, 615

\bibitem[{{Pratt} {et~al.}(2010){Pratt}, {Arnaud}, {Piffaretti},
  {B{\"o}hringer}, {Ponman}, {Croston}, {Voit}, {Borgani}, \&
  {Bower}}]{Pratt10}
{Pratt}, G.~W., {Arnaud}, M., {Piffaretti}, R., {et~al.} 2010, \aap, 511, A85

\bibitem[{{Pratt} {et~al.}(2009){Pratt}, {Croston}, {Arnaud}, \&
  {B{\"o}hringer}}]{Pratt09}
{Pratt}, G.~W., {Croston}, J.~H., {Arnaud}, M., \& {B{\"o}hringer}, H. 2009,
  \aap, 498, 361

\bibitem[{{Predehl} {et~al.}(2010){Predehl}, {Andritschke}, {B{\"o}hringer},
  {Bornemann}, {Br{\"a}uninger}, {Brunner}, {Brusa}, {Burkert}, {Burwitz},
  {Cappelluti}, {Churazov}, {Dennerl}, {Eder}, {Elbs}, {Freyberg}, {Friedrich},
  {F{\"u}rmetz}, {Gaida}, {H{\"a}lker}, {Hartner}, {Hasinger}, {Hermann},
  {Huber}, {Kendziorra}, {von Kienlin}, {Kink}, {Kreykenbohm}, {Lamer},
  {Lapchov}, {Lehmann}, {Meidinger}, {Mican}, {Mohr}, {M{\"u}hlegger},
  {M{\"u}ller}, {Nandra}, {Pavlinsky}, {Pfeffermann}, {Reiprich}, {Robrade},
  {Roh{\'e}}, {Santangelo}, {Sch{\"a}chner}, {Schanz}, {Schmid}, {Schmitt},
  {Schreib}, {Schrey}, {Schwope}, {Steinmetz}, {Str{\"u}der}, {Sunyaev},
  {Tenzer}, {Tiedemann}, {Vongehr}, \& {Wilms}}]{Predehl10}
{Predehl}, P., {Andritschke}, R., {B{\"o}hringer}, H., {et~al.} 2010, in
  Society of Photo-Optical Instrumentation Engineers (SPIE) Conference Series,
  Vol. 7732, Society of Photo-Optical Instrumentation Engineers (SPIE)
  Conference Series

\bibitem[{{Ramella} {et~al.}(2001){Ramella}, {Boschin}, {Fadda}, \&
  {Nonino}}]{Ramella01}
{Ramella}, M., {Boschin}, W., {Fadda}, D., \& {Nonino}, M. 2001, \aap, 368, 776

\bibitem[{{Reichardt} {et~al.}(2013){Reichardt}, {Stalder}, {Bleem}, {Montroy},
  {Aird}, {Andersson}, {Armstrong}, {Ashby}, {Bautz}, {Bayliss}, {Bazin},
  {Benson}, {Brodwin}, {Carlstrom}, {Chang}, {Cho}, {Clocchiatti}, {Crawford},
  {Crites}, {de Haan}, {Desai}, {Dobbs}, {Dudley}, {Foley}, {Forman}, {George},
  {Gladders}, {Gonzalez}, {Halverson}, {Harrington}, {High}, {Holder},
  {Holzapfel}, {Hoover}, {Hrubes}, {Jones}, {Joy}, {Keisler}, {Knox}, {Lee},
  {Leitch}, {Liu}, {Lueker}, {Luong-Van}, {Mantz}, {Marrone}, {McDonald},
  {McMahon}, {Mehl}, {Meyer}, {Mocanu}, {Mohr}, {Murray}, {Natoli}, {Padin},
  {Plagge}, {Pryke}, {Rest}, {Ruel}, {Ruhl}, {Saliwanchik}, {Saro}, {Sayre},
  {Schaffer}, {Shaw}, {Shirokoff}, {Song}, {Spieler}, {Staniszewski}, {Stark},
  {Story}, {Stubbs}, {{\v S}uhada}, {van Engelen}, {Vanderlinde}, {Vieira},
  {Vikhlinin}, {Williamson}, {Zahn}, \& {Zenteno}}]{Reichardt13}
{Reichardt}, C.~L., {Stalder}, B., {Bleem}, L.~E., {et~al.} 2013, \apj, 763,
  127

\bibitem[{{Reichert} {et~al.}(2011){Reichert}, {B{\"o}hringer}, {Fassbender},
  \& {M{\"u}hlegger}}]{Reichert11}
{Reichert}, A., {B{\"o}hringer}, H., {Fassbender}, R., \& {M{\"u}hlegger}, M.
  2011, \aap, 535, A4

\bibitem[{{Reiprich} \& {B{\"o}hringer}(2002)}]{Reiprich02}
{Reiprich}, T.~H. \& {B{\"o}hringer}, H. 2002, \apj, 567, 716

\bibitem[{{Romer} {et~al.}(1994){Romer}, {Collins}, {B{\"o}hringer},
  {Cruddace}, {Ebeling}, {MacGillivray}, \& {Voges}}]{Romer94}
{Romer}, A.~K., {Collins}, C.~A., {B{\"o}hringer}, H., {et~al.} 1994, \nat,
  372, 75

\bibitem[{{Romer} {et~al.}(2000){Romer}, {Nichol}, {Holden}, {Ulmer}, {Pildis},
  {Merrelli}, {Adami}, {Burke}, {Collins}, {Metevier}, {Kron}, \&
  {Commons}}]{Romer00}
{Romer}, A.~K., {Nichol}, R.~C., {Holden}, B.~P., {et~al.} 2000, \apjs, 126,
  209

\bibitem[{{Romer} {et~al.}(2001){Romer}, {Viana}, {Liddle}, \&
  {Mann}}]{Romer01}
{Romer}, A.~K., {Viana}, P.~T.~P., {Liddle}, A.~R., \& {Mann}, R.~G. 2001,
  \apj, 547, 594

\bibitem[{{Rosati} {et~al.}(2002){Rosati}, {Borgani}, \& {Norman}}]{Rosati02}
{Rosati}, P., {Borgani}, S., \& {Norman}, C. 2002, \araa, 40, 539

\bibitem[{{Rosati} {et~al.}(1998){Rosati}, {della Ceca}, {Norman}, \&
  {Giacconi}}]{Rosati98}
{Rosati}, P., {della Ceca}, R., {Norman}, C., \& {Giacconi}, R. 1998, \apjl,
  492, L21

\bibitem[{{Rykoff} {et~al.}(2008){Rykoff}, {McKay}, {Becker}, {Evrard},
  {Johnston}, {Koester}, {Rozo}, {Sheldon}, \& {Wechsler}}]{Rykoff08}
{Rykoff}, E.~S., {McKay}, T.~A., {Becker}, M.~R., {et~al.} 2008, \apj, 675,
  1106

\bibitem[{{Santos} {et~al.}(2004){Santos}, {Ellis}, {Kneib}, {Richard}, \&
  {Kuijken}}]{Santos04}
{Santos}, M.~R., {Ellis}, R.~S., {Kneib}, J.-P., {Richard}, J., \& {Kuijken},
  K. 2004, \apj, 606, 683

\bibitem[{{Sarazin}(1988)}]{Sarazin88}
{Sarazin}, C.~L. 1988, {X-ray emission from clusters of galaxies} (Cambridge,
  UK: Cambridge Univ. Press)

\bibitem[{{Scharf} {et~al.}(1997){Scharf}, {Jones}, {Ebeling}, {Perlman},
  {Malkan}, \& {Wegner}}]{Scharf97}
{Scharf}, C.~A., {Jones}, L.~R., {Ebeling}, H., {et~al.} 1997, \apj, 477, 79

\bibitem[{{Schlegel} {et~al.}(1998){Schlegel}, {Finkbeiner}, \&
  {Davis}}]{Schlegel98}
{Schlegel}, D.~J., {Finkbeiner}, D.~P., \& {Davis}, M. 1998, \apj, 500, 525

\bibitem[{{Schuecker} {et~al.}(2003){Schuecker}, {B{\"o}hringer}, {Collins}, \&
  {Guzzo}}]{Schuecker03}
{Schuecker}, P., {B{\"o}hringer}, H., {Collins}, C.~A., \& {Guzzo}, L. 2003,
  \aap, 398, 867

\bibitem[{{Schuecker} {et~al.}(2004){Schuecker}, {B{\"o}hringer}, \&
  {Voges}}]{Schuecker04}
{Schuecker}, P., {B{\"o}hringer}, H., \& {Voges}, W. 2004, \aap, 420, 61

\bibitem[{{Sehgal} {et~al.}(2008){Sehgal}, {Hughes}, {Wittman}, {Margoniner},
  {Tyson}, {Gee}, \& {dell'Antonio}}]{Sehgal08}
{Sehgal}, N., {Hughes}, J.~P., {Wittman}, D., {et~al.} 2008, \apj, 673, 163

\bibitem[{{Shectman}(1985)}]{Shectman85}
{Shectman}, S.~A. 1985, \apjs, 57, 77

\bibitem[{{Smail} {et~al.}(1998){Smail}, {Edge}, {Ellis}, \&
  {Blandford}}]{Smail98}
{Smail}, I., {Edge}, A.~C., {Ellis}, R.~S., \& {Blandford}, R.~D. 1998, \mnras,
  293, 124

\bibitem[{{Smith}(1936)}]{Smith36}
{Smith}, S. 1936, \apj, 83, 23

\bibitem[{{Solinger} \& {Tucker}(1972)}]{Solinger72}
{Solinger}, A.~B. \& {Tucker}, W.~H. 1972, \apjl, 175, L107

\bibitem[{{Stott} {et~al.}(2008){Stott}, {Edge}, {Smith}, {Swinbank}, \&
  {Ebeling}}]{Stott08}
{Stott}, J.~P., {Edge}, A.~C., {Smith}, G.~P., {Swinbank}, A.~M., \& {Ebeling},
  H. 2008, \mnras, 384, 1502

\bibitem[{{Stott} {et~al.}(2009){Stott}, {Pimbblet}, {Edge}, {Smith}, \&
  {Wardlow}}]{Stott09}
{Stott}, J.~P., {Pimbblet}, K.~A., {Edge}, A.~C., {Smith}, G.~P., \& {Wardlow},
  J.~L. 2009, \mnras, 394, 2098

\bibitem[{{Sunyaev} \& {Zeldovich}(1980)}]{Sunyaev80}
{Sunyaev}, R.~A. \& {Zeldovich}, I.~B. 1980, \araa, 18, 537

\bibitem[{{Sunyaev} \& {Zeldovich}(1970)}]{Sunyaev70}
{Sunyaev}, R.~A. \& {Zeldovich}, Y.~B. 1970, Comments on Astrophysics and Space
  Physics, 2, 66

\bibitem[{{Sunyaev} \& {Zeldovich}(1972)}]{Sunyaev72}
{Sunyaev}, R.~A. \& {Zeldovich}, Y.~B. 1972, Comments on Astrophysics and Space
  Physics, 4, 173

\bibitem[{{Szabo} {et~al.}(2011){Szabo}, {Pierpaoli}, {Dong}, {Pipino}, \&
  {Gunn}}]{Szabo11}
{Szabo}, T., {Pierpaoli}, E., {Dong}, F., {Pipino}, A., \& {Gunn}, J. 2011,
  \apj, 736, 21

\bibitem[{{Takey} {et~al.}(2011){Takey}, {Schwope}, \& {Lamer}}]{Takey11}
{Takey}, A., {Schwope}, A., \& {Lamer}, G. 2011, \aap, 534, A120

\bibitem[{{Takey} {et~al.}(2013{\natexlab{a}}){Takey}, {Schwope}, \&
  {Lamer}}]{Takey13a}
{Takey}, A., {Schwope}, A., \& {Lamer}, G. 2013{\natexlab{a}}, \aap, 558, A75

\bibitem[{{Takey} {et~al.}(2013{\natexlab{b}}){Takey}, {Schwope}, \&
  {Lamer}}]{Takey13b}
{Takey}, A., {Schwope}, A., \& {Lamer}, G. 2013{\natexlab{b}}, \aap

\bibitem[{{Takey} {et~al.}(2013{\natexlab{c}}){Takey}, {Schwope}, \&
  {Lamer}}]{Takey13c}
{Takey}, A., {Schwope}, A., \& {Lamer}, G. 2013{\natexlab{c}}, \aap

\bibitem[{{Tundo} {et~al.}(2012){Tundo}, {Moretti}, {Tozzi}, {Teng}, {Rosati},
  {Tagliaferri}, \& {Campana}}]{Tundo12}
{Tundo}, E., {Moretti}, A., {Tozzi}, P., {et~al.} 2012, \aap, 547, A57

\bibitem[{{{\v S}uhada} {et~al.}(2010){{\v S}uhada}, {Song}, {B{\"o}hringer},
  {Benson}, {Mohr}, {Fassbender}, {Finoguenov}, {Pierini}, {Pratt},
  {Andersson}, {Armstrong}, \& {Desai}}]{Suhada10}
{{\v S}uhada}, R., {Song}, J., {B{\"o}hringer}, H., {et~al.} 2010, \aap, 514,
  L3

\bibitem[{{{\v S}uhada} {et~al.}(2012){{\v S}uhada}, {Song}, {B{\"o}hringer},
  {Mohr}, {Chon}, {Finoguenov}, {Fassbender}, {Desai}, {Armstrong}, {Zenteno},
  {Barkhouse}, {Bertin}, {Buckley-Geer}, {Hansen}, {High}, {Lin},
  {M{\"u}hlegger}, {Ngeow}, {Pierini}, {Pratt}, {Verdugo}, \&
  {Tucker}}]{Suhada12}
{{\v S}uhada}, R., {Song}, J., {B{\"o}hringer}, H., {et~al.} 2012, \aap, 537,
  A39

\bibitem[{{Vanderlinde} {et~al.}(2010){Vanderlinde}, {Crawford}, {de Haan},
  {Dudley}, {Shaw}, {Ade}, {Aird}, {Benson}, {Bleem}, {Brodwin}, {Carlstrom},
  {Chang}, {Crites}, {Desai}, {Dobbs}, {Foley}, {George}, {Gladders}, {Hall},
  {Halverson}, {High}, {Holder}, {Holzapfel}, {Hrubes}, {Joy}, {Keisler},
  {Knox}, {Lee}, {Leitch}, {Loehr}, {Lueker}, {Marrone}, {McMahon}, {Mehl},
  {Meyer}, {Mohr}, {Montroy}, {Ngeow}, {Padin}, {Plagge}, {Pryke}, {Reichardt},
  {Rest}, {Ruel}, {Ruhl}, {Schaffer}, {Shirokoff}, {Song}, {Spieler},
  {Stalder}, {Staniszewski}, {Stark}, {Stubbs}, {van Engelen}, {Vieira},
  {Williamson}, {Yang}, {Zahn}, \& {Zenteno}}]{Vanderlinde10}
{Vanderlinde}, K., {Crawford}, T.~M., {de Haan}, T., {et~al.} 2010, \apj, 722,
  1180

\bibitem[{{Vikhlinin} {et~al.}(2009{\natexlab{a}}){Vikhlinin}, {Burenin},
  {Ebeling}, {Forman}, {Hornstrup}, {Jones}, {Kravtsov}, {Murray}, {Nagai},
  {Quintana}, \& {Voevodkin}}]{Vikhlinin09b}
{Vikhlinin}, A., {Burenin}, R.~A., {Ebeling}, H., {et~al.} 2009{\natexlab{a}},
  \apj, 692, 1033

\bibitem[{{Vikhlinin} {et~al.}(2006){Vikhlinin}, {Kravtsov}, {Forman}, {Jones},
  {Markevitch}, {Murray}, \& {Van Speybroeck}}]{Vikhlinin06}
{Vikhlinin}, A., {Kravtsov}, A., {Forman}, W., {et~al.} 2006, \apj, 640, 691

\bibitem[{{Vikhlinin} {et~al.}(2009{\natexlab{b}}){Vikhlinin}, {Kravtsov},
  {Burenin}, {Ebeling}, {Forman}, {Hornstrup}, {Jones}, {Murray}, {Nagai},
  {Quintana}, \& {Voevodkin}}]{Vikhlinin09a}
{Vikhlinin}, A., {Kravtsov}, A.~V., {Burenin}, R.~A., {et~al.}
  2009{\natexlab{b}}, \apj, 692, 1060

\bibitem[{{Vikhlinin} {et~al.}(1998){Vikhlinin}, {McNamara}, {Forman}, {Jones},
  {Quintana}, \& {Hornstrup}}]{Vikhlinin98}
{Vikhlinin}, A., {McNamara}, B.~R., {Forman}, W., {et~al.} 1998, \apj, 502, 558

\bibitem[{{Voges} {et~al.}(1999){Voges}, {Aschenbach}, {Boller},
  {Br{\"a}uninger}, {Briel}, {Burkert}, {Dennerl}, {Englhauser}, {Gruber},
  {Haberl}, {Hartner}, {Hasinger}, {K{\"u}rster}, {Pfeffermann}, {Pietsch},
  {Predehl}, {Rosso}, {Schmitt}, {Tr{\"u}mper}, \& {Zimmermann}}]{Voges99}
{Voges}, W., {Aschenbach}, B., {Boller}, T., {et~al.} 1999, \aap, 349, 389

\bibitem[{{Voit}(2005)}]{Voit05}
{Voit}, G.~M. 2005, Reviews of Modern Physics, 77, 207

\bibitem[{{Watson} {et~al.}(2009){Watson}, {Schr{\"o}der}, {Fyfe}, {Page},
  {Lamer}, {Mateos}, {Pye}, {Sakano}, {Rosen}, {Ballet}, {Barcons}, {Barret},
  {Boller}, {Brunner}, {Brusa}, {Caccianiga}, {Carrera}, {Ceballos}, {Della
  Ceca}, {Denby}, {Denkinson}, {Dupuy}, {Farrell}, {Fraschetti}, {Freyberg},
  {Guillout}, {Hambaryan}, {Maccacaro}, {Mathiesen}, {McMahon}, {Michel},
  {Motch}, {Osborne}, {Page}, {Pakull}, {Pietsch}, {Saxton}, {Schwope},
  {Severgnini}, {Simpson}, {Sironi}, {Stewart}, {Stewart}, {Stobbart}, {Tedds},
  {Warwick}, {Webb}, {West}, {Worrall}, \& {Yuan}}]{Watson09}
{Watson}, M.~G., {Schr{\"o}der}, A.~C., {Fyfe}, D., {et~al.} 2009, \aap, 493,
  339

\bibitem[{{Weisskopf} {et~al.}(2000){Weisskopf}, {Tananbaum}, {Van Speybroeck},
  \& {O'Dell}}]{Weisskopf00}
{Weisskopf}, M.~C., {Tananbaum}, H.~D., {Van Speybroeck}, L.~P., \& {O'Dell},
  S.~L. 2000, in Society of Photo-Optical Instrumentation Engineers (SPIE)
  Conference Series, Vol. 4012, Society of Photo-Optical Instrumentation
  Engineers (SPIE) Conference Series, ed. J.~E. {Truemper} \& B.~{Aschenbach},
  2--16

\bibitem[{{Wen} {et~al.}(2009){Wen}, {Han}, \& {Liu}}]{Wen09}
{Wen}, Z.~L., {Han}, J.~L., \& {Liu}, F.~S. 2009, \apjs, 183, 197

\bibitem[{{Wen} {et~al.}(2012){Wen}, {Han}, \& {Liu}}]{Wen12}
{Wen}, Z.~L., {Han}, J.~L., \& {Liu}, F.~S. 2012, \apjs, 199, 34

\bibitem[{{White} {et~al.}(1999){White}, {Bliton}, {Bhavsar}, {Bornmann},
  {Burns}, {Ledlow}, \& {Loken}}]{White99}
{White}, R.~A., {Bliton}, M., {Bhavsar}, S.~P., {et~al.} 1999, \aj, 118, 2014

\bibitem[{{Wilms} {et~al.}(2000){Wilms}, {Allen}, \& {McCray}}]{Wilms00}
{Wilms}, J., {Allen}, A., \& {McCray}, R. 2000, \apj, 542, 914

\bibitem[{{Wittman} {et~al.}(2006){Wittman}, {Dell'Antonio}, {Hughes},
  {Margoniner}, {Tyson}, {Cohen}, \& {Norman}}]{Wittman06}
{Wittman}, D., {Dell'Antonio}, I.~P., {Hughes}, J.~P., {et~al.} 2006, \apj,
  643, 128

\bibitem[{{Yang} {et~al.}(2007){Yang}, {Mo}, {van den Bosch}, {Pasquali}, {Li},
  \& {Barden}}]{Yang07}
{Yang}, X., {Mo}, H.~J., {van den Bosch}, F.~C., {et~al.} 2007, \apj, 671, 153

\bibitem[{{Yoon} {et~al.}(2008){Yoon}, {Schawinski}, {Sheen}, {Ree}, \&
  {Yi}}]{Yoon08}
{Yoon}, J.~H., {Schawinski}, K., {Sheen}, Y.-K., {Ree}, C.~H., \& {Yi}, S.~K.
  2008, \apjs, 176, 414

\bibitem[{{York} {et~al.}(2000){York}, {Adelman}, {Anderson}, {Anderson},
  {Annis}, {Bahcall}, {Bakken}, {Barkhouser}, {Bastian}, {Berman}, {Boroski},
  {Bracker}, {Briegel}, {Briggs}, {Brinkmann}, {Brunner}, {Burles}, {Carey},
  {Carr}, {Castander}, {Chen}, {Colestock}, {Connolly}, {Crocker}, {Csabai},
  {Czarapata}, {Davis}, {Doi}, {Dombeck}, {Eisenstein}, {Ellman}, {Elms},
  {Evans}, {Fan}, {Federwitz}, {Fiscelli}, {Friedman}, {Frieman}, {Fukugita},
  {Gillespie}, {Gunn}, {Gurbani}, {de Haas}, {Haldeman}, {Harris}, {Hayes},
  {Heckman}, {Hennessy}, {Hindsley}, {Holm}, {Holmgren}, {Huang}, {Hull},
  {Husby}, {Ichikawa}, {Ichikawa}, {Ivezi{\'c}}, {Kent}, {Kim}, {Kinney},
  {Klaene}, {Kleinman}, {Kleinman}, {Knapp}, {Korienek}, {Kron}, {Kunszt},
  {Lamb}, {Lee}, {Leger}, {Limmongkol}, {Lindenmeyer}, {Long}, {Loomis},
  {Loveday}, {Lucinio}, {Lupton}, {MacKinnon}, {Mannery}, {Mantsch}, {Margon},
  {McGehee}, {McKay}, {Meiksin}, {Merelli}, {Monet}, {Munn}, {Narayanan},
  {Nash}, {Neilsen}, {Neswold}, {Newberg}, {Nichol}, {Nicinski}, {Nonino},
  {Okada}, {Okamura}, {Ostriker}, {Owen}, {Pauls}, {Peoples}, {Peterson},
  {Petravick}, {Pier}, {Pope}, {Pordes}, {Prosapio}, {Rechenmacher}, {Quinn},
  {Richards}, {Richmond}, {Rivetta}, {Rockosi}, {Ruthmansdorfer}, {Sandford},
  {Schlegel}, {Schneider}, {Sekiguchi}, {Sergey}, {Shimasaku}, {Siegmund},
  {Smee}, {Smith}, {Snedden}, {Stone}, {Stoughton}, {Strauss}, {Stubbs},
  {SubbaRao}, {Szalay}, {Szapudi}, {Szokoly}, {Thakar}, {Tremonti}, {Tucker},
  {Uomoto}, {Vanden Berk}, {Vogeley}, {Waddell}, {Wang}, {Watanabe},
  {Weinberg}, {Yanny}, {Yasuda}, \& {SDSS Collaboration}}]{York00}
{York}, D.~G., {Adelman}, J., {Anderson}, Jr., J.~E., {et~al.} 2000, \aj, 120,
  1579

\bibitem[{{Yu} {et~al.}(2011){Yu}, {Tozzi}, {Borgani}, {Rosati}, \&
  {Zhu}}]{Yu11}
{Yu}, H., {Tozzi}, P., {Borgani}, S., {Rosati}, P., \& {Zhu}, Z.-H. 2011, \aap,
  529, A65

\bibitem[{{Zwicky}(1933)}]{Zwicky33}
{Zwicky}, F. 1933, Helvetica Physica Acta, 6, 110

\bibitem[{{Zwicky}(1937)}]{Zwicky37}
{Zwicky}, F. 1937, \apj, 86, 217

\bibitem[{{Zwicky} {et~al.}(1961){Zwicky}, {Herzog}, \& {Wild}}]{Zwicky61}
{Zwicky}, F., {Herzog}, E., \& {Wild}, P. 1961, Pasadena: California Institute
  of Technology, volume I

\end{thebibliography}


\begin{thebibliography}{17}
\expandafter\ifx\csname natexlab\endcsname\relax\def\natexlab#1{#1}\fi

\bibitem[{{Ahn} {et~al.}(2012){Ahn}, {Alexandroff}, {Allende Prieto},
  {Anderson}, {Anderton}, {Andrews}, {Aubourg}, {Bailey}, {Balbinot}, {Barnes},
  \& et~al.}]{Ahn12}
{Ahn}, C.~P., {Alexandroff}, R., {Allende Prieto}, C., {et~al.} 2012, \apjs,
  203, 21

\bibitem[{{Hao} {et~al.}(2010){Hao}, {McKay}, {Koester}, {Rykoff}, {Rozo},
  {Annis}, {Wechsler}, {Evrard}, {Siegel}, {Becker}, {Busha}, {Gerdes},
  {Johnston}, \& {Sheldon}}]{Hao10}
{Hao}, J., {McKay}, T.~A., {Koester}, B.~P., {et~al.} 2010, \apjs, 191, 254

\bibitem[{{Hilton} {et~al.}(2012){Hilton}, {Romer}, {Kay}, {Mehrtens},
  {Lloyd-Davies}, {Thomas}, {Short}, {Mayers}, {Rooney}, {Stott}, {Collins},
  {Harrison}, {Hoyle}, {Liddle}, {Mann}, {Miller}, {Sahl{\'e}n}, {Viana},
  {Davidson}, {Hosmer}, {Nichol}, {Sabirli}, {Stanford}, \& {West}}]{Hilton12}
{Hilton}, M., {Romer}, A.~K., {Kay}, S.~T., {et~al.} 2012, \mnras, 3303

\bibitem[{{Koester} {et~al.}(2007){Koester}, {McKay}, {Annis}, {Wechsler},
  {Evrard}, {Bleem}, {Becker}, {Johnston}, {Sheldon}, {Nichol}, {Miller},
  {Scranton}, {Bahcall}, {Barentine}, {Brewington}, {Brinkmann}, {Harvanek},
  {Kleinman}, {Krzesinski}, {Long}, {Nitta}, {Schneider}, {Sneddin}, {Voges},
  \& {York}}]{Koester07}
{Koester}, B.~P., {McKay}, T.~A., {Annis}, J., {et~al.} 2007, \apj, 660, 239

\bibitem[{{Mehrtens} {et~al.}(2012){Mehrtens}, {Romer}, {Hilton},
  {Lloyd-Davies}, {Miller}, {Stanford}, {Hosmer}, {Hoyle}, {Collins}, {Liddle},
  {Viana}, {Nichol}, {Stott}, {Dubois}, {Kay}, {Sahl{\'e}n}, {Young}, {Short},
  {Christodoulou}, {Watson}, {Davidson}, {Harrison}, {Baruah}, {Smith},
  {Burke}, {Mayers}, {Deadman}, {Rooney}, {Edmondson}, {West}, {Campbell},
  {Edge}, {Mann}, {Sabirli}, {Wake}, {Benoist}, {da Costa}, {Maia}, \&
  {Ogando}}]{Mehrtens12}
{Mehrtens}, N., {Romer}, A.~K., {Hilton}, M., {et~al.} 2012, \mnras, 2912

\bibitem[{{Mittal} {et~al.}(2011){Mittal}, {Hicks}, {Reiprich}, \&
  {Jaritz}}]{Mittal11}
{Mittal}, R., {Hicks}, A., {Reiprich}, T.~H., \& {Jaritz}, V. 2011, \aap, 532,
  A133

\bibitem[{{Mittal} {et~al.}(2009){Mittal}, {Hudson}, {Reiprich}, \&
  {Clarke}}]{Mittal09}
{Mittal}, R., {Hudson}, D.~S., {Reiprich}, T.~H., \& {Clarke}, T. 2009, \aap,
  501, 835

\bibitem[{{Piffaretti} {et~al.}(2011){Piffaretti}, {Arnaud}, {Pratt},
  {Pointecouteau}, \& {Melin}}]{Piffaretti11}
{Piffaretti}, R., {Arnaud}, M., {Pratt}, G.~W., {Pointecouteau}, E., \&
  {Melin}, J.-B. 2011, \aap, 534, A109

\bibitem[{{Pratt} {et~al.}(2009){Pratt}, {Croston}, {Arnaud}, \&
  {B{\"o}hringer}}]{Pratt09}
{Pratt}, G.~W., {Croston}, J.~H., {Arnaud}, M., \& {B{\"o}hringer}, H. 2009,
  \aap, 498, 361

\bibitem[{{Szabo} {et~al.}(2011){Szabo}, {Pierpaoli}, {Dong}, {Pipino}, \&
  {Gunn}}]{Szabo11}
{Szabo}, T., {Pierpaoli}, E., {Dong}, F., {Pipino}, A., \& {Gunn}, J. 2011,
  \apj, 736, 21

\bibitem[{{Takey} {et~al.}(2011){Takey}, {Schwope}, \& {Lamer}}]{Takey11}
{Takey}, A., {Schwope}, A., \& {Lamer}, G. 2011, \aap, 534, A120

\bibitem[{{Takey} {et~al.}(2013{\natexlab{a}}){Takey}, {Schwope}, \&
  {Lamer}}]{Takey13a}
{Takey}, A., {Schwope}, A., \& {Lamer}, G. 2013{\natexlab{a}}, \aap

\bibitem[{{Takey} {et~al.}(2013{\natexlab{b}}){Takey}, {Schwope}, \&
  {Lamer}}]{Takey13b}
{Takey}, A., {Schwope}, A., \& {Lamer}, G. 2013{\natexlab{b}}, \aap

\bibitem[{{Takey} {et~al.}(2013{\natexlab{c}}){Takey}, {Schwope}, \&
  {Lamer}}]{Takey13c}
{Takey}, A., {Schwope}, A., \& {Lamer}, G. 2013{\natexlab{c}}, \aap

\bibitem[{{Watson} {et~al.}(2009){Watson}, {Schr{\"o}der}, {Fyfe}, {Page},
  {Lamer}, {Mateos}, {Pye}, {Sakano}, {Rosen}, {Ballet}, {Barcons}, {Barret},
  {Boller}, {Brunner}, {Brusa}, {Caccianiga}, {Carrera}, {Ceballos}, {Della
  Ceca}, {Denby}, {Denkinson}, {Dupuy}, {Farrell}, {Fraschetti}, {Freyberg},
  {Guillout}, {Hambaryan}, {Maccacaro}, {Mathiesen}, {McMahon}, {Michel},
  {Motch}, {Osborne}, {Page}, {Pakull}, {Pietsch}, {Saxton}, {Schwope},
  {Severgnini}, {Simpson}, {Sironi}, {Stewart}, {Stewart}, {Stobbart}, {Tedds},
  {Warwick}, {Webb}, {West}, {Worrall}, \& {Yuan}}]{Watson09}
{Watson}, M.~G., {Schr{\"o}der}, A.~C., {Fyfe}, D., {et~al.} 2009, \aap, 493,
  339

\bibitem[{{Wen} {et~al.}(2009){Wen}, {Han}, \& {Liu}}]{Wen09}
{Wen}, Z.~L., {Han}, J.~L., \& {Liu}, F.~S. 2009, \apjs, 183, 197

\bibitem[{{Wen} {et~al.}(2012){Wen}, {Han}, \& {Liu}}]{Wen12}
{Wen}, Z.~L., {Han}, J.~L., \& {Liu}, F.~S. 2012, \apjs, 199, 34

\end{thebibliography}




\end{document}